\documentclass[preprint,10pt]{elsarticle}

\usepackage{graphicx,subcaption}
\usepackage{natbib}
\usepackage{colortbl}
\usepackage{amssymb}
\usepackage{moreverb} % JAHO
 \usepackage{float}   % JAHO
\usepackage{listings}
\usepackage[linesnumbered,ruled]{algorithm2e}
\SetCommentSty{mycommfont}
\usepackage{nomencl}
\usepackage[english]{babel}
\usepackage{nomencl}  % it doesn't compile !!!! See MISCELL.doc
\usepackage{leftidx} % For attaching left-hand sided scripts
 \usepackage{amsmath, amsthm, amssymb}

\usepackage[left=1.5cm,top=2cm,right=1.5cm,bottom=2.5cm]{geometry} %
\usepackage{color}
\usepackage{listings}
\usepackage{mathrsfs}

\makeatletter
\DeclareRobustCommand\bigop[2][1]{%
  \mathop{\vphantom{\sum}\mathpalette\bigop@{{#1}{#2}}}\slimits@
}
\newcommand{\bigop@}[2]{\bigop@@#1#2}
\newcommand{\bigop@@}[3]{%
  \vcenter{%
    \sbox\z@{$#1\sum$}%
    \hbox{\resizebox{\ifx#1\displaystyle#2\fi\dimexpr\ht\z@+\dp\z@}{!}{$\m@th#3$}}%
  }%
}
\makeatother
\newcommand{\bigA}{\DOTSB\bigop[0.92]{\mathrm{A}}}
\usepackage{multirow}

\definecolor{mygreen}{rgb}{0,0.6,0}
\definecolor{mygray}{rgb}{0.5,0.5,0.5}
\definecolor{mymauve}{rgb}{0.58,0,0.82}
\usepackage{hyperref}
\makenomenclature

\usepackage{lipsum}
\usepackage{pdflscape}
\usepackage{bbm}

\usepackage{import}
\usepackage{xifthen}
\usepackage{pdfpages}
\usepackage{transparent}

\newcommand{\norm}[1]{\left\lVert#1\right\rVert}

\usepackage{newfloat}

\DeclareFloatingEnvironment[
    fileext=lob,
    listname={List of Boxes},
    name=Box,
    placement=tbhp,
    within=section, % or chapter
]{mybox}

\usepackage{optidef} %added by Raul
\usepackage{algorithmic}
\usepackage{enumitem}
\usepackage{xcolor}
\newcommand{\revone}[1]{\textcolor{black}{#1}}   % Reviewer 1
\newcommand{\revtwo}[1]{\textcolor{black}{#1}}    % Reviewer 2
\newcommand{\bothrev}[1]{\textcolor{black}{#1}}    % Both
\newcommand{\own}[1]{\textcolor{black}{#1}}    % Own revisions
\newcommand{\newsec}[1]{\textcolor{black}{#1}}    % second example
\newcommand{\append}[1]{\textcolor{black}{#1}}    % appendix

\begin{document}

\begin{frontmatter}

\title{Geometrically Parametrised Reduced Order Models for Studying the Hysteresis of the Coanda Effect in \revone{Finite Element-based} Incompressible Fluid Dynamics}

\author[upc,cimne]{J.R. Bravo\corref{cor1}}
\ead{jose.raul.bravo@upc.edu}
\author[santanna]{G. Stabile}
\author[sissa]{M. Hess}
\author[upc-terrasa,cimne]{J.A. Hernandez}
\author[upc,cimne]{R. Rossi}
\author[sissa]{G. Rozza}
% \author[rvt]{}

%
\cortext[cor1]{Corresponding author}

\address[upc]{Universitat Polit\`{e}cnica de Catalunya, Department of Civil and Environmental Engineering (DECA), Barcelona, Spain}
\address[sissa]{International School for Advanced Studies, SISSA Mathematics Area (mathLab), Trieste, Italy}%
\address[santanna]{Sant'Anna School of Advanced Studies, The Biorobotics Institute, Pontedera, Pisa, Italy}%
\address[cimne]{Centre Internacional de M\`{e}todes Num\`{e}rics en Enginyeria (CIMNE), Barcelona, Spain}
\address[upc-terrasa]{Universitat Polit\`{e}cnica de Catalunya,  E.S. d'Enginyeries Industrial, Aeroespacial i Audiovisual de Terrassa (ESEIAAT), Terrassa, Spain}%

\begin{abstract}
%\lipsum[2-4]
This article presents a general reduced order model (ROM) framework for addressing fluid dynamics problems involving  time-dependent geometric parametrisations. The framework integrates Proper Orthogonal Decomposition (POD) and Empirical Cubature Method (ECM) hyper-reduction techniques to effectively approximate incompressible computational fluid dynamics simulations. To demonstrate the applicability of this framework, we investigate the behavior of a planar contraction-expansion channel geometry exhibiting bifurcating solutions known as the Coanda effect. By introducing time-dependent deformations to the channel geometry, we observe hysteresis phenomena in the solution.

The paper provides a detailed formulation of the framework, including the stabilised finite elements full order model (FOM) and ROM, with a particular focus on the considerations related to geometric parametrisation. Subsequently, we present the results obtained from the simulations, analysing the solution behavior in a \own{phase space} for the fluid velocity at a probe point, considered as the Quantity of Interest (QoI). Through qualitative and quantitative evaluations of the ROMs and hyper-reduced order models (HROMs), we demonstrate their ability to accurately reproduce the complete solution field and the QoI.

While HROMs offer significant computational speedup, enabling efficient simulations, they do exhibit some errors, particularly for testing trajectories. However, their value lies in applications where the detection of the Coanda effect holds paramount importance, even if the selected bifurcation branch is incorrect. Alternatively, for more precise results, HROMs with lower speedups can be employed.

\end{abstract}
\end{frontmatter}

\section{Introduction}
\label{sec: introduction}

\begin{figure}[h]
    \centering
    \includegraphics[width=0.4\linewidth]{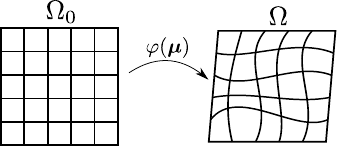}
    \caption{ Transformation of a domain $\Omega_0$ into $\Omega$ via a mapping $\varphi$}
    \label{fig:geometric mapping}
\end{figure}

This study focuses on projection-based reduced order models for fluid dynamics problems with time-dependent geometric parametrisations. We solve the Navier-Stokes equations on a parametrised domain denoted as $\Omega(\boldsymbol{\mu})\subset \mathbb{R}^d$ , with a time span $[0,T]$. Later on, in section \ref{subsec: Governing equations}, the specific form of the equations is shown. The geometric parametrisation is defined via a mapping  $\varphi(\boldsymbol{\mu}) : \Omega_0 \times \mathcal{P} \rightarrow \Omega$, such that, given a parameter $\boldsymbol{\mu} \in \mathcal{P} \subset\mathbb{R}^p$, each point in the original configuration is mapped onto a corresponding point in the deformed one as in Fig. \ref{fig:geometric mapping}.

\subsection{\own{Bifurcation and the Coanda Effect}}

\begin{figure}[h]
    \centering
    \includegraphics[width=0.35\linewidth]{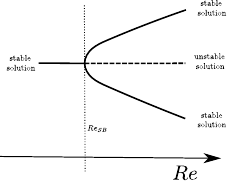}
    \caption{\own{Supercritical pitchfork bifurcation diagram for a contraction-expansion channel.} For $Re < {Re}_{\text{\tiny SB}} \ $ there exists a unique solution. For $Re \geq Re_{\text{\tiny SB}}$ three solutions coexist. \revone{In this paper, our focus lies not in constructing the bifurcation diagram, but rather on studying the application of model reduction to a dynamically morphed contraction-expansion channel operating in a neighbourhood of ${Re}_{\text{\tiny SB}}$ }}
    \label{fig:bifurcation}
\end{figure}

\revone{The Coanda effect can be described as the tendency of a flow jet to attach to solid surfaces. The term was coined after Henry Coanda, as he patented the first device taking advantage of the effect \cite{wille1965report,ahmed2019coanda}. In this paper we adopt as application case a contraction-expansion channel geometry \cite{oliveira2008simulations, mizushima2001transitions, cherdron1978asymmetric}. This case was chosen due to its tendency to exhibit complex dynamics, even at relatively low Reynolds numbers. For one, in this geometry the Coanda effect is present in the form of an asymmetric jet attaching to either the lower or upper walls, depending on the values of the Reynolds number ($Re$). From the point of view of mathematical nonlinear analysis, this application case presents a supercritical pitchfork bifurcation \cite{ambrosetti1995primer}, as shown in Fig. \ref{fig:bifurcation}. As can be seen, within a certain range of values of $Re$, a unique solution exists. However, once a critical point is surpassed, it is possible for more than one solution to exist for the same value of $Re$.}

The fluid development in a contraction-expansion channel can be described as follows:

\begin{itemize}
\item For sufficiently small \revone{$Re$}, there exists a single perfectly symmetric solution, exhibiting symmetry in both the horizontal and vertical directions.
\item As \revone{$Re$} increases, the horizontal symmetry of the solution is lost, but the solution remains symmetric about the vertical axis. This results in the formation of a symmetric jet.
\item At a critical Reynolds number value, denoted as ${Re}_{\text{\tiny SB}}$ (see Fig. \ref{fig:bifurcation}), the symmetry of the jet is disrupted due to the Coanda Effect. In this scenario, the jet attaches itself to either the upper or lower wall. While the symmetric jet solution is still mathematically possible, it becomes unstable. Numerical simulations or experiments typically yield one of the non-symmetric solutions unless advanced techniques are employed to extract the unstable solution, for instance, when constructing the complete bifurcation diagram \cite{pintore2021efficient, pichi2023artificial}.
\item With further increases in \revone{$Re$}, the presence of eddies becomes prominent. At a specific critical Reynolds number, denoted as ${Re}_{\text{\tiny H}}$, a Hopf bifurcation occurs, resulting in the emergence of an oscillating solution. However, the discussion of the Hopf bifurcation lies beyond the scope of this work. Interested readers are referred e.g. to \cite{quaini2016symmetry, sobey1986bifurcations} for further exploration of Hopf bifurcations.
\end{itemize}

In this investigation, the variation in geometry acts as the  sole driving factor for \revone{$Re$}. The viscosity and density of the fluid remain constant throughout the study, and once the fluid is initialised, the boundary conditions remain unchanged. In this setting, as the width of the channel narrows, \revone{$Re$} increases.

\subsection{\own{Related Works}}

\begin{figure}[h]
    \centering
    \includegraphics[width=0.5\linewidth]{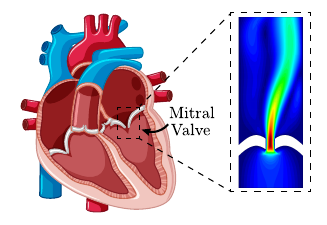}
    \caption{Mitral valve of the human heart and the contraction-expansion channel model used in this investigation. Contraction-expansion channel models can help in the study of mitral valve regurgitation disease}
    \label{fig:heart}
\end{figure}

The contraction-expansion channel has previously been utilised in research on mitral valve regurgitation disease, as demonstrated in studies such as \cite{pitton2017computational}. This disease is characterised by the improper closure of the mitral valve, as depicted in Fig \ref{fig:heart}. As a result of this faulty closure, blood is able to flow through a narrow opening, giving rise to the wall-hugging behavior mentioned earlier.

Previous numerical studies on the contraction-expansion channel have utilised an affine mapping $\varphi(\boldsymbol{\mu})$, in conjunction with the spectral element method \revone{(SEM) \cite{hess2020spectral}.} The affine nature of the geometric mapping in that study facilitated a complete online-offline decoupling of the reduced-order models, eliminating the need for additional interpolation and achieving significant speedup factors for the reduced-order models.

In a subsequent study \cite{hess2020reduced}, a nonlinear geometric mapping was introduced to account for walls with varying curvature. To effectively evaluate the reduced-order models in this scenario, the discrete empirical interpolation (DEIM) method was employed \cite{chaturantabut2010nonlinear, quarteroni2007numerical}.

More recently, the fluid-structure interaction problem has been investigated by incorporating (hyper)elastic walls \cite{khamlich2022model}. In that case, the deformation of the walls was computed from their interaction with the flow, rather than being imposed. For the present work, we will adhere to the approach of explicitly imposing the walls deformation.

Furthermore, in \cite{pichi2023artificial} a non-intrusive neural network framework, known as POD-NN, was employed for efficiently approximating the bifurcating phenomena in a contraction-expansion channel, as well as in other geometries.

In the aforementioned studies, the main objective was to construct the bifurcation diagram. To achieve this goal, steady-state solutions were computed for each parameter variation. This means that even when the geometry was parametrised, the geometric parametrisation remained unchanged over time. The focus was on capturing the steady-state behavior of the system and analyzing the bifurcation phenomena, rather than considering time-dependent variations in the geometric parametrisation.

Time dependent geometry variations for studying wall-hugging phenomena in flow jets have been investigated for example in \cite{allery2004experimental}. In that work, both numerical and experimental setups were applied to study the Coanda effect on jets interacting with inclined planes. The parameter to vary was the angle of the plate at the exit of the jet. \revone{It was observed that, as} the inclination angle of the planes was varied, the jet eventually attached, detached, or re-attached to either one of the walls, depending on whether the inclination angle was being increased or decreased. Thus, resulting on a hysteresis loop. This hysteresis phenomena in fluid problems exhibiting bifurcations had already been reported in classical works \cite{newman1961deflection, lai1996effect}, and more recently in  \cite{ pitton2017application, skotnicka2022pressure}.

\subsection{ \revone{Original Contributions}}

\begin{itemize}
    \item \revone{\textbf{Time-dependent geometric parametrisations in a reduced order model context.}} Our main contributions in this study expand on the existing literature by considering a contraction-expansion channel geometry that undergoes time-dependent deformations. By incorporating this dynamically morphed geometry, we are able to observe hysteresis phenomena in the solution, which has been previously documented in works such as \cite{pitton2017application}.
    \item \revone{\textbf{Independendence of the geometric parametrisation.}} The framework presented in this work offers broad applicability to a range of geometrically parametrised ROMs. This versatility allows for the framework to be employed in various scenarios. Unlike previous approaches that rely on assumptions about the nature of the mapping $\varphi(\boldsymbol{\mu})$ to achieve efficient offline-online decoupling, as seen in works like \cite{lassila2014model, manzoni2012reduced, quarteroni2007numerical, hess2020spectral}, we achieve efficient offline-online decoupling through the use of the empirical cubature method (ECM) \cite{hernandez2017dimensional, hernandez2020multiscale}. \revtwo{Notwithstanding, in this paper we confine our focus on finite element-based ROMs. Consequently, the geometric transformations imposed via the mapping $\varphi(\boldsymbol{\mu})$ must preserve the integrity of the elements, ensuring the positivity of the Jacobians of the isoparametric transformations. We detail the mappings employed in this study in Sec. \ref{sec: Model Description}. For geometric transformations involving more significant deformations incompatible with FEM techniques, alternative ROMs for CutFEM or shifted boundary method are available, as discussed e.g. in \cite{karatzas2020reduced, karatzas2022reduced}.}
    \item  \revone{\textbf{Hyper-reduction technology.} We note that the ECM algorithm was designed and has previously been employed for the selection of Gauss points, and the quantity to reduce pertained to internal forces. This paper presents its use in the hyper-reduction of discrete residuals and elements selection. We highlight the similarities of the method with alternative approaches. Furthermore, we emphasises that adopting ECM addresses innately what can potentially be an ill-conditioned problem.}
\end{itemize}

\subsection{Organisation of this Paper}

This paper is structured as follows. In Chapter \ref{sec: Formulation}, we present the formulation of the general framework for the fluid problem, including the formulations for the reduced and hyper-reduced order models. Section \ref{sec: Model Description} provides a detailed description of the finite element model utilised in this study, along with the two different geometric mappings employed. One mapping involves an affine transformation with straight walls, while the other employs a \own{nonaffine} mapping with curved walls. The obtained results are presented in Section \ref{sec: Results}. Finally, in Section \ref{sec: Conclusions}, we discuss the conclusions drawn from this study and provide insights into future research directions.

\section{Formulation}
\label{sec: Formulation}

In order to maintain this paper as self-contained as possible, this section presents a concise overview of the stabilised finite element discretisation of the governing equations employed in this study. We then introduce the Galerkin Proper Orthogonal Decomposition (POD-Galerkin) and the \revone{ECM} hyper-reduction techniques, highlighting the specific considerations related to the geometrically parametrised problem under investigation. We conclude the chapter by providing a summary of the simulation workflow for all the models employed in this study.

\subsection{Governing Equations}
\label{subsec: Governing equations}

As mentioned earlier in the introduction, we now proceed to explicitly state the fluids problem. The governing equations \own{considered} are the standard Differential-Algebraic incompressible Navier-Stokes equations in an Arbitrary Lagrangian Eulerian (ALE) frame of reference

\begin{equation}
    \left\{
        \begin{array}{ll}
        \frac{ \partial \boldsymbol{u}}{\partial t}+  ( \boldsymbol{c} \cdot \nabla) \boldsymbol{u}  -  \boldsymbol{b} - \nabla \cdot \boldsymbol{\sigma} = \boldsymbol{0} &  \mathrm{in \ \ } \Omega(\boldsymbol{\mu}) \times (0,T] \ , \\
    	\nabla \cdot \boldsymbol{u} = 0 &  \mathrm{in \ \ } \Omega(\boldsymbol{\mu}) \times (0,T] \ ,
    	\end{array}
    \right.
\end{equation}

\noindent where $\boldsymbol{u}$ is the velocity, $\boldsymbol{b}$ are the body forces, $\boldsymbol{\sigma} = -p \boldsymbol{I} + 2 \nu \nabla^s \boldsymbol{u}$ is the Cauchy stress tensor, $p$ is the pressure, $\nu$ is the kinematic viscosity, $\boldsymbol{c}= \boldsymbol{u} - \Hat{\boldsymbol{u}}$ is the  convective velocity, and $\Hat{\boldsymbol{u}}$ is the so-called \textit{mesh velocity} which results from the deformation of the domain and is a datum to the fluids problem. The governing equations, taking into account the provided expressions, can be written as follows:

\begin{equation}
    \left\{
        \begin{array}{ll}
        \frac{ \partial \boldsymbol{u}}{\partial t}+  \nabla \cdot (  \boldsymbol{c} \otimes \boldsymbol{u}) - \boldsymbol{b} + \nabla p  - \nabla \cdot(2 \nu \nabla^s \boldsymbol{u} ) = \boldsymbol{0} &  \mathrm{in \ \ } \Omega(\boldsymbol{\mu}) \times (0,T]  \ , \\
    	\nabla \cdot \boldsymbol{u} = 0 &  \mathrm{in \ \ } \Omega(\boldsymbol{\mu}) \times (0,T] \ .
    	\label{eq:strong form}
    	\end{array}
    \right.
\end{equation}

The initial conditions $g_0$, and Dirichlet and Neumann boundary conditions  $g_D$ and $g_N$ are case-specific and are specified by the corresponding functions,  as:

\begin{equation}
     \begin{array}{ll}
	 (\boldsymbol{u}, p)  = g_0(\boldsymbol{u}, p, \boldsymbol{x})   &  \mathrm{in \ \ } \Omega(\boldsymbol{\mu}) \times \{0\}  \ , \\
	 \left( 2 \nu \nabla^s \boldsymbol{u} -p \boldsymbol{I} \right)\boldsymbol{n}  = g_N(\boldsymbol{u}, p, \boldsymbol{x}) &  \mathrm{on \ \ } \partial  \Omega_N(\boldsymbol{\mu})\times(0,T] \ , \\
	 (\boldsymbol{u}, p)   =   g_D(\boldsymbol{u}, p, \boldsymbol{x}) &  \mathrm{on \ \ } \partial  \Omega_D(\boldsymbol{\mu})\times(0,T] \ ,
	 \end{array}
\end{equation}

\noindent with $\partial \Omega_N(\boldsymbol{\mu})\cup \partial \Omega_D(\boldsymbol{\mu}) = \partial \Omega(\boldsymbol{\mu})$ and $\partial \Omega(\boldsymbol{\mu})_N \cap \partial \Omega(\boldsymbol{\mu})_D = \emptyset$.

\subsection{\revone{Full Order Model (FOM)}}
\label{subsec: Full Order Model}

Let us introduce the functional spaces to pose the weak formulation for the space discretisation as

\begin{equation}
   \mathcal{V} := \{ \boldsymbol{v} \in H^1 (\Omega) \mid  \boldsymbol{v} = \boldsymbol{0} \text{ on } \partial \Omega_D  \} \ ,  \hspace{8mm}  \mathcal{Q} :=  L^2(\Omega) \ .
\end{equation}

In what follows, we will use the standard $L^2(\Omega)$ inner product notation $( \cdot , \cdot )_\Omega$ for brevity. The weak form of the governing equations in Eq. \ref{eq:strong form} is then given by:

\begin{equation}
    \left\{
        \begin{array}{ll}
            \left(\frac{\partial \boldsymbol{u}}{\partial t} ,  \boldsymbol{v}  \right)_\Omega - \left(   \boldsymbol{c} \otimes \boldsymbol{u}   ,   \nabla \boldsymbol{v} \right)_\Omega + 2 \nu \left( \nabla^s \boldsymbol{u} , \nabla^s \boldsymbol{v} \right)_\Omega - \left(   p  , \nabla \cdot \boldsymbol{v}   \right)_\Omega -  \left(  \boldsymbol{b}   ,  \boldsymbol{v}  \right)_\Omega = \boldsymbol{0} &  \forall \boldsymbol{\boldsymbol{v}} \in \mathcal{V}  \ , \\
            \left(  q   ,  \nabla \cdot \boldsymbol{u}  \right)_\Omega = 0 & \forall  q \in \mathcal{Q} \ .
        \end{array}
    \right.
\end{equation}

\noindent  We define the finite element spaces $\mathcal{V}^h \subset \mathcal{V}$ and $\mathcal{Q}^h \subset \mathcal{Q}$. These spaces are spanned respectively by a set of basis functions $\{\boldsymbol{\psi}_i\}_{i=1}^{N_v}$ and $\{\hat{\psi}_j\}_{j=1}^{N_p}$, being $\text{dim}\mathcal{V}^h = N_v $, and $\text{dim}\mathcal{Q}^h = N_p $. The finite elements weak form of the governing equations can be expressed as follows:

\begin{equation}
    \left\{
        \begin{array}{ll}
            \left(\frac{\partial \boldsymbol{u}^h}{\partial t} ,  \boldsymbol{v}^h \right)_\Omega - \left( \boldsymbol{c}^h  \otimes \boldsymbol{u}^h   ,   \nabla \boldsymbol{v}^h\right)_\Omega + 2 \nu \left( \nabla^s \boldsymbol{u}^h , \nabla^s \boldsymbol{v}^h\right)_\Omega - \left(   p^h  , \nabla \cdot \boldsymbol{v}^h  \right)_\Omega -  \left(  \boldsymbol{b}   ,  \boldsymbol{v}^h \right)_\Omega = \boldsymbol{0}  & \forall \boldsymbol{\boldsymbol{v}}^h \in  \mathcal{V}^h \ ,  \\
            \left(  q^h   ,  \nabla \cdot \boldsymbol{u}^h  \right)_\Omega = 0 & \forall  q^h \in \mathcal{Q}^h \ ,
        \end{array}
    \right.
\end{equation}

\noindent where $\boldsymbol{c}^h = \boldsymbol{u}^h - \Hat{\boldsymbol{u}}^h$ is the finite elements convective velocity.

To ensure the well-posedness of the problem, the selected spaces for the approximation of pressure and velocity should be compatible and satisfy the Ladyzhenskaya–Babuška–Brezzi (LBB) condition \cite{donea2003finite}. The LBB condition can be stated as follows:

\begin{equation}
   \inf_{q^h \in \mathcal{Q}^h }  \sup_{ \boldsymbol{v}^h \in \mathcal{V}^h }  \frac{ \left( q^h, \nabla \cdot \boldsymbol{v}^h \right)}{\norm{q^h}_{\mathcal{Q}^h} \norm{\boldsymbol{v}^h}_{\mathcal{V}^h} } \geq \alpha > 0   \hspace{5mm} \iff \hspace{5mm} \exists \ (\boldsymbol{u}^h \in \mathcal{V}^h, q\in \mathcal{Q}^h) \ .
\end{equation}

LBB compatible spaces necessarily comply with

\begin{equation}
    \text{dim} \mathcal{Q}^h  \leq  \text{dim} \mathcal{V}^h  \Leftarrow \hspace{5mm} \exists \ (\boldsymbol{u}^h \in \mathcal{V}^h, q\in \mathcal{Q}^h) \ ,
\end{equation}

\noindent however,  in this work, we aim to use the same order of interpolation for the finite element approximation of both variables. To overcome the limitation imposed by the LBB condition, we employ the Variational Multiscale (VMS) stabilisation technique \cite{codina2002stabilized, codina2018variational, stabile2019reduced}.

\subsubsection{\revone{Variational Multiscale (VMS) Stabilisation}}

The VMS approach introduces \textit{subgrid}  spaces $\mathcal{V}'$ and $\mathcal{Q}'$, defined as:

\begin{equation}
    \mathcal{V} =  \mathcal{V}^h  \oplus \mathcal{V}' \ , \hspace{7mm}  \mathcal{Q} = \mathcal{Q}^h  \oplus \mathcal{Q}' \ ,
\end{equation}

\noindent where $\oplus$ denotes the direct sum. The complementary velocity and pressure functions $(\boldsymbol{u}', p')$ are such that
\begin{equation}
\begin{split}
    \boldsymbol{u} = \boldsymbol{u}^h + \boldsymbol{u}'  \hspace{5mm} ,  \hspace{5mm} \boldsymbol{u}^h \in \mathcal{V}^h \ \ , \  \boldsymbol{u} ' \in  \mathcal{V}' \ ,\\
    p = p^h + p'  \hspace{5mm} ,  \hspace{5mm}  p^h \in \mathcal{Q}^h \ \ ,  \ p ' \in \mathcal{Q}' \ , \\
    \boldsymbol{u}' = \boldsymbol{0} \text{ on } \partial \Omega_D(\boldsymbol{\mu}) \ , \hspace{5mm}\boldsymbol{u}^h = g_D^h \text{ on } \partial \Omega_D(\boldsymbol{\mu}) \ .
\end{split}
\end{equation}

In VMS, the complementary functions are modelled as

\begin{equation}
    \boldsymbol{u}' = - \tau_M r_M (\boldsymbol{u}^h, p^h) \ , \hspace{7mm}
    p' = - \tau_C r_C (\boldsymbol{u}^h)  \ ,
\end{equation}

\noindent  where the terms $\tau_M$ and $\tau_C$ represent respectively the momentum and mass conservation residuals defined as
\begin{equation}
\begin{array}{l}
    r_M (\boldsymbol{u}^h, p^h) = \frac{\partial \boldsymbol{u}^h}{\partial t} + \nabla \cdot (\boldsymbol{c}^h \otimes \boldsymbol{u}^h) - \nabla \cdot (2 \nu \nabla^s \boldsymbol{u}^h) + \nabla p^h - \boldsymbol{b} \ ,
    \\
    r_C (\boldsymbol{u}^h) = \nabla \cdot \boldsymbol{u}^h \ .
\end{array}
\end{equation}

\noindent The modelling of the complementary functions with respect to the residuals ensures the consistency of the method, moreover the selection of $\tau_M$ and $\tau_C$ depends on the specific type of VMS to apply. For example, considering $\tau_M, \tau_C = 0$ leads to the Galerkin method. In the case of the quasi-static variational multiscale (QSVMS), which we employ in this study, the time evolution of the complementary functions is neglected \cite{codina2002stabilized}.

To proceed with an algebraic formulation, we consider a spatial tessellation of the domain $\Bar{\Omega}$ into $N_{el}$ finite elements, such that $\bigcup_{e=1}^{N_{el}} \Omega_e = \Bar{\Omega}$; moreover we specify standard finite elements shape functions with compact support and equal degree of interpolation. The velocity and pressure algebraic vectors are obtained by collecting the nodal solutions for both variables as $\mathbf{u} = \left(u_1 , u_2 , \dots, u_{dN_n} \right)^T$ and $\mathbf{p} = \left(p_1 , p_2 , \dots, p_{N_n} \right)^T$, where $N_n$ is the number of nodes, and $d$ is the number of spatial dimensions. The QSVMS semi-discrete system can then be expressed as follows:

\begin{equation}
    \left\{
        \begin{array}{lll}
            \boldsymbol{M} \frac{d \mathbf{u}}{d t} + \boldsymbol{A} \mathbf{u} + \boldsymbol{C}(\mathbf{c}) \mathbf{u} + \boldsymbol{D}(\mathbf{u},\mathbf{p}) + \boldsymbol{B}^T \mathbf{p} = \boldsymbol{F}  \ , \\
            \boldsymbol{B} \mathbf{u} = \boldsymbol{E}(\mathbf{u}, \mathbf{p}) \ ,
        \end{array}
    \right.
    \label{eq:semi discrete system}
\end{equation}

\noindent where the standard Galerkin finite element terms are given by

\begin{equation}
    \begin{array}{llllll}
    \boldsymbol{M}&= \bigA_{}^e \boldsymbol{M}^e  &\hspace{5mm}& M_{i j}^e  & = \left( \boldsymbol{\psi}_j, \boldsymbol{\psi}_i \right)_{\Omega^e} & \textit{the mass matrix} \ , \\
    \boldsymbol{A}&=  \bigA_{}^e \boldsymbol{A}^e && A_{i j}^e  & = \left( \nabla^s \boldsymbol{\psi}_j, \nabla^s \boldsymbol{\psi}_i \right)_{{\Omega}^e} & \textit{the diffusion matrix} \ ,   \\
    \boldsymbol{C}(\mathbf{c})&= \bigA_{}^e \boldsymbol{C}^e && C_{i j}^e  & = - \left( \boldsymbol{\psi}_j, \boldsymbol{c}^h \cdot \nabla \boldsymbol{\psi}_i \right)_{\Omega^e}  & \textit{the convection matrix}  \ ,  \\
    \boldsymbol{B}&=  \bigA_{}^e \boldsymbol{B}^e  && B_{i j}^e  & = - \left(  \hat{\psi}_j , \nabla \cdot \boldsymbol{\psi}_i  \right)_{\Omega^e} & \textit{the pressure/divergence matrix}  \ , \\
    \boldsymbol{F}&=  \bigA_{}^e \boldsymbol{F}^e  && F_{i}^e  & =\left(  b, \boldsymbol{\psi}_i \right)_{\Omega^e} & \textit{the forcing term} \ ,
    \end{array}
\end{equation}

\noindent where $\bigA_{}^e$ is the finite elements assembly operator, $1\leq i,j\leq n_{en}$, and $n_{en}$ is the number of nodes per element. The remaining terms are related to the VMS stabilisation and are defined as follows:

\begin{equation}
    \begin{array}{llllll}
    \boldsymbol{D}(\mathbf{u}, \mathbf{p}) =&\bigA_{}^e  \boldsymbol{D}^e & &\hspace{5mm}& D_{i}   = &  \left( \nabla (\boldsymbol{\psi}_i) \boldsymbol{u}^h,  \tau_M r_M (\boldsymbol{u}^h, p^h)  \right)_{\Omega^e}  \\
    &&&&&   + \left( (\nabla \boldsymbol{\psi}_i)^T \boldsymbol{u}^h ,  \tau_M r_M (\boldsymbol{u}^h, p^h)  \right)_{\Omega^e}   \\
    &&&&& - \left(  \boldsymbol{\psi}_i ,   \tau_M r_M (\boldsymbol{u}^h, p^h)  \otimes  \tau_M r_M (\boldsymbol{u}^h, p^h)  \right)_{\Omega^e}  \\
    &&&&&  + \left(  \tau_C r_C (\boldsymbol{u}^h) , \nabla \cdot \boldsymbol{\psi}_i  \right)_{\Omega^e}   \ , \\

    \boldsymbol{E}(\mathbf{u}, \mathbf{p})    =&\bigA_{}^e  \boldsymbol{E}^e &&& E_{i} = & \left(
    \nabla \hat{\psi}_i ,  \tau_M r_M (\boldsymbol{u}^h, p^h) \right)_{\Omega^e} \ .
    \end{array}
\end{equation}

\subsubsection{Time Discretisation}

We employ a generalised-$\alpha$ time integration scheme \cite{chung1993time}. In these schemes, a set of parameters $\gamma, \beta, \alpha_m, \alpha_f$ are defined and used to obtain a linear combination of the terms of the system in Eq. \ref{eq:semi discrete system} with respect to current and past time steps. Specifically, the \textit{Bossak} scheme, which is second-order accurate and unconditionally stable, is used in this work. The application of the time integration scheme leads to the fully discrete system:

\begin{equation}
    \textbf{K}_{\text{eff}}(\textbf{d}_{n+\ell}) \textbf{d}_{n+1} = \textbf{F}_{\text{eff}} \ ,
\end{equation}

\noindent where $\textbf{K}_{\text{eff}}(\textbf{d}_{n+\ell}) \in \mathbb{R}^{\mathcal{N}   \times \mathcal{N}  }$ is the fully discrete system matrix, $\mathcal{N}$ is the total number of degrees of freedom, $\textbf{d}_{n+1} \in \mathbb{R}^\mathcal{N}$ is the finite elements solution vector at time step $n+1$, that is

\begin{equation}
    \textbf{d}_{n+1} =
    \begin{bmatrix}
        \mathbf{u}_{n+1}\\
        \mathbf{p}_{n+1}
    \end{bmatrix}    \ .
\end{equation}

The dependence of the matrix $\textbf{K}_{\text{eff}}(\textbf{d}_{n+\ell})$ on the solution is due to the nonlinearity in the convective term, the subscript $n+\ell$ depends on the type of linearisation used. \revone{We now say a few words} on the linearisation technique employed.

\subsubsection{Linearisation of the Discrete System}

Given a solution at a time step $n$, the solution corresponding to time step $n+1$ is written in incremental form as

\begin{equation}
    \textbf{d}_{n+1} =  \textbf{d}_{n} + {\boldsymbol{\Delta} \textbf{d}}_{n+1} \ ,
    \label{eq: FOM representation}
\end{equation}

\noindent where $\boldsymbol{\Delta d} \in \mathbb{R}^ \mathcal{N}$ is a solution increment. We define the residual $\textbf{R}: \mathbb{R}^\mathcal{N}  \times \mathcal{P} \rightarrow \mathbb{R}^\mathcal{N}$ (highlighting the dependence of all terms of the residual on the geometric parameter $\boldsymbol{\mu} \in \mathcal{P}$), as

\begin{equation}
    \textbf{R}(\textbf{d}_{n+1} ; \boldsymbol{\mu}) := \textbf{F}_{\text{eff}}  -     \textbf{K}_{\text{eff}}(\textbf{d}_{n+ \ell}) \textbf{d}_{n+1}  \ ,
    \label{eq:fom residual}
\end{equation}

\noindent together with a fixed-point iteration method of the form

\begin{subequations}
\begin{align}
- \textbf{J} ( \textbf{d}_{n+1}^{k})  \delta \textbf{d} & = \textbf{R} (\textbf{d}_{n+1}^{k}) \ , \\
{ \boldsymbol{\Delta}\textbf{d} }_{n+1}^{k+1} & = {\boldsymbol{\Delta}\textbf{d}}^{k+1}_{n+1} + \delta \textbf{d} \ ,
\end{align}
\label{eq: fom_iterative}
\end{subequations}

\noindent where the Jacobian matrix $\textbf{J} = \partial \textbf{R} / \partial \textbf{d}$, and $k$ is the current iterate index. We use a Picard method \cite{donea2003finite}, which amounts to choosing $\ell = 0$ for the matrix   $\textbf{K}_{\text{eff}}(\textbf{d}_{n+ \ell})$ in Eq. \ref{eq:fom residual}. A Newton-Raphson method would amount to choosing the subindex $\ell = 1$, and therefore taking the tangent in the current iterate.

\subsubsection{\revone{Quantity of Interest (QoI)}}

It is not uncommon that, rather than the complete solution field $\textbf{d}$, a quantity of interest is computed through a given operator $\Gamma$ on the solution field as:

\begin{equation}
    \textbf{z} = \Gamma(\textbf{d}), \hspace{5mm} \textbf{z} \in \mathbb{R}^{\alpha} \subseteq \mathbb{R}^{\mathcal{N}} \ .
\end{equation}

Examples of this \textit{QoI} are the lift and drag coefficients. For the case at hand, such a QoI will be given by the vertical component of the velocity at a selected position indicating the presence of the Coanda effect.

\subsection{\revone{Reduced Order Model (ROM)}}
\label{subsec: Reduced Order Model}

Projection-based reduced order models involve constructing an optimal basis that spans a subspace where a high-dimensional system can be projected and solved. Various techniques in the literature fall under this classification, and for our research, we have chosen the \revone{POD-Galerkin} technique.

POD, like other projection-based reduced order modeling techniques, comprises two stages:

\begin{itemize}
    \item \textit{Offline stage}: In this stage, a set of simulations is performed using the computationally expensive FOM, and the resulting solutions are stored in matrix form. These matrices are then analysed to obtain the aforementioned basis. Ideally, the produced ROMs should not require the evaluation of full dimensional variables. We accomplish this decoupling through a hyper-reduction training.
    \item  \textit{Online stage}: With the basis and additional hyper-reduction data available, the hyper-reduced order models (HROMs) can be efficiently launched for unexplored parameters at a fraction of the cost associated with the FOMs.
\end{itemize}

In the subsequent sections, we will delve into the details primarily concerning the offline stage of ROMs and HROMs.

\subsubsection{\revone{Proper Orthogonal Decomposition (POD-Galerkin)}}
\label{subsec: Proper Orthogonal Decomposition POD}

We define the discrete solution manifold as the set of solution vectors $\mathbf{d}$ for all possible values of the parameters vector, that is

\begin{equation}
    \mathcal{M}^h = \{ \textbf{d}(\boldsymbol{\mu})\mid  \boldsymbol{\mu} \in \mathcal{P} , t \in [0,T] \}  \ \subset \ \mathbb{R}^\mathcal{N} \ ,
    \label{eq: solution manifold discrete}
\end{equation}

%\gs{is $\mathcal{P}$ the parameter space or the discrete parameter space?}

\noindent We employ a \textit{linear} approximation that, given a reduced order model solution at a time step $n$, $\tilde{\textbf{d}}_n \in \mathbb{R}^\mathcal{N}$, the solution corresponding to the step $n+1$ is

\begin{equation}
   \tilde{\textbf{d}}_{n+1} = \tilde{\textbf{d}}_n + \boldsymbol{\Phi}\boldsymbol{\Delta q}_{n+1} \ ,
    \label{eq: rom representation}
\end{equation}

\noindent where $\boldsymbol{\Delta q} \in \mathbb{R}^ N$ is the reduced solution increment, with $N \leq \mathcal{N}$ (usually $N \ll \mathcal{N}$ is expected) , and $\boldsymbol{\Phi} = [\boldsymbol{\phi}_1, \dots, \boldsymbol{\phi}_ N ] \in \mathbb{R}^{ \mathcal{N} \times N}$, is the reduced basis matrix, obtained by employing the \revone{\_}POD method \cite{sirovich1987turbulence, hesthaven2016certified}.

The procedure consists in taking $m$ samples (FOM solutions) of the discrete solution manifold, and store them in a snapshots matrix $\boldsymbol{S} = [\textbf{d}_1, \cdots, \textbf{d}_m] \in \mathbb{R}^{\mathcal{N} \times m}$. Here, each of the $m$ samples corresponds to a time step. For this, we consider a function

% \gs{ok maybe I understood, $\mathcal{M}^h$ is discrete in space but continuous in the parameter space? Am I right? In that case you can neglect the previous comments}

\begin{equation}
\begin{aligned}
   f_{\mu}: \ & \mathbb{R}_+  \rightarrow \mathbb{R}^p  \ , \\
     &  t  \mapsto \boldsymbol{\mu} \ ,\\
\end{aligned}
\label{eq: time-mu mapping}
\end{equation}

\noindent defining a \textit{trajectory} over the discrete solution manifold with $m$ cases corresponding to the pairs (t, $ f_{\mu}(t) $). The specific trajectories used in our investigation are shown later in section \ref{subsec: Trajectories}.

Having at one's disposal the snapshots matrix $\boldsymbol{S}$, we apply the truncated singular value decomposition with a truncation tolerance $0\leq \epsilon_{\text{\tiny SOL}} \leq 1$, as $\boldsymbol{S} = \boldsymbol{U}_N \boldsymbol{\Sigma}_N\boldsymbol{V}_N^T + \boldsymbol{E}$ where,
\begin{equation}
    \boldsymbol{U}_N\in \mathbb{R}^{  \mathcal{N} \times N} \ \  , \ \ \boldsymbol{\Sigma}_N = \text{diag}(\sigma_1, \sigma_2, \dots, \sigma_N) \in \mathbb{R}^{N \times N} \ \  , \ \ \boldsymbol{V}_N^T\in \mathbb{R}^{ N \times m} \ \ ,  \ \
    \norm{\boldsymbol{E}} \leq \epsilon_{\text{\tiny SOL}} \norm{\boldsymbol{S}} \ .
\end{equation}

The optimal \own{$N$-dimensional basis} \cite{eckart1936approximation} is obtained as the truncated matrix of left singular vectors  $\boldsymbol{\Phi}:= \boldsymbol{U}_N$.

Substitution of Eq. \ref{eq: rom representation} into Eq. \ref{eq:fom residual}, and subsequent projection of the over-determined system of equations onto $\boldsymbol{\Phi}$ (Galerkin projection), results in

\begin{equation}
    \boldsymbol{\Phi}^T \textbf{R}(\tilde{\textbf{d}}_{n+1}; \boldsymbol{\mu} ) = \boldsymbol{0} \ .
    \label{eq: galerkin rom}
\end{equation}

The fixed-point method for solving for the reduced increment $\boldsymbol{\Delta q} \in \mathbb{R}^N$ is given as

\begin{subequations}
\begin{align}
- \boldsymbol{\Phi}^T  \textbf{J}(\tilde{\textbf{d}}_{n+1}^{k}) \boldsymbol{\Phi} \delta \textbf{q} & = \boldsymbol{\Phi}^T \textbf{R} (\tilde{\textbf{d}}_{n+1}^{k}) \ ,\\
{ \boldsymbol{\Delta}\textbf{q} }^{k+1}_{n+1} & = {\boldsymbol{\Delta}\textbf{q}}_{n+1}^{k} + \delta \textbf{q} \ .
\end{align}
\label{eq: rom_iterative}
\end{subequations}

\noindent The same operator \footnote{Although for a ROM the same operator can be employed, for an HROM not containing all elemental variables, an equivalent operator might be required} defining the QoI for the FOM, can be used for the reduced order model solution vector as

\begin{equation}
    \tilde{\textbf{z}} = \Gamma(\tilde{\textbf{d}}), \hspace{5mm} \tilde{\textbf{z}} \in \mathbb{R}^{\alpha} \subseteq \mathbb{R}^{\mathcal{N}} \ .
\end{equation}

\subsubsection{Hyper-Reduction via Empirical Cubature}
\label{subsec: Hyper-Reduction}

Taking into account the finite elements discretisation employed, Eq. \ref{eq: galerkin rom} can be represented as

\begin{equation}
     \boldsymbol{\Phi}^T \textbf{R}(\tilde{\textbf{d}}; \boldsymbol{\mu} )  = \sum_{e=1}^{N_{el}} \boldsymbol{\Phi}^T \boldsymbol{L}_e^T \textbf{R}_e (\boldsymbol{L}_e \tilde{\textbf{d}}; \boldsymbol{\mu}  ) = \sum_{e=1}^{N_{el}} \boldsymbol{\Phi}_e^T \textbf{R}_e (\tilde{\textbf{d}}_e; \boldsymbol{\mu}  )  \ ,
    \label{eq: galerkin rom assembly}
\end{equation}

\noindent where $\boldsymbol{L}_e \in \{0,1\}^{e_{dof} \times \mathcal{N}}$ is the Boolean operator localising the high dimensional vector of dimension $\mathcal{N}$ to the degrees of freedom associated to element $e$. Consequently, $\boldsymbol{\Phi}_e$ and $\tilde{\textbf{d}}_e$ \own{are, respectively,} the entries of the basis and reduced solution vector associated to element $e$\own{, and $\textbf{R}_e : \mathbb{R}^{e_{dofs}} \times \mathcal{P} \rightarrow \mathbb{R}^{e_{dofs}}$ is the elemental residual.}

\revone{Hyper-reduction techniques within finite elements are primarily categorised into two distinct groups, based on their approach to constructing efficient approximations to Eq. \ref{eq: galerkin rom assembly}. On the one hand, there are techniques that focus on approximating (interpolating) the full-dimensional variable $\textbf{R} \in \mathbb{R}^{\mathcal{N}}$, followed by projecting the approximated vector onto a reduced space.  This category, known as the \textit{approximate-then-project}  methods \cite{farhat2020computational}, require an orthogonal basis for the nonlinear term, potentially in an unassembled form \cite{tiso2013modified}. The process requires employing a greedy algorithm \cite{chaturantabut2010nonlinear} or a QR factorisation with pivoting \cite{drmac2016new} to identify critical sampling points, often referred to as ``magic points" \cite{stabile2020efficient}.}

\revone{On the other hand, there are strategies that attempt to approximate directly the projected vector $\boldsymbol{\Phi}^T \boldsymbol{R} \in \mathbb{R}^{N}$. These techniques are classified as \textit{project-then-approximate} \cite{farhat2020computational} methods. Here, the approximation of Eq. \ref{eq: galerkin rom assembly} is performed not by sampling the full set of elements, but rather a selected subset $\mathbb{E} \subset \{ 1, 2, \dots , N_{el}\}$. Each considered projected elemental contribution is in turn adjusted using a corresponding weight factor $\omega_e$.}

\revone{We discussed this classification in detail in \cite{hernandez2017dimensional}, where we used the concepts of \textit{nodal vector approximation approaches} and \textit{integral approximation approaches}. The methodology adopted in the current paper can be classified as an integral approximation or a project-then-approximate method. Our preference for this type of methods is motivated by their demonstrated stability and robustness, which are discussed in more depth in the subsequent paragraphs.}

\revone{Having summarised the classification of the hyper-reduction techniques, we proceed to detail the necessary steps to obtain the reduced set of elements and weights. We begin by storing the elemental contributions in Eq. \ref{eq: galerkin rom assembly} for each of the $m$ studied cases.} Let the projected residual for element $e$ for parameter case $i$ (here \own{$\boldsymbol{\mu}_i = f_{\mu}(t_i) $}, see Eq. \ref{eq: time-mu mapping}) be defined as
\begin{equation}
     \mathbb{R}^N \ni \boldsymbol{\mathcal{R}}_{ie}   =  \boldsymbol{\Phi}_e^T \textbf{R}_e (\tilde{\textbf{d}}_e; \boldsymbol{\mu}_i  ) \ .
\end{equation}
\noindent We construct then the matrix of \own{projected} residuals for all $N_{el}$ elements and $m$ studied parameters, as

\begin{equation}
  \mathbb{R}^{N \cdot m \times N_{el}} \ni \boldsymbol{S_r} =
    \begin{bmatrix}
    \boldsymbol{\mathcal{R}}_{11}  &  \dots & \boldsymbol{\mathcal{R}}_{1N_{el}} \\
    \vdots & \ddots & \vdots \\
    \boldsymbol{\mathcal{R}}_{m1}  &  \dots & \boldsymbol{\mathcal{R}}_{mN_{el}}
    \end{bmatrix}    \ .
\end{equation}

The \textit{exact assembly} of the \own{projected} residuals, for the $m$ studied parameters variations, is given as

\begin{equation}
    \mathbb{R}^{N \cdot m} \ni \boldsymbol{d} \ = \ \boldsymbol{S_r} \ \mathbbm{1}  \ ,
    \label{eq: exact assembly of projected residuals}
\end{equation}

\noindent where $\mathbbm{1} := \{1\}^{N_{el}}$.

\own{The optimisation problem for finding the reduced set of elements $\mathbb{E}$ and corresponding weights $\boldsymbol{\omega}$ is then posed as}

\begin{equation}
    \begin{aligned}
        (\mathbb{E}, \boldsymbol{\omega}) = arg \ min \quad & \norm{ \boldsymbol{\zeta} }_0\\
        \textrm{s.t.} \quad & \norm{ \boldsymbol{d} - \boldsymbol{S_r}\boldsymbol{\zeta} }_2 < \epsilon \norm{ \boldsymbol{d} }_2    \\
        & \boldsymbol{\zeta} \succeq \boldsymbol{0} \ ,
    \end{aligned}
    \label{eq:optimization_problem_np_hard_1}
\end{equation}

\noindent \own{where $\boldsymbol{\zeta} \in \mathbb{R}^{N_{el}}$ is a sparse vector with non-zero values $\boldsymbol{\omega}$ at indices $\mathbb{E}$, $\norm{\cdot}_0$ represents the zero pseudo norm (counting the number of non-zero entries of its argument), $\epsilon$ is a user-defined tolerance, and we use the symbol $\cdot \succeq \cdot $ to represent inequality with respect to the non-negative orthant (this constraint requires the non-negative entries of $\boldsymbol{\zeta}$ to be positive).} \revone{This non-negative constraint was considered critical for maintaining the positive definiteness of the involved operators in works such as \cite{farhat2014dimensional,hernandez2017dimensional}. Moreover, a proof that solutions to the optimisation problem in Eq. \ref{eq:optimization_problem_np_hard_1} are stable was provided in \cite{farhat2015structure}, showing that such solutions preserve the Lagrangian structure of the problem at HROM level.}

\revone{Given the optimisation problem, the question arises as to which algorithm should be used to obtain a solution. It is widely recognized \cite{boyd2004convex} that the problem, as formulated in Eq. \ref{eq:optimization_problem_np_hard_1}, is NP-hard, hence computationally intractable. Therefore, one must resort to suboptimal greedy heuristics or convex relaxation. Among the alternatives is the Energy-Conserving Sampling and Weighting (ECSW) algorithm \cite{farhat2014dimensional,farhat2015structure}, which employs a non-negative least squares (NNLS) procedure with an early stopping criterion to account for the user-defined tolerance. Another alternative involves modifying the cost function to include the $\norm{\cdot}_1$ norm, thus convexifying the problem and making it amenable to established convex optimisation algorithms, as suggested e.g. in \cite{pan2015subspace, yano2019lp}.}

\revone{The ECM algorithm, that we employ in this study, addresses a modified version of the optimisation problem by initially applying a truncated SVD, with truncation tolerance $\epsilon_{\text{\tiny RES}}$ to matrix $\boldsymbol{S_r}$, as}

\begin{equation}
    \boldsymbol{S_r} = \boldsymbol{U}_\beta \boldsymbol{\Sigma}_\beta \boldsymbol{G} + \boldsymbol{E}  \hspace{4mm} \, \hspace{4mm}  \norm{\boldsymbol{E}} \leq \epsilon_{\text{\tiny RES}} \norm{\boldsymbol{S}_r}   \ ,
    \label{eq: truncated svd residual}
\end{equation}

\noindent \own{where $\boldsymbol{U} \in \mathbb{R}^{N\cdot m \times \beta}$, $\boldsymbol{\Sigma} \in \mathbb{R}^{\beta \times \beta}$, $\boldsymbol{G} \in \mathbb{R}^{\beta \times N_{el}}$. }

We use the truncated matrix of right singular vectors $\boldsymbol{G}$ to define a vector containing the \own{\textit{assembly of the modes of the projected residuals} (in contrast to the exact assembly of the projected residuals in Eq. \ref{eq: exact assembly of projected residuals})}, as

\begin{equation}
    \mathbb{R}^{\beta} \ni \boldsymbol{b} \ = \ \boldsymbol{G} \ \mathbbm{1} \ .
\end{equation}

\revone{In this way, the optimisation problem tackled by the ECM algorithm is}

\revone{
\begin{equation}
    \begin{aligned}
        (\mathbb{E}, \boldsymbol{\omega}) = arg \ min \quad & \norm{ \boldsymbol{\zeta} }_0\\
        \textrm{s.t.} \quad & \boldsymbol{b} - \boldsymbol{G} \boldsymbol{\zeta}  = \boldsymbol{0}  \\
        & \boldsymbol{\zeta} \succeq \boldsymbol{0} \ .
    \end{aligned}
    \label{eq:optimization_problem_np_hard_2}
\end{equation}
}

\revone{By operating on a basis for the row space of \(\boldsymbol{S_r}\), the ECM algorithm circumvents the need to employ a NNLS algorithm and instead uses a standard least-squares algorithm in every greedy iteration. Admittedly, the computational burden is now passed to the SVD algorithm, which could be required to tackle large matrices. Fortunately, efficient SVD algorithms have been proposed; for example, a sequential-randomized SVD for dealing with matrices larger than the available memory on a local PC was presented in \cite{HERNANDEZ2024116552}. Parallel algorithms are also available for the SVD, for example leveraging the tall-and-skinny matrices involved \cite{tomas2020tall}. }

\revone{An additional advantage of employing the ECM is that it readily incorporates a treatment for ill-posed problems, consisting of assembled projected residuals close to or numerically zero, i.e., when $\boldsymbol{S_r} \ \mathbbm{1} \approx \boldsymbol{0}$ \cite{tezaur2022robust}. This condition was discussed in our paper \cite{hernandez2017dimensional} for self-equilibrium problems, where the equivalent quantity for the assembled projected residuals is \textit{exactly} zero by construction. As a result of this prevision, the application of the ECM algorithm leads to the selection of $\beta +1 $ elements, one more than the number of rows in $\boldsymbol{G}$, as}

\revone{
\begin{equation}
    (\mathbb{E}, \boldsymbol{\omega} ) \leftarrow \texttt{ECM} (\boldsymbol{G}, \mathbbm{1}) \ .
\end{equation}
}

\revone{Such solution exactly integrates the assembled modes of the projected residuals as}

\revone{
\begin{equation}
    \boldsymbol{G} \boldsymbol{\zeta} - \boldsymbol{b} = \boldsymbol{0}  \ .
\end{equation}
}

\revone{A discussion on the relation between the truncation tolerance \(\epsilon_{\text{\tiny RES}}\) in Eq. \ref{eq: truncated svd residual}, and the user-defined tolerance \(\epsilon\) in the original optimisation problem in Eq. \ref{eq:optimization_problem_np_hard_1} has been recently presented in \cite{HERNANDEZ2024116552}. One can readily check that a solution obtained by the ECM algorithm also complies with}

\revone{
\begin{equation}
    \| \boldsymbol{S_r} \boldsymbol{\zeta} - \boldsymbol{d} \|_2 \leq \epsilon_{\text{\tiny RES}} \| \boldsymbol{d} \|_2 \ .
\end{equation}
}

\revone{
Indeed, when using the ECM algorithm, the truncation tolerance for the SVD of the projected residuals, $\epsilon_{\text{\tiny RES}}$, is the parameter dictating the sparsity of the reduced mesh, and in turn the precision of the HROM integration with respect to the ROM. If no truncation is performed, the number of selected elements will equal \texttt{Rank}($\boldsymbol{S_r}$)+1.
}

\own{Having at one's diposal the HROM elements and weights, the reduced terms in Eq. \ref{eq: rom_iterative} are assembled by looping over the reduced set of elements multiplied by the corresponding elemental weight, as}

\own{
\begin{equation}
     \boldsymbol{\Phi}^T \textbf{R}(\tilde{\textbf{d}} ) \approx  \sum_{e\in \mathbb{E}}{\boldsymbol{\Phi}}_e^T \textbf{R}_e (\tilde{\textbf{d}}_e; \boldsymbol{\mu}  ) \omega_e  \  ,
     \label{eq: hrom right hand side construction}
\end{equation}
}

\own{
\begin{equation}
     \boldsymbol{\Phi}^T  \textbf{J}(\tilde{\textbf{d}})\boldsymbol{\Phi} \approx \sum_{e\in \mathbb{E}} {\boldsymbol{\Phi}}_e^T  \textbf{J}_e(\tilde{\textbf{d}}_e ; \boldsymbol{\mu}){\boldsymbol{\Phi}}_e  \omega_e  \  .
     \label{eq: hrom left hand side construction}
\end{equation}
}

\subsection{Global Workflow}
\label{subsec: Global Workflow}

\own{Having introduced all the components, we now straightforwardly assemble them in this section as a summary. The following box presents the summarised workflow for both the offline and online stages of the POD-Galerkin with ECM Hyper-reduction.}

\begin{mybox}
\centering
\revone{\fbox{\begin{minipage}{\textwidth}
\textbf{Offline stage}\\
1. Define the $m$ training scenarios $\{\boldsymbol{\mu}_i\}_{i=1}^m$\\
2. Launch the corresponding FOM simulations and build the snapshots matrix $\boldsymbol{S} \in \mathbb{R}^{\mathcal{N} \times m}$\\
3. Compute the SVD of $\boldsymbol{S}$, with truncation tolerance $ \epsilon_{\text{\tiny SOL}}$ to get the basis $\boldsymbol{\Phi}\in \mathbb{R}^{\mathcal{N} \times N}$\\
4. Construct the snapshots of residuals projected $\boldsymbol{S_r} \in \mathbb{R}^{ N\cdot m  \times N_{el}} $\\
5. Compute the SVD of $\boldsymbol{S_r}$, with truncation tolerance $\epsilon_{\text{\tiny RES}}$, to get the basis for its rowspace $\boldsymbol{G}\in \mathbb{R}^{ \beta \times N_{el}} $ \\
6. Obtain the set of elements and positive weights using the ECM algorithm \cite{hernandez2020multiscale}
\[
(\mathbb{E}, \boldsymbol{\omega}) \leftarrow \texttt{ECM}(\boldsymbol{G}, \mathbbm{1})
\]
\textbf{Online stage}\\
1. Given a new set of parameters $\{\boldsymbol{\mu}^*_i\}_{i=1}^\ell$\\
2. Construct and solve the reduced system by looping over the subset of elements $\mathbb{E}$ multiplied by their corresponding weight $\boldsymbol{\omega}$
\end{minipage}}
\caption{\revone{Workflow of POD-Galerkin with ECM hyper-reduction}} % Add your caption here
\label{box:workflow-box}} % This is the label you can use to reference the box
\end{mybox}

\revone{The $m$ training scenarios described in step 1 of the offline stage in Box~\ref{box:workflow-box} are defined in this work using a specific function $f_{\mu}$ defining a \textit{training trajectory}, as formulated in Eq.~\ref{eq: time-mu mapping}. It is possible to employ more than one training trajectory; furthermore, incorporating other parameter variations can be achieved without significant complications. Similarly, in this work, the parameters introduced in step 1 of the online stage,  denote \textit{testing trajectories}. These trajectories are distinct from those used during the training stage.}

\revone{Alg. \ref{alg: global workflow} shows the speudocode for running the $\boldsymbol{\mu}-$geometrically parametrised simulation, where the steps to be followed by either a FOM, ROM or HROM simulation are delineated}. The main difference among these models occurs in point \ref{alg: ROM enters here}.

\begin{algorithm}[H]
\caption{Pseudo code for the geometrically parametrised simulations}
\label{alg: global workflow}
\hspace*{\algorithmicindent} \textbf{Input:} final time $T$, time increment size $\Delta t$, trajectory function $f_{\mu}(t):t\mapsto \boldsymbol{\mu}$\revone{, for a ROM also the basis $\boldsymbol{\Phi}\in\mathbb{R}^{\mathcal{N} \times N}$, and for an HROM, both the basis $\boldsymbol{\Phi}\in\mathbb{R}^{\mathcal{N} \times N}$ and the HROM elements and weights $\mathbb{E}, \boldsymbol{\omega}$} \\
\hspace*{\algorithmicindent} \textbf{Output:} State variable $\mathbf{u}_n \in \mathbb{R}^{\mathcal{N}}$ for all time steps ($\boldsymbol{S}$) and/or QoI $\boldsymbol{z}_n \in \mathbb{R}^{\alpha}$ for all time steps ($\boldsymbol{Z}$)
\begin{algorithmic}[1]
    \STATE \textbf{Initialisations} \\
    inital time $t \leftarrow 0$ \\
    initial geometry $\Omega_0  \leftarrow \varphi(\boldsymbol{\mu}_0)$  \\
    initial solution fields $(\boldsymbol{u}^h, p^h)  \leftarrow g_0( \boldsymbol{x}, \boldsymbol{u}, p)$
    \WHILE{$t \leq T$}
        \STATE update time: $t \leftarrow t + \Delta t$
        \STATE update the geometric parameter: $\boldsymbol{\mu}_{n+1} \leftarrow f_{\mu}(t)$
        \STATE morph the geometry $ \Omega_{n+1} \leftarrow \varphi(\boldsymbol{\mu}_{n+1})$
        \STATE compute the mesh velocity $\hat{\boldsymbol{u}}^h  \leftarrow
 g_{\text{\tiny mesh}}(\Omega_{n+1})$
        \STATE get the solution increment $\Delta \textbf{d}_{n+1}$ by

        \begin{itemize}
            \item \revone{FOM: solve until convergence system in Eq. \ref{eq: fom_iterative}}
            \item \revone{ROM: solve until convergence system in Eq. \ref{eq: rom_iterative}}
            \item \revone{HROM: solve until convergence system in Eq. \ref{eq: rom_iterative}, construct the system as in Eqs. \ref{eq: hrom right hand side construction} and \ref{eq: hrom left hand side construction}}
        \end{itemize}
        \label{alg: ROM enters here}
        \STATE store required data $\boldsymbol{S}[:,n+1] \leftarrow \textbf{d}_{n+1}$; $\boldsymbol{Z}[:,n+1] \leftarrow  \textbf{z}_{n+1}$
    \ENDWHILE
\end{algorithmic}
\end{algorithm}

It is worth noting that for this work, step 4 of Box \ref{box:workflow-box} was performed by running a ROM following Alg. \ref{alg: global workflow}, while storing the projected residuals as outlined in Section \ref{subsec: Hyper-Reduction}. This way of performing the HROM training can be expensive (in particular in terms of memory). In other works, the authors have chosen to project the readily available snapshots matrices onto the column space of the basis as $\mathbf{d}_{\text{\tiny projected}} = \boldsymbol{\Phi} \boldsymbol{\Phi}^T \mathbf{d}$, to finally obtain the residuals with respect to these \textit{projected} snapshots. Our decision to run Alg. \ref{alg: global workflow} to obtain the residuals projected is completely consistent with the theory presented in section \ref{subsec: Hyper-Reduction}.

\section{Model Description}
\label{sec: Model Description}

\begin{figure}[H]
\centering
\resizebox{0.6\linewidth}{!}{\includegraphics{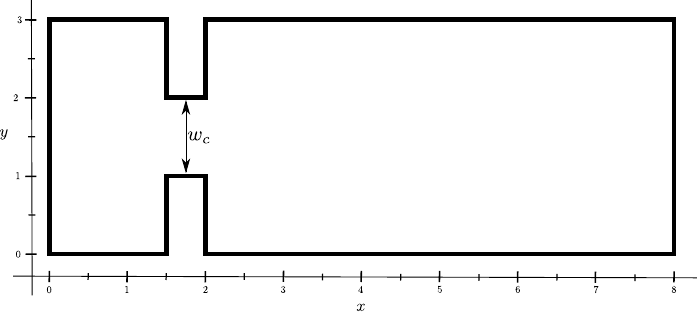}}
\caption{Base model of contraction-expansion channel}
\label{fig:Geometry}
\end{figure}

The plane contraction-expansion channel used in this study is shown in \revone{Fig. \ref{fig:Geometry}. This geometry serves as the base or reference configuration $\Omega_0$. To effectively morph this base geometry into a deformed configuration $\Omega$, as depicted in Fig. \ref{fig:geometric mapping}, we employ two distinct geometric mappings. This approach is employed to showcase the framework's versatility, irrespective of the nature of the mappings.} Each geometric mapping depends on a single scalar geometric parameter $\mu \in \mathbb{R}$, and are in turn an affine mapping denoted as $\boldsymbol{\varphi}_{\text{\tiny AFFINE}}$ and a \own{nonaffine} mapping that combines Free Form Deformation (FFD) and Radial Basis Functions (RBF) denoted as $\boldsymbol{\varphi}_{\text{\tiny FFD + RBF}}$.

The affine mapping $\boldsymbol{\varphi}_{\text{\tiny AFFINE}}$ changes the narrowing width $w_c$ in a symmetric way while preserving the walls straight. On the other hand, the \own{nonaffine} mapping $\boldsymbol{\varphi}_{\text{\tiny FFD + RBF}}$ allows to obtain deformed configurations that undergo more complex transformations and lead to curved walls of the narrowing.

\subsection{Affine \own{M}apping}
\label{subsec: Affine mapping}

For the affine mapping, the only parameter that dictates the shape of the geometry is the narrowing width $\mu = w_c \in [0.1, 2.9]$. Following \cite{hess2020spectral}, the affine mapping

\begin{equation}
\begin{aligned}
    \boldsymbol{\varphi}_{\text{\tiny AFFINE}}: \ & \Omega_0 \times [0.1, 2.9] \rightarrow \Omega \ , \\
    & (\boldsymbol{x}_0 , \mu ) \mapsto \boldsymbol{x} \ ,\\
\end{aligned}
\end{equation}

\noindent is defined by decomposing the original domain $\Omega_0$ into $N_{dom}$ non-overlapping subdomains as,

\begin{equation}
    \Omega_0 = \bigcup_{i=1}^{N_{dom}} \Omega_0^i  \hspace{4mm} \text{with }, \ \  \Omega_0^i \cap \Omega_0^j = \emptyset  \ \ \ \forall i\neq j \ ,
\end{equation}

\noindent and defining for each subdomain

\begin{equation}
    \boldsymbol{x}_0 = \boldsymbol{T}_i(\mu) (\boldsymbol{x} -\boldsymbol{g}_i ) + \boldsymbol{g}_i \ , \hspace{5mm} \boldsymbol{x}_0 \in \Omega_0 \ \ \boldsymbol{x} \in \Omega \ , \hspace{5mm} \own{\boldsymbol{T}_i} \in \mathbb{R}^{2 \times 2} \ \ \boldsymbol{g}_i \in \mathbb{R}^2 \ .
\end{equation}

\begin{figure}[H]
\centering
\resizebox{0.6\linewidth}{!}{\includegraphics{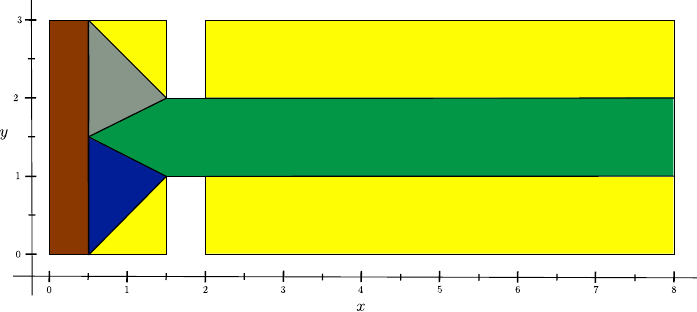}}
\caption{Reference subdomains for affine mapping}
\label{fig: different jacobians}
\end{figure}

\noindent For the geometry at hand, five distinct regions can be identified, as shown in Fig. \ref{fig: different jacobians}. For each of them, a particular \revone{transformation matrix} allows to map the point in the original configuration to the deformed configuration, using as parameter the narrowing width $w_c$. Eq. \ref{eq:jacobians} shows the \revone{transformation matrices} corresponding to each of the coloured regions.

\begin{equation}
    \boldsymbol{T}_{brown} =
    \begin{bmatrix}
        1 & 0 \\
        0 & 1
    \end{bmatrix}  \hspace{3mm}
    \boldsymbol{T}_{yellow} =
    \begin{bmatrix}
        1 & 0 \\
        0 & \frac{2}{3 - w_c}
    \end{bmatrix}  \hspace{3mm}
    \boldsymbol{T}_{green} =
    \begin{bmatrix}
        1 & 0 \\
        0 & \frac{1}{w_c}
    \end{bmatrix}    \hspace{3mm}
    \boldsymbol{T}_{grey} =
    \begin{bmatrix}
        1 & 0 \\
        \frac{1- w_c}{2} & 1
    \end{bmatrix}   \hspace{3mm}
    \boldsymbol{T}_{blue} =
    \begin{bmatrix}
        1 & 0 \\
        \frac{w_c - 1}{2} & 1
    \end{bmatrix}  \ .
    \label{eq:jacobians}
\end{equation}

\subsection{\own{Nonaffine} Mapping}
\label{subsec: Nonlinear mapping}

We perform the nonaffine mapping in two stages. In the first stage, an \revone{FFD} technique is applied to the boundary points of the 2D geometry. The second stage involves moving the points in the interior of the geometry. This is accomplished using an \revone{RBF} interpolator.

\begin{figure}[H]
  \centering
  \begin{subfigure}[b]{0.47\textwidth}
    \includegraphics[width=0.9\linewidth]{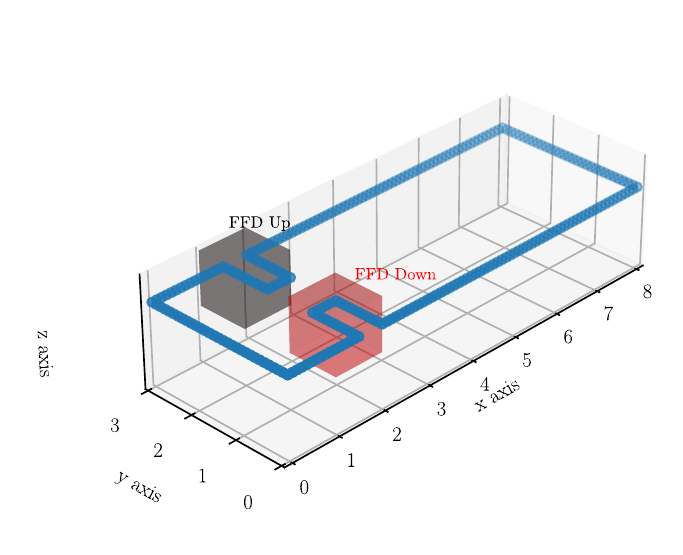}
    \caption{Bounding boxes for FFD}
  \end{subfigure}
  \hfill
  \begin{subfigure}[b]{0.47\textwidth}
    \includegraphics[width=0.9\linewidth]{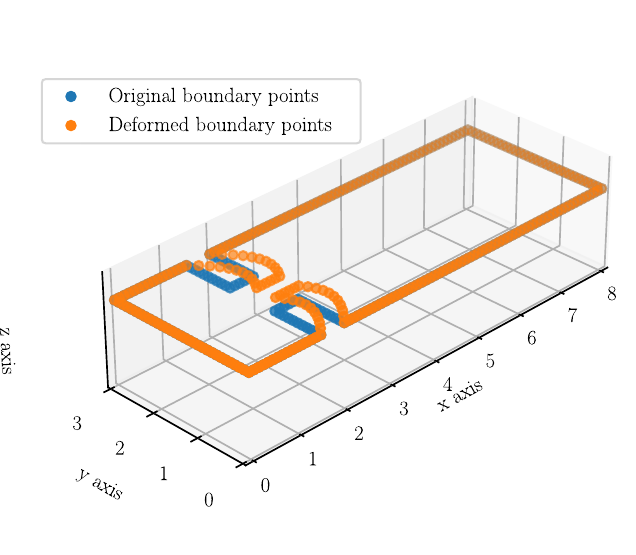}
    \caption{Boundary points deformed by changing control points of the FFD}
  \end{subfigure}
  \caption{\revone{FFD} part of the nonaffine mapping. For the second part of the nonaffine mapping, the interior points are moved according to \revone{an RBF} interpolator}
  \label{fig:FFD}
\end{figure}

In FFD, the geometry to morph is surrounded by a box of control points, as can be seen in Fig. \ref{fig:FFD}. Let $\boldsymbol{P}_0$ represent the set containing the original position of the control points defining the bounding box. For our particular case, the set of deformed control points' positions $\boldsymbol{P}(\mu)$ depends on a single scalar $\mu \in [0,1]$. The deformed position $\boldsymbol{x} \in  \partial \Omega$ of a point $\boldsymbol{x}_0 \in \partial \Omega_0$ \own{is then given as,}

\begin{equation}
    \boldsymbol{x} = \text{FFD}(\boldsymbol{x}_0, \boldsymbol{P}_0, \boldsymbol{P}(\mu))    \hspace{5mm}  , \forall \ \boldsymbol{x}_0 \in \partial \Omega_0 \ , \ \boldsymbol{x} \in \partial\Omega \ .
\end{equation}

For information on the specific expression of the mappings and blending functions used, the interested reader is directed to \cite{sederberg1986free, lombardi2012numerical, tezzele2018dimension}. We employ the implementation of FFD from the Python library PyGeM \cite{TezzeleDemoMolaRozza2020PyGeM}.

After the deformed boundary points have been computed, the deformed position of the points in the interior of the geometry $\boldsymbol{x} \in \Omega$ can be obtained by applying an RBF interpolator.

Let $\hat{\boldsymbol{x}}_i$ represent the $i$-th undeformed boundary mesh point (whose deformed position is obtained via the FFD mapping). The deformed position $\boldsymbol{x} \in \Omega$ of a point $\boldsymbol{x}_0 \in \Omega_0$ is then given as,

\begin{equation}
    \boldsymbol{x} = \text{RBF}(\boldsymbol{x}_0) = \sum_{i=1}^{N_b} \beta_i \phi \left( \norm{\boldsymbol{x}_0 - \hat{\boldsymbol{x}}_i  } \right) + q(\boldsymbol{x}_0) \ ,
\end{equation}

\noindent where $N_b$ is the number of points in the deformed boundary, $\phi(\cdot)$ is a given basis function depending on the distance from the desired point in the interior to a mesh point in the boundary, and $q(\boldsymbol{x}_0)$ is a polynomial. The coefficients $\beta_i$ and the polynomial $q(\boldsymbol{x}_0)$, are obtained as a function of the deformed boundary points by imposing the interpolation conditions

\begin{equation}
    \text{RBF}( \hat{\boldsymbol{x}}_i) = \text{FFD}(\hat{\boldsymbol{x}}_i, \boldsymbol{P}_0, \boldsymbol{P}(\mu)) \hspace{5mm}, \ i=1, \dots, N_b \ .
\end{equation}

Further details about \revone{RBF} interpolation can be found in \cite{stabile2020efficient, beckert2001multivariate}. Successive application of the FFD mapping for all mesh points on the boundary of the geometry, followed by RBF parameters computation and application for all mesh points in the interior, completely defines the mapping
\begin{equation}
\begin{aligned}
   \boldsymbol{\varphi}_{\text{\tiny FFD+RBF}}: \ & \Omega_0 \times [0,1] \rightarrow \Omega \ , \\
    & (\boldsymbol{x}_0 ,  \mu ) \mapsto \boldsymbol{x} \ .\\
\end{aligned}
\end{equation}

\subsection{Mesh}
\label{subsec: Mesh}

\begin{figure}[H]
  \centering
  \begin{subfigure}[b]{0.47\textwidth}
    \includegraphics[width=\linewidth]{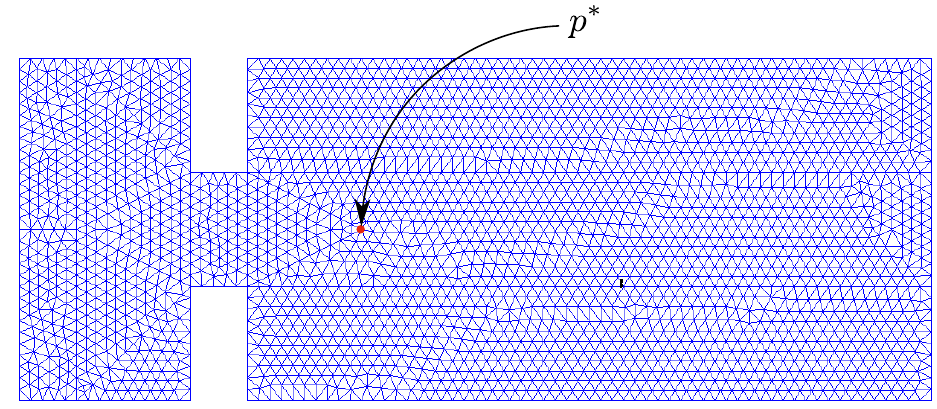}
    \caption{Mesh for mapping $\varphi_{\text{\tiny AFFINE}}$}
  \end{subfigure}
  \hfill
  \begin{subfigure}[b]{0.47\textwidth}
    \includegraphics[width=\linewidth]{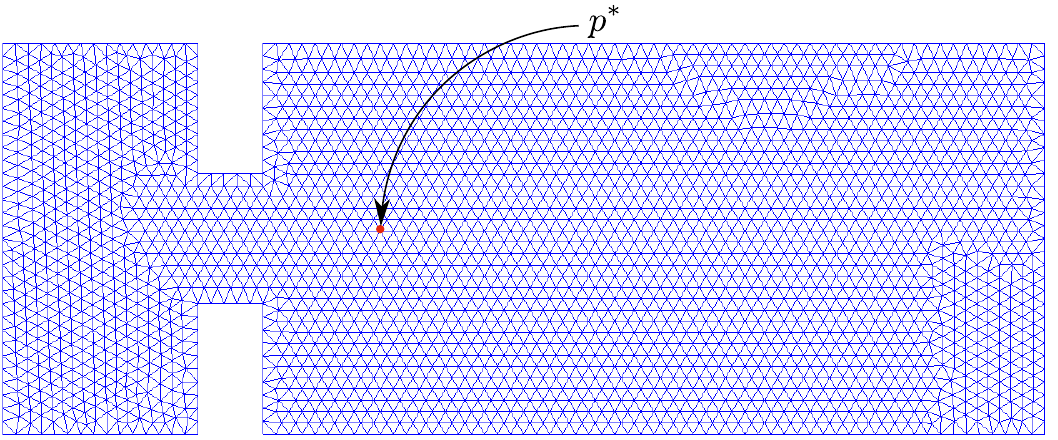}
    \caption{Mesh for mapping $\varphi_{\text{\tiny FFD+RBF}}$}
  \end{subfigure}
  \caption{\own{Finite elements meshes used for the models. The point $p^*$ refers to the probe position in the mesh whose information is used for the hysteresis plot}}
  \label{fig: meshes}
\end{figure}

The two geometric mappings require different control surfaces with a corresponding label to exist. For example, the affine mapping requires to clearly identify the coloured regions in Fig. \ref{fig: different jacobians} to apply the respective transformations on them. Therefore, a conforming mesh fitted to each of them was generated. On the other hand, the \own{nonaffine} mapping only required to identify the upper and lower rectangles defining the narrowing. \revone{Moreover, while the nature of the affine mapping allows to preserve the positive Jacobians of the isoparametric finite elements under relatively large deformations, the \own{nonaffine} mapping can lead to distorted elements which intersect, or completely swap, whenever deformations are too large. These differences on the mappings forced us to use a different mesh for each of them, while keeping the same element size in both. We have used unstructured meshes in both cases because they are known to work better for the specific example under consideration \cite{quaini2016symmetry}.} The information of the meshes used\footnote{ In KratosMultiphysics, the mesh entities are separated in 2D triangles and 1D boundary elements known in Kratos as \textit{Conditions}. Given this distinction there are  4916 Elements +231 Conditions for the affine mapping, and  5152 elements + 260 Conditions for the \own{nonaffine} mapping} can be seen in Table \ref{table:mesh_data}, while Fig. \ref{fig: meshes} show the generated meshes. In both figures, the presence of the probe point $p^*$ allows to monitor the evolution of the QoI, and therefore the occurrence of the Coanda effect. As the geometry morphs in time following a circular trajectory, a hysteresis plot can be created. The position of the probe point is selected to be located at a node in the corresponding finite element mesh that is close to the \revone{$y$}-direction centre line, and slightly behind the narrowing. The generation of the meshes was performed using the pre- and post-processor GiD \cite{gidhome}.

\begin{table}[h]
\centering
\begin{tabular}{ccc}
\hline
\rowcolor[gray]{0.8}
\textbf{Type of mapping} & \textbf{Number of Nodes} & \textbf{Number of Elements} \\ \hline
affine                   & 2589                     & 5147                        \\ \hline
\own{nonaffine}                  & 2707                     & 5412                        \\ \hline
\end{tabular}
\caption{ \revone{Information of the mesh employed in this study for each geometric mapping.}}
\label{table:mesh_data}
\end{table}

The general purpose software KratosMultiphysics \cite{vicente_mataix_ferrandiz_2023_7681287} was used for launching the FOM, ROM and HROM simulations, by taking advantage of the KratosRomApplication. In Kratos, the finite elements discretisation is closely linked to the physics to be simulated. For the physics in these simulations, Navier-Stokes in an Arbitrary Lagrangian Eulerian framework, the only available option was to use P1-P1 triangular elements.

\subsection{Boundary and Initial Conditions}
\label{subsec: Boundary Conditions}

Both models \revone{prescribe} the initial and boundary conditions shown next

\begin{equation}
 \boldsymbol{u} = \boldsymbol{0} \ , \ p = 0  \ \ , \ \boldsymbol{x} = \begin{bmatrix}
     x \\ y
 \end{bmatrix} \in \Omega_0  \ \ t = 0   \ ,
\end{equation}

\begin{equation}
    \begin{array}{ll}
    \boldsymbol{u} =
    \begin{bmatrix}
    u \\
    v
    \end{bmatrix} & =
    \left\{
        \begin{array}{ll}
                \begin{bmatrix}
    y(3-y) \ sin\left( \frac{t \pi}{2} \right) \\
    0
    \end{bmatrix}   & , x = 0 \ \ t = [0,1] \ , \\
    \\
    \begin{bmatrix}
    y(3-y)  \\
    0
    \end{bmatrix}  & , x = 0 \ \ t = (1,T] \ ,
        \end{array}
    \right. \\ \\
    p = 0 & \ \  \ \ , x = 8 \ \ t = [0,T] \ .
    \end{array}
\end{equation}

\noindent As can be seen, the maximum horizontal velocity at the inlet is set to 3. Moreover, the inlet velocity has been allowed to undergo an initialisation period for one second following a smooth path. \own{Finally, both upper and lower walls prescribe non-slip conditions, and }\revone{we are not considering body forces}.

%While in our case the total time $T = 244 s$,
% and the time steps increment is set to $\Delta t = 0.1 s$.

% REYNOLDS NUMBER NECESSARY? (RATHER THAN SPECIFYING THE PARAMETES), THE SIMULATION IS SUPER MESH-DEPENDENT!!

\subsection{Material Properties}
\label{subsec: Material properties}

The values of the viscosity and density for the fluid have been selected so that, given the distortion allowed by the finite element discretisations under the respective geometric mappings, the results were qualitatively comparable to the ones reported in similar works, e.g. \cite{quaini2016symmetry,oliveira2008simulations,pitton2017computational, hess2020reduced,hess2020spectral,hess2019localized}. Moreover, the material properties are held constant for all simulations.

For  both models, we used the values of dynamic viscosity and density reported in the following table.

\begin{table}[h]
\centering
\begin{tabular}{cc}
\rowcolor[gray]{0.8}
\textbf{fluid property} & \textbf{value} \\ \hline
dynamic viscosity $\tilde{\mu}$ & 0.1 \\ \hline
density $\rho$          & 1.0   \\ \hline
\end{tabular}
\end{table}

\section{Results}
\label{sec: Results}

In this section, we begin by providing a qualitative overview of the \revone{FOM}. We present the different geometrical configurations and showcase the observed Coanda effect along with their corresponding hysteresis loops. Additionally, we offer insights into the training process of the \revone{ROMs} and \revone{HROMs}. We discuss the \own{different trajectories} and present the results obtained from all models \own{providing both a qualitative and} a quantitative comparison including error and speedup factors \own{for each of the two examples studied, which are indented to highlight the performance of ROMs and HROMs in challenging scenarios.}

The models' files will be made available via the Examples repository of \revone{KratosMultiphysics \cite{vicente_mataix_ferrandiz_2023_7681287}} at \newline \revtwo{ \url{https://kratosmultiphysics.github.io/Examples/}}.

\subsection{FOM Results}
\label{subsec: FOM Results}

Before delving into the study of the hysteresis of the Coanda effect for the contraction-expansion channel under scrutiny, we first examine the behavior of the \revone{FOM} and how variations in the geometry impact the fluid solution.

\begin{figure}[h]
  \centering
  \begin{subfigure}[b]{0.45\textwidth}
    \includegraphics[width=\linewidth]{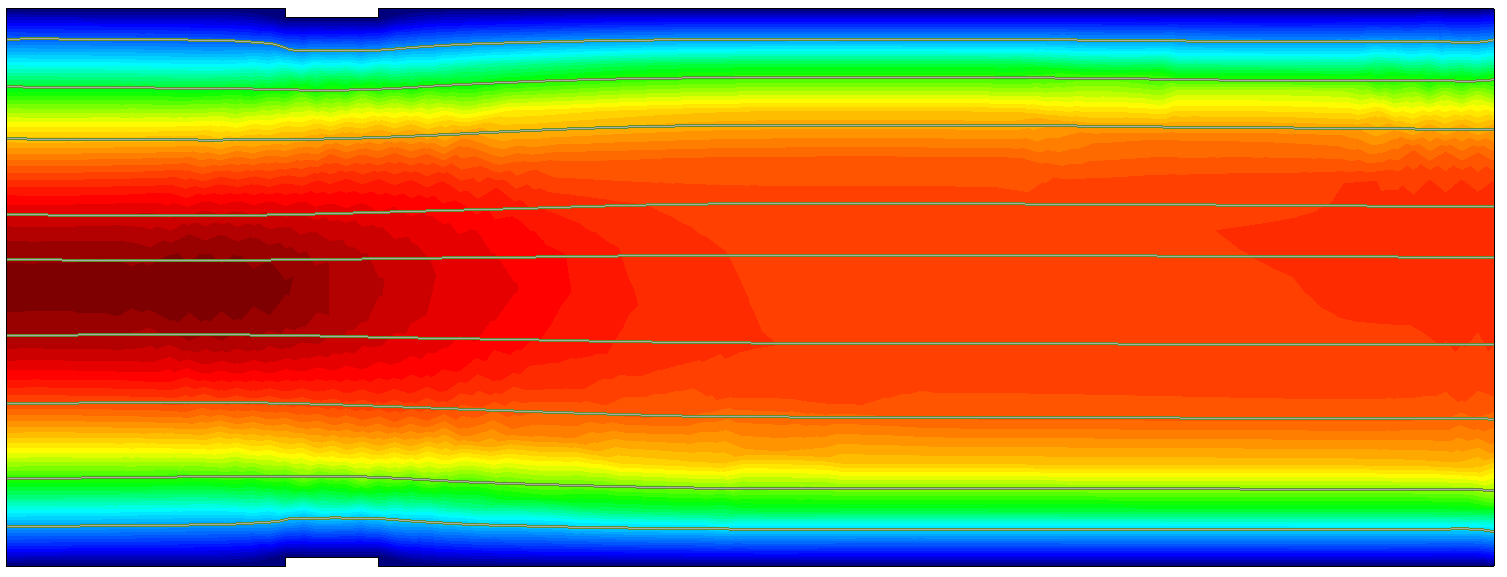}
    \caption{$\boldsymbol{\varphi}_{\text{\tiny AFFINE}}(\cdot, 2.9), \ \  w_c = 2.9 $}
    \label{fig:FOM solutions a}
  \end{subfigure}
  \hfill
  \begin{subfigure}[b]{0.45\textwidth}
    \includegraphics[width=\linewidth]{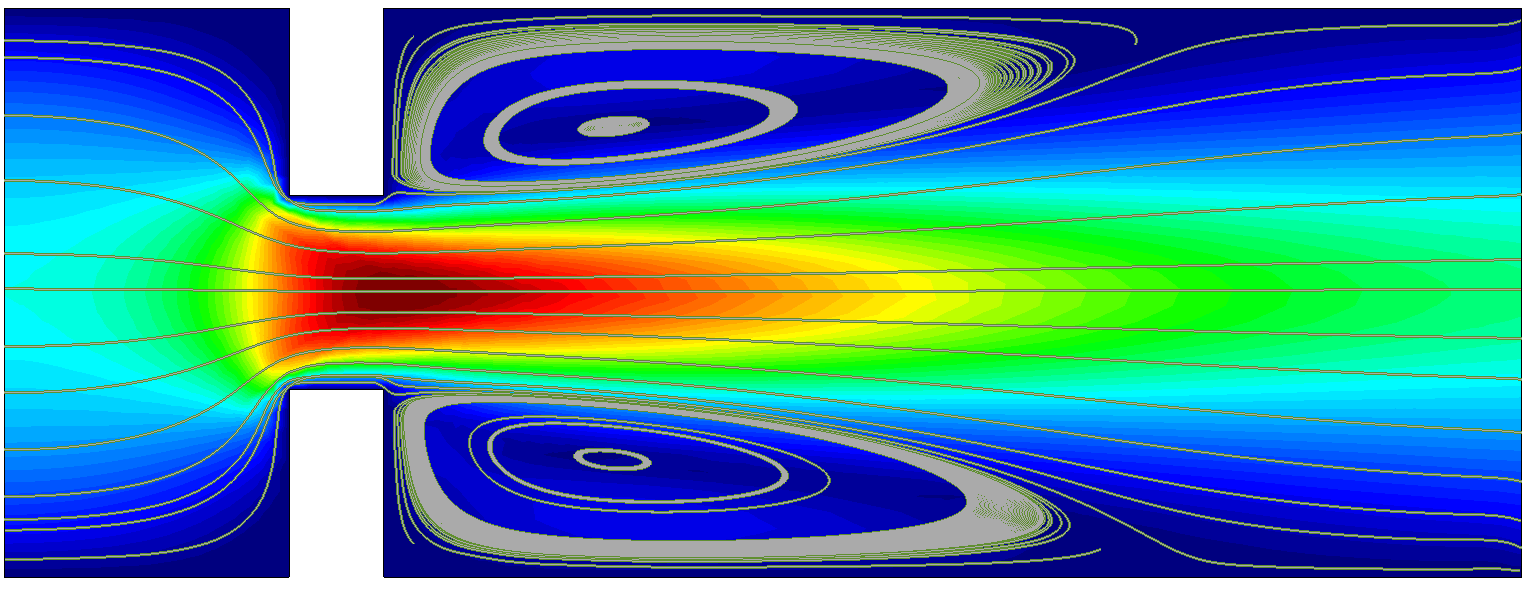}
    \caption{$\boldsymbol{\varphi}_{\text{\tiny AFFINE}}(\cdot,1), \ \ w_c = 1$ }
    \label{fig:FOM solutions b}
  \end{subfigure}

  \begin{subfigure}[b]{0.45\textwidth}
    \includegraphics[width=\linewidth]{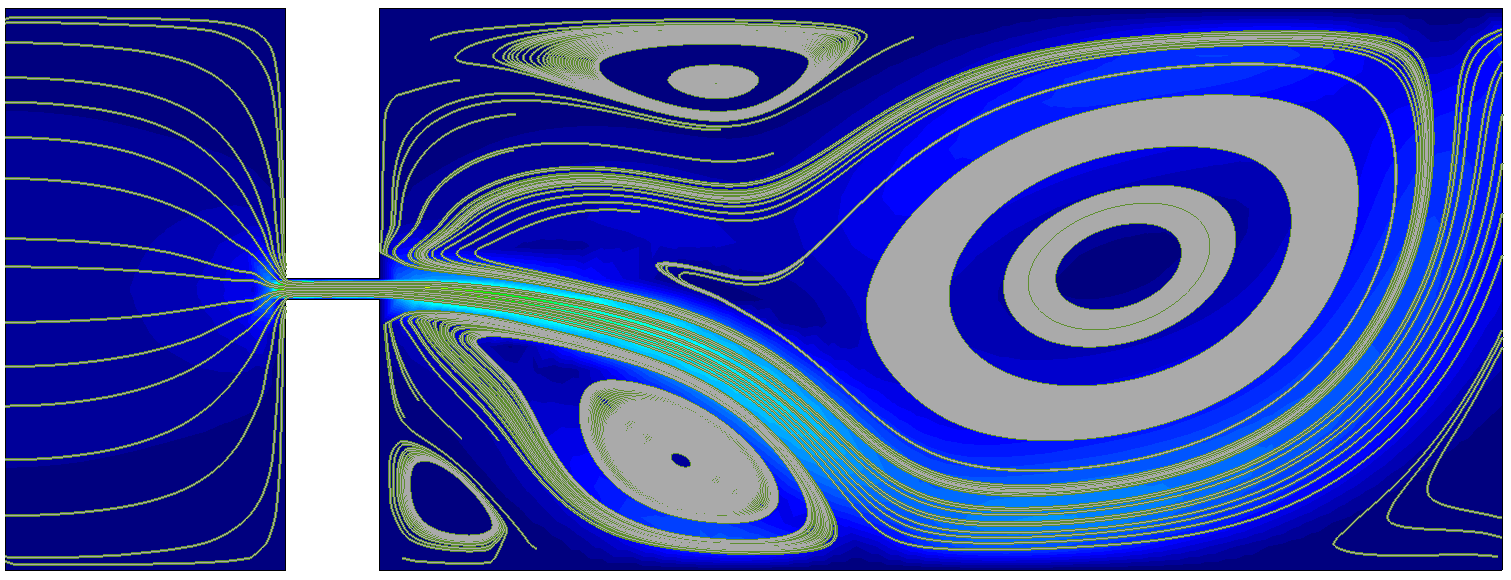}
    \caption{$\boldsymbol{\varphi}_{\text{\tiny AFFINE}}(\cdot,0.1), \ \ w_c = 0.1$ }
    \label{fig:FOM solutions c}
  \end{subfigure}
  \hfill
  \begin{subfigure}[b]{0.45\textwidth}
    \includegraphics[width=\linewidth]{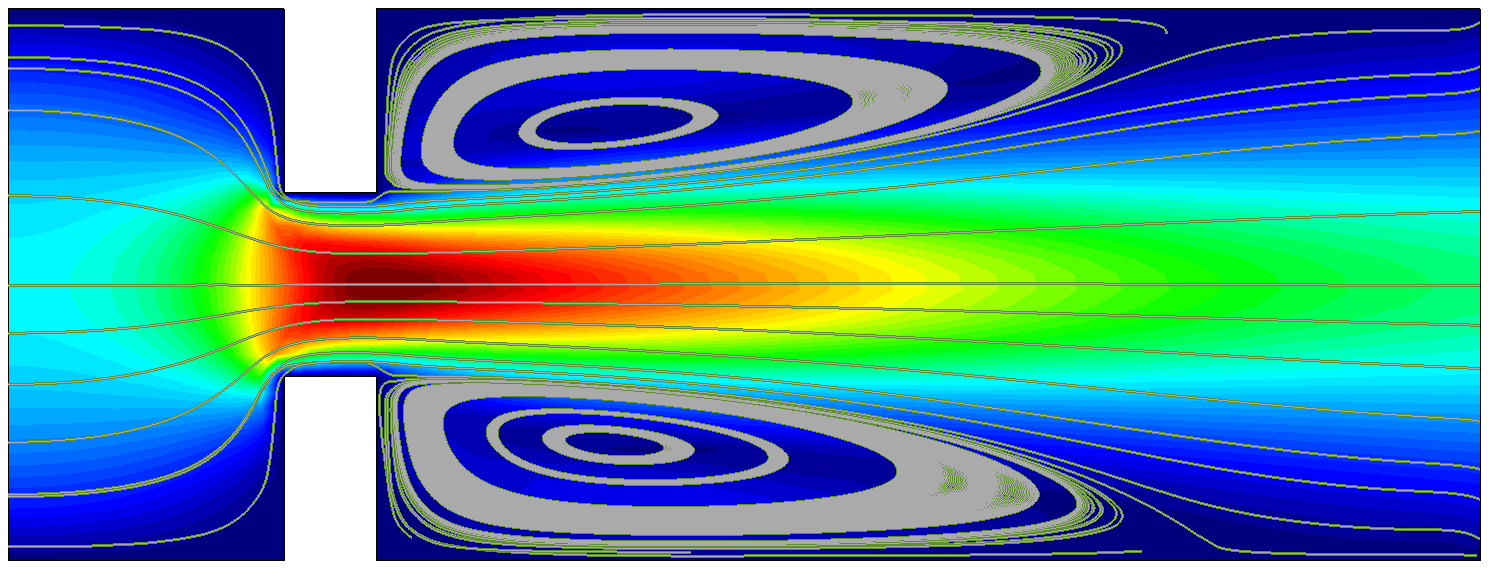}
    \caption{$\boldsymbol{\varphi}_{\text{\tiny FFD+RBF}}(\cdot,0), \ \ w_c = 1 $}
    \label{fig:FOM solutions d}
  \end{subfigure}

  \begin{subfigure}[b]{0.45\textwidth}
    \includegraphics[width=\linewidth]{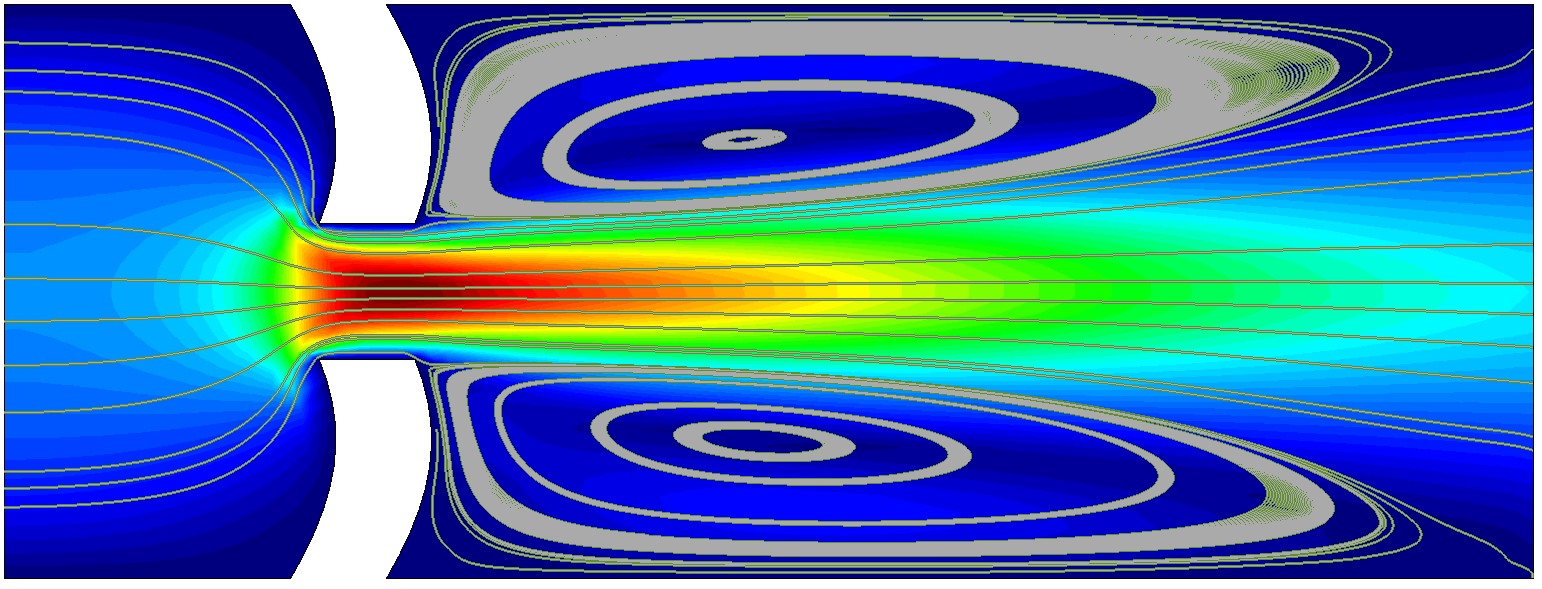}
    \caption{$\boldsymbol{\varphi}_{\text{\tiny FFD+RBF}}(\cdot,0.5), \ \ w_c = 0.7113 $}
    \label{fig:FOM solutions e}
  \end{subfigure}
  \hfill
  \begin{subfigure}[b]{0.45\textwidth}
    \includegraphics[width=\linewidth]{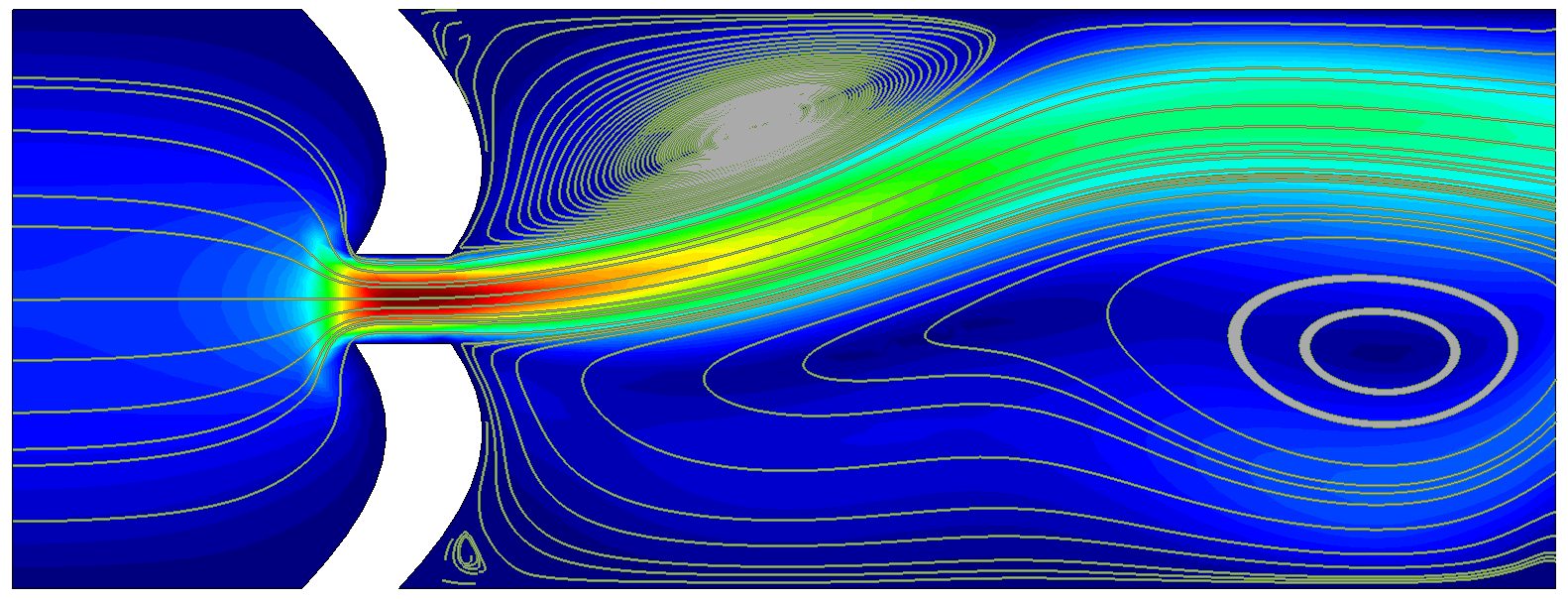}
    \caption{$\boldsymbol{\varphi}_{\text{\tiny FFD+RBF}}(\cdot,1), \ \ w_c = 0.422$ }
    \label{fig:FOM solutions f}
  \end{subfigure}
  \caption{FOM solutions for different deformed configurations}
  \label{fig:FOM solutions}
\end{figure}

Fig. \ref{fig:FOM solutions} illustrates the flow field for extreme values of the geometric parameter $\mu$ for both the affine \own{geometric} mapping (ranging from 0.1 to 2.9) and the \own{nonaffine geometric} mapping (ranging from 0 to 1) in the contraction-expansion channel.

For the case where the narrowing width $w_c$ is set to 1 (Figures \ref{fig:FOM solutions b} and \ref{fig:FOM solutions d}), both mappings yield qualitatively similar solutions, characterized by a horizontally symmetric jet. However, as the narrowing width decreases, the Coanda effect becomes present, and small eddies appear on the asymmetric jets (Figures \ref{fig:FOM solutions c} and \ref{fig:FOM solutions f}).

Fig. \ref{fig:FOM solutions c} exhibits a complex flow pattern. Notice the relatively small value of the narrowing width made possible by the affine \own{geometric} mapping thanks to its bijectivity. On the other hand, in Fig. \ref{fig:FOM solutions f}, the maximum distortion allowed by employing the \own{nonaffine geometric} mapping produces a narrowing width of $w_c = 0.422$. Although this value is larger than the minimum narrowing width for the affine \own{geometric} mapping, it still produces flow patterns relevant to the current investigation.

\begin{figure}[H]
  \centering
  \begin{subfigure}[b]{0.45\textwidth}
    \includegraphics[width=\linewidth]{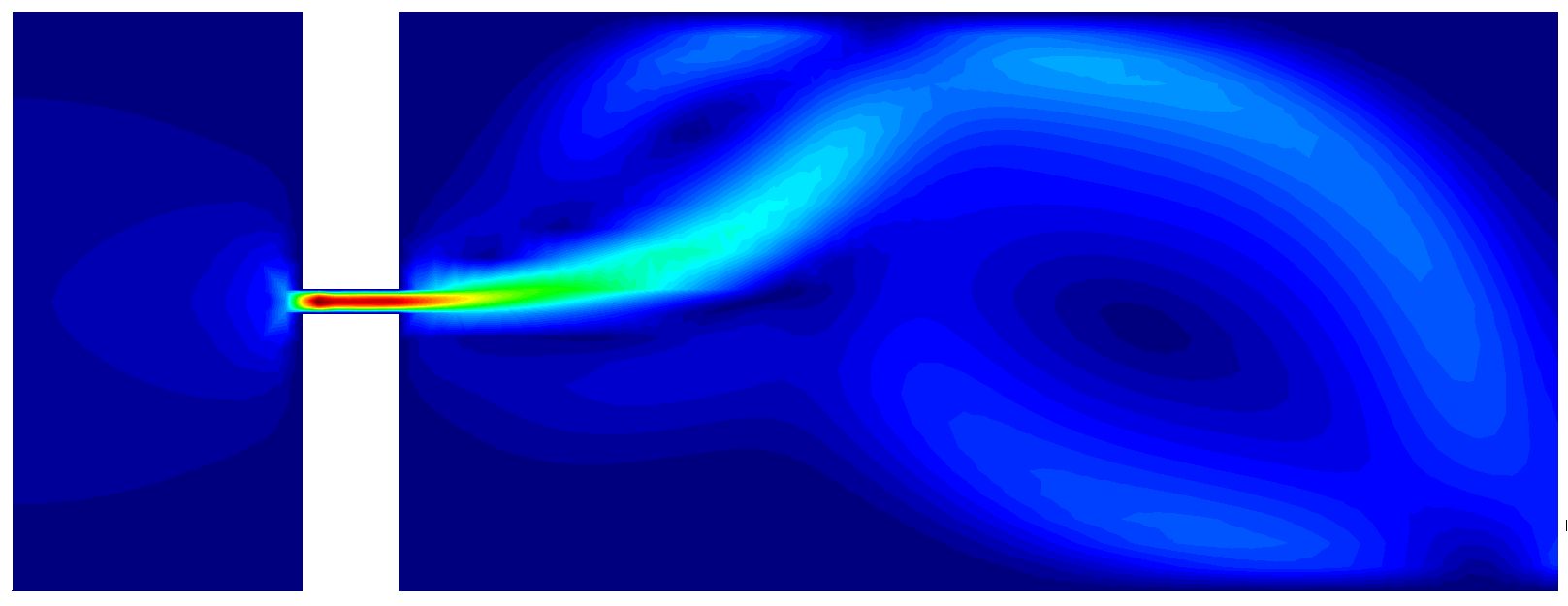}
    \caption{$\boldsymbol{\varphi}_{\text{\tiny AFFINE}}(\cdot,0.1), \ \ w_c = 0.1$ }
    \label{fig:FOM stable solutions a}
  \end{subfigure}
  \hfill
  \begin{subfigure}[b]{0.45\textwidth}
    \includegraphics[width=\linewidth]{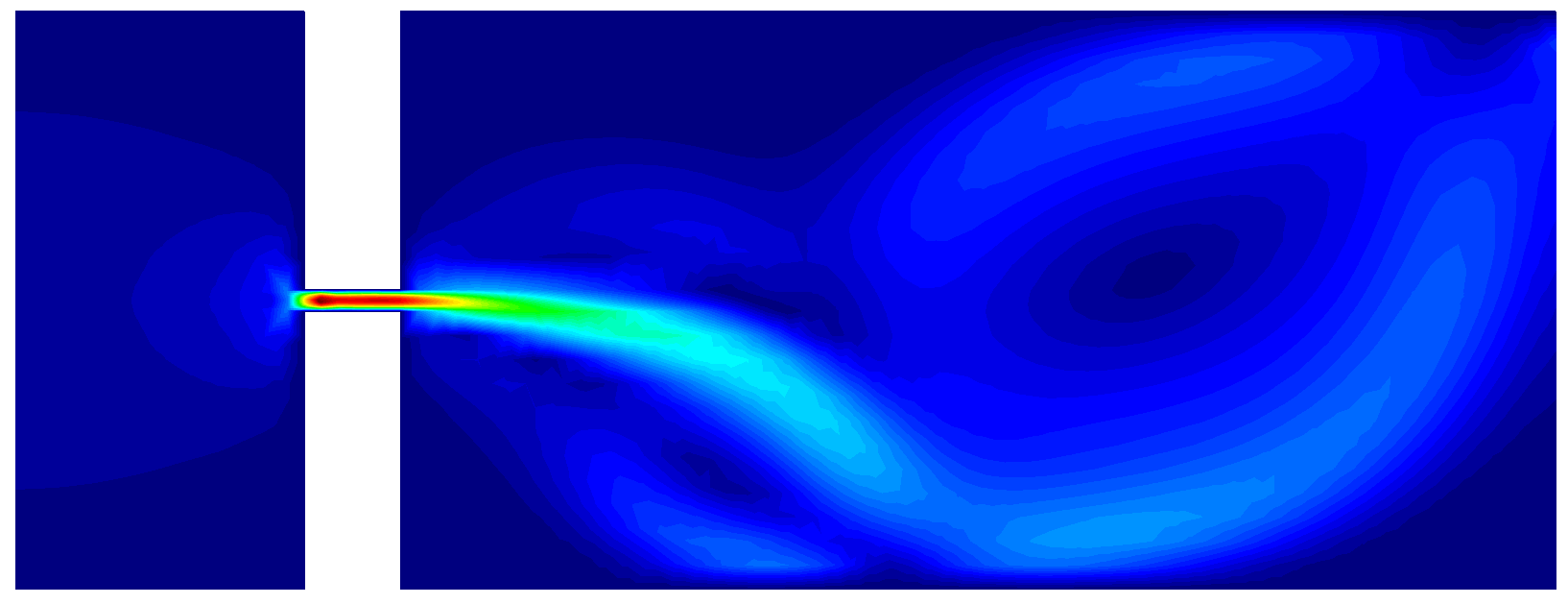}
    \caption{$\boldsymbol{\varphi}_{\text{\tiny AFFINE}}(\cdot,0.1), \ \ w_c = 0.1$ }
    \label{fig:FOM stable solutions b}
  \end{subfigure}

  \begin{subfigure}[b]{0.45\textwidth}
    \includegraphics[width=\linewidth]{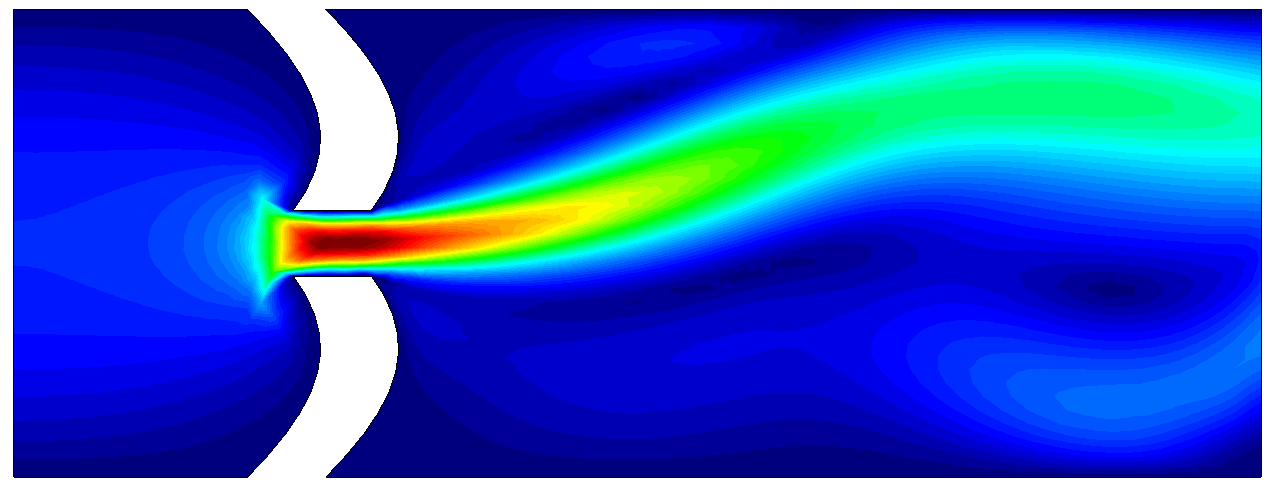}
    \caption{$\boldsymbol{\varphi}_{\text{\tiny FFD+RBF}}(\cdot,1), \ \ w_c = 0.422$ }
    \label{fig:FOM stable solutions c}
  \end{subfigure}
  \hfill
  \begin{subfigure}[b]{0.45\textwidth}
    \includegraphics[width=\linewidth]{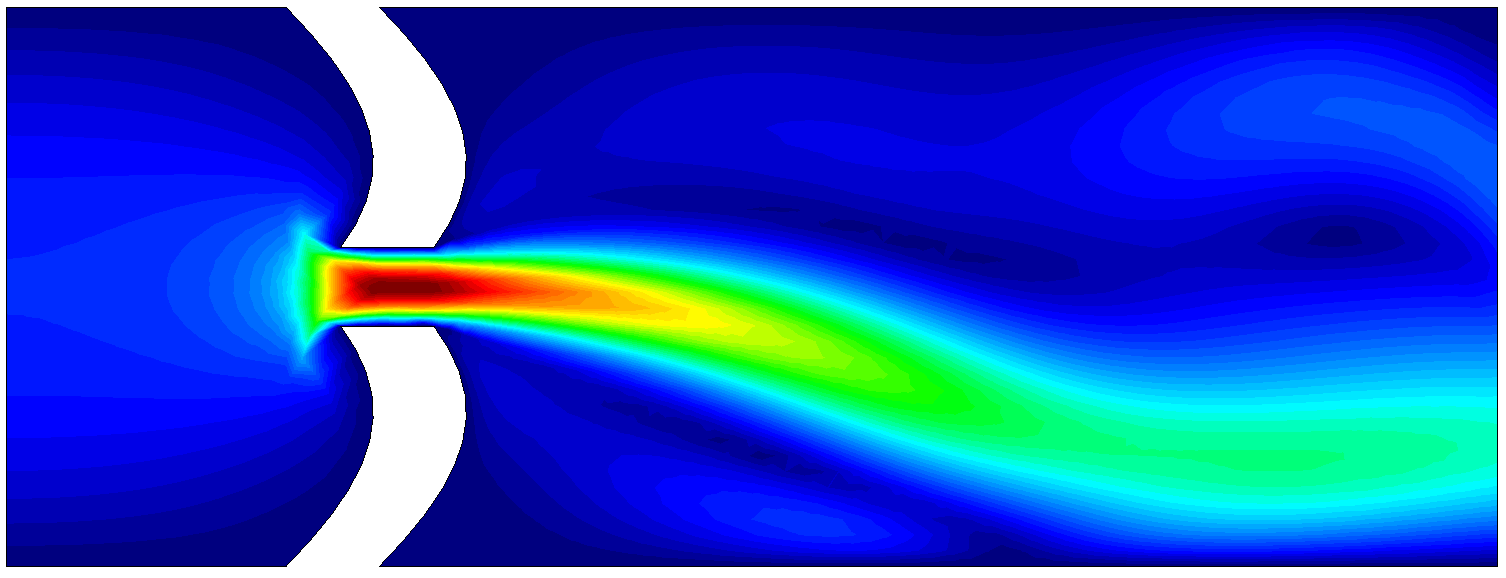}
    \caption{$\boldsymbol{\varphi}_{\text{\tiny FFD+RBF}}(\cdot,1), \ \ w_c = 0.422$ }
    \label{fig:FOM stable solutions d}
  \end{subfigure}

  \caption{Stable solutions attach to the upper or lower walls. The symmetric jet solution is unstable and therefore not achieved}
  \label{fig:FOM stable solutions}
\end{figure}

In Fig. \ref{fig:FOM stable solutions}, we show the possible solutions obtained for the same values of the geometric parameter $\mu$ for both mappings. Advanced methods can be used for obtaining one or the other of the stable solutions, e.g. \cite{pintore2021efficient}. For our specific case, we have observed that given a specific trajectory (earlier mentioned to be defined by $f_\mu$), that is, given an initial value of the parameter $\mu$, and then deforming the geometry in a closed loop, one of the stable solutions is obtained consistently, and a hysteresis loop can be observed.

\subsection{\newsec{Trajectories}}
\label{subsec: Trajectories}

\newsec{Here we highlight the different values of the parameter $\mu$ imposed to both models, through the description of the trajectory functions $f_{\mu}: t \mapsto \mu$.}

\newsec{In this work, we are working with three distinct trajectories. The first two trajectories are straightforward and represent a contraction followed by an expansion, and an expansion followed by a contraction respectively. Both models are therefore subjected to a morphing going from an initial geometry $\Omega_0$, going to a maximally deformed geometry and from there back to the initial geometric configuration. These first two trajectories are therefore}

\begin{figure}[H]
\begin{minipage}{0.5\textwidth}
\newsec{
\[
f_{\mu}(t) =
\left\{
\begin{array}{ll}
    \mu_{max}                &, \text{ if } t<T_{1} \\
    \mu_{max} - (t- T_1)c  &, \text{ if } t \geq T_{1} \text{ and } t<T_{2} \\
    \mu_{min}                &, \text{ if } t \geq T_{2} \text{ and } t<T_{3} \\
    \mu_{min} + (t - T_3)c  &, \text{ if } t \geq T_{3} \text{ and } t<T_{4} \\
    \mu_{max}                &, \text{ if } t \geq T_{4} \\
\end{array}
\right.
\]
}
\caption*{\newsec{Trajectory 1: Contract then expand}}
\end{minipage}
\begin{minipage}{0.5\textwidth}
\newsec{
\[
f_{\mu}(t) =
\left\{
\begin{array}{ll}
    \mu_{min}                &, \text{ if } t<T_{1} \\
    \mu_{min} + (t- T_1)c  &, \text{ if } t \geq T_{1} \text{ and } t<T_{2} \\
    \mu_{max}                &, \text{ if } t \geq T_{2} \text{ and } t<T_{3} \\
    \mu_{max} - (t - T_3)c  &, \text{ if } t \geq T_{3} \text{ and } t<T_{4} \\
    \mu_{min}                &, \text{ if } t \geq T_{4} \\
\end{array}
\right.
\]
}
\caption*{\newsec{Trajectory 2: Expand then contract}}
\end{minipage}
\end{figure}

\newsec{The third trajectory is a more complicated function defined as follows}

\begin{figure}[H]
\begin{minipage}{\textwidth}
\newsec{
\[
    f_{\mu}(t) = \left\{ \begin{array}{ll}
         \mu_0 \hspace{4mm} & , \text{ if } t<T_{1} \\
        \frac{1}{2}(1+ sin(\frac{2\pi t-T_1 }{T} - \frac{\pi}{2}) cos(\frac{6\pi}{T})) \hspace{4mm} & ,\text{ if } t  \geq T_{1}
    \end{array}
    \right.
\]
}
\caption*{\newsec{Trajectory 3: Trigonometric function}}
\end{minipage}
\end{figure}

\newsec{For all of the trajectories, we keep the initial geometry static for a period of time $T_1$ to allow the flow to develop. Moreover, for trajectories 1 and 2, we treat differently time instances $T_2$ and $T_3$. One the one hand, for the nonaffine mapping we keep the value of the parameter $\mu$ static during enough time for the Coanda effect to fully develop. On the other hand, in the case of the affine mapping the asymmetric jet is already strongly present in the solution by the time the maximum deformation is reached, therefore, for this mapping we immediately start the progressive deformation back to the initial configuration.}

\begin{figure}[H]
  \centering
  \begin{subfigure}[b]{0.55\textwidth}
    \includegraphics[width=\linewidth]{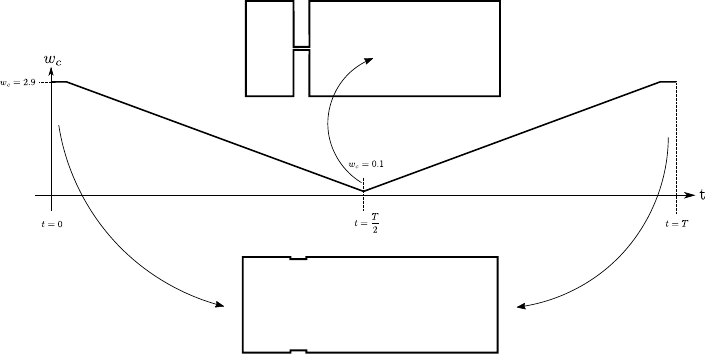}
    \caption{$\boldsymbol{\varphi}_{\text{\tiny AFFINE}}(\cdot, \cdot), \ \ w_c \in [0.1, 2.9] $ }
  \end{subfigure}
  \hfill
  \begin{subfigure}[b]{0.40\textwidth}
    \includegraphics[width=\linewidth]{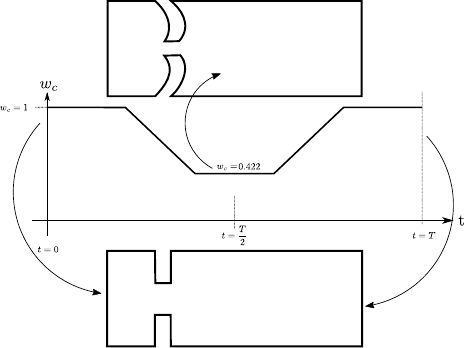}
    \caption{$\boldsymbol{\varphi}_{\text{\tiny FFD+RBF}}(\cdot, \cdot), \ \ w_c \in [0.422, 1] $ }
  \end{subfigure}

  \caption{Trajectory 1 for both mappings}
  \label{fig:Trajectory 1 both mappings}
\end{figure}

\begin{figure}[H]
  \centering
  \begin{subfigure}[b]{0.55\textwidth}
    \includegraphics[width=\linewidth]{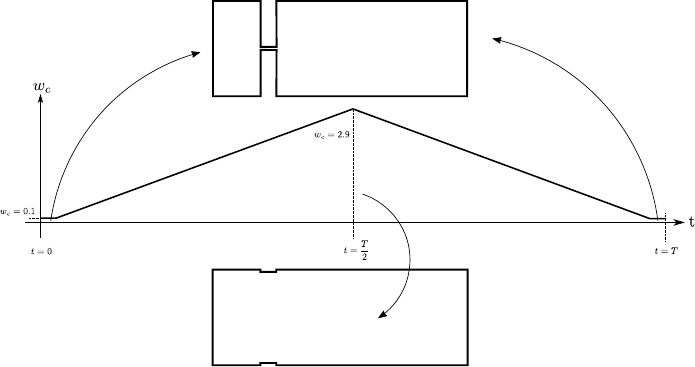}
    \caption{$\boldsymbol{\varphi}_{\text{\tiny AFFINE}}(\cdot, \cdot), \ \ w_c \in [0.1, 2.9] $ }
  \end{subfigure}
  \hfill
  \begin{subfigure}[b]{0.40\textwidth}
    \includegraphics[width=\linewidth]{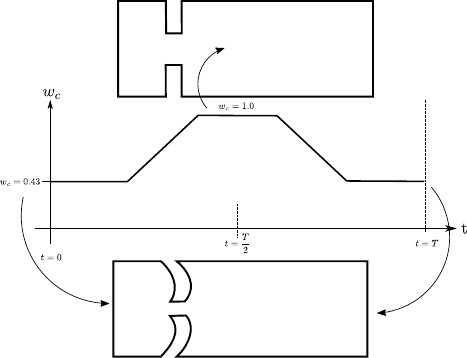}
    \caption{$\boldsymbol{\varphi}_{\text{\tiny FFD+RBF}}(\cdot, \cdot), \ \ w_c \in [0.422, 1] $ }
  \end{subfigure}

  \caption{Trajectory 2 for both mappings}
  \label{fig:Trajectory 2 both mappings}
\end{figure}

\begin{figure}[H]
  \centering
  \begin{subfigure}[b]{0.55\textwidth}
    \includegraphics[width=\linewidth]{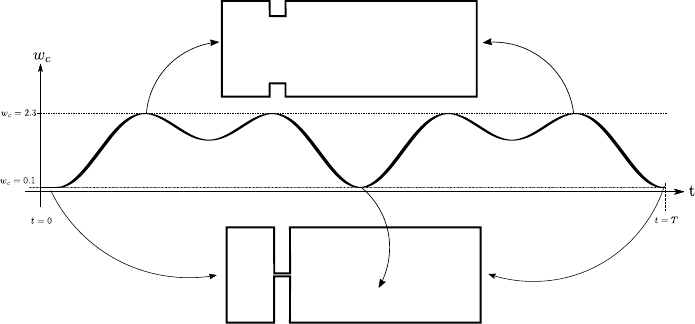}
    \caption{$\boldsymbol{\varphi}_{\text{\tiny AFFINE}}(\cdot, \cdot), \ \ w_c \in [0.1, 2.3] $ }
  \end{subfigure}
  \hfill
  \begin{subfigure}[b]{0.40\textwidth}
    \includegraphics[width=\linewidth]{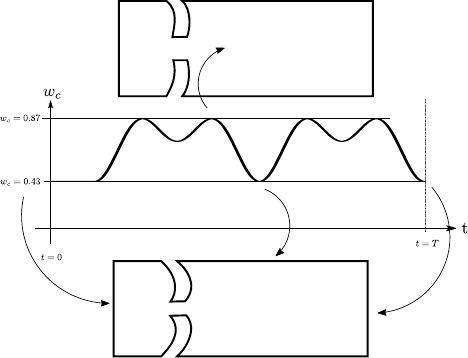}
    \caption{$\boldsymbol{\varphi}_{\text{\tiny FFD+RBF}}(\cdot, \cdot), \ \ w_c \in [0.43, 0.87] $ }
  \end{subfigure}

  \caption{\newsec{Trajectory 3 for both mappings}}
  \label{fig:Trajectory 3 both mappings}
\end{figure}

\newsec{A graphical representation of these functions applied to each of the two models is shown in Figs. \ref{fig:Trajectory 1 both mappings}, \ref{fig:Trajectory 2 both mappings}, and \ref{fig:Trajectory 3 both mappings}.}

\subsection{\newsec{Description of the Methodology}}

\newsec{Having at hand two models defined by their own geometric mappings, and three trajectories, we have devised a strategy to test the efficacy of the ROMs and HROMs to reconstruct a \textit{training} and a \textit{testing} trajectory. We have therefore come up with two different examples as follows:}
\newsec{
\begin{itemize}
    \item \textbf{Example 1}: Use Trajectory 1 as training trajectory, and Trajectory 2 as testing trajectory
    \item \textbf{Example 2}: Use both Trajectory 1 and Trajectory 2 as training trajectories, and Trajectory 3 as testing trajectory
\end{itemize}
}

\newsec{For each of the two models we have constructed ROMs employing four different truncation tolerances for the SVD of the snapshots matrices of the solution, namely $ \epsilon_{\text{\tiny SOL}} = \{ 1e-3, 1e-4, 1e-5, 1e-6 \}$, such that}

\begin{equation}
    \norm{\boldsymbol{S} - \boldsymbol{\Phi}\boldsymbol{\Phi}^T \boldsymbol{S}}_F \leq \epsilon_{\text{\tiny SOL}} \norm{\boldsymbol{S}}_F  \ ,
\end{equation}

\noindent \newsec{where the snapshots matrix $\boldsymbol{S}\in \mathbb{R}^{\mathcal{N}\times m}$, and the basis matrix $\boldsymbol{\Phi} \in \mathbb{R}^{\mathcal{N} \times N}$, as described in Sec. \ref{subsec: Proper Orthogonal Decomposition POD}.}

\newsec{Moreover, for each of the two models  and for each of the four ROMs, we construct HROMs employing four different truncation tolerances for the SVD of projected residuals, namely $\epsilon_{\text{\tiny RES}} = \{ 1e-3, 1e-4, 1e-5, 1e-6 \}$, such that}

\begin{equation}
\norm{\boldsymbol{S}_r - \boldsymbol{G}^T\boldsymbol{G} \boldsymbol{S}_r}_F \leq \epsilon_{\text{\tiny RES}} \norm{\boldsymbol{S}_r}_F \ ,
\end{equation}

\noindent \newsec{where the snapshots matrix of projected residuals $\boldsymbol{S}_r \in \mathbb{R}^{N \cdot m \times N_{el}}$, and the basis for its rowspace $\boldsymbol{G} \in \mathbb{R}^{\beta \times N_{el}}$. Finally, as described in \ref{subsec: Hyper-Reduction}, the ECM algorithm is used to obtain the HROM elements and weights, as}

\begin{equation}
(\mathbb{E}, \boldsymbol{\omega}) \leftarrow \texttt{ECM}(\boldsymbol{G} , \mathbbm{1} ) \ .
\end{equation}

\newsec{Therefore, for each of the two examples, 8 ROMs and 32 HROMs have been constructed, half of them for the model employing the affine geometric mapping and the other half corresponding to the model employing the nonaffine mapping.}

\newsec{To asses these ROMs and HROMs, we employ three strategies:}

\begin{itemize}
    \item \newsec{\textbf{Qualitative assessment through phase space plot}}

    \newsec{Throughout this section we use a phase space plot for the QoI, which as mentioned before is the horizontal velocity at the probe point $p^*$ (as shown in Fig. \ref{fig: meshes}). The plot consists then on a comparison between the horizontal velocity $\boldsymbol{v}^*_y$, against the narrowing width $w_c$. This plot straightforwardly provides insight on the performance of the ROMs for reconstructing the behavior of the FOMs, as well as of the performance of the HROMs for reconstructing the respective ROMs and FOMs.}

    \item \newsec{\textbf{Quantitative assessment through measure of errors}}

    To quantitatively evaluate the results obtained for the different models, we define a percentage error operator as follows:

    \begin{equation}
    \begin{aligned}
        e : \ & ( \boldsymbol{a}, \boldsymbol{b} ) \mapsto \frac{\norm{ \boldsymbol{a} - \boldsymbol{b} }}{\norm{\boldsymbol{a}}} \cdot 100  \ .
    \end{aligned}
    \label{eq:error definition}
    \end{equation}

    Here, $\boldsymbol{a}$ and $\boldsymbol{b}$ can be either matrices or column vectors. In our case, we will use this operator to compare the snapshots matrices of the solution fields for the FOM, ROM, and HROM, denoted respectively as $\boldsymbol{S}_{\text{\tiny FOM}}$, $\boldsymbol{S}_{\text{\tiny ROM}}$, and $\boldsymbol{S}_{\text{\tiny HROM}}$. \own{As well as for the QoI, denoted as ${\boldsymbol{v}^*_y}_{\text{\tiny FOM}}$, ${\boldsymbol{v}^*_y}_{\text{\tiny ROM}}$, and ${\boldsymbol{v}^*_y}_{\text{\tiny HROM}}$} for each of the considered models.

    \item \newsec{\textbf{Speedup}}

    In order to quantify the speedup factor, a comparison was conducted between the ROM and the HROM in relation to the FOM. This analysis considered the time required for the construction and solution of the linear system of equations. The time necessary for visualising the complete solution field was not taken into account.

    To evaluate the speedup factor, we introduced a speedup operator denoted as $s$, as follows:

    \begin{equation}
    \begin{aligned}
    s : \ & (T_{ \text{\tiny a }}, T_{ \text{\tiny FOM }}) \mapsto \frac{ T_{\text{\tiny FOM } }}{ T_{\text{\tiny a } }} \ ,
    \end{aligned}
    \label{eq:speedup definition}
    \end{equation}

    where $T_{ \text{\tiny a }}$ represents the time required for either the ROM or the HROM, while $T_{\text{\tiny FOM }}$ corresponds to the time taken by the FOM. The output of this operator indicates the number of times the reduced order model is faster in comparison to the FOM.

\end{itemize}

\subsection{\newsec{Example 1}}

\begin{figure}[H]
  \centering
  \begin{subfigure}[b]{0.47\textwidth}
    \includegraphics[width=\linewidth]{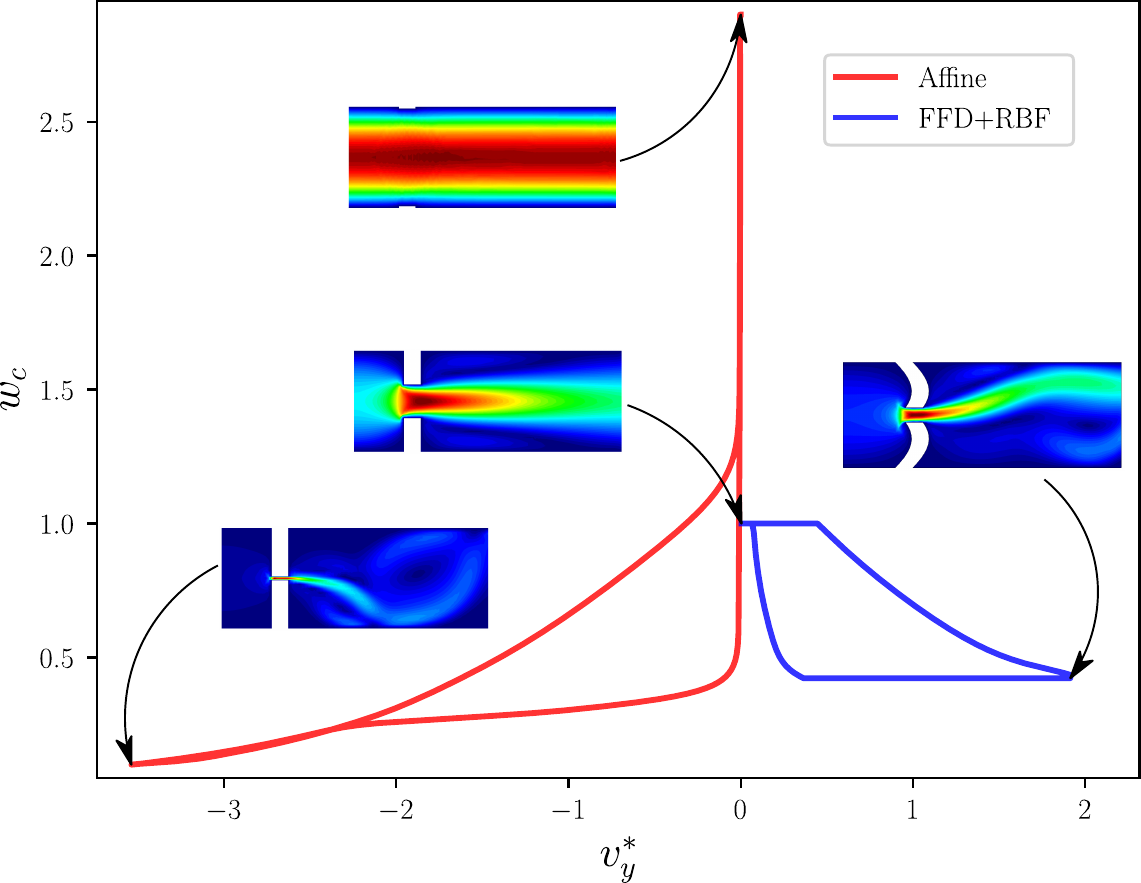}
    \caption{Trajectory 1}
  \end{subfigure}
  \hfill
  \begin{subfigure}[b]{0.47\textwidth}
    \includegraphics[width=\linewidth]{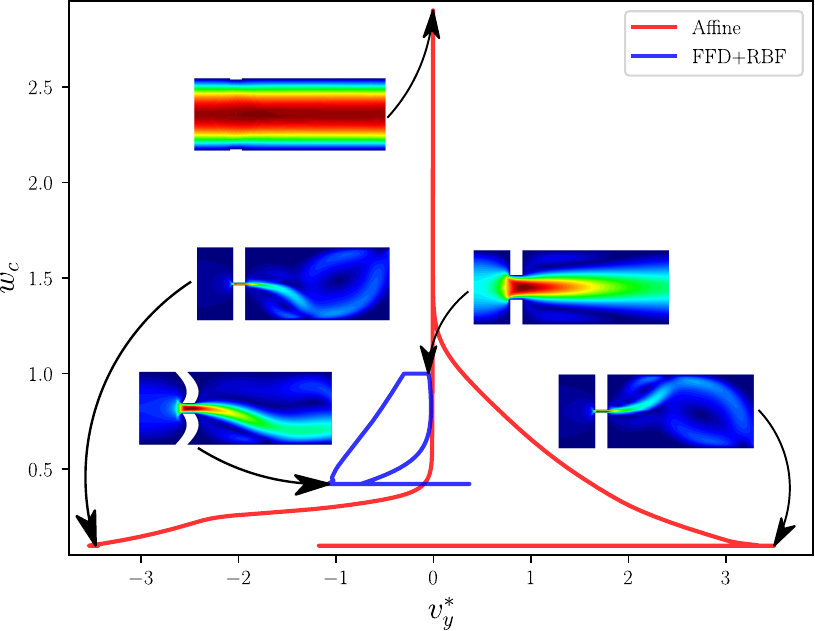}
    \caption{Trajectory 2}
  \end{subfigure}
    \caption{ \revone{QoI phase space plots depicting the impact of the two trajectories in the FOMs for both geometric mappings. a) When subjected to Trajectory 1, the FOM for the affine mapping produces a solution with a jet attaching to the lower wall of the channel, while applying Trajectory 1 to the FOM of the nonaffine mapping results in a jet attaching to the upper wall. b) When subjected to Trajectory 2, the FOM for the affine geometric mapping produces a jet attaching, first to the upper wall and subsequently to the lower wall. On the other hand, when applying Trajectory 2 to the FOM for the nonaffine mapping, the jet produced attaches only to the lower wall, before going back to the symmetric position}}
  \label{fig: Ex1 hysteresis FOM}
\end{figure}

\newsec{For the first example, Trajectory 1 serves as the training trajectory, while Trajectory 2 is taken as the testing trajectory. Fig. \ref{fig: Ex1 hysteresis FOM} displays the QoI phase space plot for both models and for both trajectories. The expectation is that ROMs whose modes favor one branch in the bifurcation encounter difficulties accurately capturing the behavior of the opposite branch. Consequently, this testing approach offers valuable insights into the performance and limitations of the ROMs and HROMs in capturing the full range of behavior of the solution.}

\subsubsection{\newsec{ROM}}

\begin{figure}[H]
  \centering
  \begin{subfigure}[b]{0.53\textwidth}
    \includegraphics[width=\linewidth]{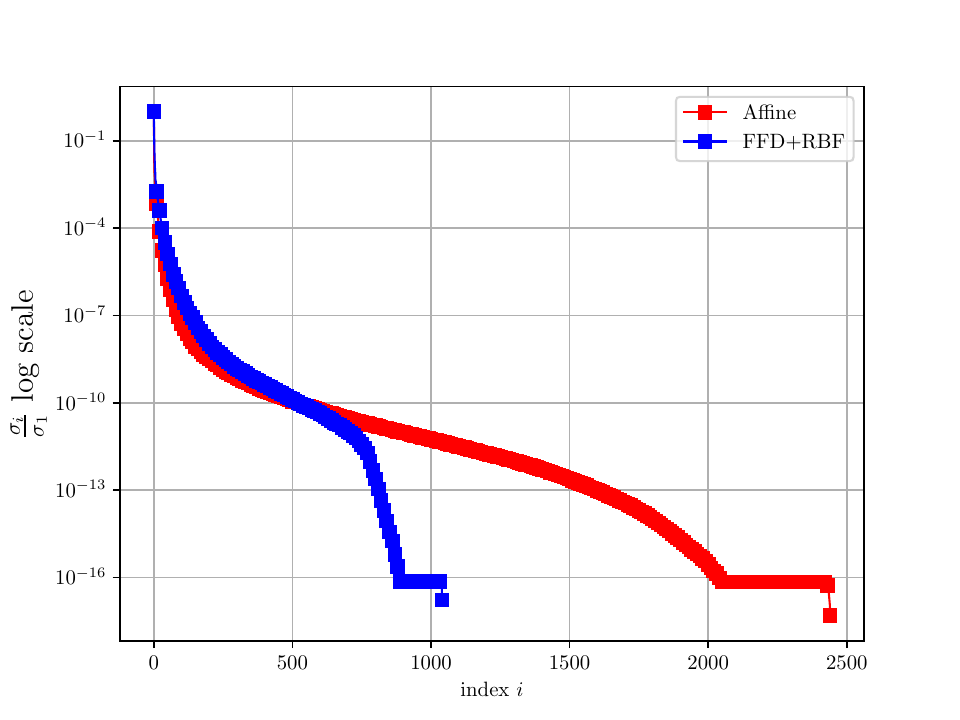}
  \end{subfigure}
  \hfill
  \begin{subfigure}[b]{0.45\textwidth}
    \includegraphics[width=\linewidth]{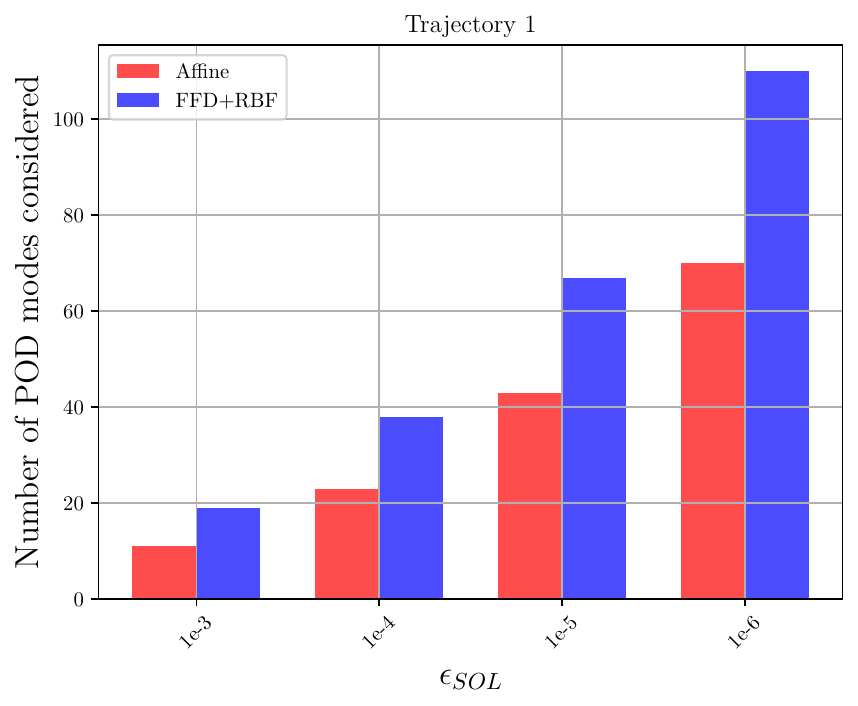}
  \end{subfigure}

    \caption{\revone{Singular values of solution snapshots and the corresponding number of POD modes for both geometric mappings under Trajectory 1}}
  \label{fig: Example 1 singular values S and Number of POD modes}
\end{figure}

\newsec{Fig. \ref{fig: Example 1 singular values S and Number of POD modes} illustrates the singular values of the solution snapshots and the corresponding number of POD modes required by both models. These results are presented for various truncation tolerances $\epsilon_{\text{\tiny SOL}}$. Notably, the nonaffine geometric mapping consistently demands a greater number of modes to achieve the same tolerance compared to the affine mapping. We generated four ROMs for each geometric mapping, accounting for the respective number of POD modes.}

\begin{figure}[H]
  \centering
  \begin{subfigure}[b]{0.45\textwidth}
    \includegraphics[width=\linewidth]{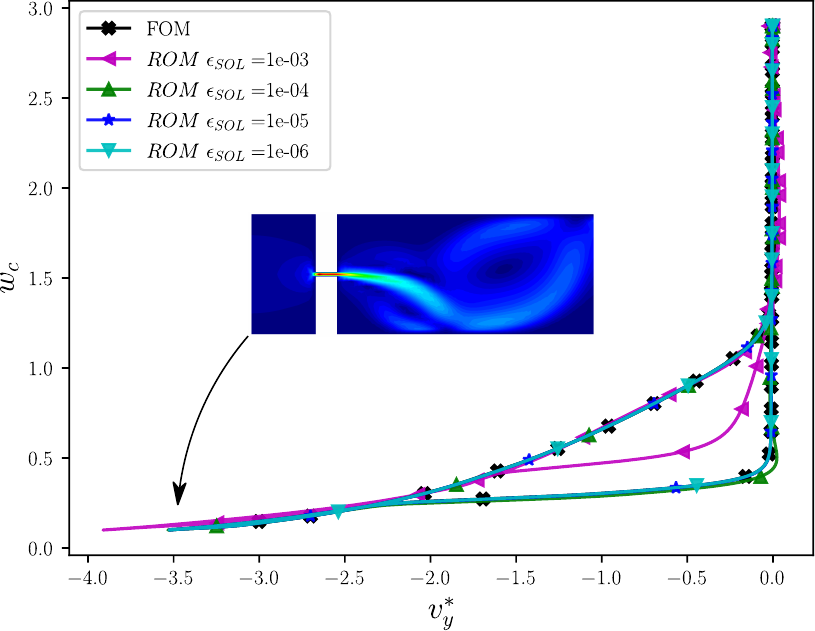}
    \caption{$\boldsymbol{\varphi}_{\text{\tiny AFFINE}}$ }
    %\label{fig:XXX}
  \end{subfigure}
  \hfill
  \begin{subfigure}[b]{0.45\textwidth}
    \includegraphics[width=\linewidth]{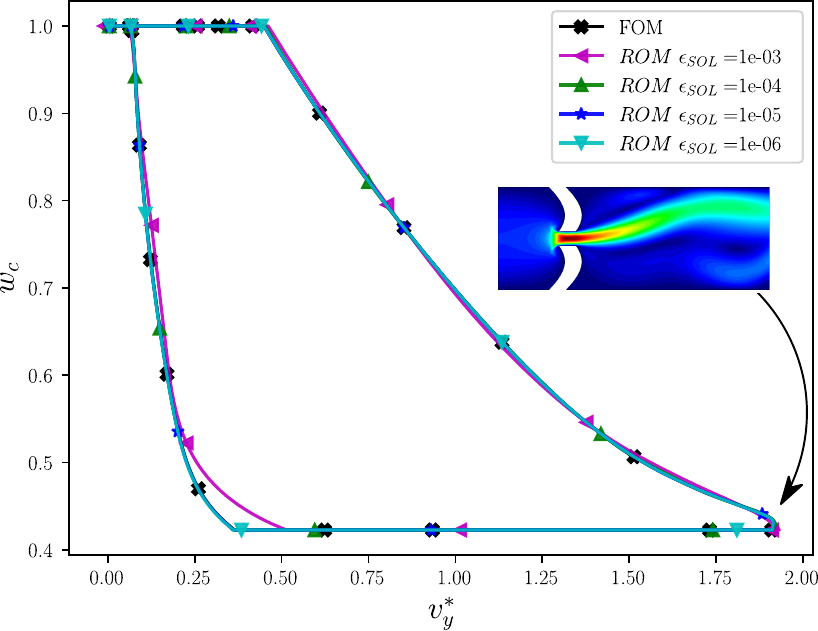}
    \caption{$\boldsymbol{\varphi}_{\text{\tiny FFD+RBF}} $ }
    %\label{fig:XXX}
  \end{subfigure}

  % \caption{FOM vs ROM hysteresis for the training trajectories for both mappings}
\caption{QoI phase space plot for the FOM against various ROMs for the training trajectory (Trajectory 1) for both geometric mappings}
  \label{fig: Example 1 train QoI ROM vs FOM}
\end{figure}

\begin{figure}[H]
  \centering
  \begin{subfigure}[b]{0.45\textwidth}
    \includegraphics[width=\linewidth]{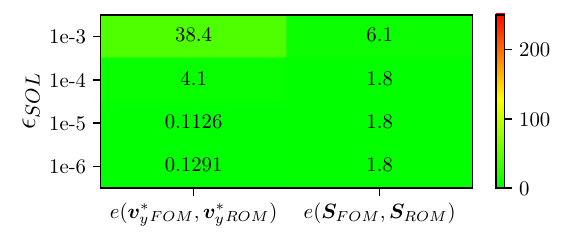}
    \caption{$\boldsymbol{\varphi}_{\text{\tiny AFFINE}}$ }
    %\label{}
  \end{subfigure}
  \hfill
  \begin{subfigure}[b]{0.45\textwidth}
    \includegraphics[width=\linewidth]{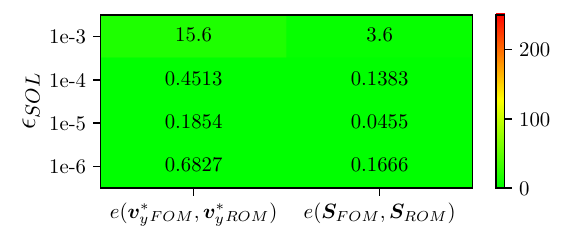}
    \caption{$\boldsymbol{\varphi}_{\text{\tiny FFD+RBF}}$ }
    %\label{}
  \end{subfigure}
  \caption{\bothrev{Percentage error on QoI and solution field of FOM against various ROMs for the training trajectory (Trajectory 1) for both geometric mappings}}
  \label{fig: Example 1 errors train ROM vs FOM}
\end{figure}

\newsec{We now check the performance of the ROMs for reconstructing the training trajectory, both qualitatively and quantitatively. Fig. \ref{fig: Example 1 train QoI ROM vs FOM} shows the QoI phase space plot of the FOMs and ROMs, demonstrating the consistent behaviour of the ROMs: as the truncation tolerance $\epsilon_{\text{\tiny SOL}}$ decreases, the discrepancy between the QoI of the ROMs and that of the FOMs also decreases. Additionally, Fig. \ref{fig: Example 1 errors train ROM vs FOM} illustrates the percentage errors for the reconstruction of the QoI and the complete solution fields, quantifying the observed trend in the phase space plots.}


\begin{figure}[H]
  \centering
  \begin{subfigure}[b]{0.45\textwidth}
    \includegraphics[width=\linewidth]{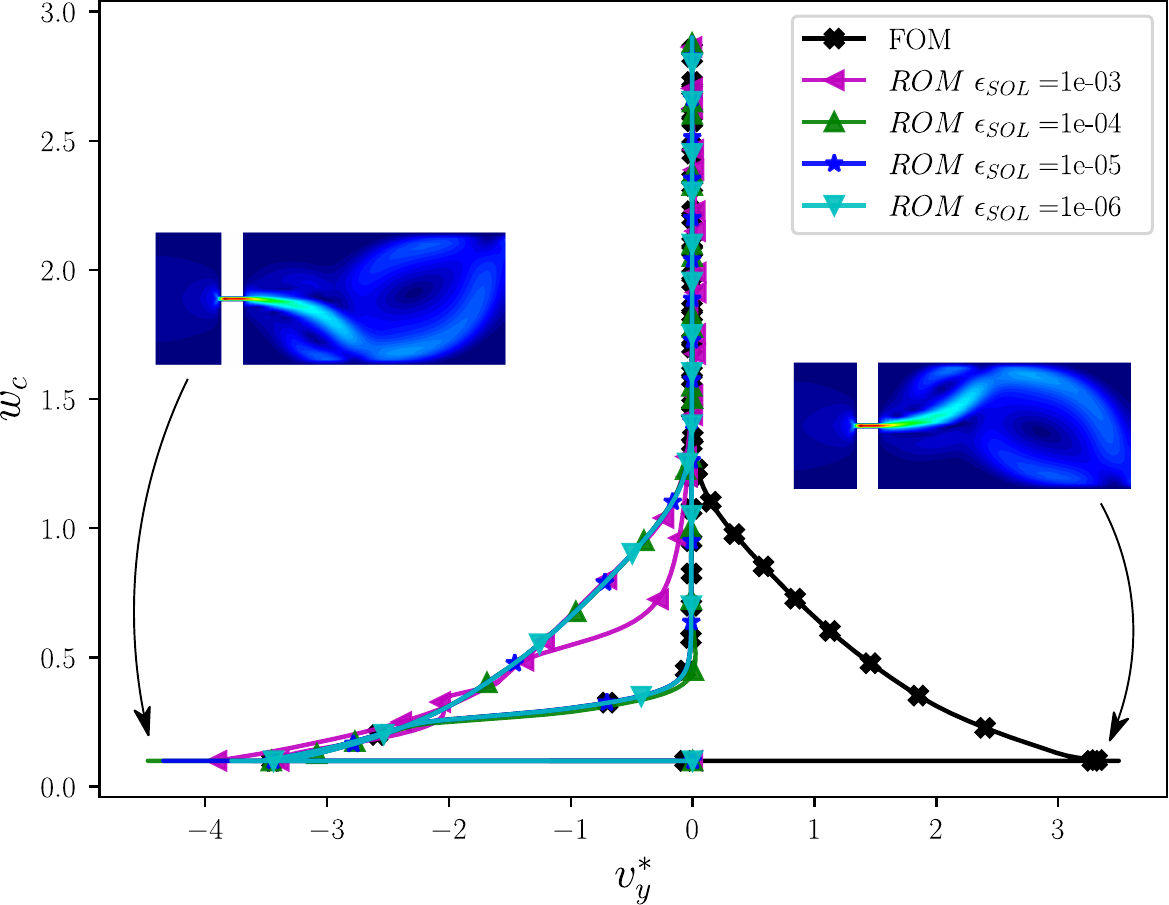}
    \caption{$\boldsymbol{\varphi}_{\text{\tiny AFFINE}}$ }
    %\label{fig: QoI ROM test a}
  \end{subfigure}
  \hfill
  \begin{subfigure}[b]{0.45\textwidth}
    \includegraphics[width=\linewidth]{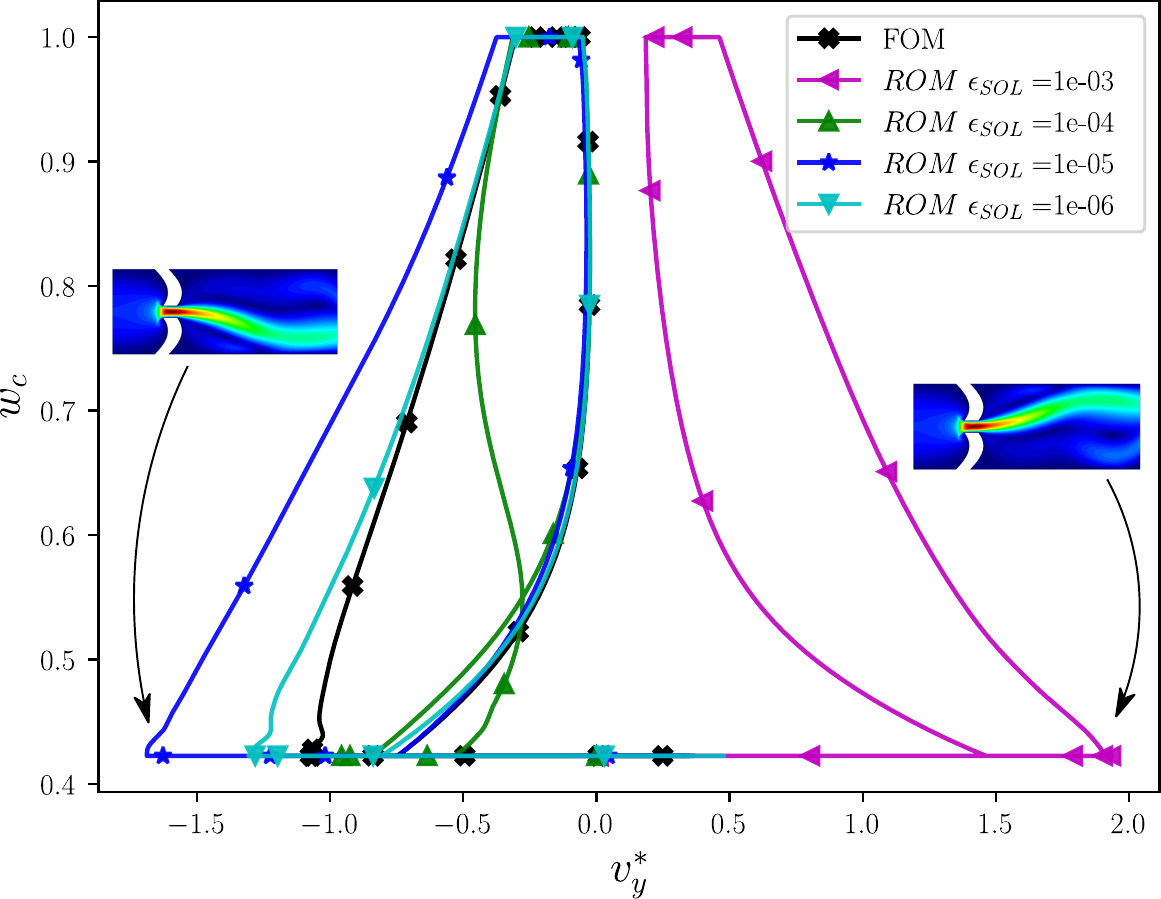}
    \caption{$\boldsymbol{\varphi}_{\text{\tiny FFD+RBF}}$ }
    %\label{fig: QoI ROM test b}
  \end{subfigure}
  \caption{QoI phase space plot for the FOM against various ROMs for the testing trajectory (Trajectory 2) for both geometric mappings}
  \label{fig: Example 1 test QoI ROM vs FOM}
\end{figure}

\newsec{We now turn our focus to the performance of the ROMs for reconstructing the testing trajectory. Fig. \ref{fig: Example 1 test QoI ROM vs FOM} presents the QoI phase space plots for FOMs and ROMs. We note that none of the ROMs for the affine mapping accurately selects the ``correct" branch of the bifurcation, while the ROMs for the nonaffine mapping progressively converge towards the FOM solution.}

\begin{figure}[H]
  \centering
  \begin{subfigure}[b]{0.45\textwidth}
    \includegraphics[width=\linewidth]{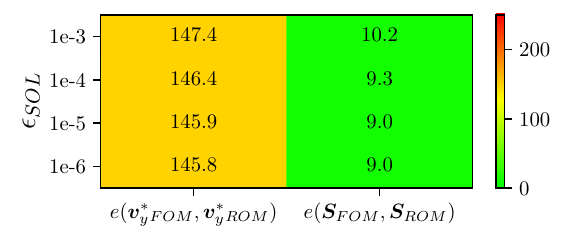}
    \caption{$\boldsymbol{\varphi}_{\text{\tiny AFFINE}}$ }
    %%\label{}
  \end{subfigure}
  \hfill
  \begin{subfigure}[b]{0.45\textwidth}
    \includegraphics[width=\linewidth]{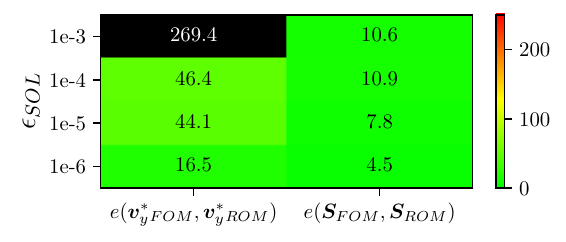}
    \caption{$\boldsymbol{\varphi}_{\text{\tiny FFD+RBF}}$ }
    %%\label{}
  \end{subfigure}

  \caption{\bothrev{Percentage error on QoI and solution field of FOM against various ROMs for the testing trajectory (Trajectory 2) for both geometric mappings}}
  \label{fig: Example 1 errors test ROM vs FOM}
\end{figure}

\newsec{Fig. \ref{fig: Example 1 errors test ROM vs FOM} shows the percentage errors incurred by both models during the reconstruction of the QoI and the complete solution fields of the respective FOM. We note that the errors in the QoI are significantly larger than those in the complete solution fields. For the affine mapping, the errors in the QoI of the ROMs reach around 145\%, whereas the errors in the complete solution fields are approximately 9\%. In the case of the nonaffine mapping, the ROMs result in smaller errors in the QoI, except for the model with a truncation tolerance $\epsilon_{\text{\tiny SOL}} = 1e-3$, which exhibits the opposite branch in the bifurcation compared to the FOM.}

\subsubsection{\newsec{HROM}}

\begin{figure}[H]
  \centering
  \begin{subfigure}[b]{0.55\textwidth}
    \includegraphics[width=\linewidth]{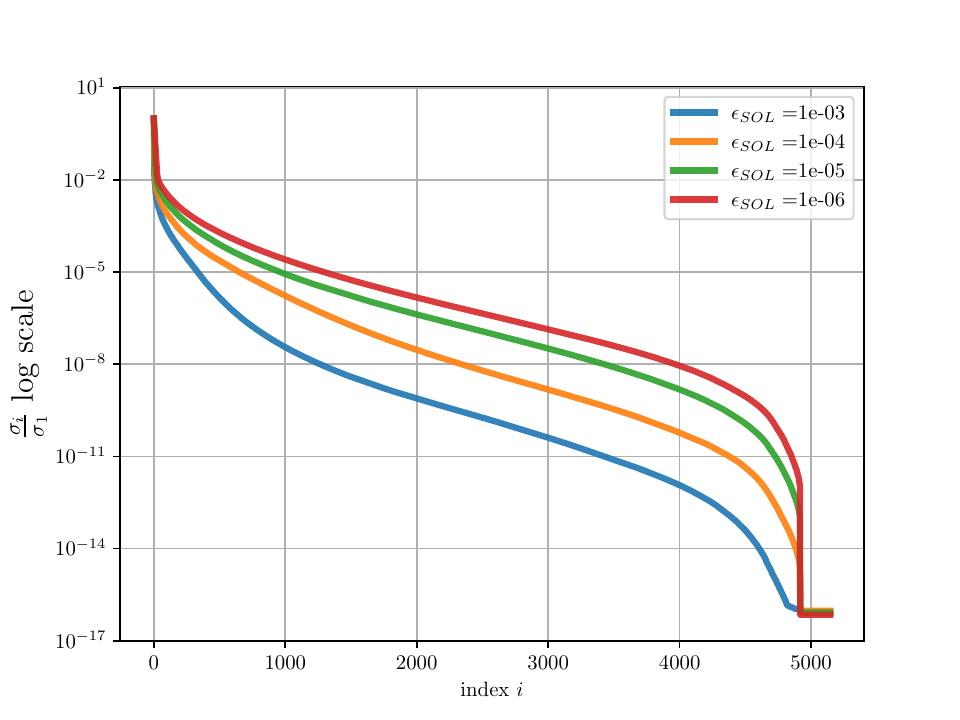}
    \caption{Singular values for matrices $\boldsymbol{S_r}(\epsilon_{\text{\tiny SOL}})$}
    %\label{}
  \end{subfigure}
  \hfill
  \begin{subfigure}[b]{0.40\textwidth}
    \includegraphics[width=\linewidth]{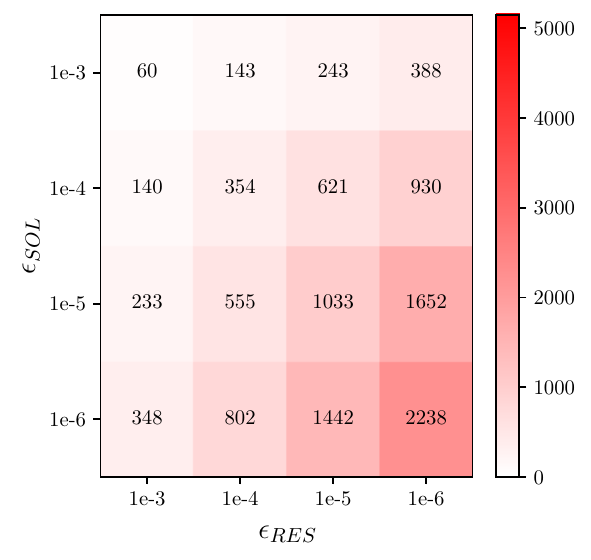}
    \caption{Number of selected elements for all combinations of $\epsilon_{\text{\tiny SOL}}$ and $\epsilon_{\text{\tiny RES}}$}
    %\label{}
  \end{subfigure}

  \caption{\revone{Singular values decay profile of matrices of projected residuals, and number of selected elements by ECM algorithm for geometric mapping  $\boldsymbol{\varphi}_{\text{\tiny AFFINE}}$}}
  \label{fig: Example 1 Sr affine num elements}
\end{figure}

\begin{figure}[H]
  \centering
  \begin{subfigure}[b]{0.55\textwidth}
    \includegraphics[width=\linewidth]{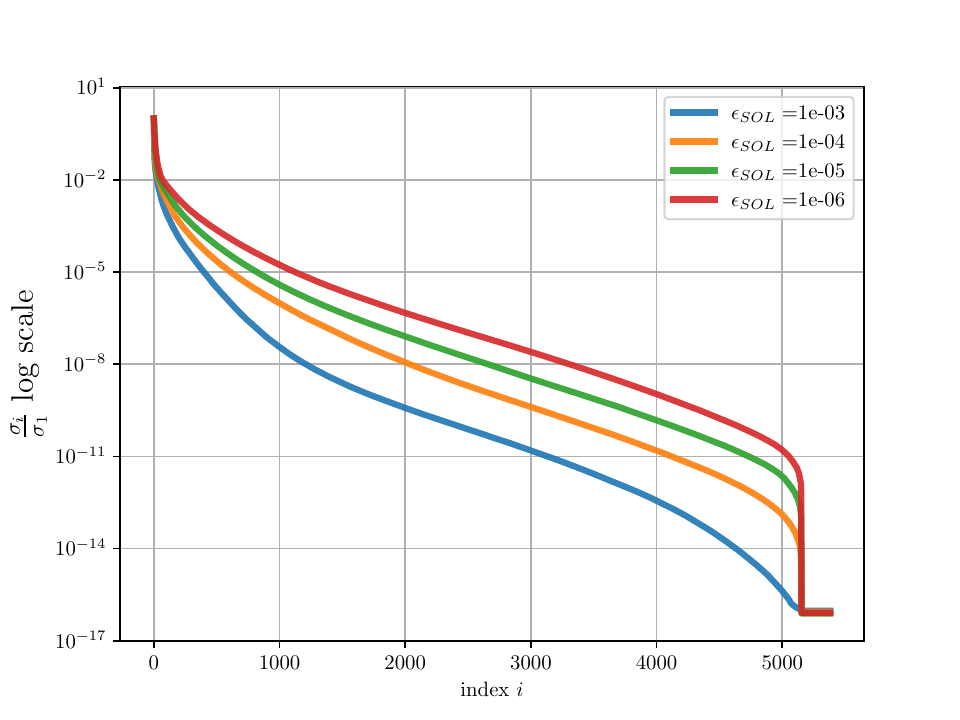}
    \caption{Singular values for matrices $\boldsymbol{S_r}(\epsilon_{\text{\tiny SOL}})$}
    %%\label{}
  \end{subfigure}
  \hfill
  \begin{subfigure}[b]{0.40\textwidth}
    \includegraphics[width=\linewidth]{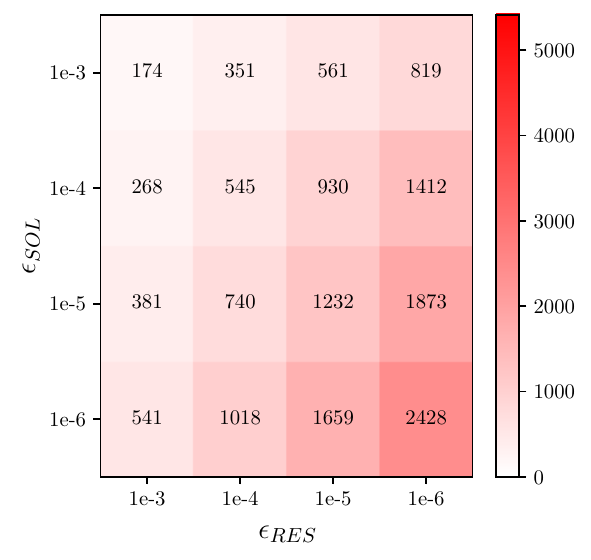}
    \caption{Number of selected elements for all combinations of $\epsilon_{\text{\tiny SOL}}$ and $\epsilon_{\text{\tiny RES}}$}
    %%\label{}
  \end{subfigure}

  \caption{\revone{Singular values decay profile of matrices of projected residuals, and number of selected elements by ECM algorithm for geometric mapping  $\boldsymbol{\varphi}_{\text{\tiny FFD+RBF}}$}
  \label{fig: Example 1 Sr nonlinear num elements}}
\end{figure}

\newsec{Since the ECM algorithm computes rules featuring one element more than the number of components in the basis for the rowspace of the matrix of projected residuals (See Sec. \ref{subsec: Hyper-Reduction}), the decay profile of its singular values provides an indication of the potential to hyper-reduce the case at hand. Fig. \ref{fig: Example 1 Sr affine num elements} shows the singular values decay profile of $\boldsymbol{S_r}$ and the number of selected elements for of the HROMs created for the affine geometric mapping. Moreover, Fig. \ref{fig: Example 1 Sr nonlinear num elements} shows it for the nonaffine mapping. In both cases, we can observe that the larger the value of the truncation tolerance $\epsilon_{\text{\tiny SOL}}$, the more pronounced the decay profile. This affects the required number of components of the SVD of $\boldsymbol{S_r}$ to comply with the four truncation tolerances $\epsilon_{\text{\tiny RES}}$, and therefore directly dictates the number of selected elements by the ECM algorithm.}

\begin{figure}[H]
\begin{subfigure}{.5\textwidth}
  \centering
  % include first image
  \includegraphics[width=.8\linewidth]{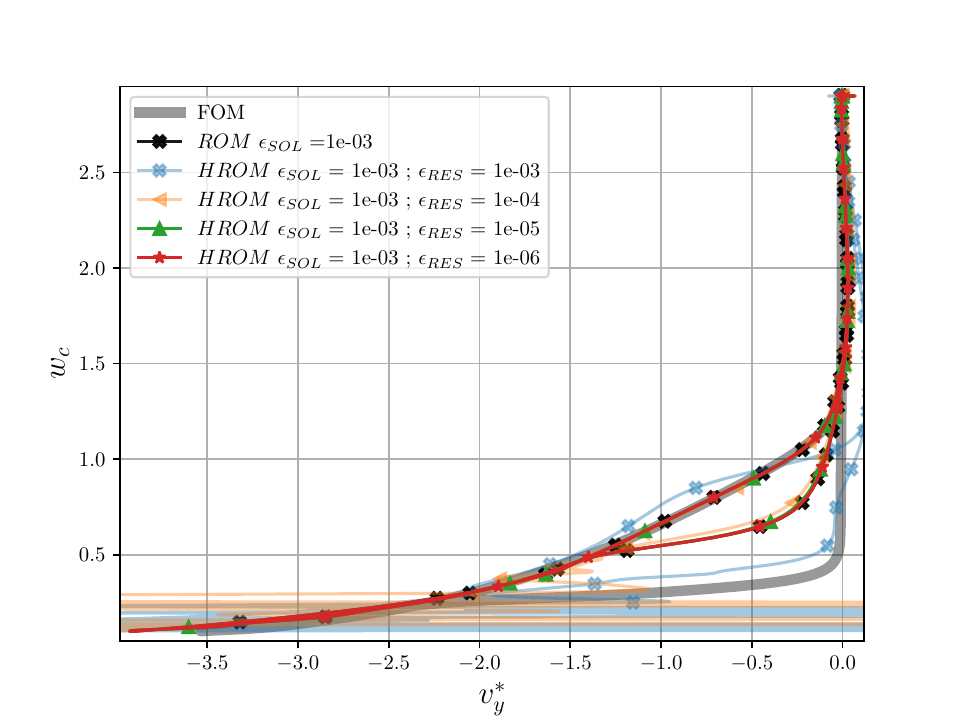}
  \caption{ROM $1e-3$ vs HROM}
  %\label{fig:hysteresis train affine_a}
\end{subfigure}
\begin{subfigure}{.5\textwidth}
  \centering
  % include second image
  \includegraphics[width=.8\linewidth]{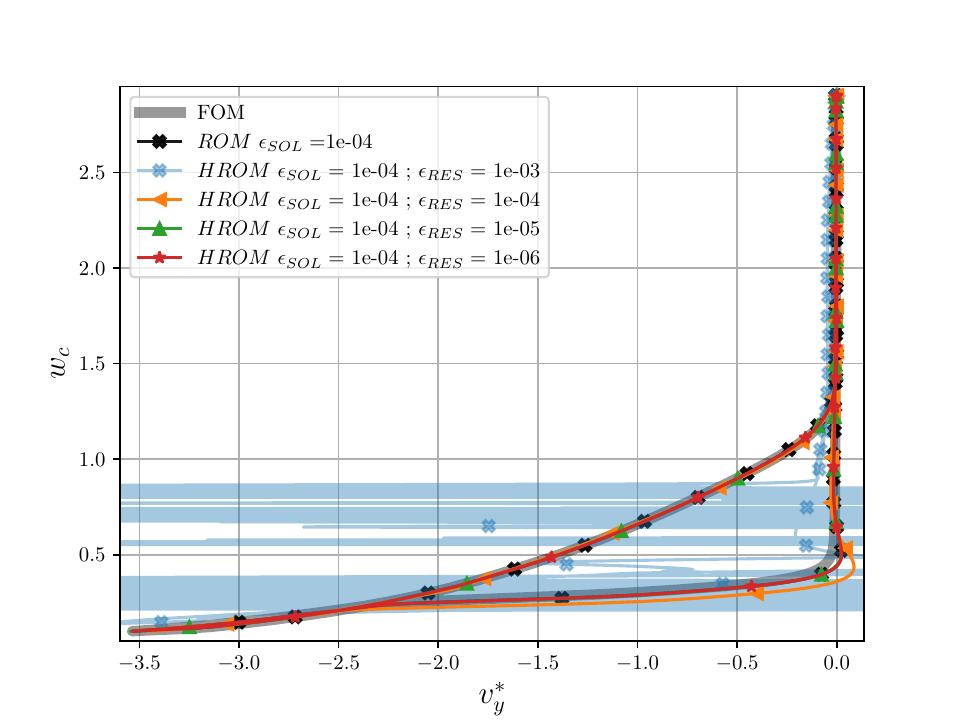}
  \caption{ROM $1e-4$ vs HROM}
  %\label{fig:hysteresis train affine_b}
\end{subfigure}
\begin{subfigure}{.5\textwidth}
  \centering
  % include third image
  \includegraphics[width=.8\linewidth]{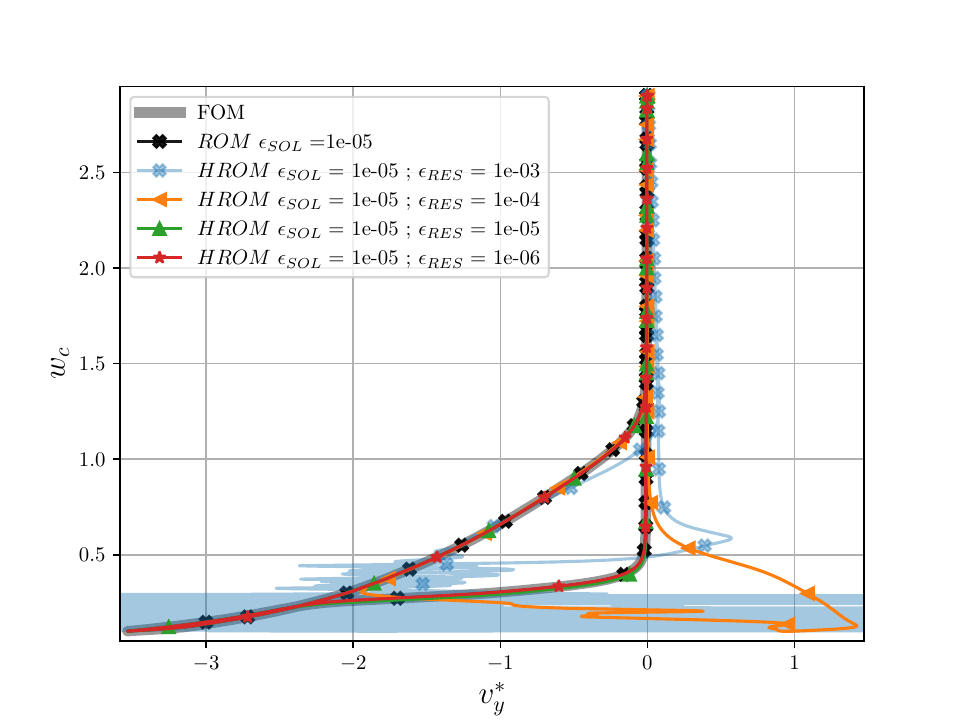}
  \caption{ROM $1e-5$ vs HROM}
  %\label{fig:hysteresis train affine_c}
\end{subfigure}
\begin{subfigure}{.5\textwidth}
  \centering
  % include fourth image
  \includegraphics[width=.8\linewidth]{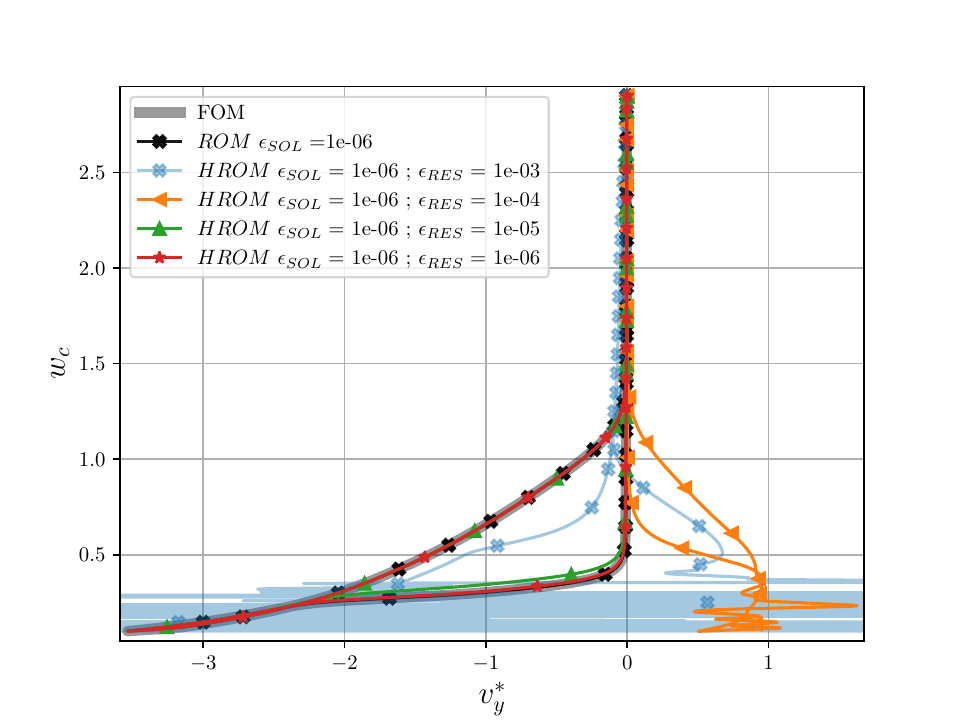}
  \caption{ROM $1e-6$ vs HROM}
  %\label{fig:hysteresis train affine_d}
\end{subfigure}
\caption{\bothrev{QoI phase space plot for the ROMs against various HROMs for the training trajectory (Trajectory 1) for mapping $\varphi_{\text{\tiny AFFINE}}$}}
\label{fig: Example 1 train QoI HROM affine}
\end{figure}

\newsec{For each geometric mapping, we have constructed 16 HROMs containing the number of elements specified on letter (b) in the above figures. \append{Appendix \ref{sec: appendix} shows the location of the selected elements in the meshes employed.} We now proceed to check their performance to reconstruct the training trajectories. Fig. \ref{fig: Example 1 train QoI HROM affine} shows the QoI phase space plot for each of the ROMs against their corresponding HROMs for the affine mapping. We recall that the error committed by the HROM is bounded by the corresponding ROM used for its construction, however, in these plots we have also included the FOM with a ghosted line for reference. We immediately observe that some of the models are numerically unstable. In the plots, exploding solutions were also ghosted to allow visualisation of the rest. In particular, all of the HROMs constructed with a truncation tolerance of $\epsilon_{\text{\tiny RES}} = 1e-3$, as well as the HROM constructed with the combination $\epsilon_{\text{\tiny SOL}} = 1e-3, \epsilon_{\text{\tiny RES}} = 1e-4$ were not numerically stable. The rest of the HROMs behave as expected, approaching the corresponding ROM solution, as the tolerance $\epsilon_{\text{\tiny RES}}$ gets smaller.}

\begin{figure}[H]
  \centering
  \begin{subfigure}[b]{0.245\textwidth}
    \includegraphics[width=\linewidth]{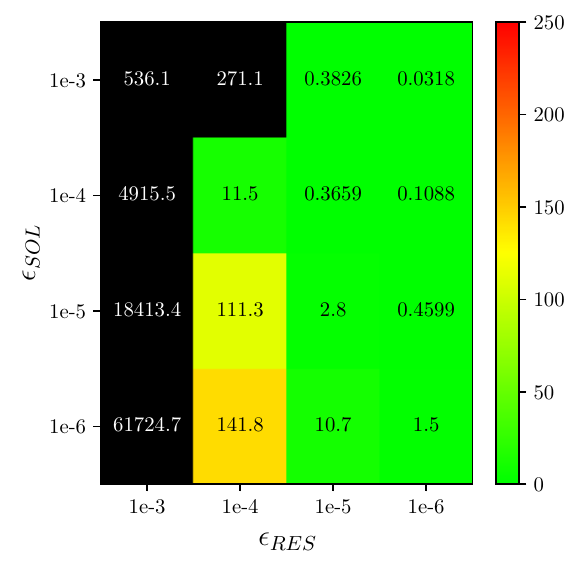}
    \caption{$e({\boldsymbol{v}^*_y}_{\text{\tiny ROM}}, {\boldsymbol{v}^*_y}_{\text{\tiny HROM}} )$}
    %%\label{}
  \end{subfigure}
  \hfill
  \begin{subfigure}[b]{0.245\textwidth}
    \includegraphics[width=\linewidth]{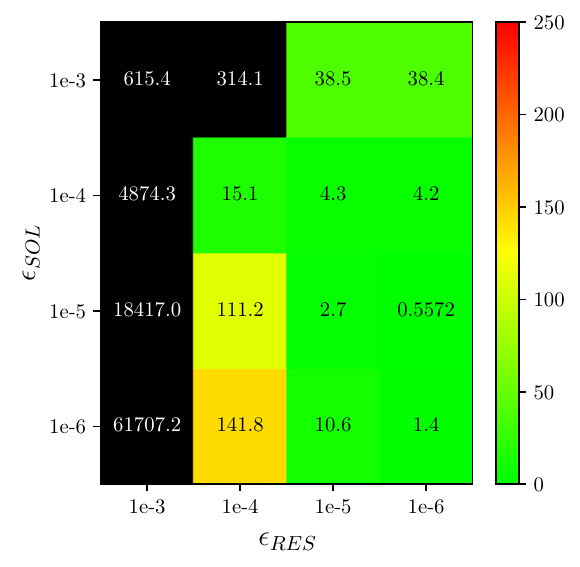}
    \caption{ $e({\boldsymbol{v}^*_y}_{\text{\tiny FOM}}, {\boldsymbol{v}^*_y}_{\text{\tiny HROM}} )$}
    %%\label{}
  \end{subfigure}
  \hfill
  \begin{subfigure}[b]{0.245\textwidth}
    \includegraphics[width=\linewidth]{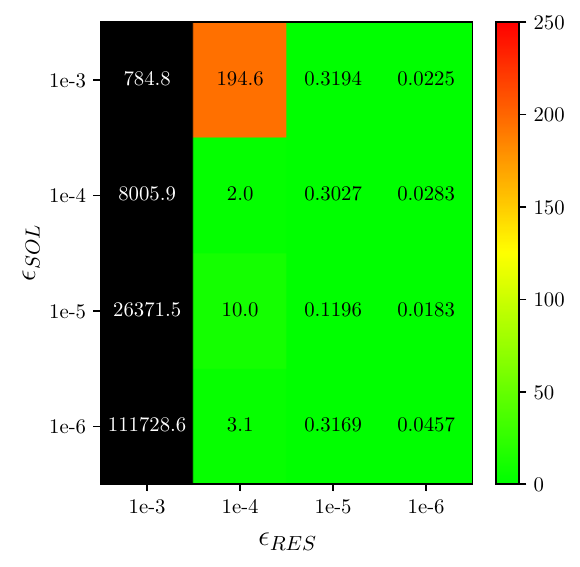}
    \caption{ $e(\boldsymbol{S}_{\text{\tiny ROM}},  \boldsymbol{S}_{\text{\tiny HROM}})$}
    %%\label{}
  \end{subfigure}
  \hfill
  \begin{subfigure}[b]{0.245\textwidth}
    \includegraphics[width=\linewidth]{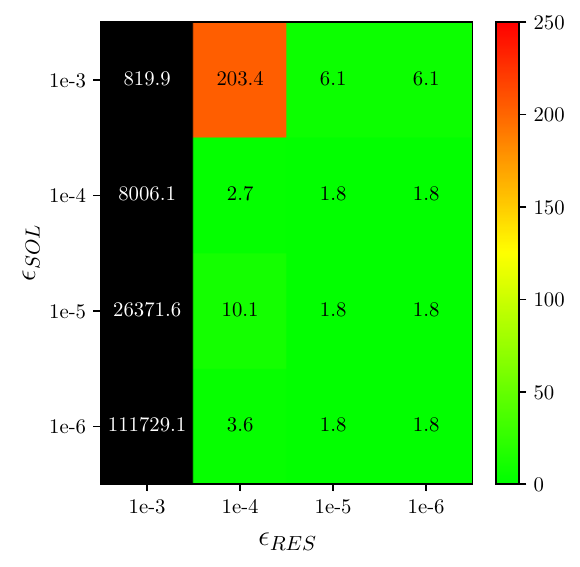}
    \caption{  $e(\boldsymbol{S}_{\text{\tiny FOM}},  \boldsymbol{S}_{\text{\tiny HROM}})$ }
    %%\label{}
  \end{subfigure}
  \caption{ \bothrev{Percentage error on QoI and solution field of ROM and FOM against various HROMs for the training trajectory (Trajectory 1) for mapping $\varphi_{\text{\tiny AFFINE}}$}}
  \label{fig: Example 1 errors train HROM affine}
\end{figure}

\newsec{Fig. \ref{fig: Example 1 errors train HROM affine} further confirms the trends observed in the phase space plots, by displaying the percentage errors incurred by the HROMs in the complete solution field and the QoI, when compared to the FOM and ROMs.}

\begin{figure}[H]
  \begin{subfigure}{.5\textwidth}
    \centering
    % include first image
    \includegraphics[width=.8\linewidth]{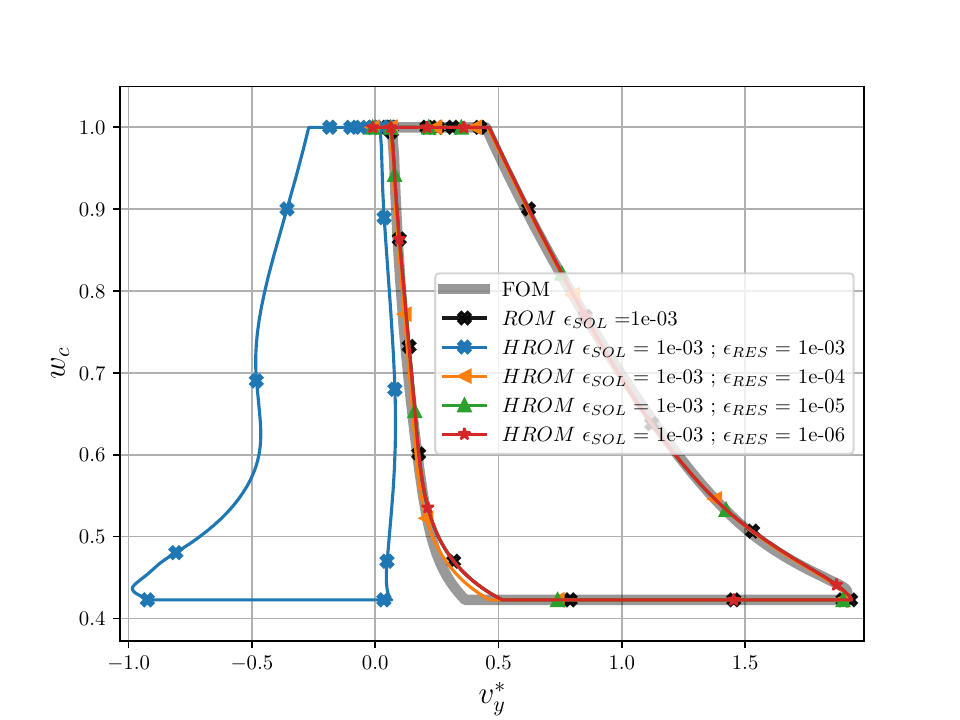}
    \caption{ROM $1e-3$ vs HROM}
  \end{subfigure}
  \begin{subfigure}{.5\textwidth}
    \centering
    % include second image
    \includegraphics[width=.8\linewidth]{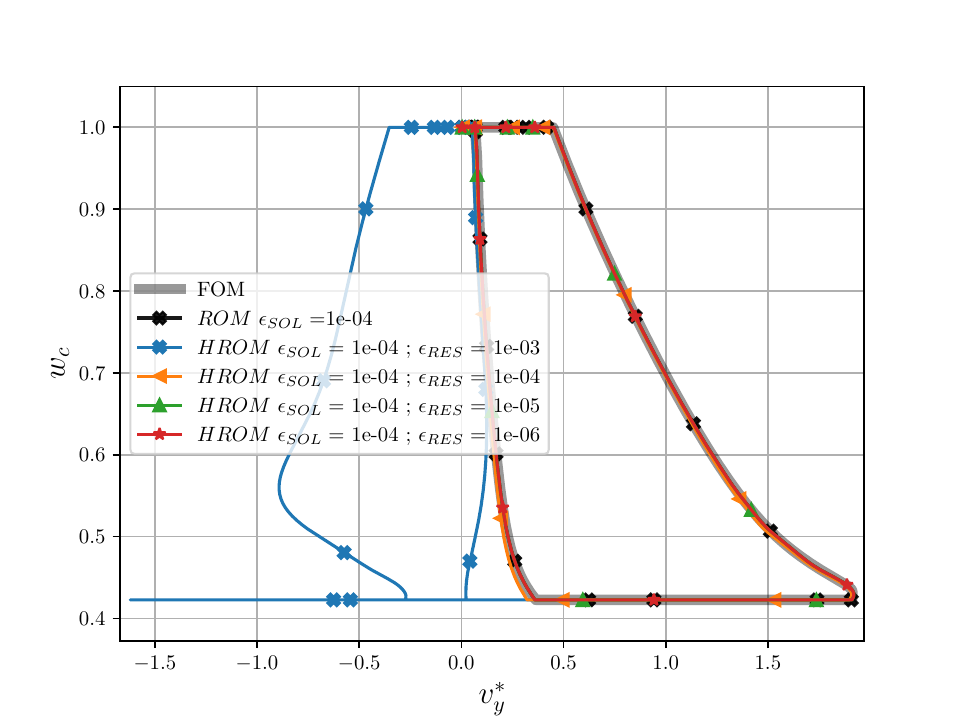}
    \caption{ROM $1e-4$ vs HROM}
  \end{subfigure}
  \begin{subfigure}{.5\textwidth}
    \centering
    % include third image
    \includegraphics[width=.8\linewidth]{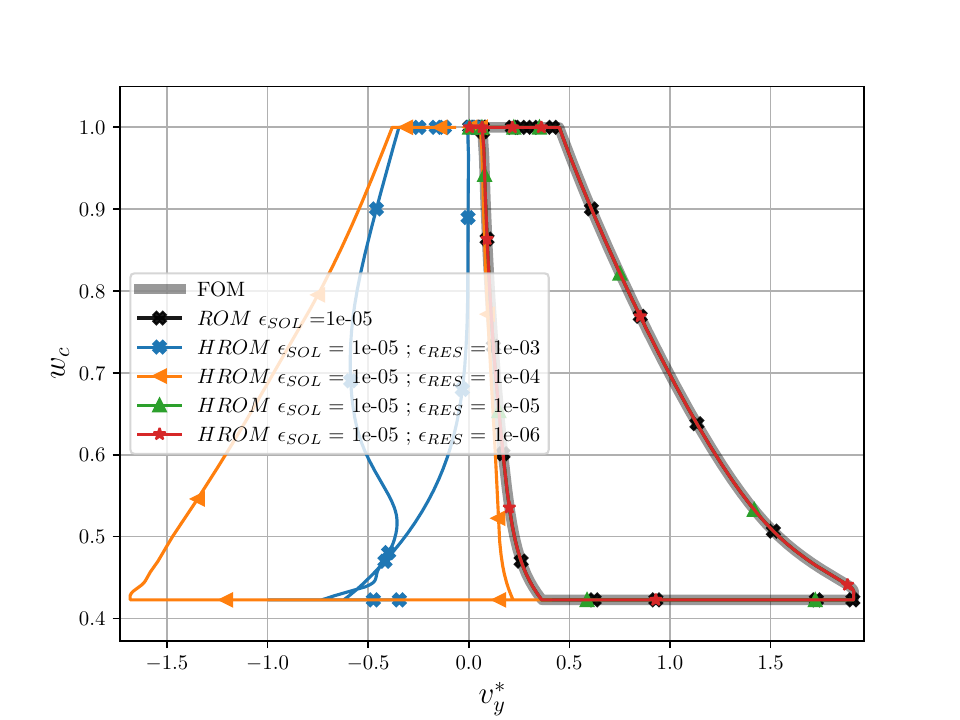}
    \caption{ROM $1e-5$ vs HROM}
  \end{subfigure}
  \begin{subfigure}{.5\textwidth}
    \centering
    % include fourth image
    \includegraphics[width=.8\linewidth]{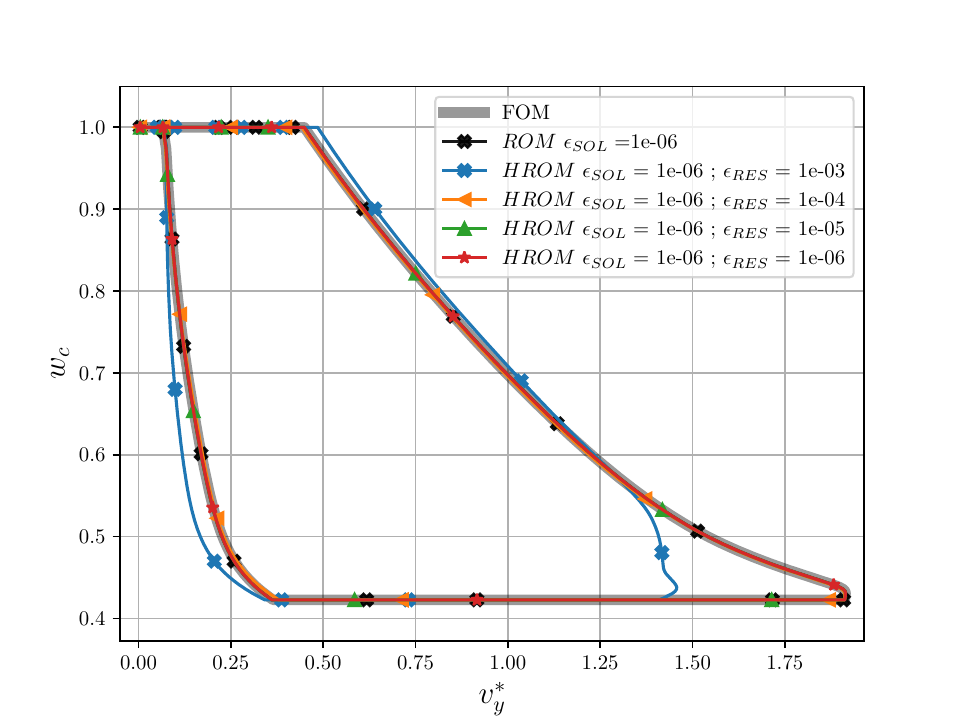}
    \caption{ROM $1e-6$ vs HROM}
  \end{subfigure}
  \caption{  \bothrev{QoI phase space plot for the ROMs against various HROMs for the training trajectory (Trajectory 1) for mapping $\varphi_{\text{\tiny FFD+RBF}}$ }}
  \label{fig: Example 1 train QoI HROM nonlinear}
  \end{figure}

\newsec{Turning our attention to the HROMs associated with the nonaffine geometric mapping, we evaluate their performance for reconstructing the training trajectory by first observing  Fig. \ref{fig: Example 1 train QoI HROM nonlinear}. This figure shows the QoI phase space plots of the ROMs against their respective HROMs. We observe that all of the models are numerically stable, but some of the HROMs choose the opposite stable branch in the bifurcation with respect to the ROM.}

\begin{figure}[H]
  \centering
  \begin{subfigure}[b]{0.245\textwidth}
    \includegraphics[width=\linewidth]{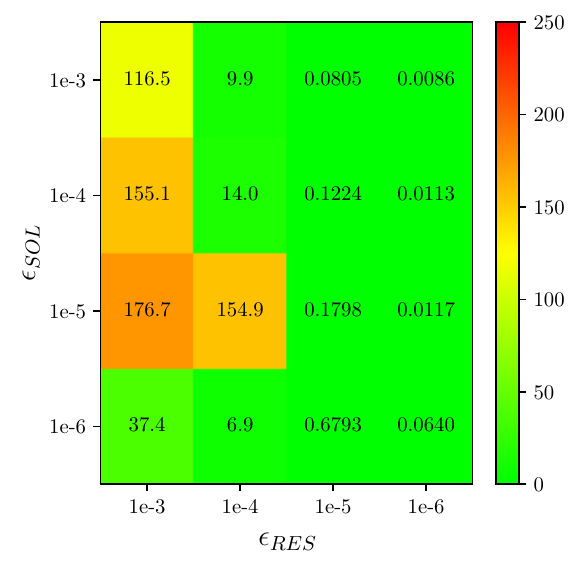}
    \caption{$e({\boldsymbol{v}^*_y}_{\text{\tiny ROM}}, {\boldsymbol{v}^*_y}_{\text{\tiny HROM}} )$}
    %%\label{}
  \end{subfigure}
  \hfill
  \begin{subfigure}[b]{0.245\textwidth}
    \includegraphics[width=\linewidth]{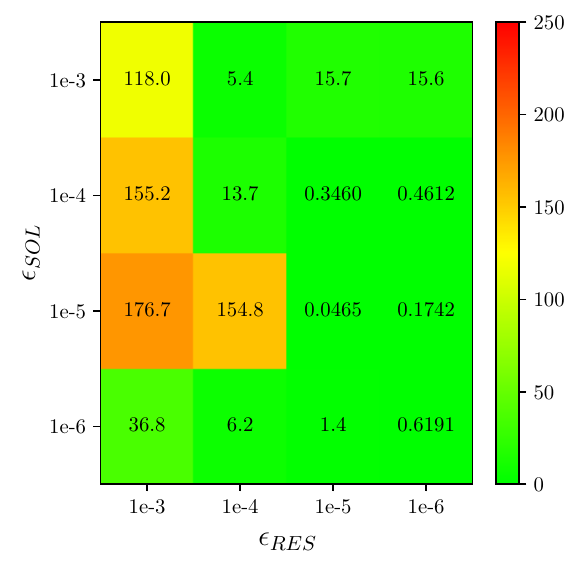}
    \caption{ $e({\boldsymbol{v}^*_y}_{\text{\tiny FOM}}, {\boldsymbol{v}^*_y}_{\text{\tiny HROM}} )$}
    %%\label{}
  \end{subfigure}
  \hfill
  \begin{subfigure}[b]{0.245\textwidth}
    \includegraphics[width=\linewidth]{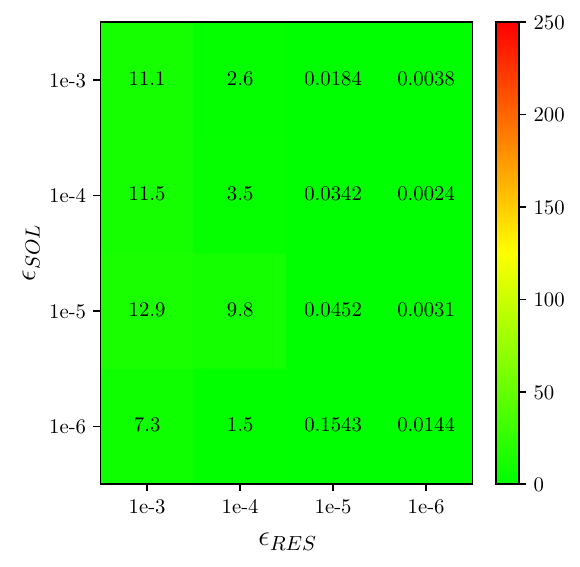}
    \caption{ $e(\boldsymbol{S}_{\text{\tiny ROM}},  \boldsymbol{S}_{\text{\tiny HROM}})$}
    %%\label{}
  \end{subfigure}
  \hfill
  \begin{subfigure}[b]{0.245\textwidth}
    \includegraphics[width=\linewidth]{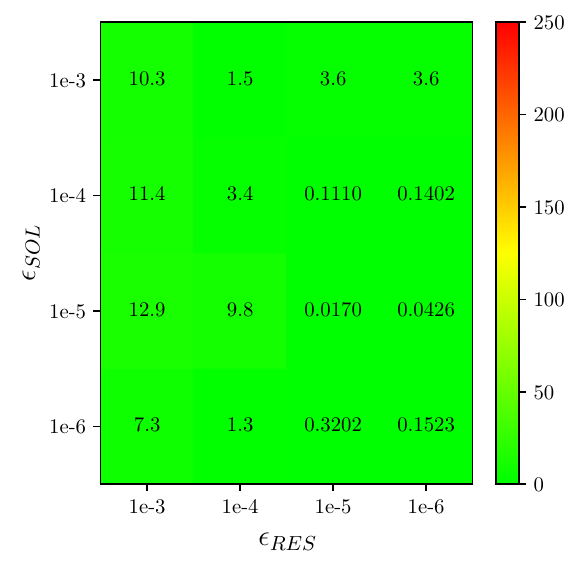}
    \caption{  $e(\boldsymbol{S}_{\text{\tiny FOM}},  \boldsymbol{S}_{\text{\tiny HROM}})$ }
    %%\label{}
  \end{subfigure}
  \caption{ \bothrev{ Percentage error on QoI and solution field of ROM and FOM against various HROMs for the training trajectory (Trajectory 1) for mapping $\varphi_{\text{\tiny FFD+RBF}}$}}
  \label{fig: Example 1 errors train HROM nonlinear}
\end{figure}

\newsec{Fig. \ref{fig: Example 1 errors train HROM nonlinear} quantifies the trends observed in the phase space plots, by displaying the percentage errors incurred by the HROMs in the complete solution field and the QoI, when compared to the FOM and ROMs.}

\newsec{Having assessed the performance of the HROMs for reproducing the training trajectory, we now turn our attention to their capability to handle the testing trajectory.}

\begin{figure}[H]
\begin{subfigure}{.5\textwidth}
  \centering
  % include first image
  \includegraphics[width=.8\linewidth]{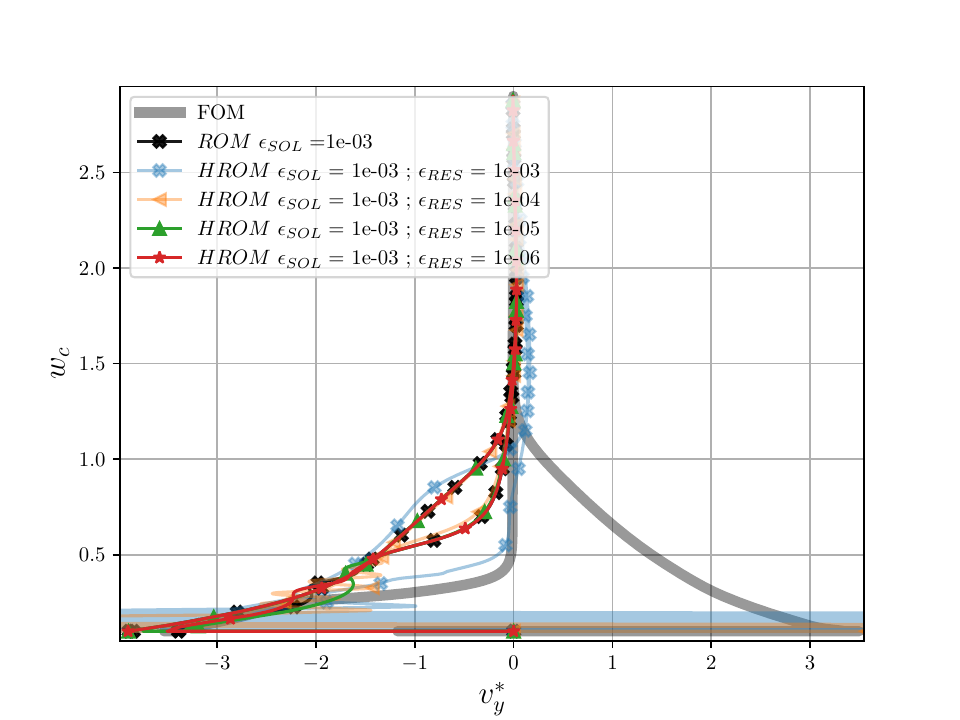}
  \caption{ROM $1e-3$ vs HROM}
\end{subfigure}
\begin{subfigure}{.5\textwidth}
  \centering
  % include second image
  \includegraphics[width=.8\linewidth]{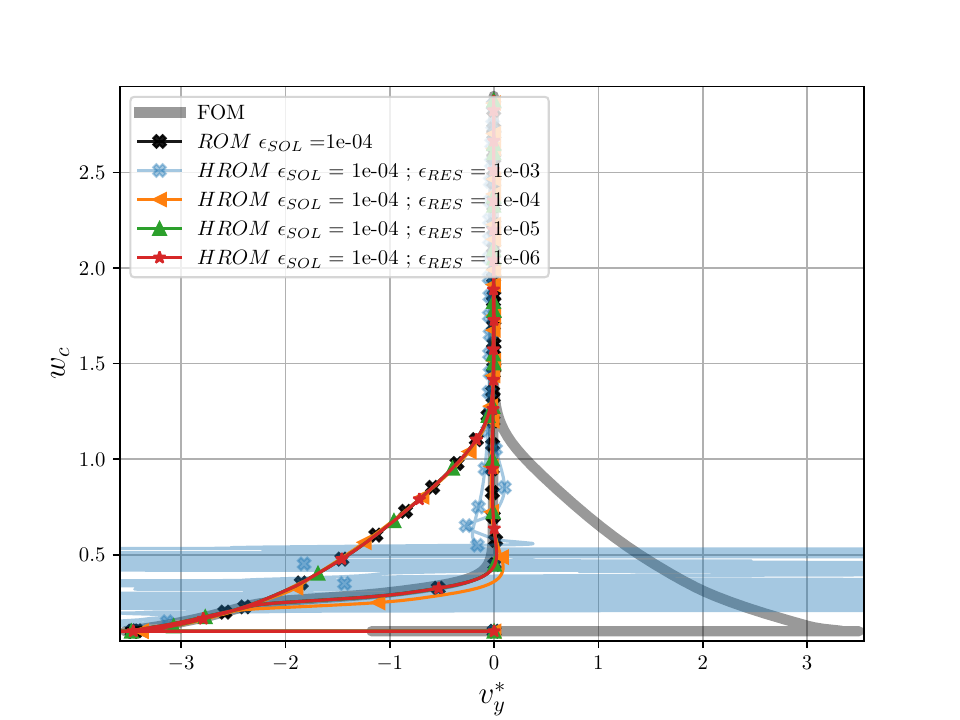}
  \caption{ROM $1e-4$ vs HROM}
\end{subfigure}
\begin{subfigure}{.5\textwidth}
  \centering
  % include third image
  \includegraphics[width=.8\linewidth]{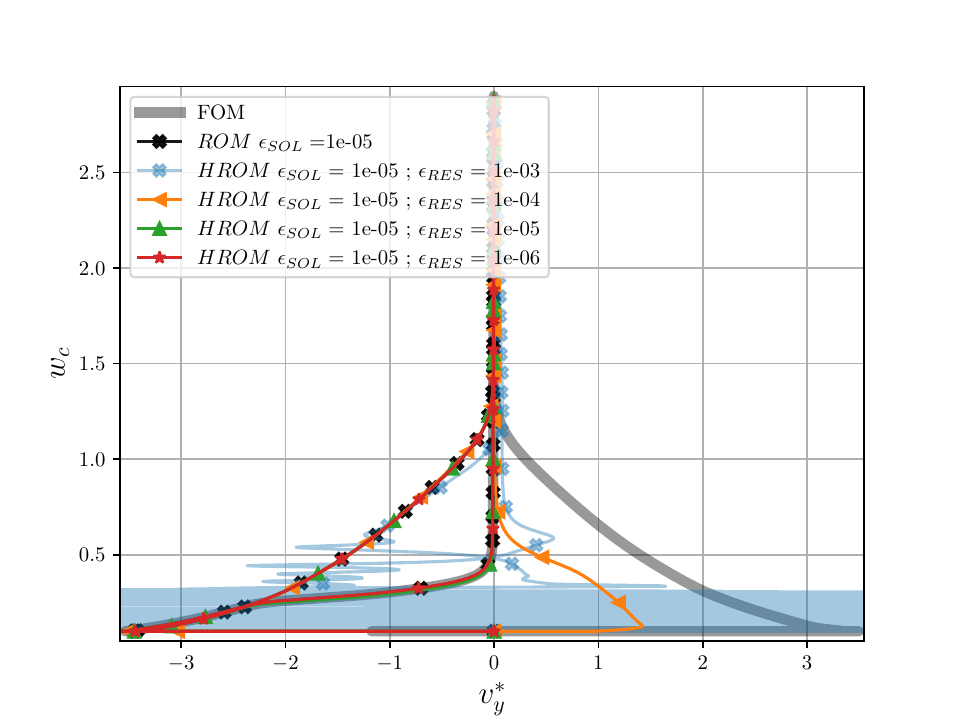}
  \caption{ROM $1e-5$ vs HROM}
\end{subfigure}
\begin{subfigure}{.5\textwidth}
  \centering
  % include fourth image
  \includegraphics[width=.8\linewidth]{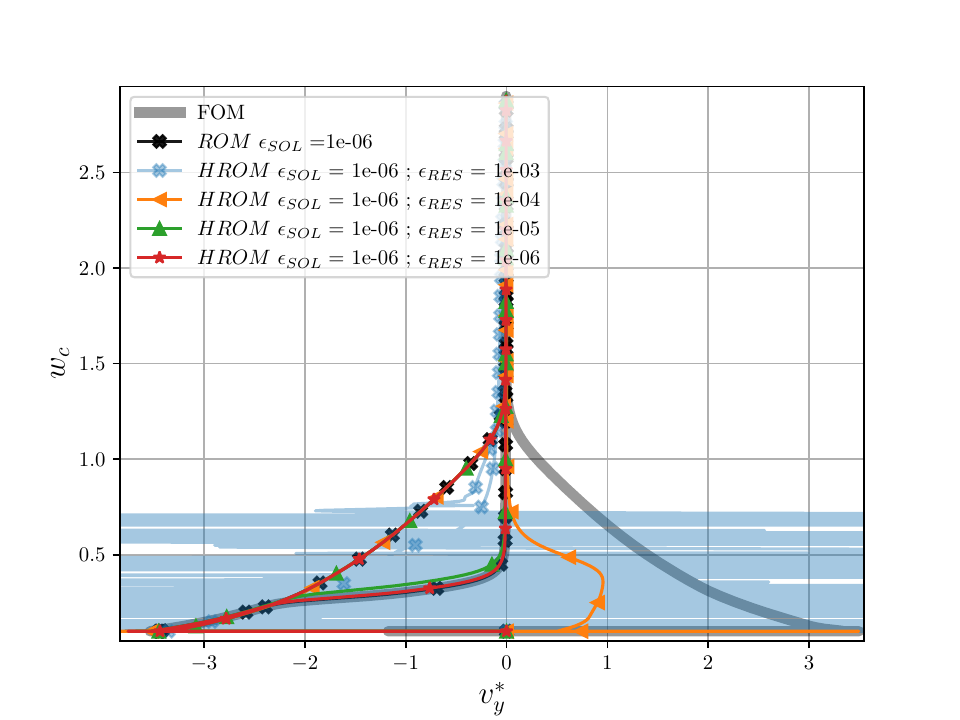}
  \caption{ROM $1e-6$ vs HROM}
\end{subfigure}
\caption{ \bothrev{ QoI phase space plot for the ROMs against various HROMs for the testing trajectory (Trajectory 2) for mapping $\varphi_{\text{\tiny AFFINE}}$ }}
\label{fig: Example 1 test QoI HROM affine}
\end{figure}

\begin{figure}[H]
  \centering
  \begin{subfigure}[b]{0.245\textwidth}
    \includegraphics[width=\linewidth]{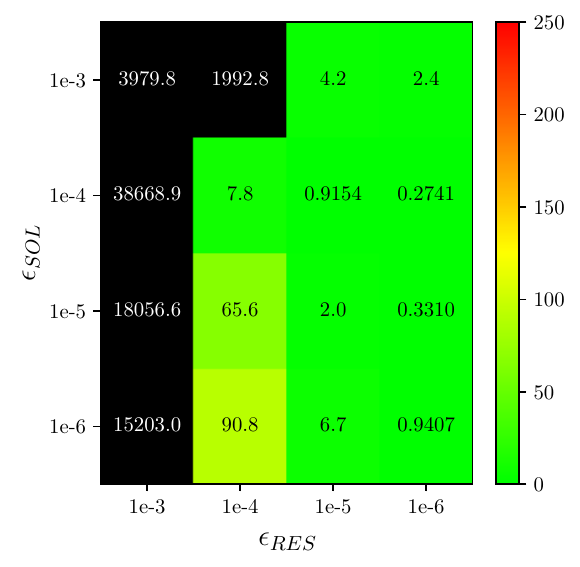}
    \caption{$e({\boldsymbol{v}^*_y}_{\text{\tiny ROM}}, {\boldsymbol{v}^*_y}_{\text{\tiny HROM}} )$}
    %%\label{}
  \end{subfigure}
  \hfill
  \begin{subfigure}[b]{0.245\textwidth}
    \includegraphics[width=\linewidth]{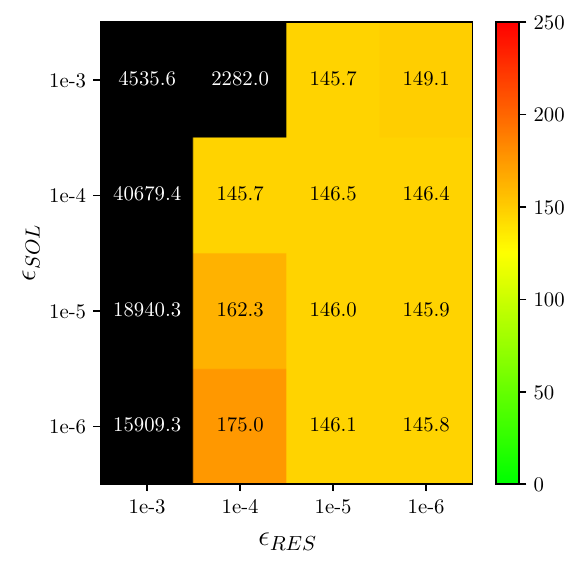}
    \caption{ $e({\boldsymbol{v}^*_y}_{\text{\tiny FOM}}, {\boldsymbol{v}^*_y}_{\text{\tiny HROM}} )$}
    %%\label{}
  \end{subfigure}
  \hfill
  \begin{subfigure}[b]{0.245\textwidth}
    \includegraphics[width=\linewidth]{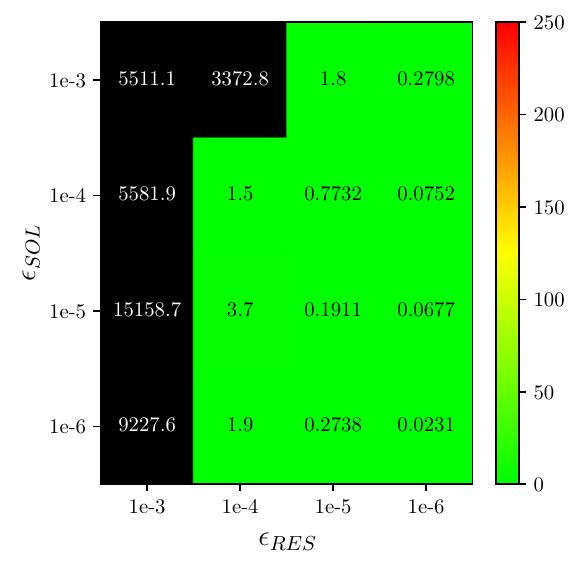}
    \caption{ $e(\boldsymbol{S}_{\text{\tiny ROM}},  \boldsymbol{S}_{\text{\tiny HROM}})$}
    %%\label{}
  \end{subfigure}
  \hfill
  \begin{subfigure}[b]{0.245\textwidth}
    \includegraphics[width=\linewidth]{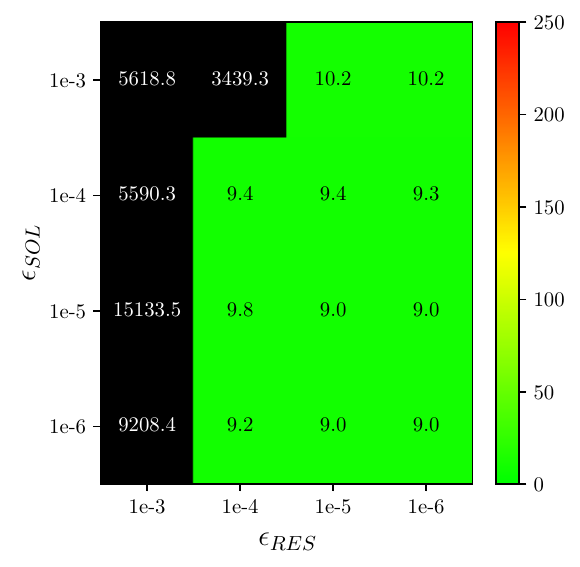}
    \caption{  $e(\boldsymbol{S}_{\text{\tiny FOM}},  \boldsymbol{S}_{\text{\tiny HROM}})$ }
    %%\label{}
  \end{subfigure}
  \caption{\bothrev{Percentage error on QoI and solution field of ROM and FOM against various HROMs for the testing trajectory (Trajectory 2) for mapping $\varphi_{\text{\tiny AFFINE}}$ }}
  \label{fig: Example 1 errors test HROM affine}
\end{figure}

\newsec{We begin by examining the HROMs employing the affine geometric mapping. Fig. \ref{fig: Example 1 test QoI HROM affine} presents the QoI phase space plot of each of the four ROMs against their corresponding HROMs. As in the case of the training trajectory, all of the HROMs constructed with a tolerance $\epsilon_{\text{\tiny RES}}=1e-3$ are numerically unstable, and as  $\epsilon_{\text{\tiny RES}}$ gets smaller, the HROMs approach their corresponding ROMs. For this case, however, the FOM selects the opposite stable branch in the bifurcation diagram with respect to both ROMs and HROMs. Fig. \ref{fig: Example 1 errors test HROM affine} provides quantification of these trends, for example although the errors of the HROMs with respect to their ROMs can be less than 1\% in the QoI, the errors with respect to the FOM are no less than 145\%.}

\begin{figure}[H]
  \begin{subfigure}{.5\textwidth}
    \centering
    % include first image
    \includegraphics[width=.8\linewidth]{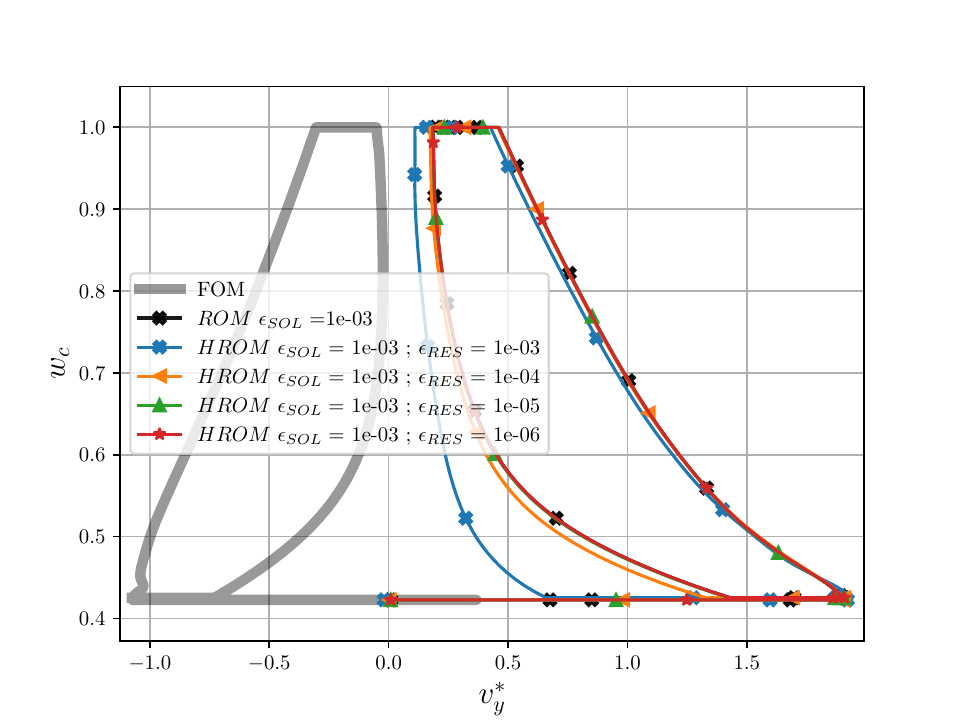}
    \caption{ ROM $1e-3$ vs HROM }
    %\label{fig:hysteresis test ffd rbf_a}
  \end{subfigure}
  \begin{subfigure}{.5\textwidth}
    \centering
    % include second image
    \includegraphics[width=.8\linewidth]{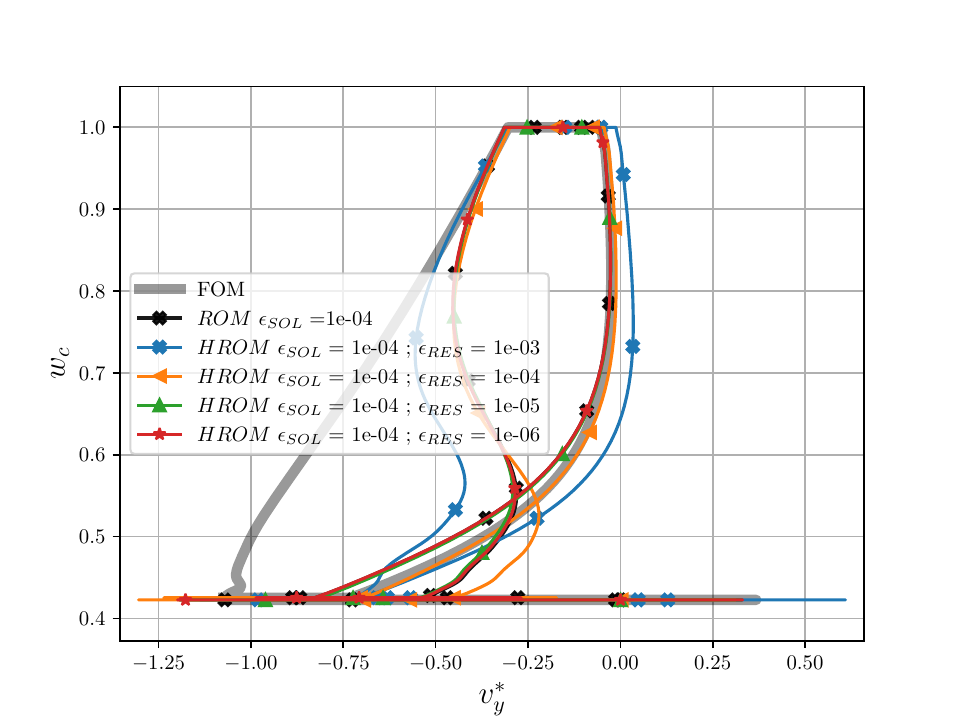}
    \caption{ROM $1e-4$ vs HROM}
    %\label{fig:hysteresis test ffd rbf_b}
  \end{subfigure}
  \begin{subfigure}{.5\textwidth}
    \centering
    % include third image
    \includegraphics[width=.8\linewidth]{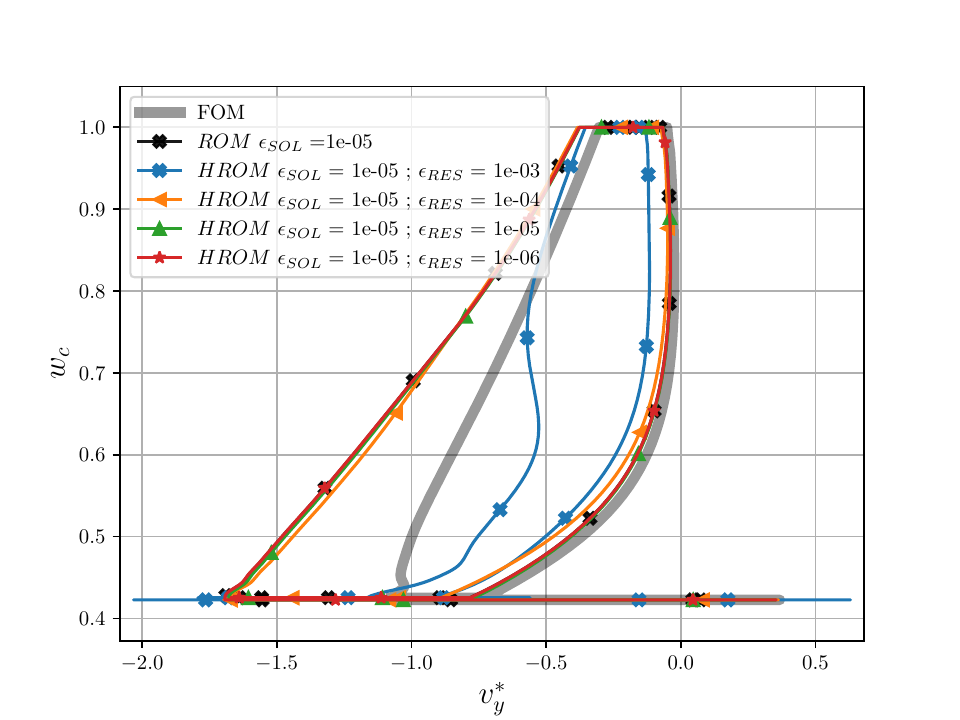}
    \caption{ROM $1e-5$ vs HROM}
    %\label{fig:hysteresis test ffd rbf_c}
  \end{subfigure}
  \begin{subfigure}{.5\textwidth}
    \centering
    % include fourth image
    \includegraphics[width=.8\linewidth]{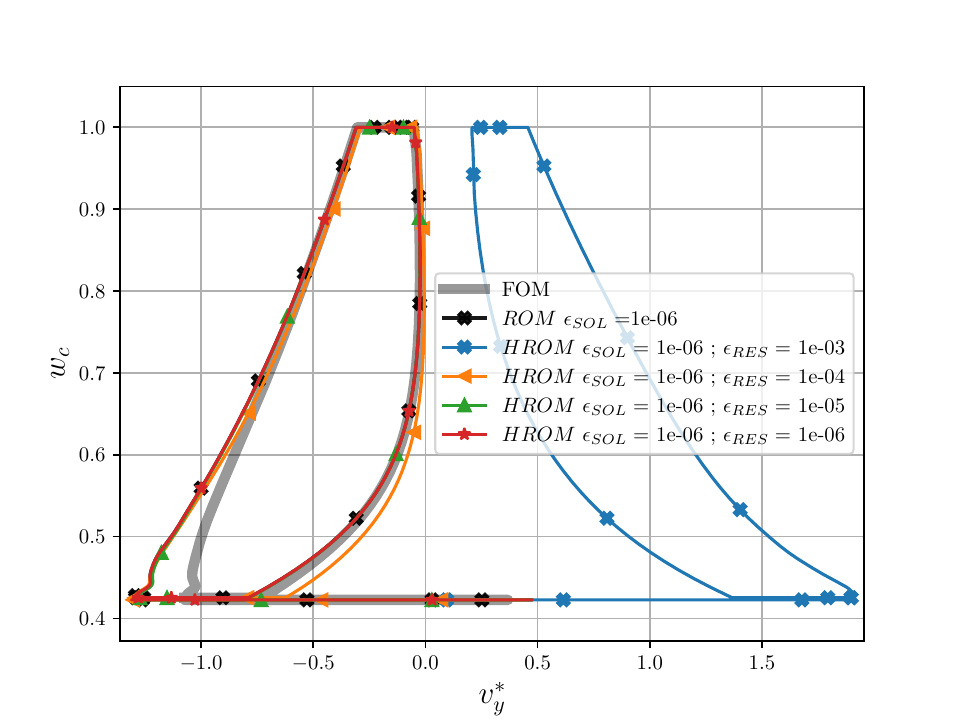}
    \caption{ROM $1e-6$ vs HROM}
    %\label{fig:hysteresis test ffd rbf_d}
  \end{subfigure}
  \caption{ \bothrev{QoI phase space plot for the ROMs against various HROMs for the testing trajectory (Trajectory 2) for mapping $\varphi_{\text{\tiny FFD+RBF}}$ }}
  \label{fig: Example 1 test QoI HROM nonlinear}
  \end{figure}

\begin{figure}[H]
  \centering
  \begin{subfigure}[b]{0.245\textwidth}
    \includegraphics[width=\linewidth]{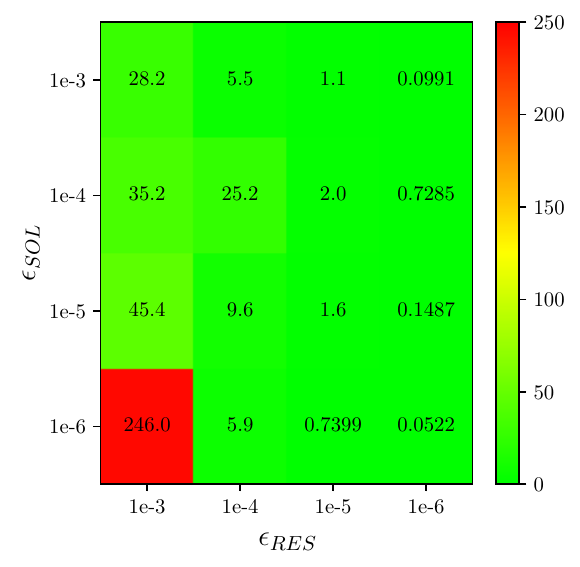}
    \caption{$e({\boldsymbol{v}^*_y}_{\text{\tiny ROM}}, {\boldsymbol{v}^*_y}_{\text{\tiny HROM}} )$}
    %%\label{}
  \end{subfigure}
  \hfill
  \begin{subfigure}[b]{0.245\textwidth}
    \includegraphics[width=\linewidth]{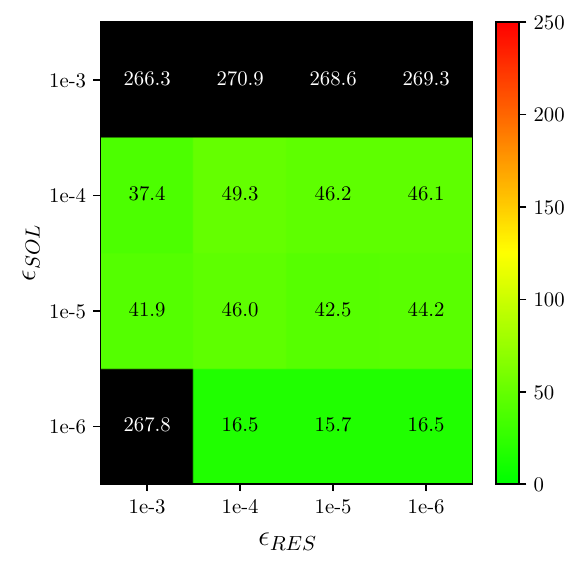}
    \caption{ $e({\boldsymbol{v}^*_y}_{\text{\tiny FOM}}, {\boldsymbol{v}^*_y}_{\text{\tiny HROM}} )$}
    %%\label{}
  \end{subfigure}
  \hfill
  \begin{subfigure}[b]{0.245\textwidth}
    \includegraphics[width=\linewidth]{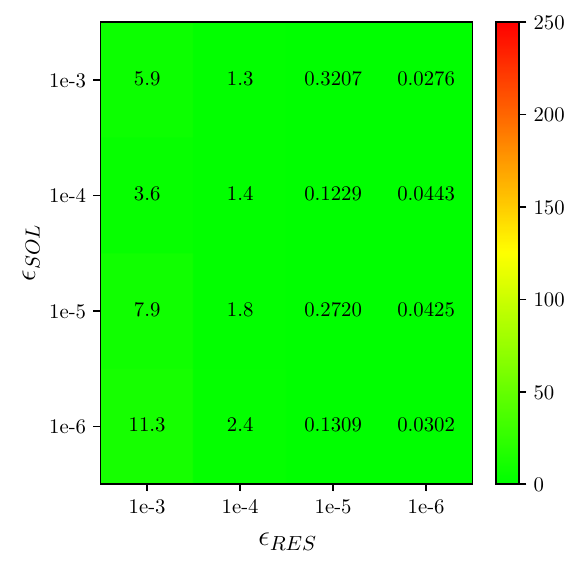}
    \caption{ $e(\boldsymbol{S}_{\text{\tiny ROM}},  \boldsymbol{S}_{\text{\tiny HROM}})$}
    %%\label{}
  \end{subfigure}
  \hfill
  \begin{subfigure}[b]{0.245\textwidth}
    \includegraphics[width=\linewidth]{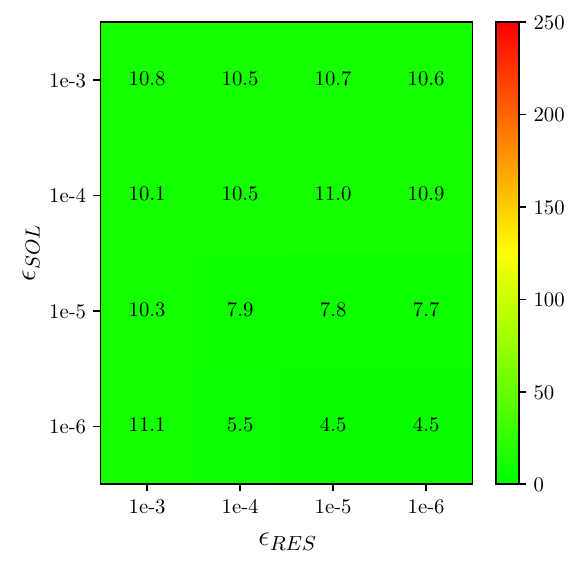}
    \caption{  $e(\boldsymbol{S}_{\text{\tiny FOM}},  \boldsymbol{S}_{\text{\tiny HROM}})$ }
    %%\label{}
  \end{subfigure}
  \caption{ \bothrev{Percentage error on QoI and solution field of ROM and FOM against various HROMs for the testing trajectory (Trajectory 2) for mapping $\varphi_{\text{\tiny FFD+RBF}}$ }}
  \label{fig: Example 1 errors test HROM nonlinear}
\end{figure}

\newsec{Turning our attention to the performance of the HROMs for reconstructing the testing trajectory with the nonaffine geometric mapping, Fig. \ref{fig: Example 1 test QoI HROM nonlinear} displays the QoI phase space plot of the ROMs against their corresponding HROMs. Remarkably, all HROMs are numerically stable, progressively approaching their corresponding ROMs as the tolerance $\epsilon_{\text{\tiny RES}}$ decreases. Fig. \ref{fig: Example 1 errors test HROM nonlinear} offers quantification of the errors for QoI and solution fields of the HROMs against ROMs and FOM.}

\subsubsection{\newsec{Speedup}}

\begin{figure}[H]
  \centering
  \begin{subfigure}[b]{0.44\textwidth}
    \includegraphics[width=\linewidth]{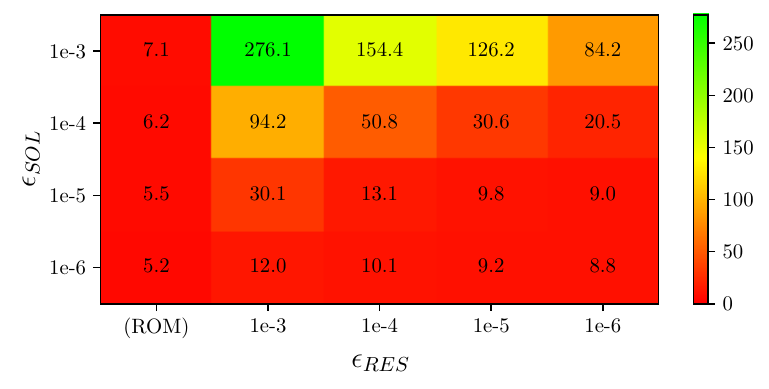}
    \caption{$\boldsymbol{\varphi}_{\text{\tiny AFFINE}}$ }
    %%\label{}
  \end{subfigure}
  \hfill
  \begin{subfigure}[b]{0.44\textwidth}
    \includegraphics[width=\linewidth]{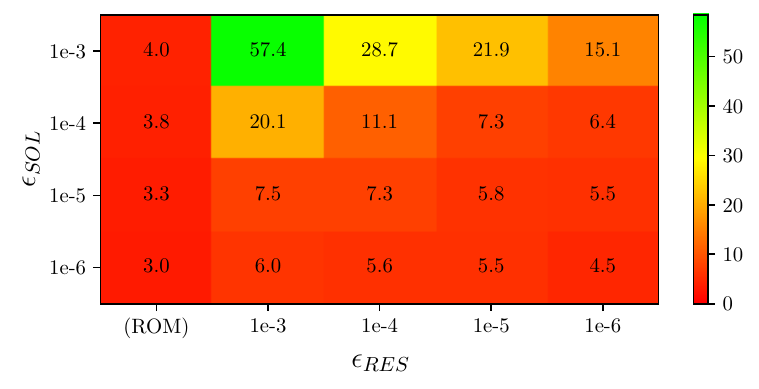}
    \caption{$\boldsymbol{\varphi}_{\text{\tiny FFD+RBF}}$ }
    %%\label{}
  \end{subfigure}

  \caption{\revone{Speedup factors for the ROMs and HROMs presented in this example. The first column displays the performance of the ROMs, while the subsequent columns depict the HROMs' performance for each combination of truncation tolerances $\epsilon_{\text{\tiny SOL}}$ and $\epsilon_{\text{\tiny RES}}$ considered}}
  \label{fig: Speedups Example 1}
\end{figure}

\newsec{Fig. \ref{fig: Speedups Example 1} shows the speedup factors obtained for all models with respect to the FOMs of each geometric mapping, as computed using Eq. \ref{eq:speedup definition}. Notably, the ROMs without hyper-reduction achieve a speedup of less than an order of magnitude. In contrast, the HROMs exhibit significantly larger speedup factors. It is important to exercise caution, considering that the HROMs constructed considering the less stringent truncation tolerances resulted invariably in high errors.}

\subsubsection{\newsec{Discussion}}

\newsec{The results presented for this first example highlight the challenges of accurately capturing the Coanda effect and its hysteresis behavior using reduced order models whose testing trajectories require capturing the behavior of the opposite stable branch in the bifurcation, with respect of the information contained in the training trajectory. While in the case of the affine geometric mapping, none of four ROMs was capable of accurately reproducing the FOM trend in the Coanda effect for the testing trajectory, ROMs for the nonaffine geometric mapping were able to capture the main patterns of this unseeen brach when sufficiently decreasing the truncation tolerances of the solution and residuals. In this case, although none of the ROMs perfectly matched the FOM solution, they progressively approached it.}

\newsec{To delve deeper on the potential of ROMs and HROMs to deal with the characterisation of the Coanda effect in this geometry, we proceed to consider both Trajectory 1 and Trajectory 2 as training trajectories in Example 2.}

\subsection{\newsec{Example 2}}

\newsec{In the second example, we consider the use of two training trajectories: Trajectory 1 and Trajectory 2. The presence of jets attaching to both, the upper and lower walls in these training trajectories anticipates that the ROMs and HROMs will be robust. We subject all models to Trajectory 3 as the testing trajectory and asses their performance to check this hypothesis.}

\subsubsection{\newsec{ROM}}

\begin{figure}[H]
  \centering
  \begin{subfigure}[b]{0.53\textwidth}
    \includegraphics[width=\linewidth]{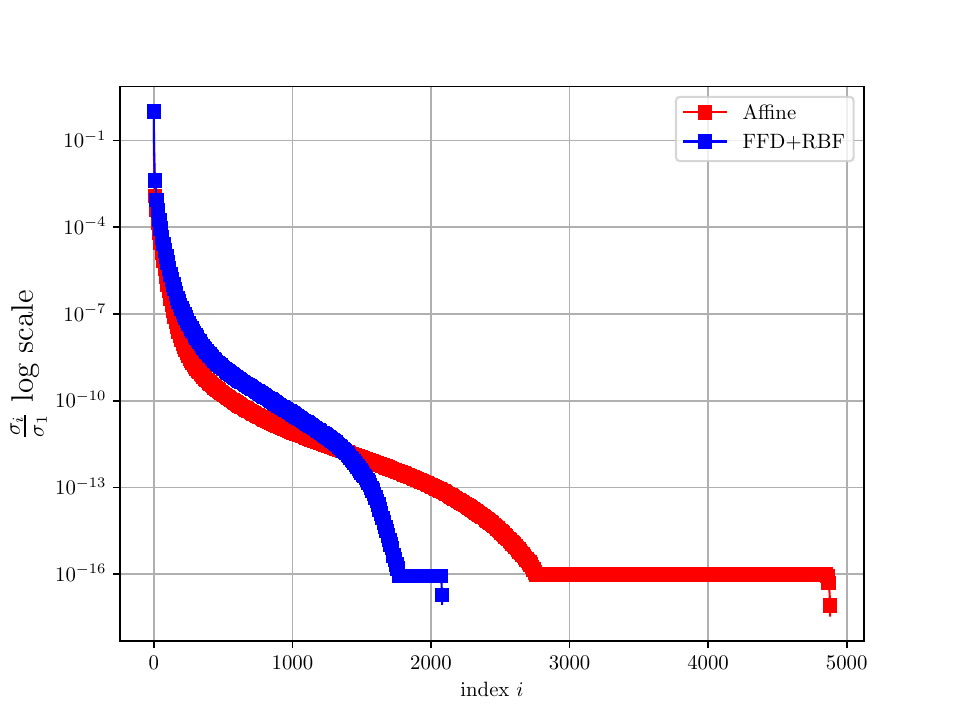}
    %\caption{}
    %\label{}
  \end{subfigure}
  \hfill
  \begin{subfigure}[b]{0.45\textwidth}
    \includegraphics[width=\linewidth]{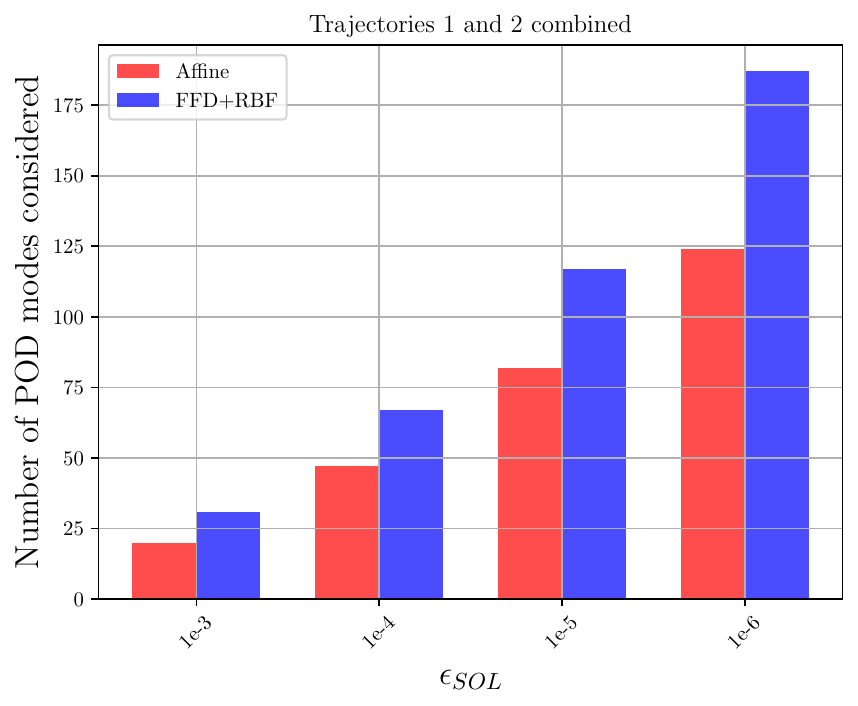}
    %\caption{}
    %\label{}
  \end{subfigure}
\caption{\revone{Singular values of solution snapshots and the corresponding number of POD modes for both geometric mappings under Trajectory 1 and Trajectory 2 combined}}
\label{fig: Example 3 singular values S and Number of POD modes}
\end{figure}

\newsec{Fig. \ref{fig: Example 3 singular values S and Number of POD modes} illustrates the singular values of the solution snapshots and the corresponding number of POD modes required by both models. As in the previous example, the nonaffine geometric mapping requires a greater number of modes to achieve the same tolerance compared to the affine mapping. We generated four ROMs for each geometric mapping, accounting for the respective number of POD modes.}

\begin{figure}[H]
  \centering
  \begin{subfigure}[b]{0.45\textwidth}
    \includegraphics[width=\linewidth]{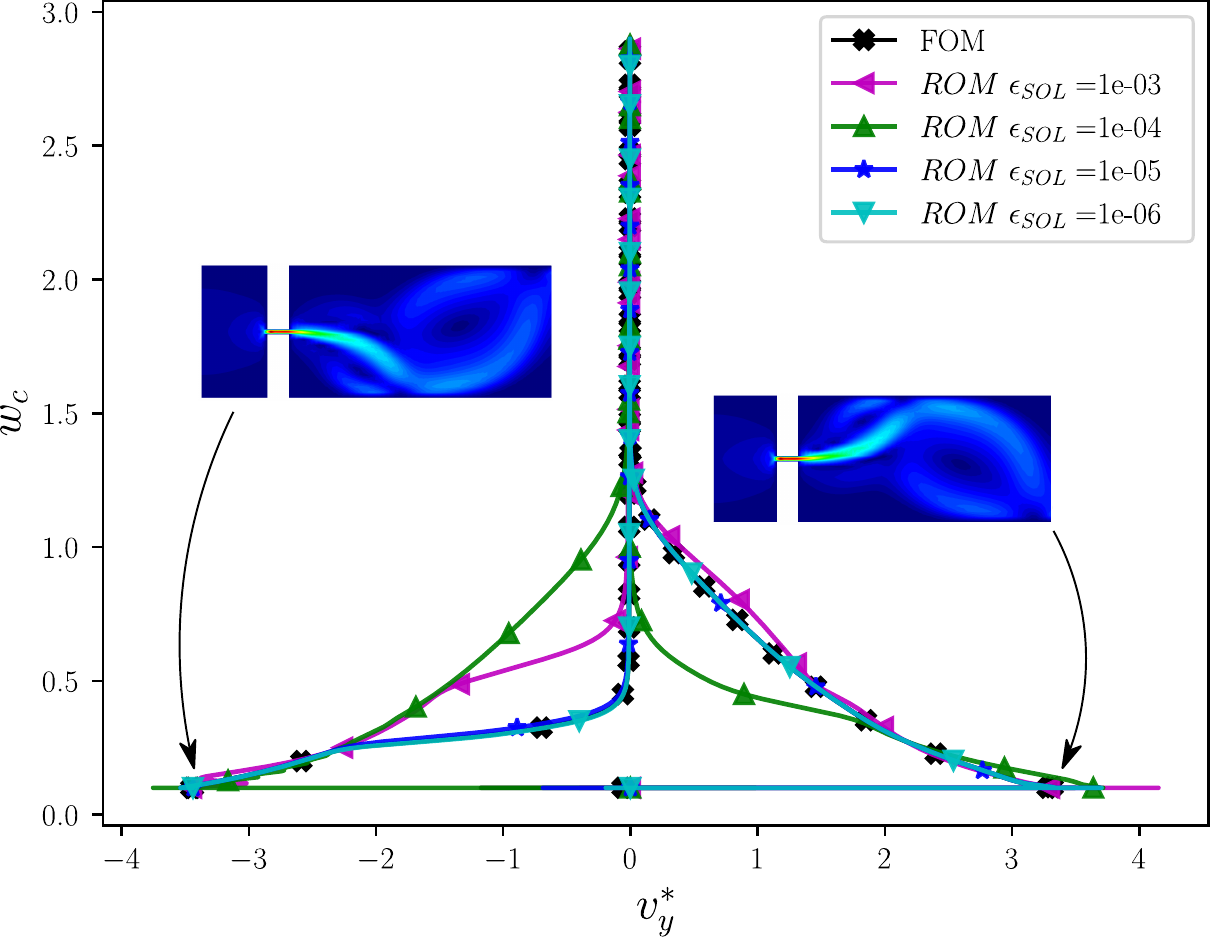}
    \caption{$\boldsymbol{\varphi}_{\text{\tiny AFFINE}}$ }
    %\label{fig:XXX}
  \end{subfigure}
  \hfill
  \begin{subfigure}[b]{0.45\textwidth}
    \includegraphics[width=\linewidth]{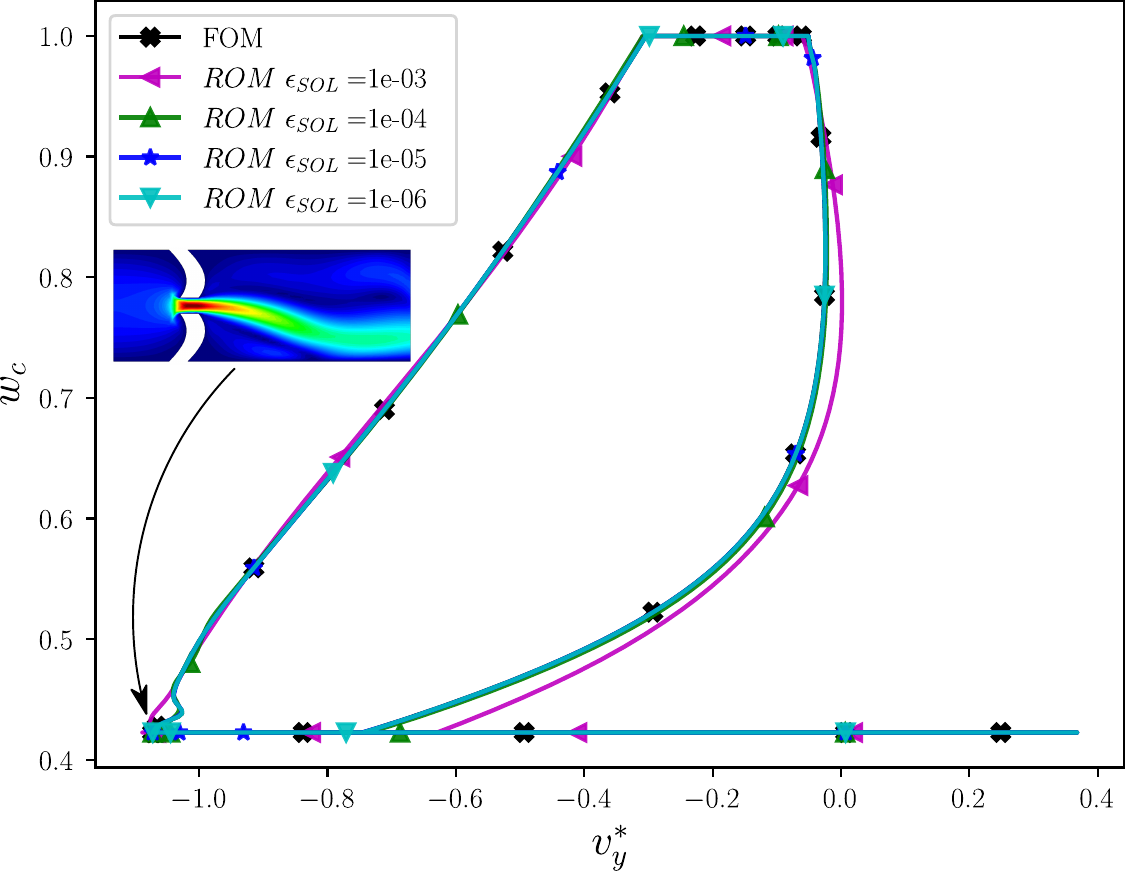}
    \caption{$\boldsymbol{\varphi}_{\text{\tiny FFD+RBF}}$ }
    %\label{fig:XXX}
  \end{subfigure}
\caption{QoI phase space plot for the FOM against various ROMs for the first training trajectory (Trajectory 2) for both geometric mappings}
  \label{fig: Example 3 train QoI ROM vs FOM}
\end{figure}

\begin{figure}[H]
  \centering
  \begin{subfigure}[b]{0.45\textwidth}
    \includegraphics[width=\linewidth]{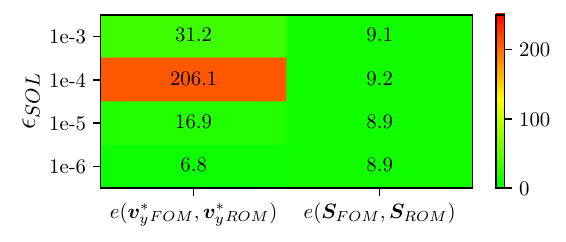}
    \caption{$\boldsymbol{\varphi}_{\text{\tiny AFFINE}}$ }
    %\label{}
  \end{subfigure}
  \hfill
  \begin{subfigure}[b]{0.45\textwidth}
    \includegraphics[width=\linewidth]{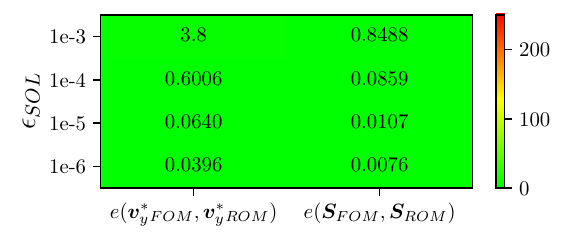}
    \caption{$\boldsymbol{\varphi}_{\text{\tiny FFD+RBF}}$ }
    %\label{}
  \end{subfigure}
  \caption{\bothrev{Percentage error on QoI and solution field of FOM against various ROMs for the first training trajectory (Trajectory 2) for both geometric mappings}}
  \label{fig: Example 3 errors train ROM vs FOM}
\end{figure}

\newsec{We asses the capability of the constructed ROMs to replicate the training trajectories, starting with Trajectory 2. Fig. \ref{fig: Example 3 train QoI ROM vs FOM} and Fig. \ref{fig: Example 3 errors train ROM vs FOM} show the QoI phase space plots and the percentage errors for the ROMs against the FOMs for both geometric mappings. The ROMs for both geometric mappings exhibit the expected behaviour that, as the truncation tolerance $\epsilon_{\text{\tiny SOL}}$ decreases, the discrepancy between the QoI of the ROMs and that of the FOMs also decreases. }

\begin{figure}[H]
  \centering
  \begin{subfigure}[b]{0.45\textwidth}
    \includegraphics[width=\linewidth]{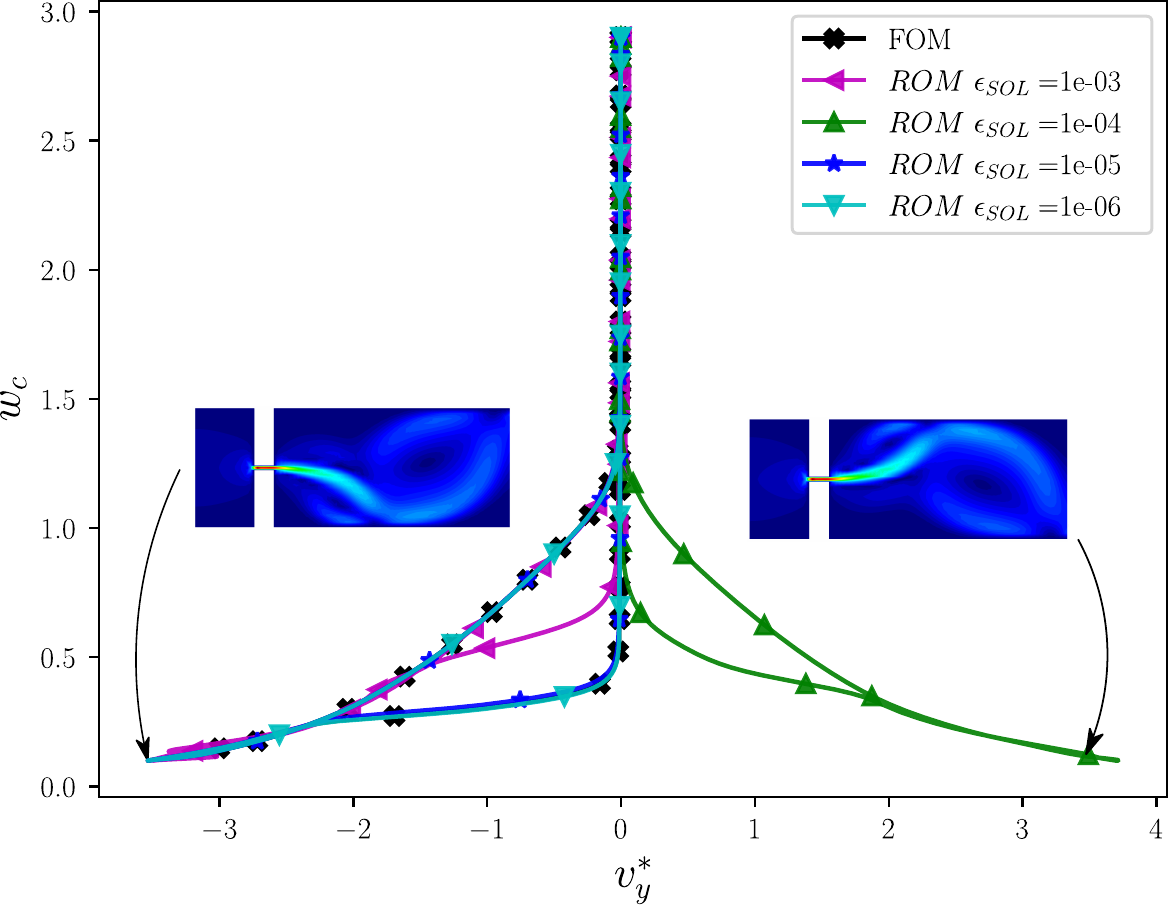}
    \caption{$\boldsymbol{\varphi}_{\text{\tiny AFFINE}}$ }
    %\label{fig: QoI ROM test a}
  \end{subfigure}
  \hfill
  \begin{subfigure}[b]{0.45\textwidth}
    \includegraphics[width=\linewidth]{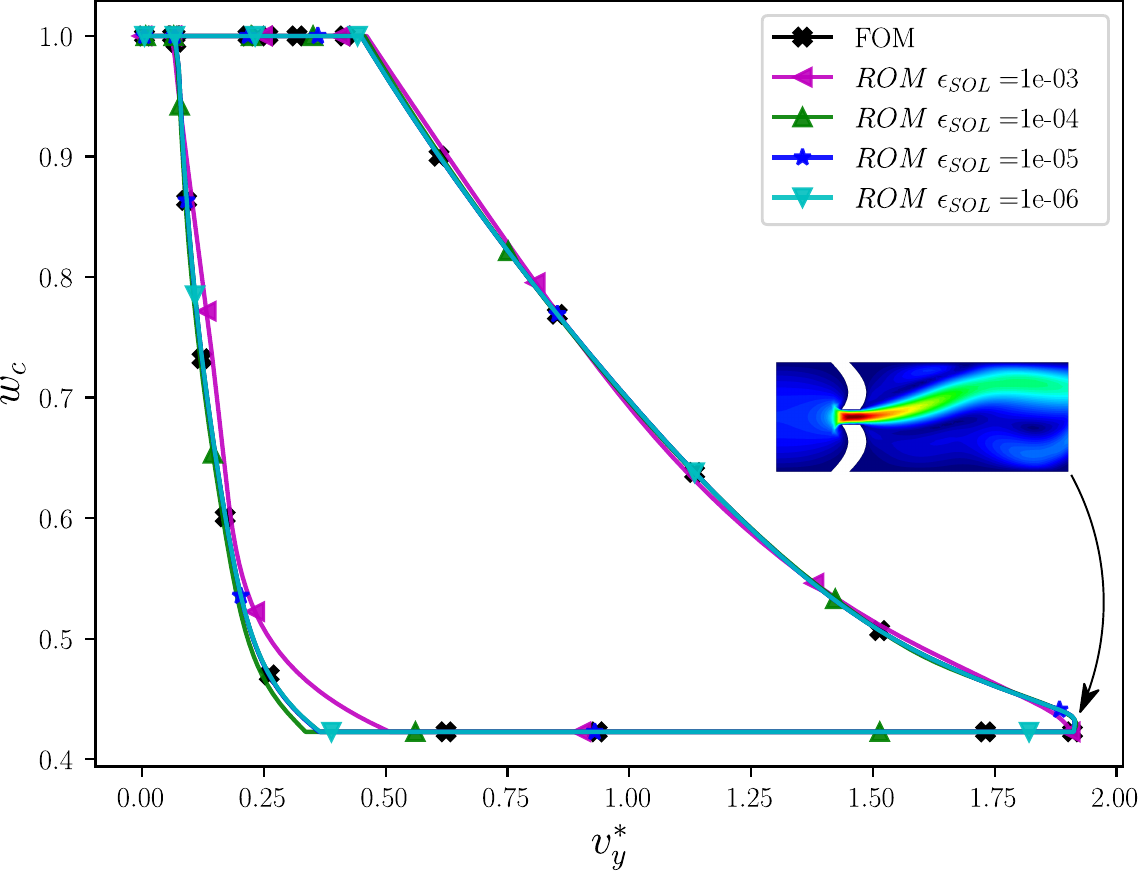}
    \caption{$\boldsymbol{\varphi}_{\text{\tiny FFD+RBF}}$ }
    %\label{fig: QoI ROM test b}
  \end{subfigure}
\caption{QoI phase space plot for the FOM against various ROMs for the second training trajectory (Trajectory 1) for both geometric mappings}
\label{fig: Example 3 test QoI ROM vs FOM}
\end{figure}

\begin{figure}[H]
  \centering
  \begin{subfigure}[b]{0.45\textwidth}
    \includegraphics[width=\linewidth]{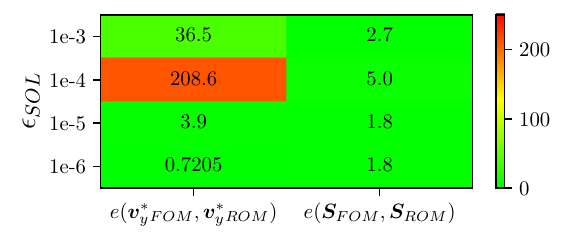}
    \caption{$\boldsymbol{\varphi}_{\text{\tiny AFFINE}}$ }
    %\label{}
  \end{subfigure}
  \hfill
  \begin{subfigure}[b]{0.45\textwidth}
    \includegraphics[width=\linewidth]{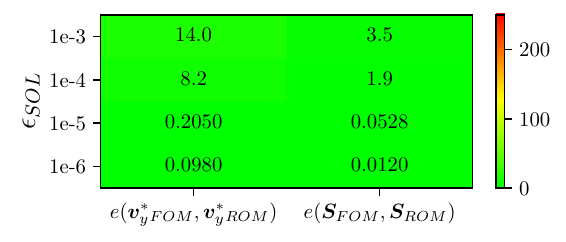}
    \caption{$\boldsymbol{\varphi}_{\text{\tiny FFD+RBF}}$ }
    %\label{}
  \end{subfigure}
  \caption{\bothrev{Percentage error on QoI and solution field of FOM against various ROMs for the second training trajectory (Trajectory 1) for both geometric mappings}}
  \label{fig: Example 3 errors test ROM vs FOM}
\end{figure}

\newsec{Moving on, we examine the performance of the ROMs for reproducing Trajectory 1, which here we consider as the second training trajectory. Fig. \ref{fig: Example 3 test QoI ROM vs FOM} and Fig. \ref{fig: Example 3 errors test ROM vs FOM} show the QoI phase space plots and percentage errors incurred by the ROMs against their respective FOMs for this training trajectory. We observe the resemblance of this results with those obtained in Example 1 for reproducing this same training trajectory.}

\subsubsection{\newsec{HROM}}

\begin{figure}[H]
  \centering
  \begin{subfigure}[b]{0.55\textwidth}
    \includegraphics[width=\linewidth]{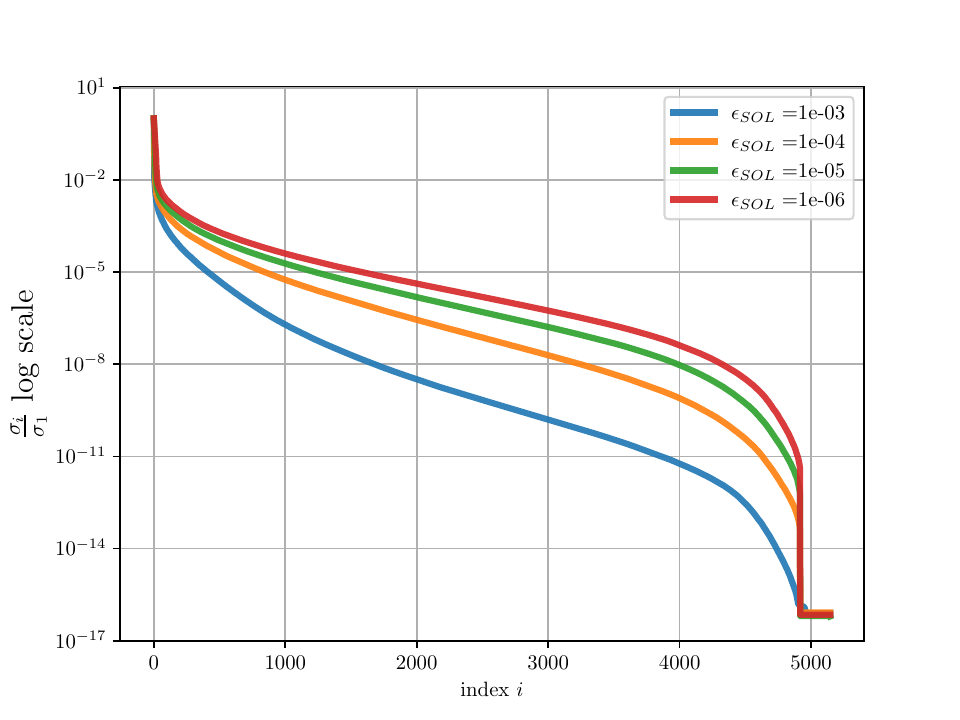}
    \caption{Singular values for matrices $\boldsymbol{S_r}(\epsilon_{\text{\tiny SOL}})$}
  \end{subfigure}
  \hfill
  \begin{subfigure}[b]{0.40\textwidth}
    \includegraphics[width=\linewidth]{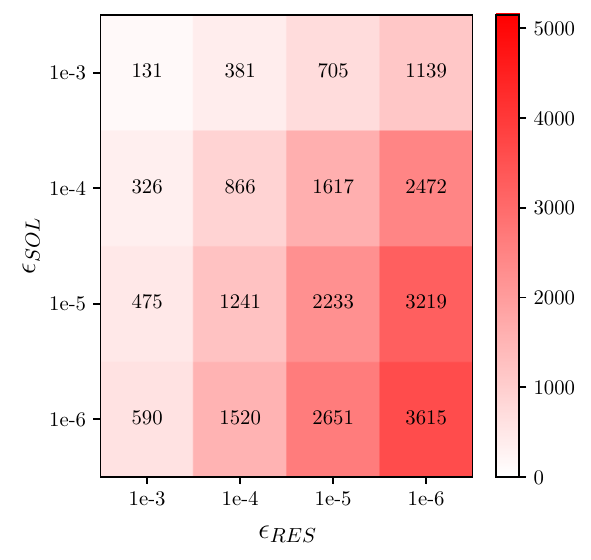}
    \caption{Number of selected elements for all combinations of $\epsilon_{\text{\tiny SOL}}$ and $\epsilon_{\text{\tiny RES}}$}
  \end{subfigure}
  \caption{\revone{Singular values decay profile of matrices of projected residuals, and number of selected elements by ECM algorithm for geometric mapping  $\boldsymbol{\varphi}_{\text{\tiny AFFINE}}$}}
  \label{fig: Example 3 Sr affine num elements}
\end{figure}

\begin{figure}[H]
  \centering
  \begin{subfigure}[b]{0.55\textwidth}
    \includegraphics[width=\linewidth]{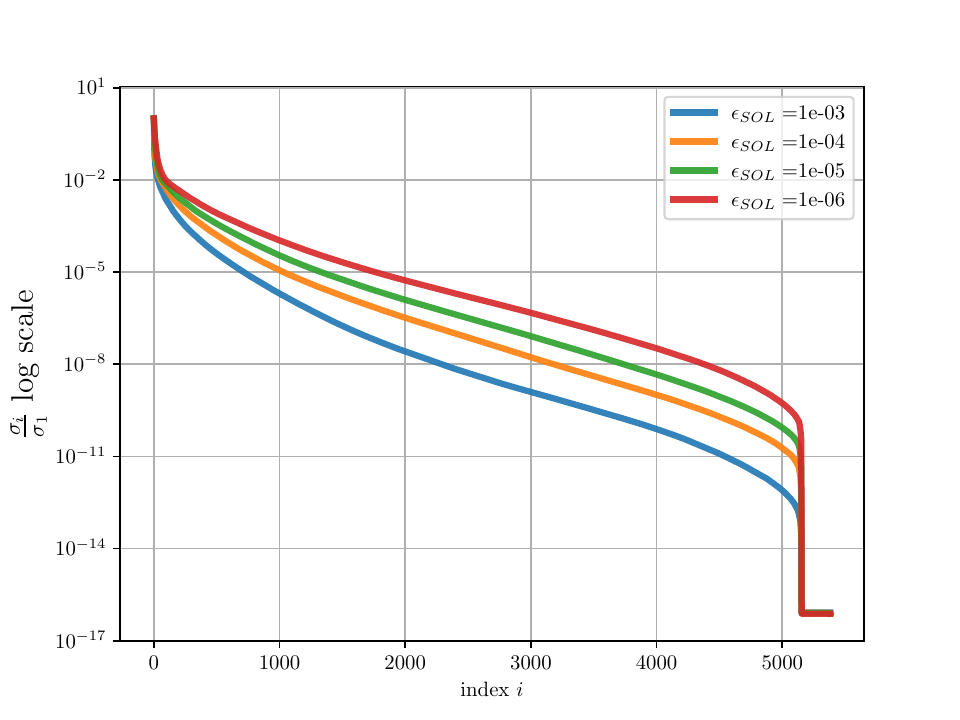}
    \caption{Singular values for matrices $\boldsymbol{S_r}(\epsilon_{\text{\tiny SOL}})$}
  \end{subfigure}
  \hfill
  \begin{subfigure}[b]{0.40\textwidth}
    \includegraphics[width=\linewidth]{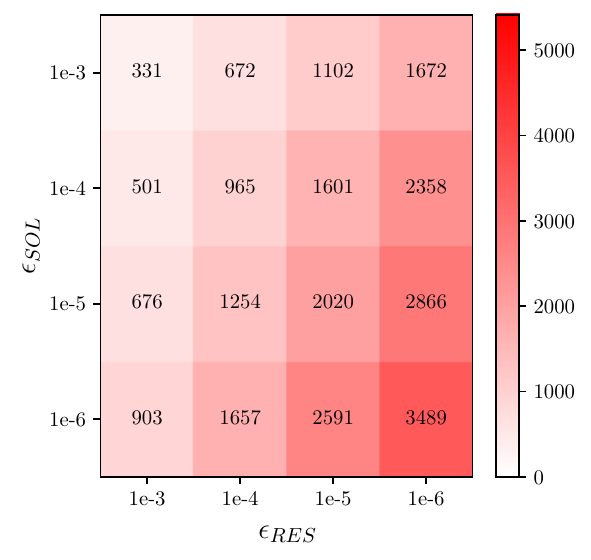}
    \caption{Number of selected elements for all combinations of $\epsilon_{\text{\tiny SOL}}$ and $\epsilon_{\text{\tiny RES}}$}
  \end{subfigure}
  \caption{\revone{Singular values decay profile of matrices of projected residuals, and number of selected elements by ECM algorithm for geometric mapping  $\boldsymbol{\varphi}_{\text{\tiny FFD+RBF}}$}}
  \label{fig: Example 3 Sr nonlinear num elements}
\end{figure}

\newsec{Fig. \ref{fig: Example 3 Sr affine num elements} and Fig. \ref{fig: Example 3 Sr nonlinear num elements} show the singular values decay profile of the matrices of projected residuals, and the count of selected elements by the ECM algorithm across the 32 constructed HROMs. \append{Appendix \ref{sec: appendix} shows the location of the selected elements in the meshes employed.} As anticipated, the number of HROMs surpasses those in the preceding example. This is a consequence of the richer POD bases, and the subsequent slower decay of singular values of matrices $\boldsymbol{S_r}(\epsilon_{\text{\tiny SOL}})$.}

\begin{figure}[H]
\begin{subfigure}{.5\textwidth}
  \centering
  % include first image
  \includegraphics[width=.8\linewidth]{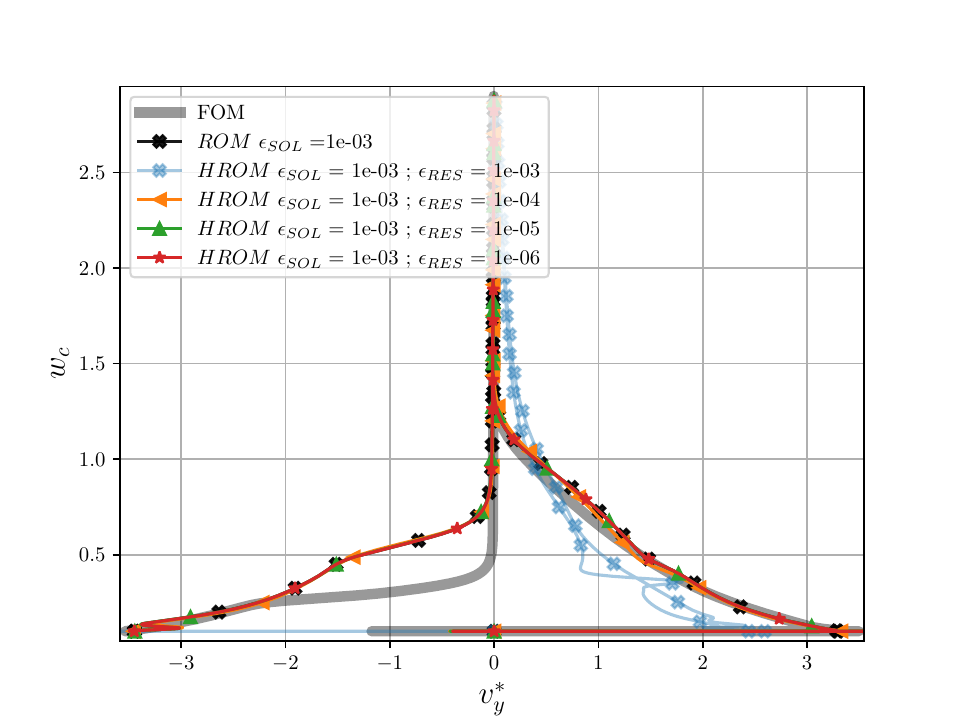}
  \caption{ROM $1e-3$ vs HROM}
  %\label{fig: example3 hysteresis train affine_a}
\end{subfigure}
\begin{subfigure}{.5\textwidth}
  \centering
  % include second image
  \includegraphics[width=.8\linewidth]{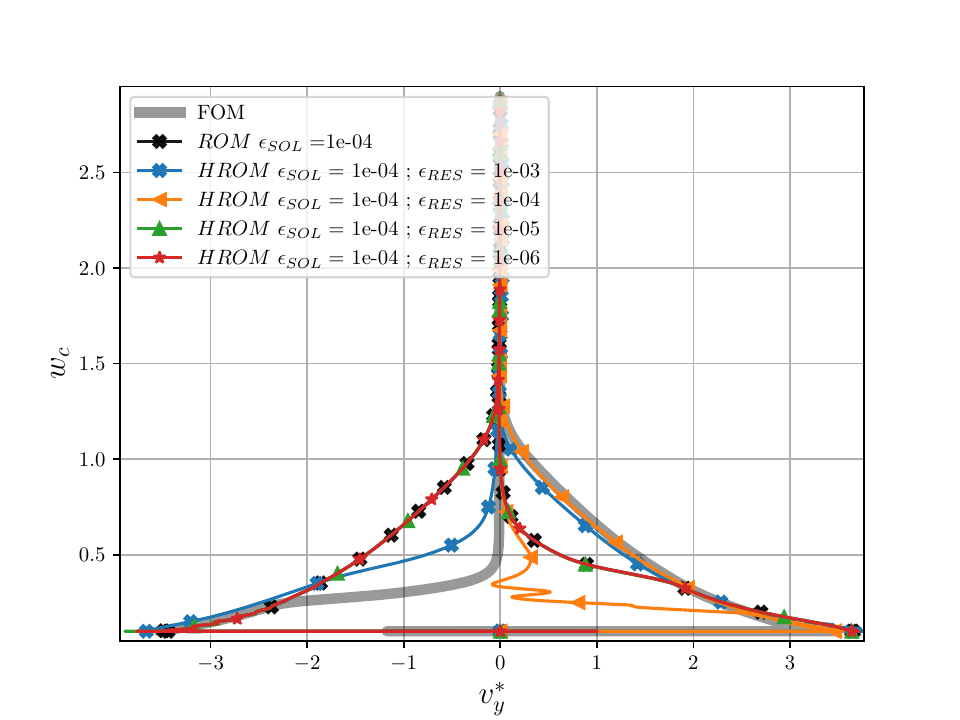}
  \caption{ROM $1e-4$ vs HROM}
  %\label{fig: example3 hysteresis train affine_b}
\end{subfigure}
\begin{subfigure}{.5\textwidth}
  \centering
  % include third image
  \includegraphics[width=.8\linewidth]{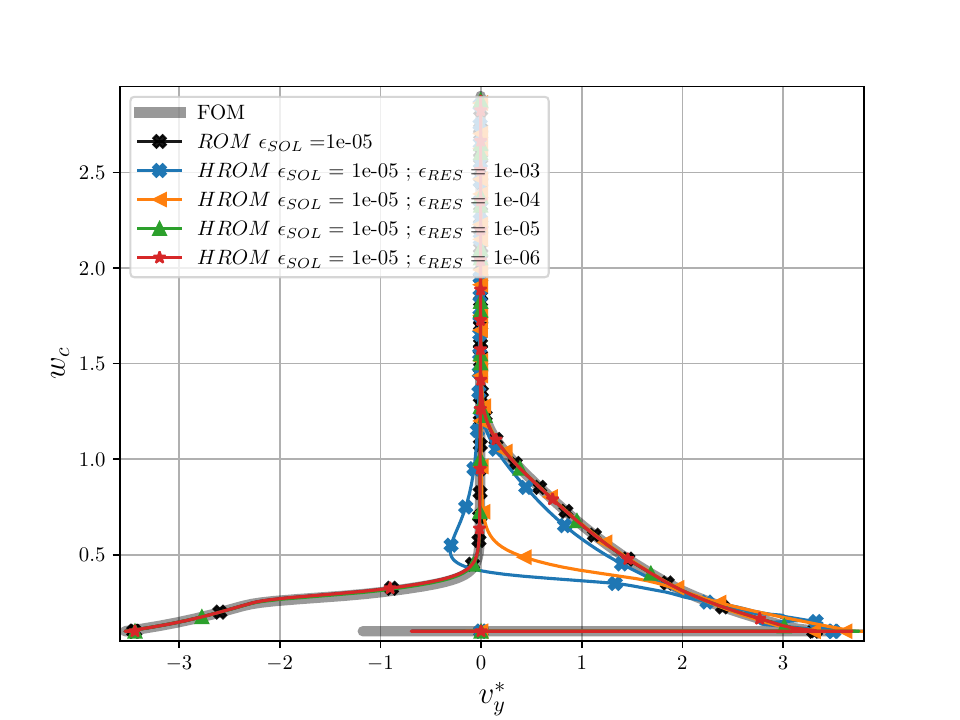}
  \caption{ROM $1e-5$ vs HROM}
  %\label{fig: example3 hysteresis train affine_c}
\end{subfigure}
\begin{subfigure}{.5\textwidth}
  \centering
  % include fourth image
  \includegraphics[width=.8\linewidth]{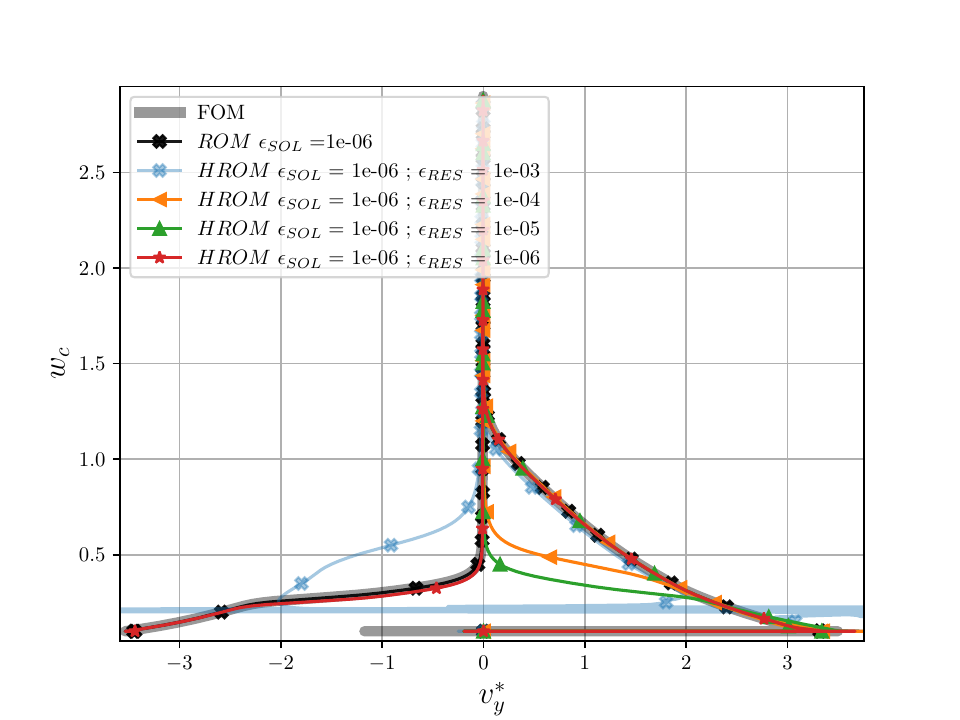}
  \caption{ROM $1e-6$ vs HROM}
  %\label{fig: example3 hysteresis train affine_d}
\end{subfigure}
\caption{ \bothrev{QoI phase space plot for the ROMS against various HROMs for the first training trajectory (Trajectory 2) for mapping $\varphi_{\text{\tiny AFFINE}}$ }}
\label{fig: Example 3 train QoI HROM affine}
\end{figure}

\begin{figure}[H]
  \centering
  \begin{subfigure}[b]{0.245\textwidth}
    \includegraphics[width=\linewidth]{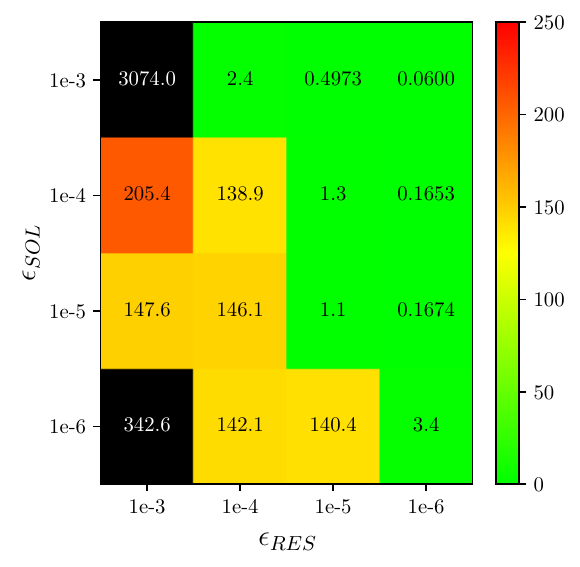}
    \caption{$e({\boldsymbol{v}^*_y}_{\text{\tiny ROM}}, {\boldsymbol{v}^*_y}_{\text{\tiny HROM}} )$}
    %%\label{}
  \end{subfigure}
  \hfill
  \begin{subfigure}[b]{0.245\textwidth}
    \includegraphics[width=\linewidth]{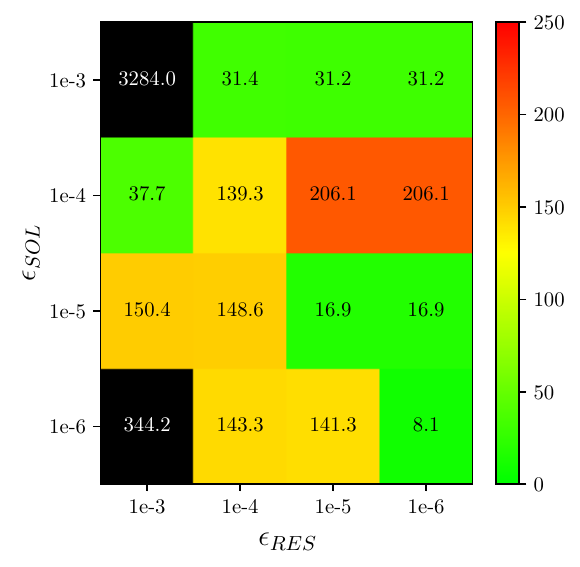}
    \caption{ $e({\boldsymbol{v}^*_y}_{\text{\tiny FOM}}, {\boldsymbol{v}^*_y}_{\text{\tiny HROM}} )$}
    %%\label{}
  \end{subfigure}
  \hfill
  \begin{subfigure}[b]{0.245\textwidth}
    \includegraphics[width=\linewidth]{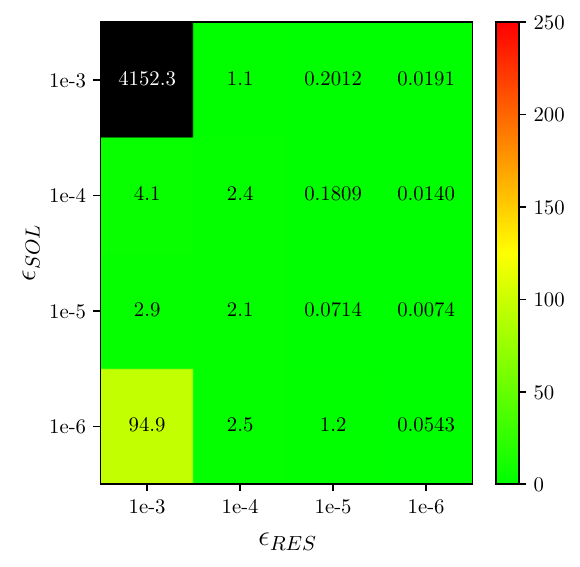}
    \caption{ $e(\boldsymbol{S}_{\text{\tiny ROM}},  \boldsymbol{S}_{\text{\tiny HROM}})$}
    %%\label{}
  \end{subfigure}
  \hfill
  \begin{subfigure}[b]{0.245\textwidth}
    \includegraphics[width=\linewidth]{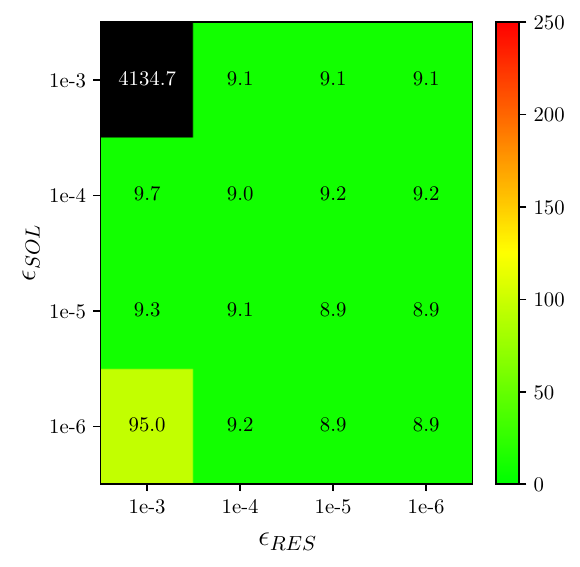}
    \caption{  $e(\boldsymbol{S}_{\text{\tiny FOM}},  \boldsymbol{S}_{\text{\tiny HROM}})$ }
    %%\label{}
  \end{subfigure}
  \caption{ \bothrev{ Percentage error on QoI and solution field of ROM and FOM against various HROMs for the first training trajectory (Trajectory 2) for mapping $\varphi_{\text{\tiny AFFINE}}$}}
  \label{fig: Example 3 errors train HROM affine}
\end{figure}

\newsec{We proceed to investigate the performance of the constructed HROMs for reproducing the training trajectories. Let us begin focusing on the first training trajectory, which for this example is Trajectory 2. Fig. \ref{fig: Example 3 train QoI HROM affine} and Fig. \ref{fig: Example 3 errors train HROM affine} show the performance of the HROMs for the affine geometric mapping against their respective ROMs and FOM.}

\begin{figure}[H]
\begin{subfigure}{.5\textwidth}
  \centering
  % include first image
  \includegraphics[width=.8\linewidth]{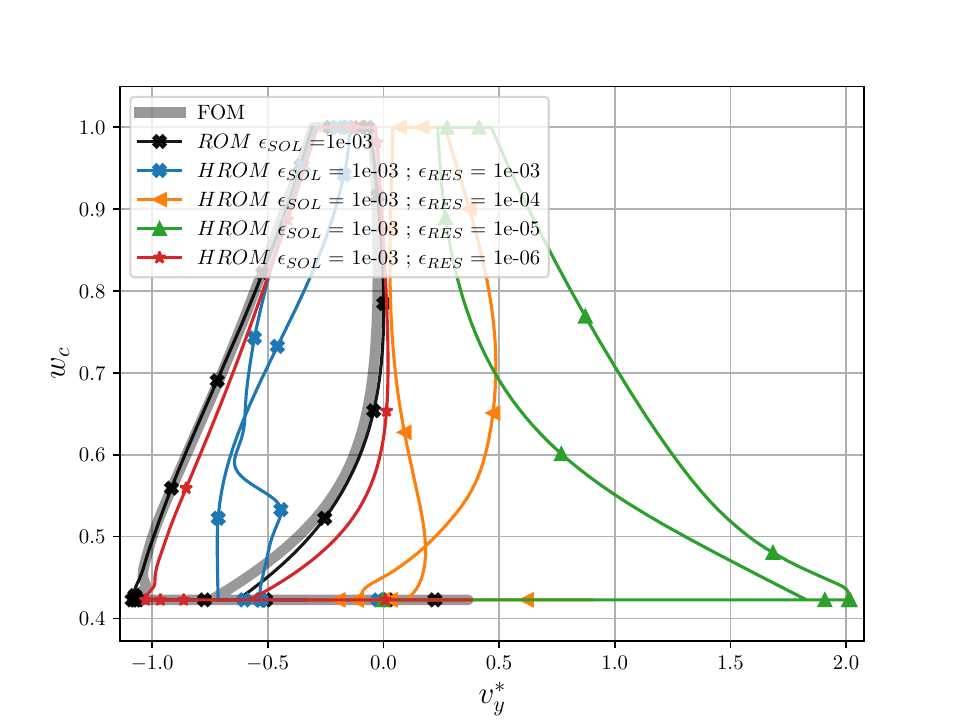}
  \caption{ROM $1e-3$ vs HROM}
\end{subfigure}
\begin{subfigure}{.5\textwidth}
  \centering
  % include second image
  \includegraphics[width=.8\linewidth]{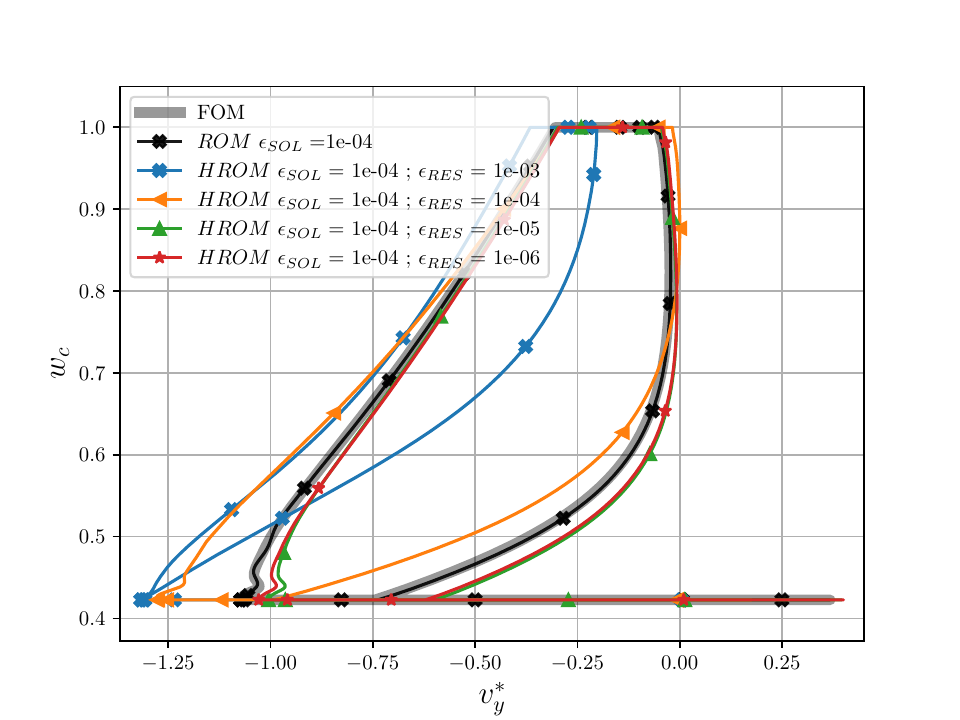}
  \caption{ROM $1e-4$ vs HROM}
\end{subfigure}
\begin{subfigure}{.5\textwidth}
  \centering
  % include third image
  \includegraphics[width=.8\linewidth]{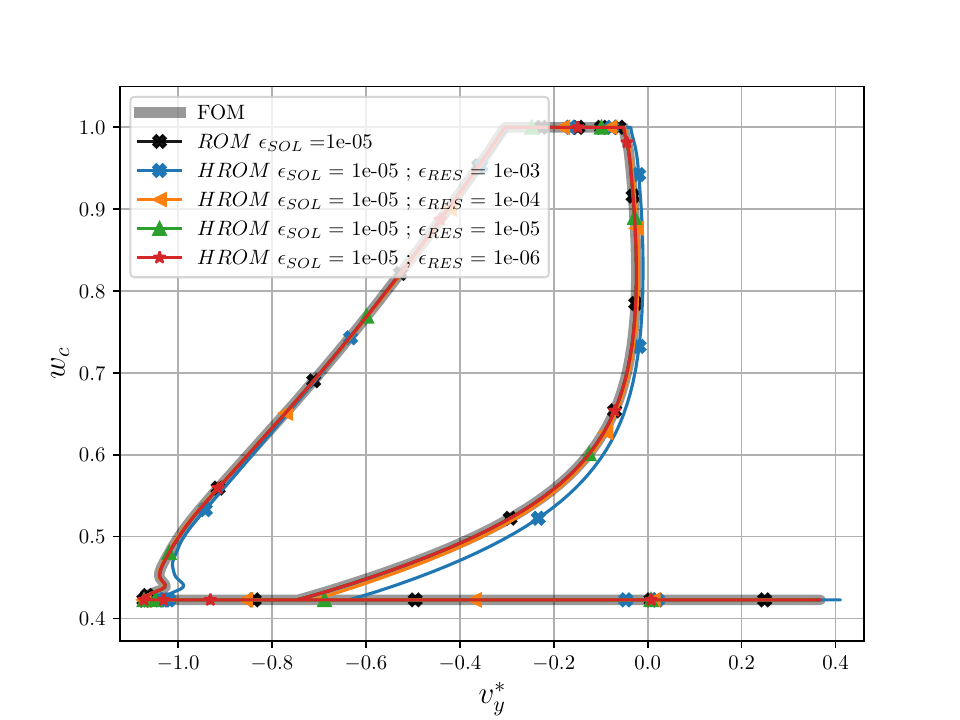}
  \caption{ROM $1e-5$ vs HROM}
\end{subfigure}
\begin{subfigure}{.5\textwidth}
  \centering
  % include fourth image
  \includegraphics[width=.8\linewidth]{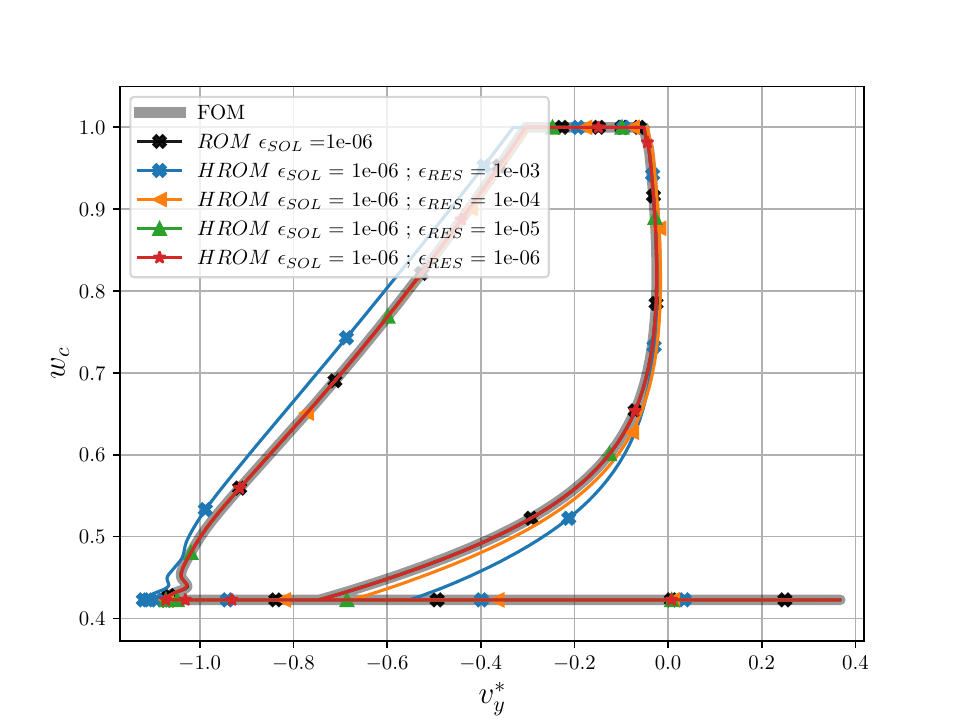}
  \caption{ROM $1e-6$ vs HROM}
\end{subfigure}
\caption{ \bothrev{ QoI phase space plot for the ROMs against various HROMs for the first training trajectory (Trajectory 2) for mapping $\varphi_{\text{\tiny FFD+RBF}}$   }}
\label{fig: Example 3 train QoI HROM nonlinear}
\end{figure}

\begin{figure}[H]
  \centering
  \begin{subfigure}[b]{0.245\textwidth}
    \includegraphics[width=\linewidth]{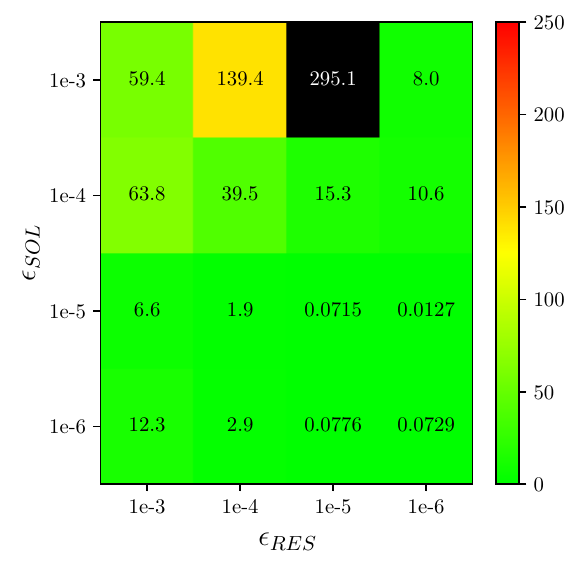}
    \caption{$e({\boldsymbol{v}^*_y}_{\text{\tiny ROM}}, {\boldsymbol{v}^*_y}_{\text{\tiny HROM}} )$}
    %%\label{}
  \end{subfigure}
  \hfill
  \begin{subfigure}[b]{0.245\textwidth}
    \includegraphics[width=\linewidth]{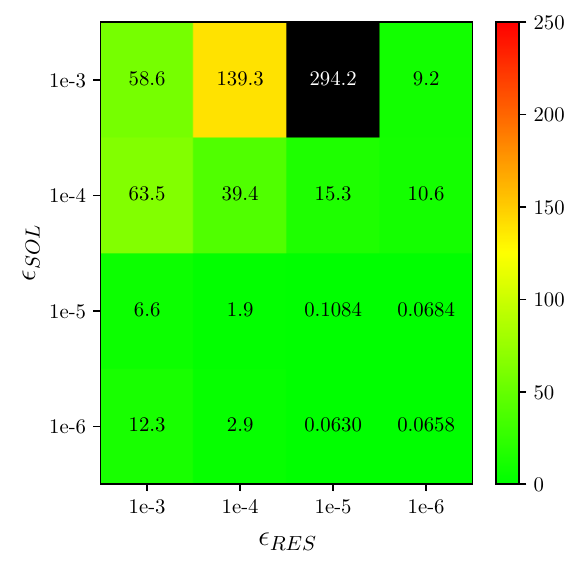}
    \caption{ $e({\boldsymbol{v}^*_y}_{\text{\tiny FOM}}, {\boldsymbol{v}^*_y}_{\text{\tiny HROM}} )$}
    %%\label{}
  \end{subfigure}
  \hfill
  \begin{subfigure}[b]{0.245\textwidth}
    \includegraphics[width=\linewidth]{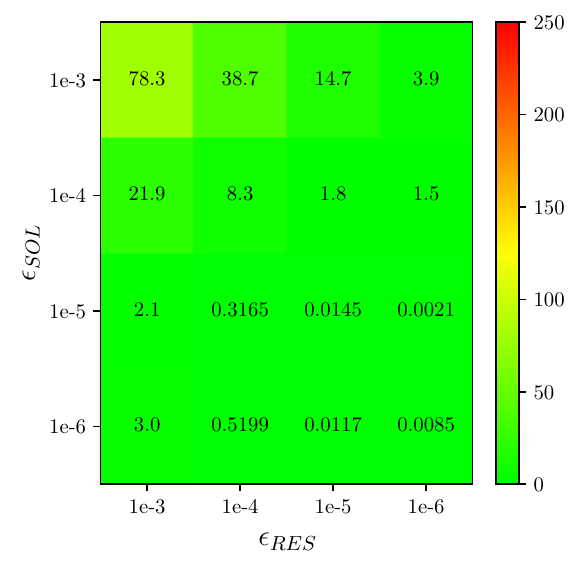}
    \caption{ $e(\boldsymbol{S}_{\text{\tiny ROM}},  \boldsymbol{S}_{\text{\tiny HROM}})$}
    %%\label{}
  \end{subfigure}
  \hfill
  \begin{subfigure}[b]{0.245\textwidth}
    \includegraphics[width=\linewidth]{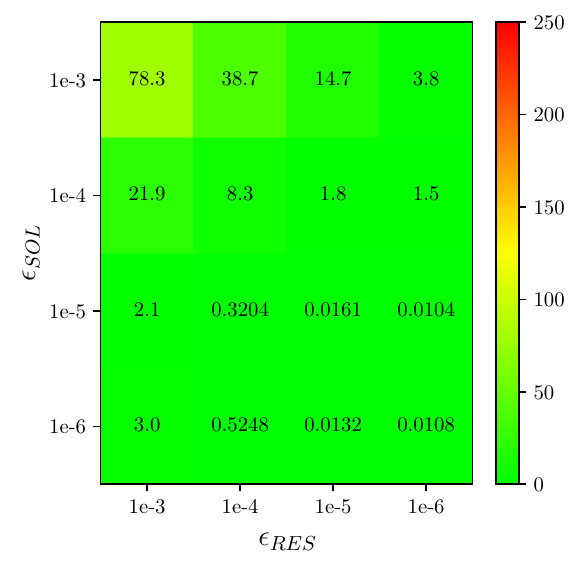}
    \caption{  $e(\boldsymbol{S}_{\text{\tiny FOM}},  \boldsymbol{S}_{\text{\tiny HROM}})$ }
    %%\label{}
  \end{subfigure}
  \caption{  \bothrev{Percentage error on QoI and solution field of ROM and FOM against various HROMs for the first training trajectory (Trajectory 2) for mapping $\varphi_{\text{\tiny FFD+RBF}}$}}
  \label{fig: Example 3 errors train HROM nonlinear}
\end{figure}

\newsec{Staying with the first training trajectory, Fig. \ref{fig: Example 3 train QoI HROM nonlinear} and Fig. \ref{fig: Example 3 errors train HROM nonlinear} show the performance of the HROMs for the nonaffine geometric mapping against their respective ROMs and FOM.}

\begin{figure}[H]
  \begin{subfigure}{.5\textwidth}
    \centering
    % include first image
    \includegraphics[width=.8\linewidth]{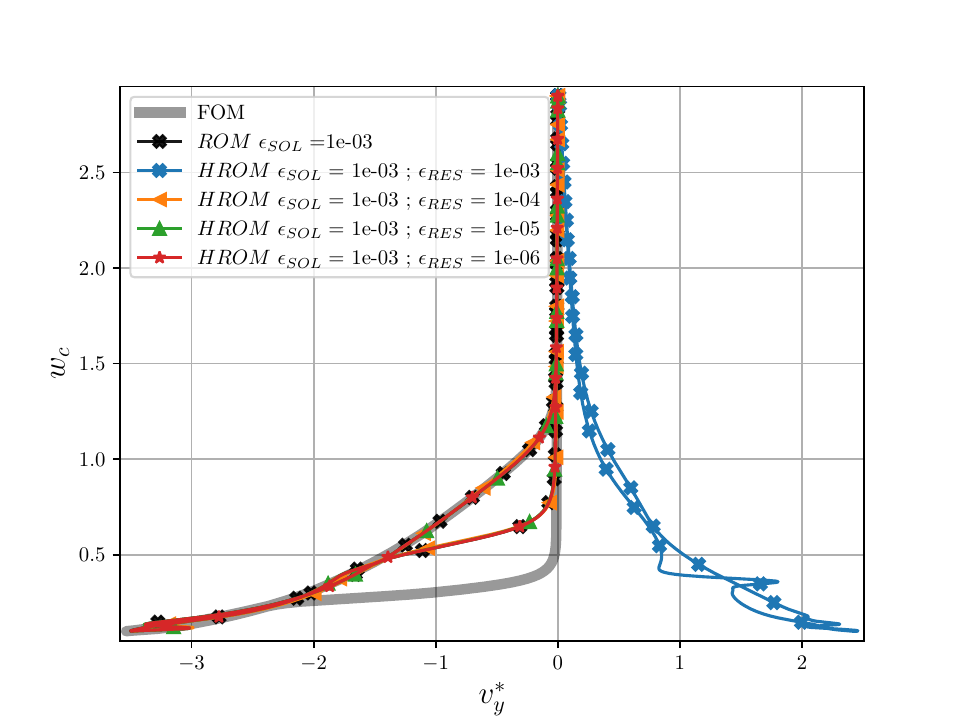}
    \caption{ROM $1e-3$ vs HROM}
  \end{subfigure}
  \begin{subfigure}{.5\textwidth}
    \centering
    % include second image
    \includegraphics[width=.8\linewidth]{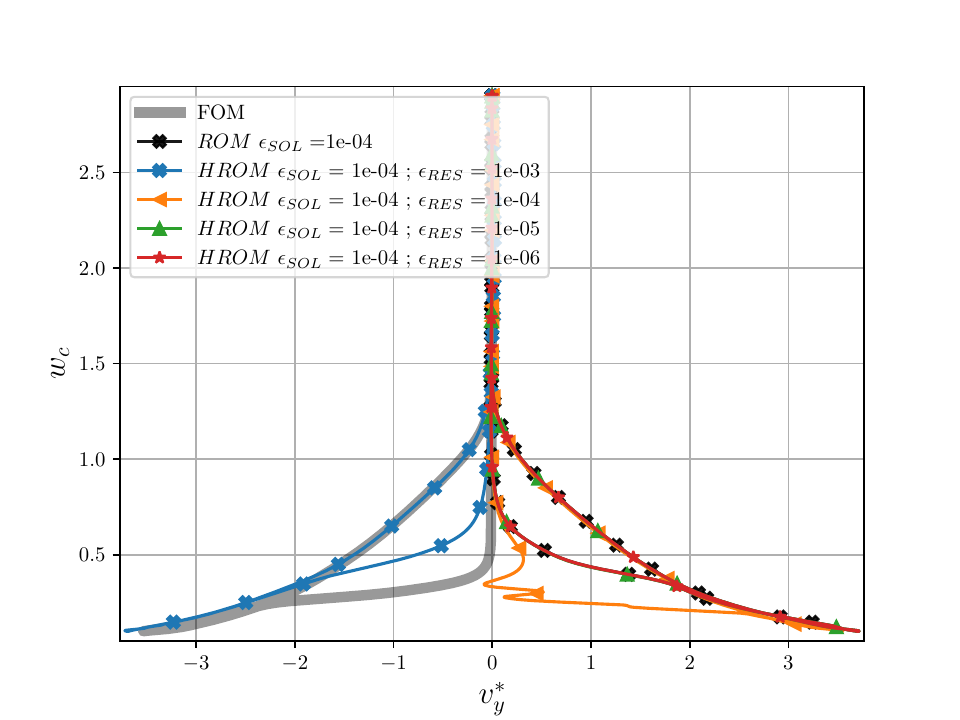}
    \caption{ROM $1e-4$ vs HROM}
  \end{subfigure}
  \begin{subfigure}{.5\textwidth}
    \centering
    % include third image
    \includegraphics[width=.8\linewidth]{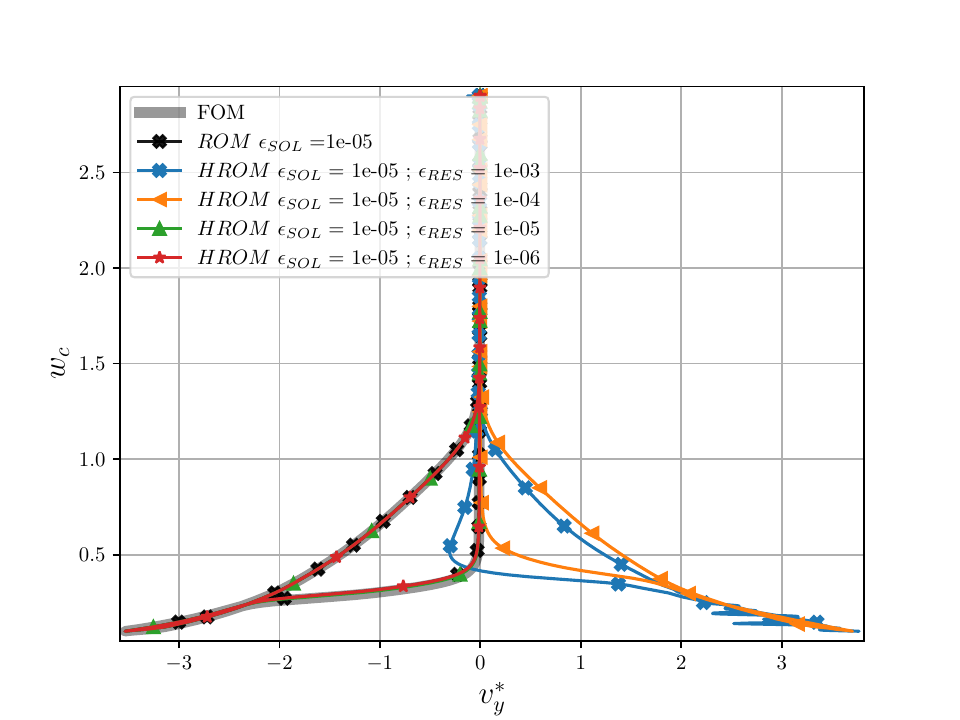}
    \caption{ROM $1e-5$ vs HROM}
  \end{subfigure}
  \begin{subfigure}{.5\textwidth}
    \centering
    % include fourth image
    \includegraphics[width=.8\linewidth]{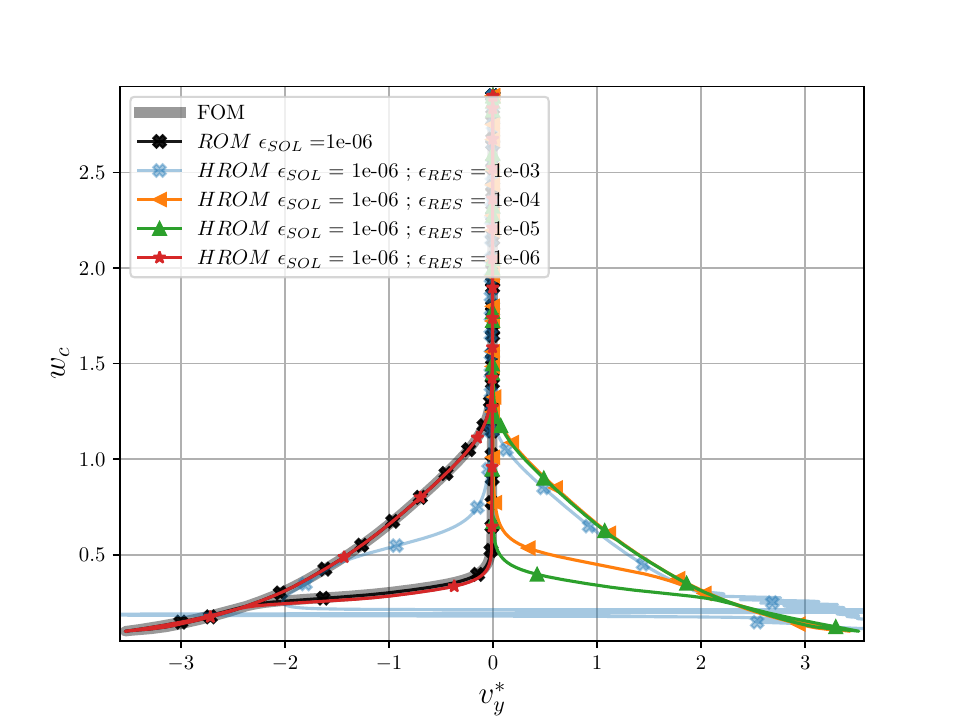}
    \caption{ROM $1e-6$ vs HROM}
  \end{subfigure}
\caption{ \bothrev{QoI phase space plot for the ROMs against various HROMs for the second training trajectory (Trajectory 1) for mapping $\varphi_{\text{\tiny AFFINE}}$}}
\label{fig: Example 3 test QoI HROM affine}
  \end{figure}

\begin{figure}[H]
  \centering
  \begin{subfigure}[b]{0.245\textwidth}
    \includegraphics[width=\linewidth]{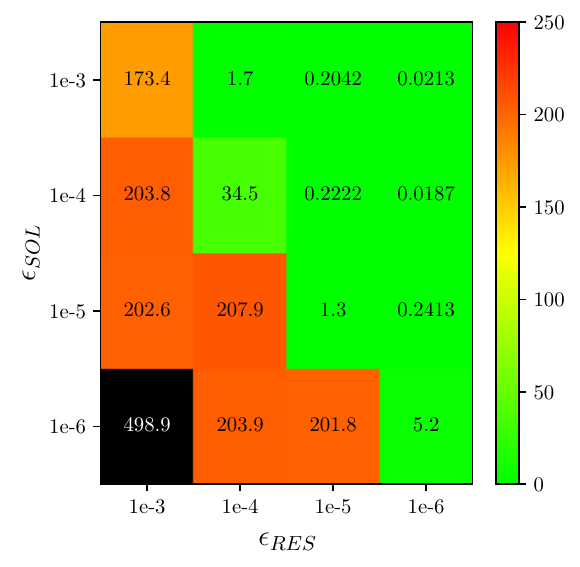}
    \caption{$e({\boldsymbol{v}^*_y}_{\text{\tiny ROM}}, {\boldsymbol{v}^*_y}_{\text{\tiny HROM}} )$}
    %%\label{}
  \end{subfigure}
  \hfill
  \begin{subfigure}[b]{0.245\textwidth}
    \includegraphics[width=\linewidth]{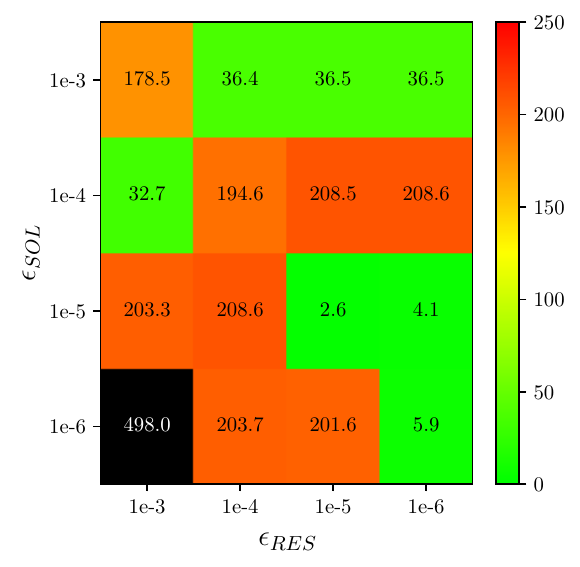}
    \caption{ $e({\boldsymbol{v}^*_y}_{\text{\tiny FOM}}, {\boldsymbol{v}^*_y}_{\text{\tiny HROM}} )$}
    %%\label{}
  \end{subfigure}
  \hfill
  \begin{subfigure}[b]{0.245\textwidth}
    \includegraphics[width=\linewidth]{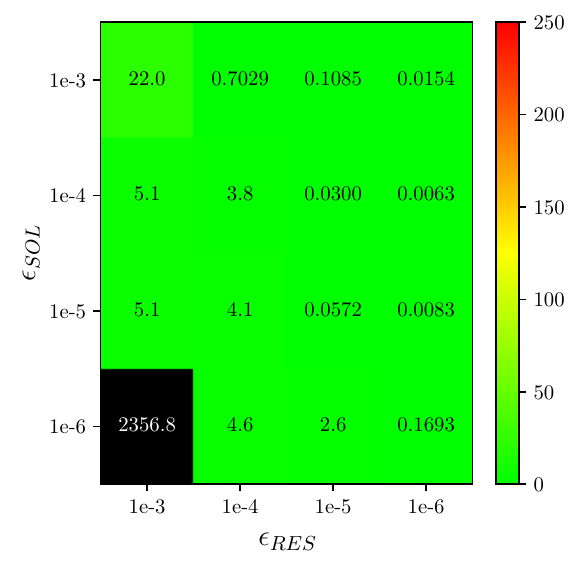}
    \caption{ $e(\boldsymbol{S}_{\text{\tiny ROM}},  \boldsymbol{S}_{\text{\tiny HROM}})$}
    %%\label{}
  \end{subfigure}
  \hfill
  \begin{subfigure}[b]{0.245\textwidth}
    \includegraphics[width=\linewidth]{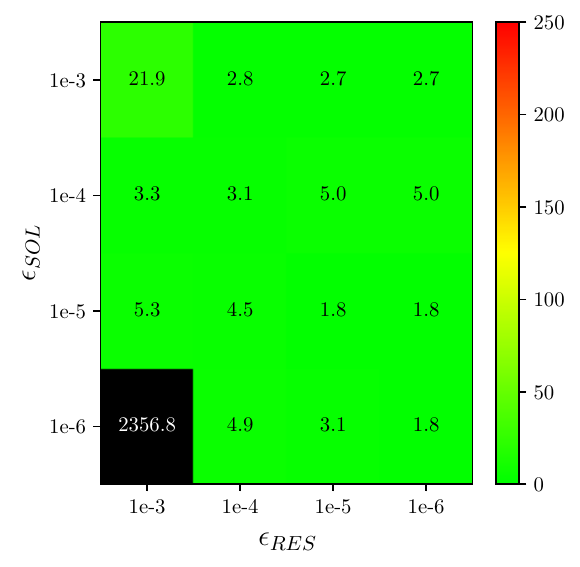}
    \caption{  $e(\boldsymbol{S}_{\text{\tiny FOM}},  \boldsymbol{S}_{\text{\tiny HROM}})$ }
    %%\label{}
  \end{subfigure}
  \caption{ \bothrev{Percentage error on QoI and solution field of ROM and FOM against various HROMs for the second training trajectory (Trajectory 1) for mapping $\varphi_{\text{\tiny AFFINE}}$ }}
  \label{fig: Example 3 errors test HROM affine}
\end{figure}

\newsec{Moving on to the second training trajectory, which for this example is Trajectory 1, Fig. \ref{fig: Example 3 test QoI HROM affine} and Fig. \ref{fig: Example 3 errors test HROM affine} show the performance of the HROMs for the affine geometric mapping aginst their respective ROMs and FOM.}

\begin{figure}[H]
  \begin{subfigure}{.5\textwidth}
    \centering
    % include first image
    \includegraphics[width=.8\linewidth]{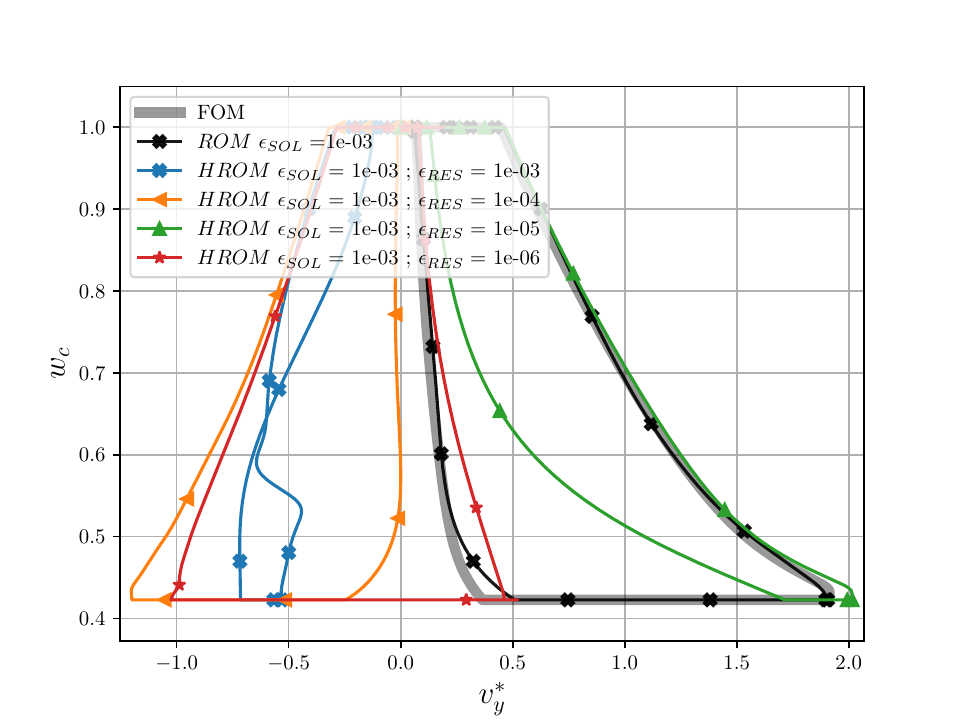}
    \caption{ ROM $1e-3$ vs HROM }
    %\label{fig: example 3 hysteresis test ffd rbf_a}
  \end{subfigure}
  \begin{subfigure}{.5\textwidth}
    \centering
    % include second image
    \includegraphics[width=.8\linewidth]{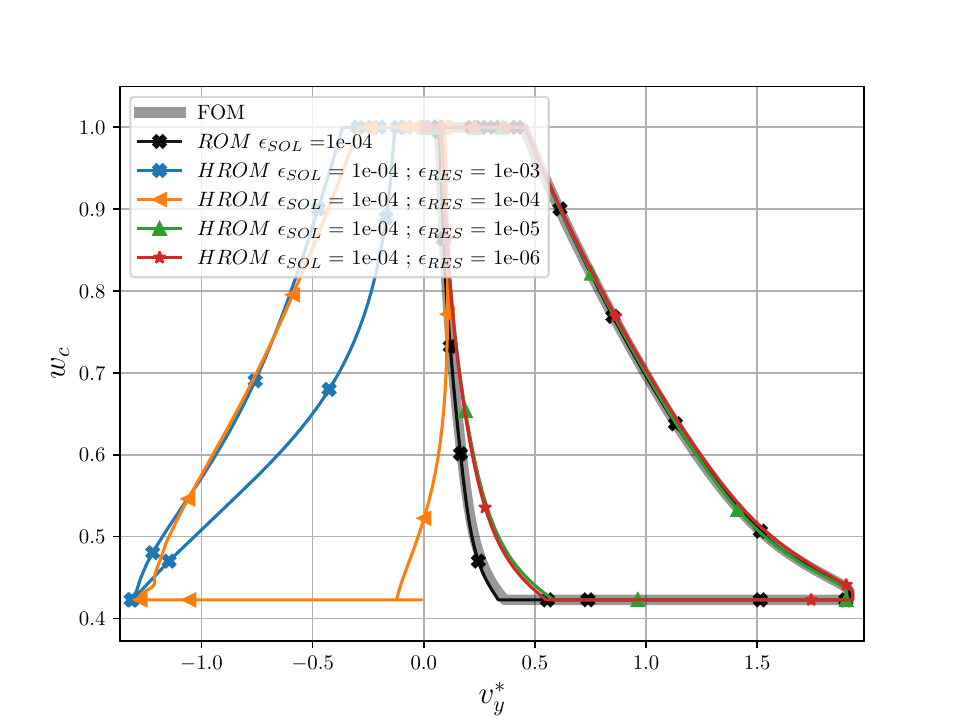}
    \caption{ROM $1e-4$ vs HROM}
    %\label{fig: example 3 hysteresis test ffd rbf_b}
  \end{subfigure}
  \begin{subfigure}{.5\textwidth}
    \centering
    % include third image
    \includegraphics[width=.8\linewidth]{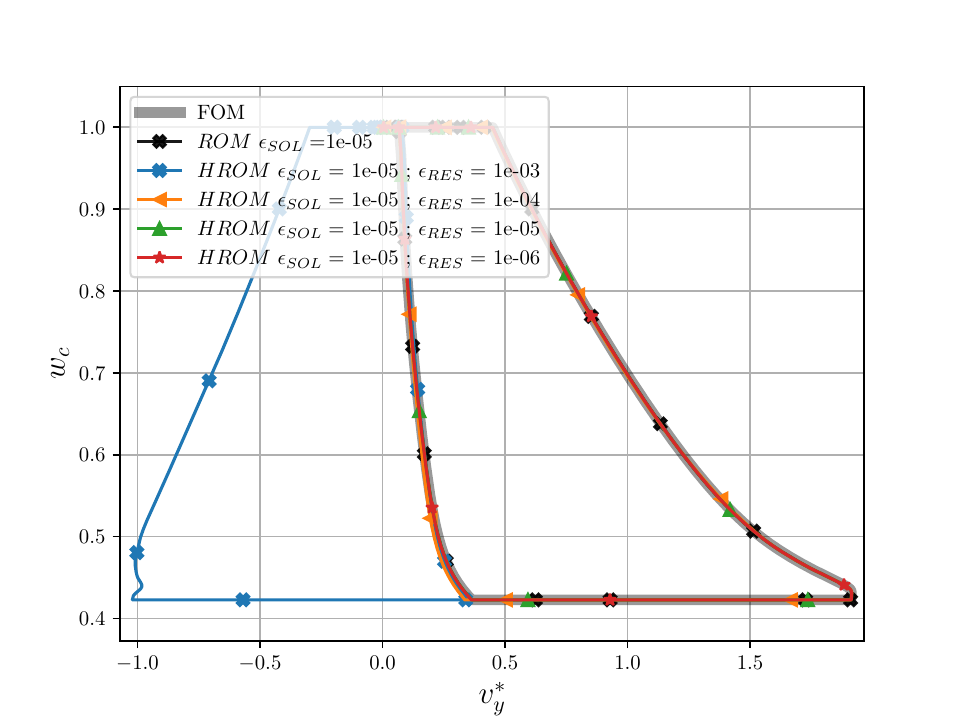}
    \caption{ROM $1e-5$ vs HROM}
    %\label{fig: example 3 hysteresis test ffd rbf_c}
  \end{subfigure}
  \begin{subfigure}{.5\textwidth}
    \centering
    % include fourth image
    \includegraphics[width=.8\linewidth]{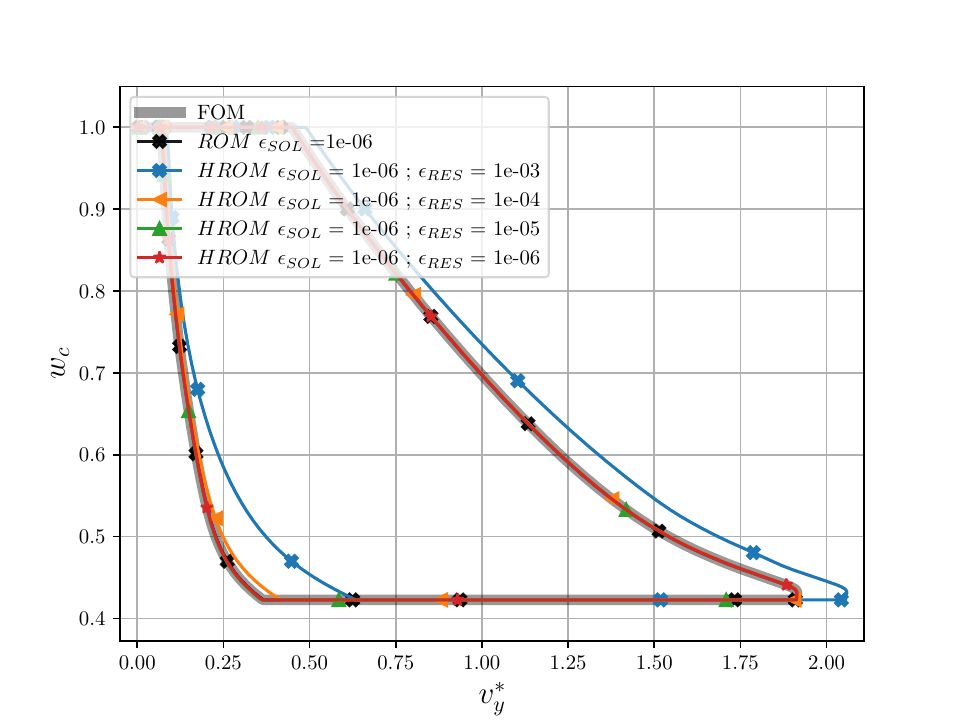}
    \caption{ROM $1e-6$ vs HROM}
    %\label{fig: example 3 hysteresis test ffd rbf_d}
  \end{subfigure}
\caption{\bothrev{ QoI phase space plot for the ROMs against various HROMs for the second training trajectory (Trajectory 1) for mapping $\varphi_{\text{\tiny FFD+RBF}}$  }}
\label{fig: Example 3 test QoI HROM nonlinear}
\end{figure}

\begin{figure}[H]
  \centering
  \begin{subfigure}[b]{0.245\textwidth}
    \includegraphics[width=\linewidth]{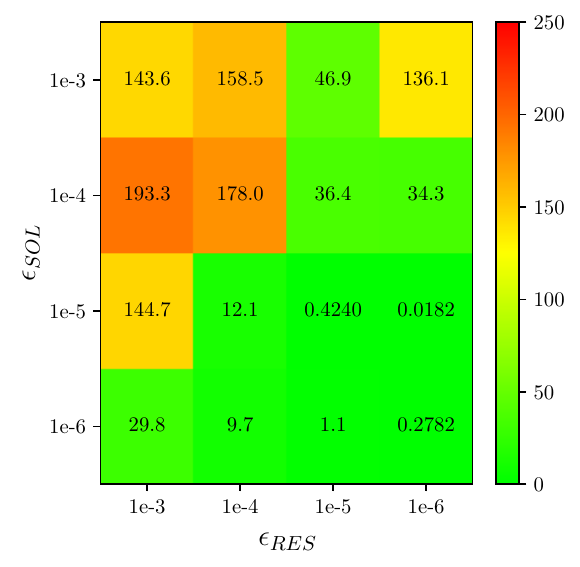}
    \caption{$e({\boldsymbol{v}^*_y}_{\text{\tiny ROM}}, {\boldsymbol{v}^*_y}_{\text{\tiny HROM}} )$}
    %%\label{}
  \end{subfigure}
  \hfill
  \begin{subfigure}[b]{0.245\textwidth}
    \includegraphics[width=\linewidth]{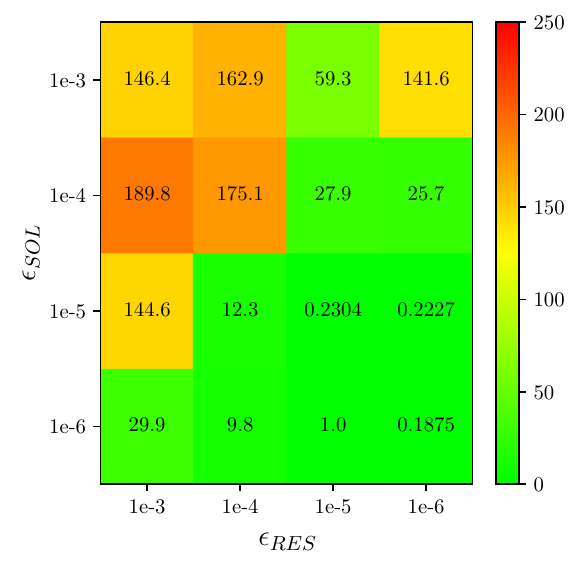}
    \caption{ $e({\boldsymbol{v}^*_y}_{\text{\tiny FOM}}, {\boldsymbol{v}^*_y}_{\text{\tiny HROM}} )$}
    %%\label{}
  \end{subfigure}
  \hfill
  \begin{subfigure}[b]{0.245\textwidth}
    \includegraphics[width=\linewidth]{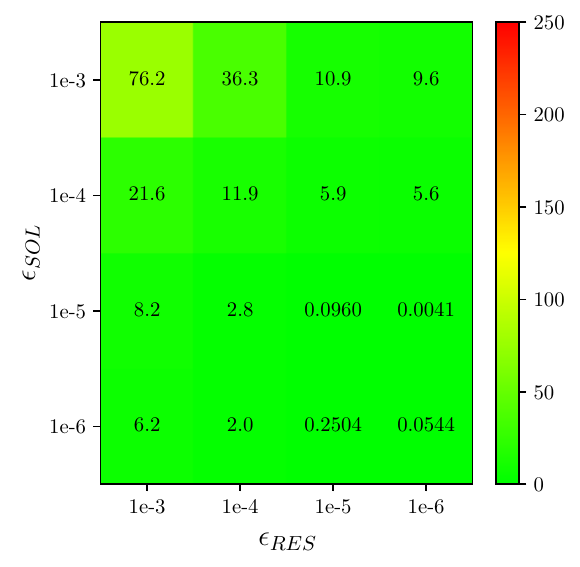}
    \caption{ $e(\boldsymbol{S}_{\text{\tiny ROM}},  \boldsymbol{S}_{\text{\tiny HROM}})$}
    %%\label{}
  \end{subfigure}
  \hfill
  \begin{subfigure}[b]{0.245\textwidth}
    \includegraphics[width=\linewidth]{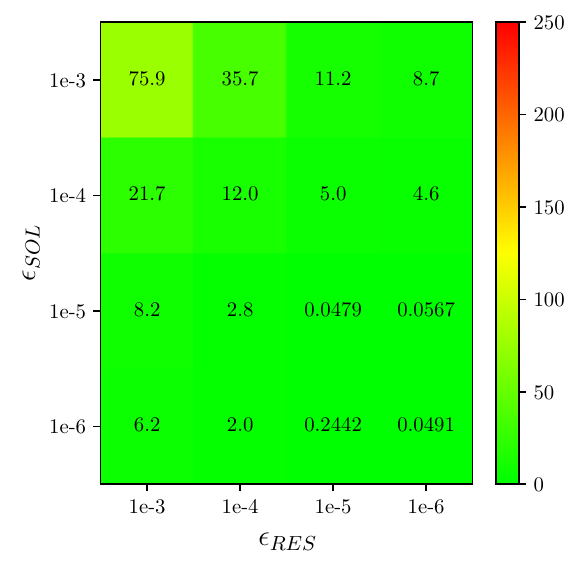}
    \caption{  $e(\boldsymbol{S}_{\text{\tiny FOM}},  \boldsymbol{S}_{\text{\tiny HROM}})$ }
    %%\label{}
  \end{subfigure}
  \caption{ \bothrev{Percentage error on QoI and solution field of ROM and FOM against various HROMs for the second training trajectory (Trajectory 1) for mapping $\varphi_{\text{\tiny FFD+RBF}}$ }}
  \label{fig: Example 3 errors test HROM nonlinear}
\end{figure}

\newsec{We finish the assesment of the training trajectories by examining the performance of the HROMs for the nonaffine geometric mapping for reproducing the second training trajectory on Fig. \ref{fig: Example 3 test QoI HROM nonlinear} and Fig. \ref{fig: Example 3 errors test HROM nonlinear}.}

\newsec{In this example, the results obtained for both the ROMs and HROMs in reproducing the training trajectories are consistent that is, as the tolerances are reduced, they converge towards their respective ROMs and FOM. Moreover, in the case of Trajectory 1, the results are comparable to those from the previous example for the same trajectory. This similarity is anticipated, given that the ROMs and HROMs are subjected in both cases to the trajectories used for their construction. To thoroughly assess their capabilities, we proceed to subject the ROMs and HROMs described in this example to Trajectory 3.}

\subsubsection{\newsec{Trajectory 3}}

\begin{figure}[H]
  \centering
  \begin{subfigure}[b]{0.45\textwidth}
    \includegraphics[width=\linewidth]{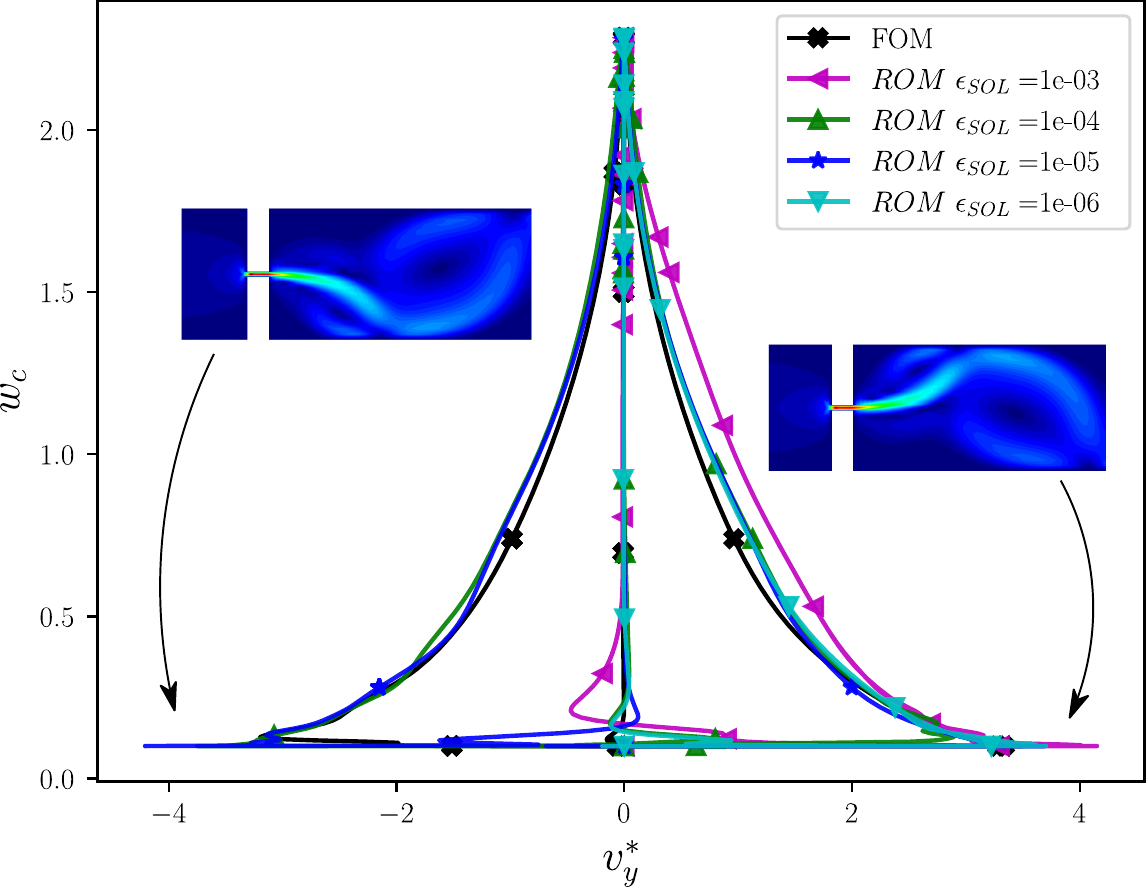}
    \caption{$\boldsymbol{\varphi}_{\text{\tiny AFFINE}}$ }
    %\label{fig: QoI ROM test a}
  \end{subfigure}
  \hfill
  \begin{subfigure}[b]{0.45\textwidth}
    \includegraphics[width=\linewidth]{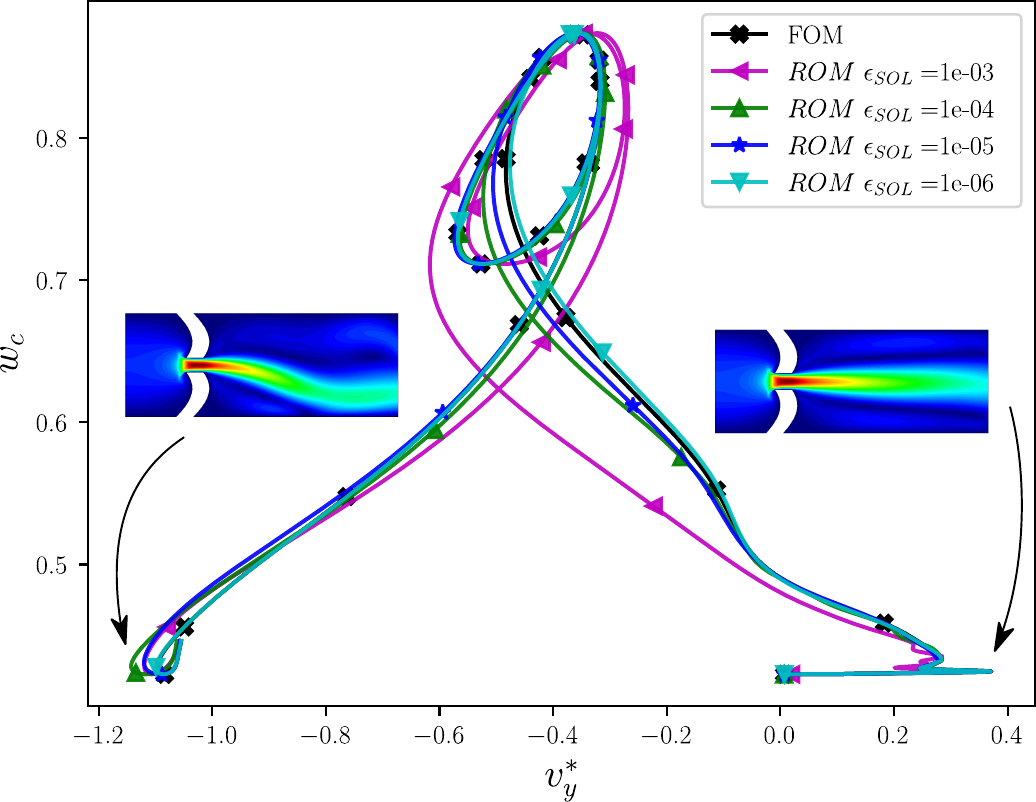}
    \caption{$\boldsymbol{\varphi}_{\text{\tiny FFD+RBF}}$ }
    %\label{fig: QoI ROM test b}
  \end{subfigure}
  \caption{\newsec{QoI phase space plot for the FOM against various ROMs for the testing trajectory (Trajectory 3) for both geometric mappings}}
  \label{fig: Example 4 train QoI ROM vs FOM}
\end{figure}

\begin{figure}[H]
  \centering
  \begin{subfigure}[b]{0.45\textwidth}
    \includegraphics[width=\linewidth]{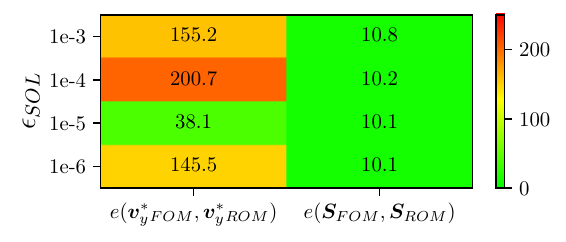}
    \caption{$\boldsymbol{\varphi}_{\text{\tiny AFFINE}}$ }
    %\label{}
  \end{subfigure}
  \hfill
  \begin{subfigure}[b]{0.45\textwidth}
    \includegraphics[width=\linewidth]{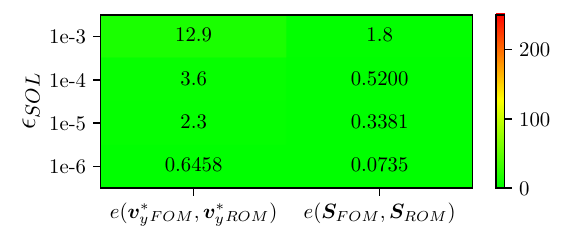}
    \caption{$\boldsymbol{\varphi}_{\text{\tiny FFD+RBF}}$ }
    %\label{}
  \end{subfigure}

  \caption{\bothrev{Percentage error on QoI and solution field of FOM against various ROMs for the testing trajectory (Trajectory 3) for both geometric mappings}}
  \label{fig: Example 2 errors test ROM vs FOM}
\end{figure}

\newsec{We now examine the performance of the ROMs for each geometric mapping in reproducing Trajectory 3, considered as the testing trajectory for this example. The results are depicted in Fig. \ref{fig: Example 4 train QoI ROM vs FOM} and Fig. \ref{fig: Example 2 errors test ROM vs FOM}.}

\newsec{For the affine geometric mapping, Trajectory 3 induces solutions attaching to both the upper and lower walls, posing a challenge for the ROMs to capture accurately. Notably, the ROM with the smallest error in the QoI is not associated with the smallest truncation tolerance $\epsilon_{\text{\tiny SOL}}$, but rather to the second smallest. However, the percentage error in the solution field is the same for both, at 10.1\%.}

\newsec{In contrast, the ROMs for the nonaffine geometric mapping effectively capture the behavior of the solutions, demonstrating errors in both the QoI and the solution field of less than 1\% for the smallest truncation tolerances.}

\begin{figure}[H]
  \begin{subfigure}{.5\textwidth}
    \centering
    % include first image
    \includegraphics[width=.8\linewidth]{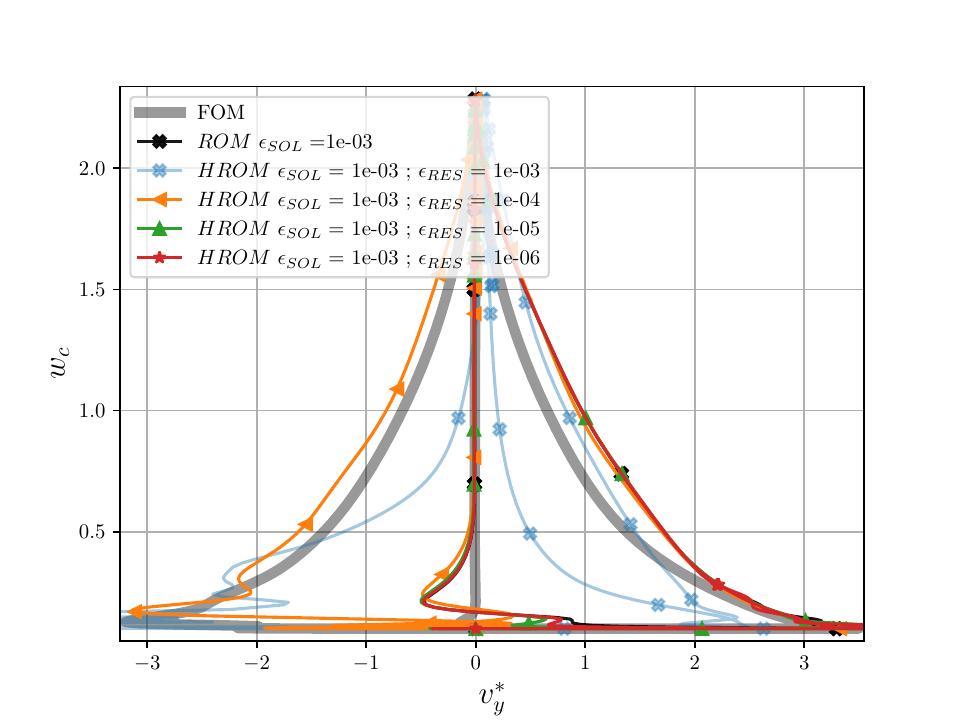}
    \caption{ROM $1e-3$ vs HROM}
  \end{subfigure}
  \begin{subfigure}{.5\textwidth}
    \centering
    % include second image
    \includegraphics[width=.8\linewidth]{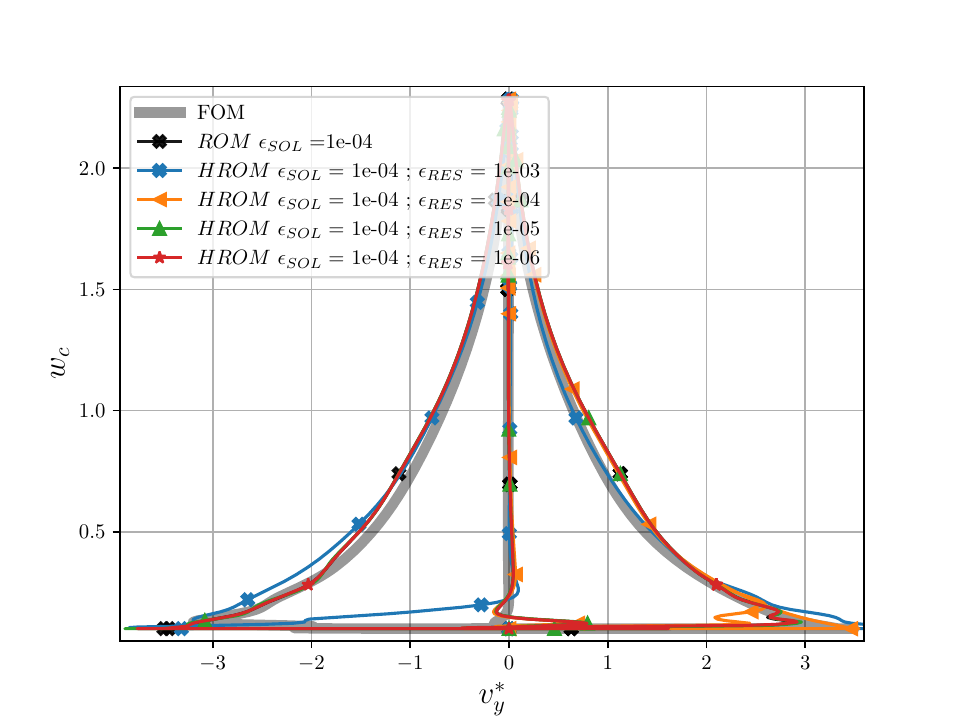}
    \caption{ROM $1e-4$ vs HROM}
  \end{subfigure}
  \begin{subfigure}{.5\textwidth}
    \centering
    % include third image
    \includegraphics[width=.8\linewidth]{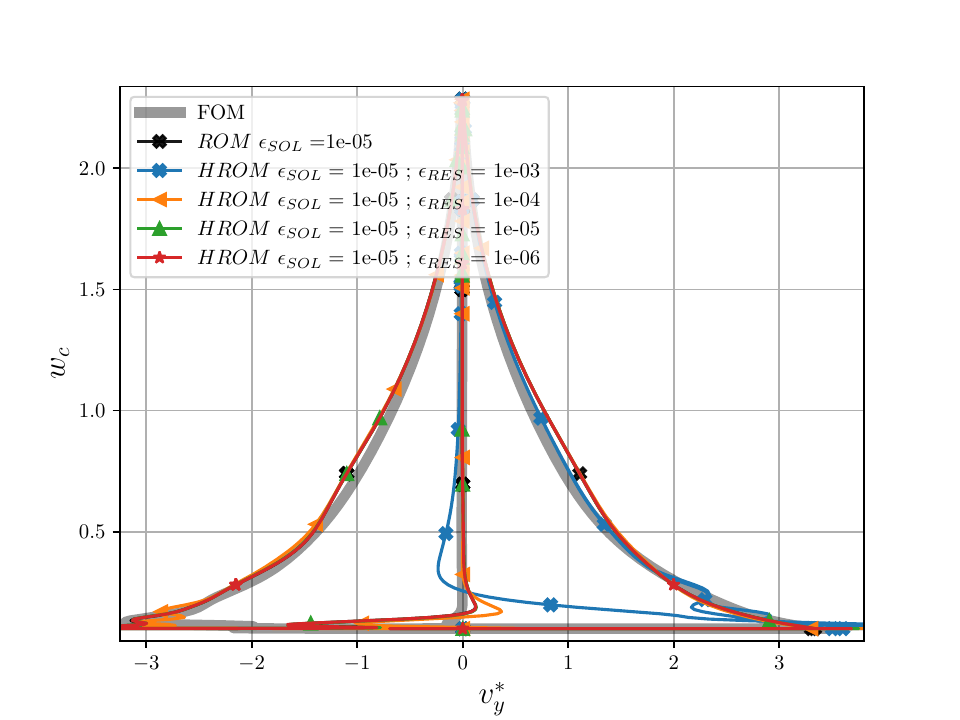}
    \caption{ROM $1e-5$ vs HROM}
  \end{subfigure}
  \begin{subfigure}{.5\textwidth}
    \centering
    % include fourth image
    \includegraphics[width=.8\linewidth]{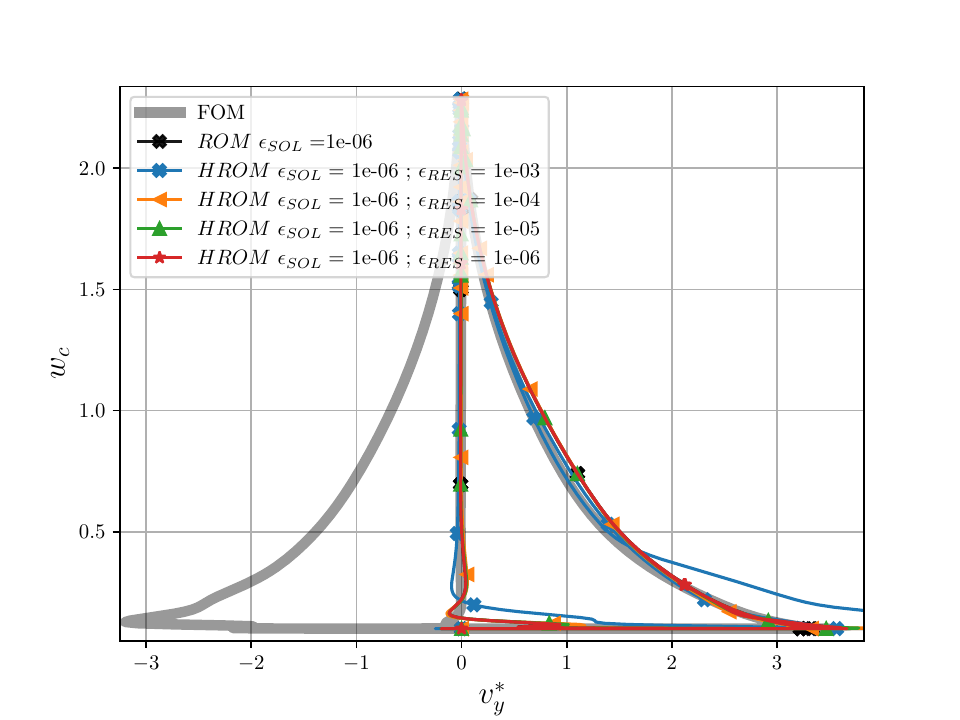}
    \caption{ROM $1e-6$ vs HROM}
  \end{subfigure}
\caption{ \bothrev{QoI phase space plot for the ROMs against various HROMs for the testing trajectory (Trajectory 3) for mapping $\varphi_{\text{\tiny AFFINE}}$  }}
\label{fig: Example 4 test QoI HROM affine}
\end{figure}

\begin{figure}[H]
  \centering
  \begin{subfigure}[b]{0.245\textwidth}
    \includegraphics[width=\linewidth]{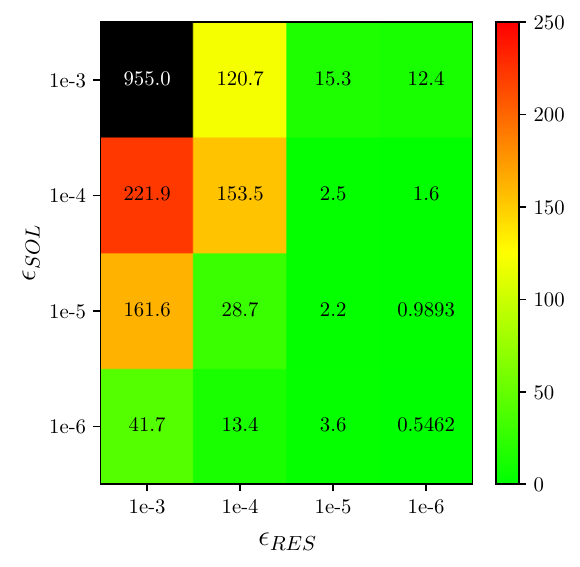}
    \caption{$e({\boldsymbol{v}^*_y}_{\text{\tiny ROM}}, {\boldsymbol{v}^*_y}_{\text{\tiny HROM}} )$}
    %%\label{}
  \end{subfigure}
  \hfill
  \begin{subfigure}[b]{0.245\textwidth}
    \includegraphics[width=\linewidth]{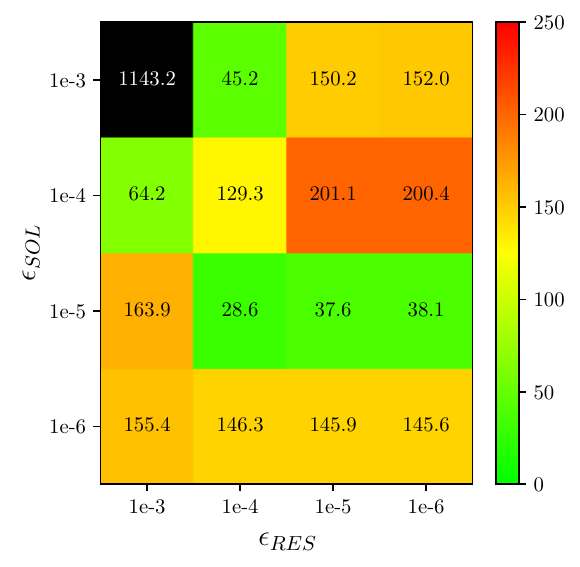}
    \caption{ $e({\boldsymbol{v}^*_y}_{\text{\tiny FOM}}, {\boldsymbol{v}^*_y}_{\text{\tiny HROM}} )$}
    %%\label{}
  \end{subfigure}
  \hfill
  \begin{subfigure}[b]{0.245\textwidth}
    \includegraphics[width=\linewidth]{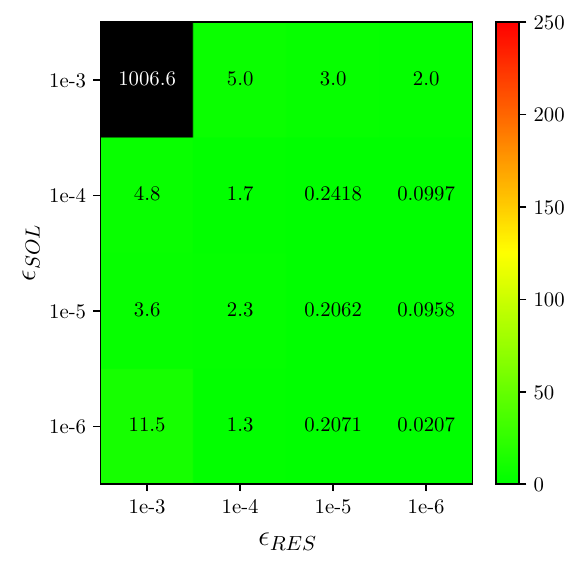}
    \caption{ $e(\boldsymbol{S}_{\text{\tiny ROM}},  \boldsymbol{S}_{\text{\tiny HROM}})$}
    %%\label{}
  \end{subfigure}
  \hfill
  \begin{subfigure}[b]{0.245\textwidth}
    \includegraphics[width=\linewidth]{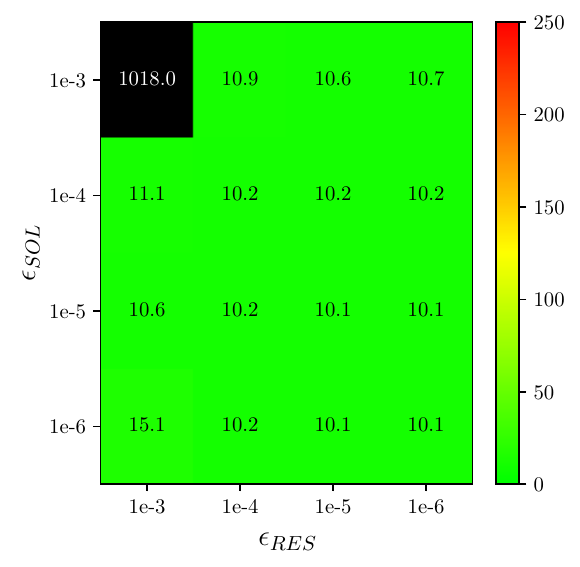}
    \caption{  $e(\boldsymbol{S}_{\text{\tiny FOM}},  \boldsymbol{S}_{\text{\tiny HROM}})$ }
    %%\label{}
  \end{subfigure}
  \caption{\bothrev{Percentage error on QoI and solution field of ROM and FOM against various HROMs for the testing trajectory (Trajectory 3) for mapping $\varphi_{\text{\tiny AFFINE}}$ }}
  \label{fig: Example 4 errors test HROM affine}
\end{figure}

\newsec{We proceed toevaluate the performance of the HROMs for reproducing the testing trajectory. Fig. \ref{fig: Example 4 test QoI HROM affine} and Fig. \ref{fig: Example 4 errors test HROM affine} illustrate the performance of the HROMs for the affine geometric mapping in comparison to their respective ROMs and FOM. The figures show that except for the model with the least stringent truncation tolerances, the remaining HROMs are capable of producing numerically stable solutions. Clearly, the HROMs with the second smallest truncation tolerance for the solution demonstrate the smallest errors in the QoI with respect to the FOM. This outcome follows from the ROM with a truncation tolerance $\epsilon_{\text{\tiny SOL}} = 1e-5$ being the best performing in terms of the QoI. In terms of the solution filed, all numerically stable HROMs deliver solutions around 10\% accurate with respect to the FOM. }

\begin{figure}[H]
  \begin{subfigure}{.5\textwidth}
    \centering
    % include first image
    \includegraphics[width=.8\linewidth]{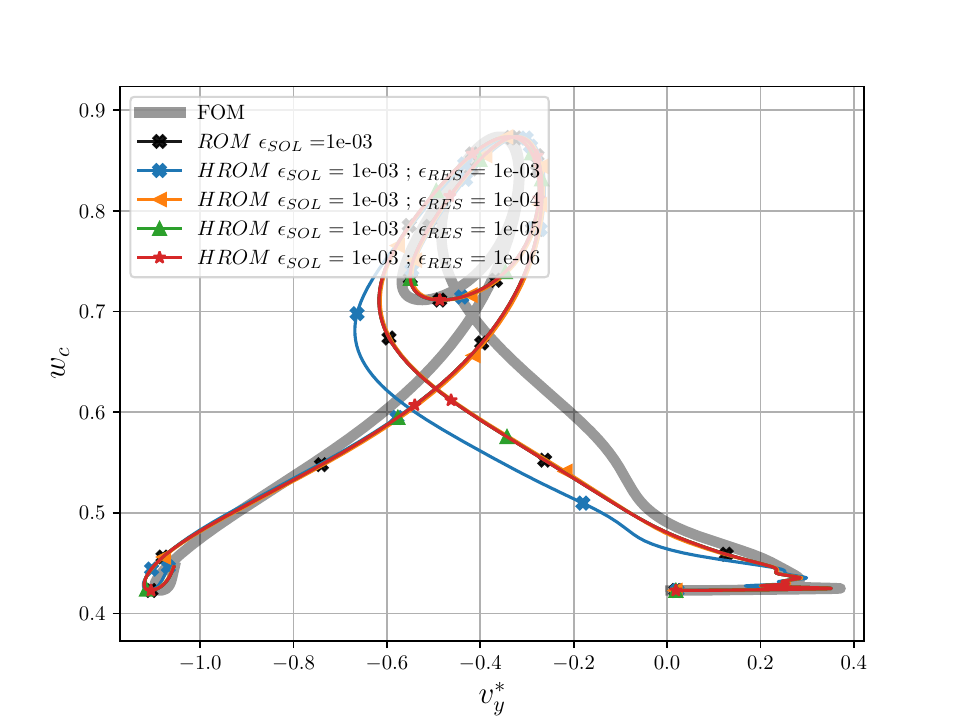}
    \caption{ ROM $1e-3$ vs HROM }
    %\label{fig: example 3 hysteresis test ffd rbf_a}
  \end{subfigure}
  \begin{subfigure}{.5\textwidth}
    \centering
    % include second image
    \includegraphics[width=.8\linewidth]{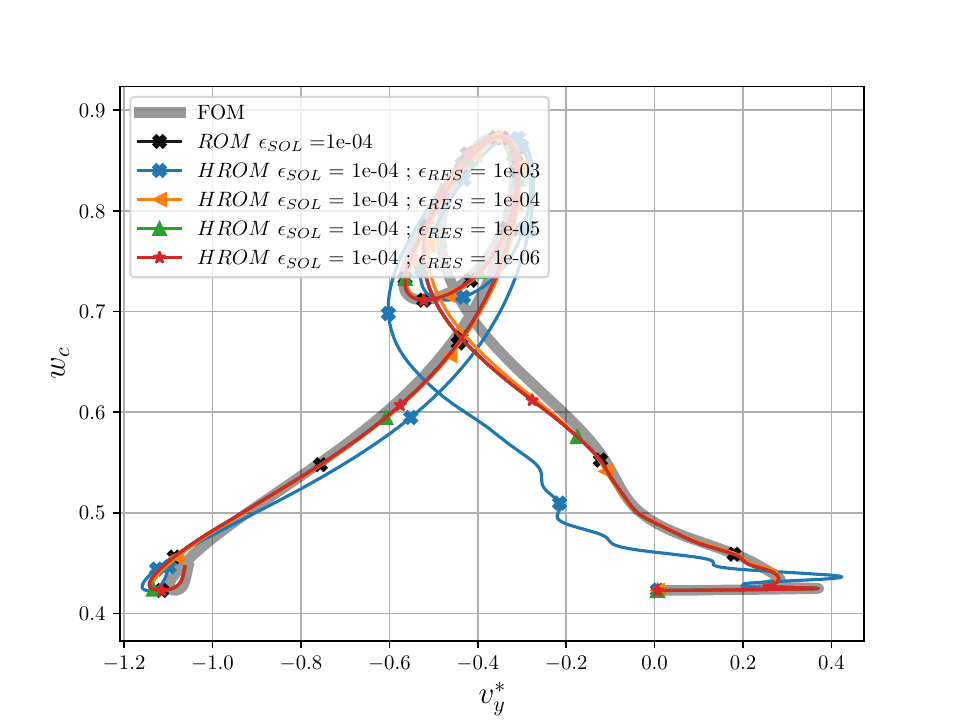}
    \caption{ROM $1e-4$ vs HROM}
    %\label{fig: example 3 hysteresis test ffd rbf_b}
  \end{subfigure}
  \begin{subfigure}{.5\textwidth}
    \centering
    % include third image
    \includegraphics[width=.8\linewidth]{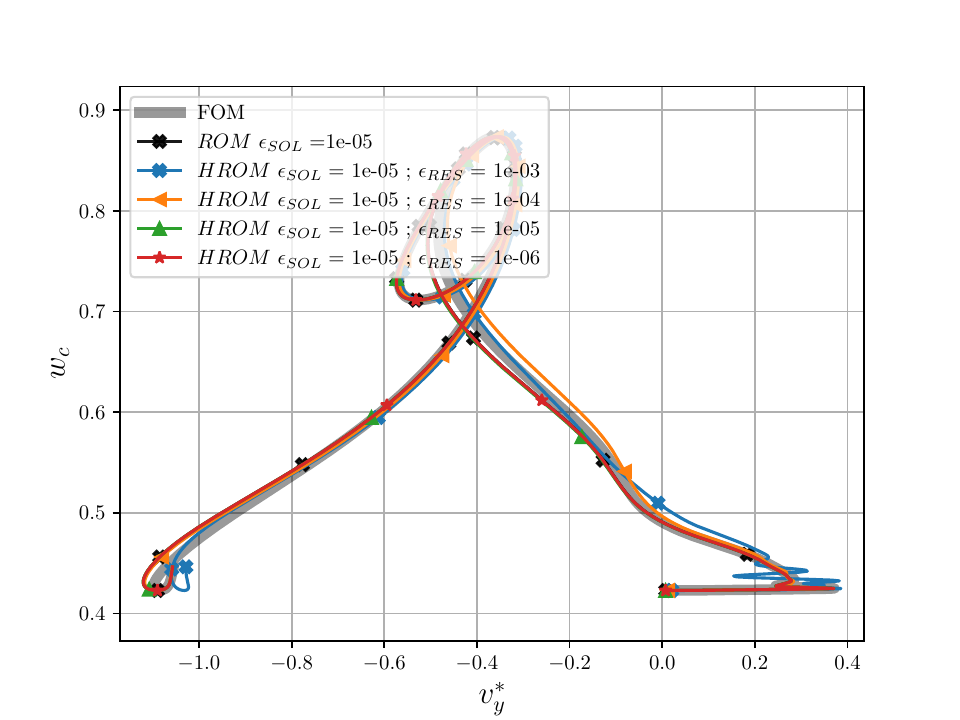}
    \caption{ROM $1e-5$ vs HROM}
    %\label{fig: example 3 hysteresis test ffd rbf_c}
  \end{subfigure}
  \begin{subfigure}{.5\textwidth}
    \centering
    % include fourth image
    \includegraphics[width=.8\linewidth]{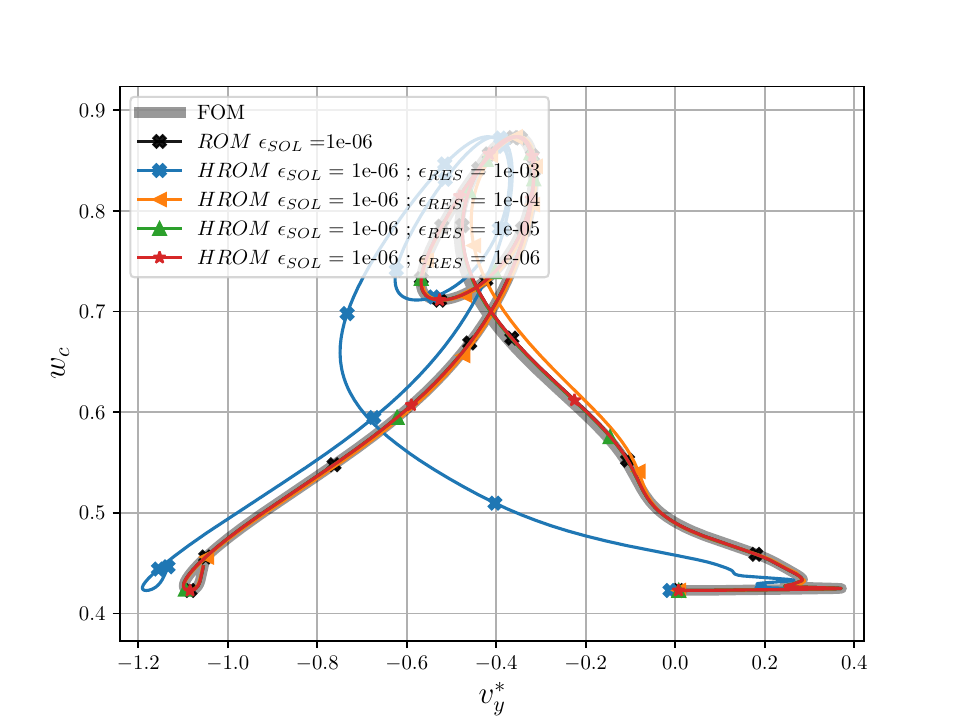}
    \caption{ROM $1e-6$ vs HROM}
    %\label{fig: example 3 hysteresis test ffd rbf_d}
  \end{subfigure}
  \caption{ \bothrev{Percentage error on QoI and solution field of ROM and FOM against various HROMs for the testing trajectory (Trajectory 3) for mapping $\varphi_{\text{\tiny FFD+RBF}}$  }}
  \label{fig: Example 4 errors test HROM linear}
\end{figure}

\begin{figure}[H]
  \centering
  \begin{subfigure}[b]{0.245\textwidth}
    \includegraphics[width=\linewidth]{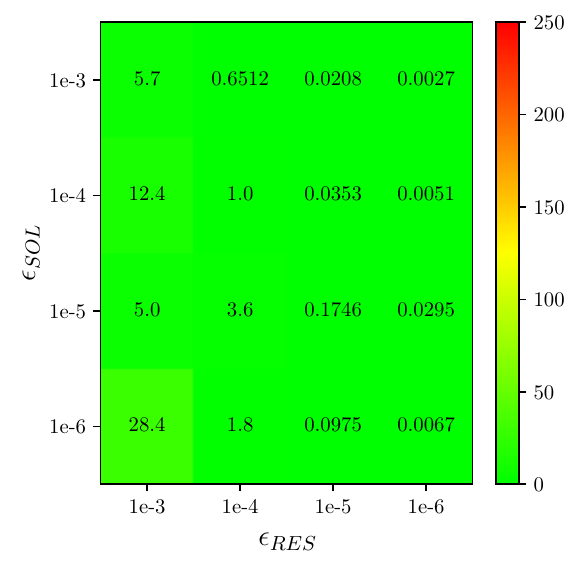}
    \caption{$e({\boldsymbol{v}^*_y}_{\text{\tiny ROM}}, {\boldsymbol{v}^*_y}_{\text{\tiny HROM}} )$}
    %%\label{}
  \end{subfigure}
  \hfill
  \begin{subfigure}[b]{0.245\textwidth}
    \includegraphics[width=\linewidth]{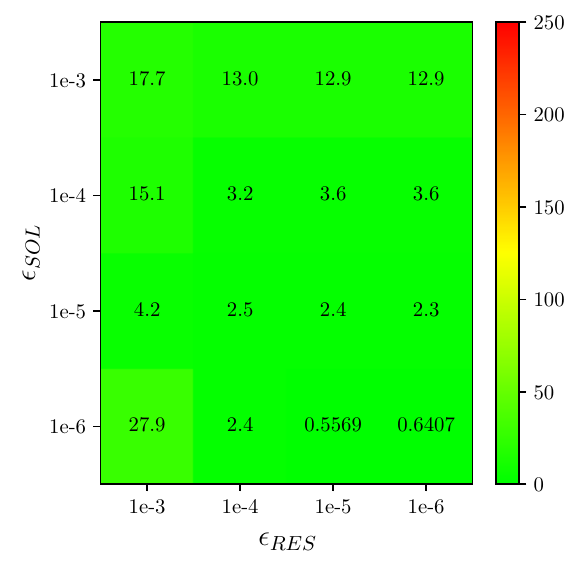}
    \caption{ $e({\boldsymbol{v}^*_y}_{\text{\tiny FOM}}, {\boldsymbol{v}^*_y}_{\text{\tiny HROM}} )$}
    %%\label{}
  \end{subfigure}
  \hfill
  \begin{subfigure}[b]{0.245\textwidth}
    \includegraphics[width=\linewidth]{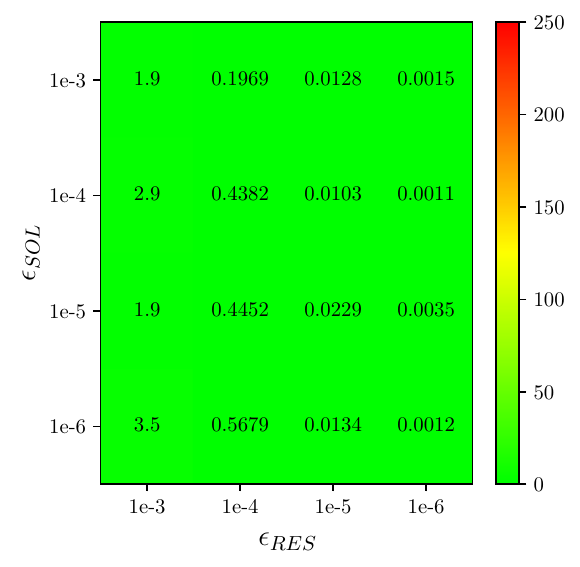}
    \caption{ $e(\boldsymbol{S}_{\text{\tiny ROM}},  \boldsymbol{S}_{\text{\tiny HROM}})$}
    %%\label{}
  \end{subfigure}
  \hfill
  \begin{subfigure}[b]{0.245\textwidth}
    \includegraphics[width=\linewidth]{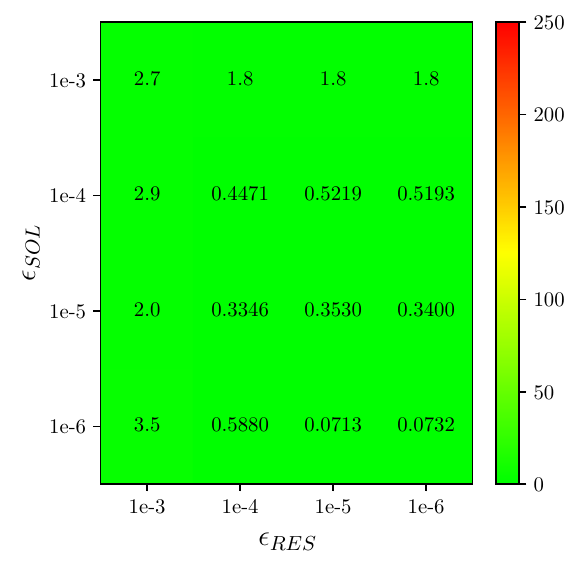}
    \caption{  $e(\boldsymbol{S}_{\text{\tiny FOM}},  \boldsymbol{S}_{\text{\tiny HROM}})$ }
    %%\label{}
\end{subfigure}
\caption{ \bothrev{Percentage error on QoI and solution field of ROM and FOM against various HROMs for the testing trajectory (Trajectory 3) for mapping $\varphi_{\text{\tiny FFD+RBF}}$  }}
\label{fig: Example 4 errors test HROM nonlinear}
\end{figure}

\newsec{We conclude our evaluation of the HROMs' performance in reconstructing the testing trajectory by examining Fig. \ref{fig: Example 4 errors test HROM linear} and Fig. \ref{fig: Example 4 errors test HROM nonlinear}. These figures show the HROMs for the nonaffine geometric mapping in reproducing Trajectory 3. Notably, for this trajectory all models exhibit relatively low errors compared to their performance when subjected to the two training trajectories (Trajectory 1 and Trajectory 2).}

\subsubsection{\newsec{Speedup}}

\begin{figure}[H]
  \centering
  \begin{subfigure}[b]{0.44\textwidth}
    \includegraphics[width=\linewidth]{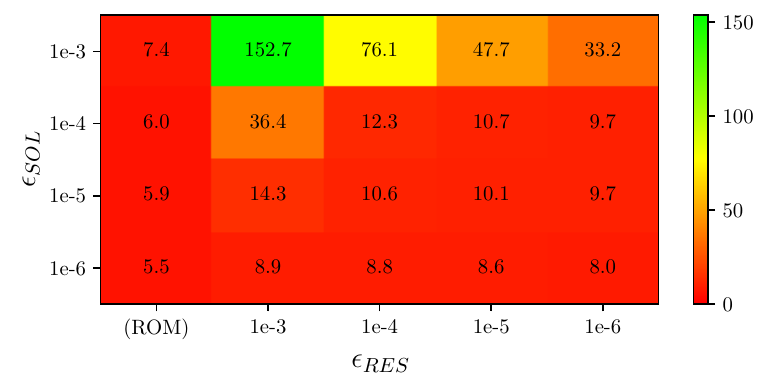}
    \caption{$\boldsymbol{\varphi}_{\text{\tiny AFFINE}}$ }
    %%\label{}
  \end{subfigure}
  \hfill
  \begin{subfigure}[b]{0.44\textwidth}
    \includegraphics[width=\linewidth]{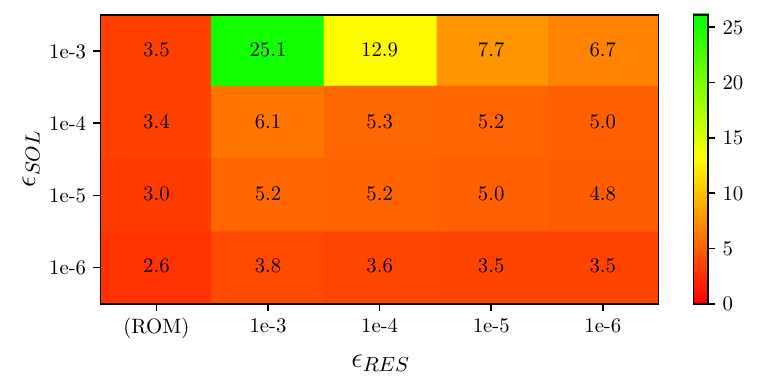}
    \caption{$\boldsymbol{\varphi}_{\text{\tiny FFD+RBF}}$ }
    %%\label{}
  \end{subfigure}

  \caption{\revone{Speedup factors for the ROMs and HROMs presented in this example. The first column displays the performance of the ROMs, while the subsequent columns depict the HROMs' performance for each combination of truncation tolerances $\epsilon_{\text{\tiny SOL}}$ and $\epsilon_{\text{\tiny RES}}$ considered}}
  \label{fig: Speedups Example 3}
\end{figure}

\newsec{Fig. \ref{fig: Speedups Example 3} shows an overview of the speedup factors achieved by all ROMs and HROMs with respect to the FOMs for both geometric mappings. The calculations are conducted according to Eq. \ref{eq:speedup definition}. A reduction in speedup is observed compared to Example 1, particularly for the models with less stringent truncation tolerances. However, as the tolerances are made tighter, the speedup factors closely resemble those reported in Example 1, with the observation that the results in this example exhibit greater robustness and precision.}

\subsubsection{\newsec{Discussion}}

\newsec{In this example, we examined the performance of ROMs and HROMs created using a training set that includes data from both stable branches of the bifurcation. Specifically, with snapshots containing a jet attaching to the upper and lower walls. The results showed that these models can accurately represent not only the two training trajectories but also a more complex testing trajectory.}

\subsection{\newsec{Summary of Results}}
\label{sec: summary results}

\newsec{As a way of summary, we can highlight three main points:}

\begin{itemize}
    \item \textbf{\newsec{On the richness of the training sets.}} \newsec{In order to develop robust ROMs and HROMs capable of adapting to unseen trajectories, it is essential to include snapshots from both stable branches of the bifurcation in the training sets. This necessity becomes evident when comparing the results from both examples. In example 1, the testing trajectories resulted in significant accuracy issues, because the FOM for the testing trajectory presented solutions on the opposite branch with respect to the training trajectory. In contrast, Example 2, which included both stable branches in its training trajectories, demonstrated improved accuracy for unseen testing trajectories.}
    \item \newsec{\textbf{On the truncation tolerance of the solutions} $\boldsymbol{\epsilon_{\text{\tiny SOL}}.}$ The truncation tolerance of the snapshots matrix of solutions sets an upper bound on the achievable accuracy of the ROM, and subsequently, its derived HROMs. Opting for larger truncation tolerances does not inherently result in numerically unstable models, as observed in all studied ROMs, where numerical stability was maintained regardless of the incurred errors. However, in scenarios involving both stable branches of the bifurcation, and therefore not limited by this fact to preferring a given solution, such as in Example 2, selecting excessively large values for $\epsilon_{\text{\tiny SOL}}$ caused the ROMs to choose the opposite stable branch compared to the FOM. While this deviation is still physically permissible, it may be significant enough to render some models unsuitable depending on the specific application.}
    \item \newsec{\textbf{On the truncation tolerance of the residuals} $\boldsymbol{\epsilon_{\text{\tiny RES}}.}$} \newsec{HROMs constructed with a truncation tolerance of $\epsilon_{\text{\tiny RES}} = 1e-3$ exhibited considerable deviations from their corresponding ROMs, to the extent of becoming numerically unstable. This observation holds true, especially in the context of Example 1, but it remains applicable even in scenarios where the POD bases were enriched with information from both stable branches of the bifurcation, as in Example 2. Opting for a more restrictive tolerance, such as $\epsilon_{\text{\tiny RES}} \leq 1e-4$, appears necessary to ensure that the HROMs provide accurate solutions with respect to their ROMs.}
\end{itemize}

\newsec{Based on these observations, the adoption of the presented HROM framework for characterising the hysteresis of the Coanda effect appears to involve a trade-off between accuracy and efficiency. Fast HROMs, while capable of capturing the overall solution trend, may struggle to select the same branch as the FOM, even when such a solution is achievable with the POD basis. On the other hand, more accurate models that reproduce intricate solutions can be obtained by choosing sufficiently stringent truncation tolerances, albeit resulting in smaller speedup factors.}

Finally, we \own{acknowledge} that we detected a significant increase in assembly time in our implementation as the number of POD modes increased. In scenarios where a \own{relatively} large number of modes are involved, we deviated from the element-by-element approach outlined in Section \ref{subsec: Reduced Order Model} for assembling the system of equations. Instead, \own{for these cases} we adopted a ``global" approach. This entailed assembling the sparse system matrix in a manner similar to the \revone{FOM}, followed by sparse-dense, and posterior dense-dense matrix product with the complete basis matrix $\boldsymbol{\Phi}$. We observed that the element-by-element formulations consistently outperformed \own{this} global approach for cases considering \newsec{less than 60 POD modes, and a reduced mesh containing less than one fifth of the total elements. The speedup factors reported at the end of each example} show the superior formulation between the two options.

\section{Conclusions and Perspectives}
\label{sec: Conclusions}

In this paper, our focus was on investigating a general ROM framework for addressing fluid dynamics problems with \own{time-dependent} geometric parametrisations. This framework encompasses the utilisation of two powerful techniques: \revone{POD-Galerkin together with ECM} hyperreduction. By employing these techniques, we aimed to effectively capture the intricate fluid behavior inherent in the contraction-expansion channel geometry. While this geometry offers a relatively straightforward setting, it still presents complex fluid dynamics phenomena, such as a bifurcating solution known as Coanda effect.

By utilising ROMs and \revone{HROMs}, we have successfully constructed accurate models capable of capturing both the training trajectories, which represent a specific deformation of the geometry over time, and \newsec{ challenging testing trajectories, which either trigger the opposite branch of the bifurcation compared to the training trajectory, or introduce a more complicated deformation sequence.}

We have analysed the solution behavior in a \own{phase space}, specifically focusing on a \revone{QoI}, which is the velocity in the \own{$y$}-direction at a probe point. This QoI allows for the detection and characterisation of phenomena such as the Coanda effect and its hysteresis. By qualitatively assessing the outputs of the ROMs and HROMs in this \own{phase space} plot, we gain insights into the performance of the models. Additionally, quantitative evaluations have been conducted to assess the accuracy of the complete solution field and the QoI.

As discussed in \own{Sec. \ref{sec: summary results}, the behaviour of the ROMs and HROMs suggest} a trade-off between accuracy and computational speedup. The HROM models exhibit significant speedups, while still providing physically acceptable and bounded solutions\own{. However,} these models incur relatively large errors in reproducing the complete solution field and the QoI \revone{of FOMs}, particularly for testing trajectories. Despite these errors, these models can still be useful in applications where the detection of the Coanda effect is crucial, even if the selected bifurcation branch is incorrect. \newsec{This is evident, for instance, in Example 2, where the HROMs successfully capture the overall trend of the solution for unseen testing trajectories, despite the fact that the obtained solution does not match the FOM solution.} For more accurate results, HROMs offering \own{moderate} speedups while maintaining low errors can be employed. \newsec{For applications demanding very high levels of accuracy for unseen trajectories, along with very high speedup factors, the incorporation of nonlinear ROM techniques could be considered, as outlined in Future Work.}

\subsection{Future Work}

There are several promising avenues for further advancing this research.

Firstly, it is important to acknowledge that our study focused on a single parameter variation within a mesh comprising a relatively small number of elements, which was suitable for our academic objectives. However, in scenarios where higher resolution and increased accuracy are required, it becomes imperative to employ a larger number of elements in the base model. Additionally, an effective reduced order model should be trained by exploring a multidimensional parameter space. Addressing these considerations necessitates addressing certain challenges within the framework presented in this paper. Specifically, the launch and analysis of simulations, as well as the management of the generated matrices via singular value decomposition, become computationally demanding on a single machine.

As mentioned in Section \ref{subsec: Hyper-Reduction}, the size of the snapshots matrices increases with the number of elements and the number of POD modes.  To alleviate this challenge, we are actively engaged in the development of parallelisation techniques for the entire workflow. This includes parallel simulation orchestration, efficient data management strategies, and the implementation of parallel algorithms for computing the singular value decomposition. These parallelisation efforts aim to significantly enhance the computational efficiency and scalability of the training process, enabling the exploration of larger parameter spaces and higher fidelity models.

Furthermore, simulations involving qualitatively distinct solutions, such as the ones demonstrated in this paper, often require a large number of modes to capture the intricate behavior of the solutions. To address this challenge, we are exploring alternative strategies. One approach involves utilising multiple piece-wise linear bases that effectively capture the specific behavior in the vicinity of a particular region in the parametric space, as demonstrated in \cite{hess2019localized, HessQuainiRozza_ICCS_2022}. Additionally, we are investigating the utilisation of nonlinear manifolds, particularly quadratic approximations as proposed in \cite{jain2017quadratic, barnett2022quadratic}, to mitigate the requirement of a high number of modes. Lastly, we are actively exploring the application of autoencoder neural networks as a form of generic manifold Galerkin approximation \cite{lee2020model, romor2022non}. We anticipate that the results of these advancements will be reported in subsequent papers, expanding upon the findings presented in this study.

\section*{Acknowledgements}

This project has received funding from the European High-Performance Computing Joint Undertaking (JU) under grant agreement No 955558. The JU receives support from the European Union’s Horizon 2020 research and innovation programme and Spain, Germany, France, Italy, Poland, Switzerland, Norway.

This publication is part of the R\&D project PCI2021-121944, financed by MICIU/AEI/10.13039/501100011033/ and by the ``European Union NextGenerationEU/PRTR”.

J.R. Bravo acknowledges the financial support of the Departament de Recerca i Universitats de la Generalitat de Catalunya through the FI-SDUR 2020 scholarship.

J.A Hern\'{a}ndez acknowledges project PID2021-122518OB-I00 financed by MICIU/AEI/10.13039/501100011033/ and by FEDER, EU.

\bibliographystyle{abbrv}
\bibliography{sample.bib}

\section{\append{Appendix}}
\label{sec: appendix}

\subsection{\append{Visualisation of the POD Modes and the Selected Elements}}

\append{In this Appendix, we present the POD modes obtained in both examples and compare them with the elements selected by the ECM algorithm. Due to the Galerkin projection employed in this paper, the location of the selected elements by the ECM algorithm matches the patterns present in the POD modes.}

\subsubsection{\append{Example 1. Affine Mapping}}

\begin{figure}[H]
  \centering
  \begin{subfigure}[b]{0.245\textwidth}
    \includegraphics[width=\linewidth]{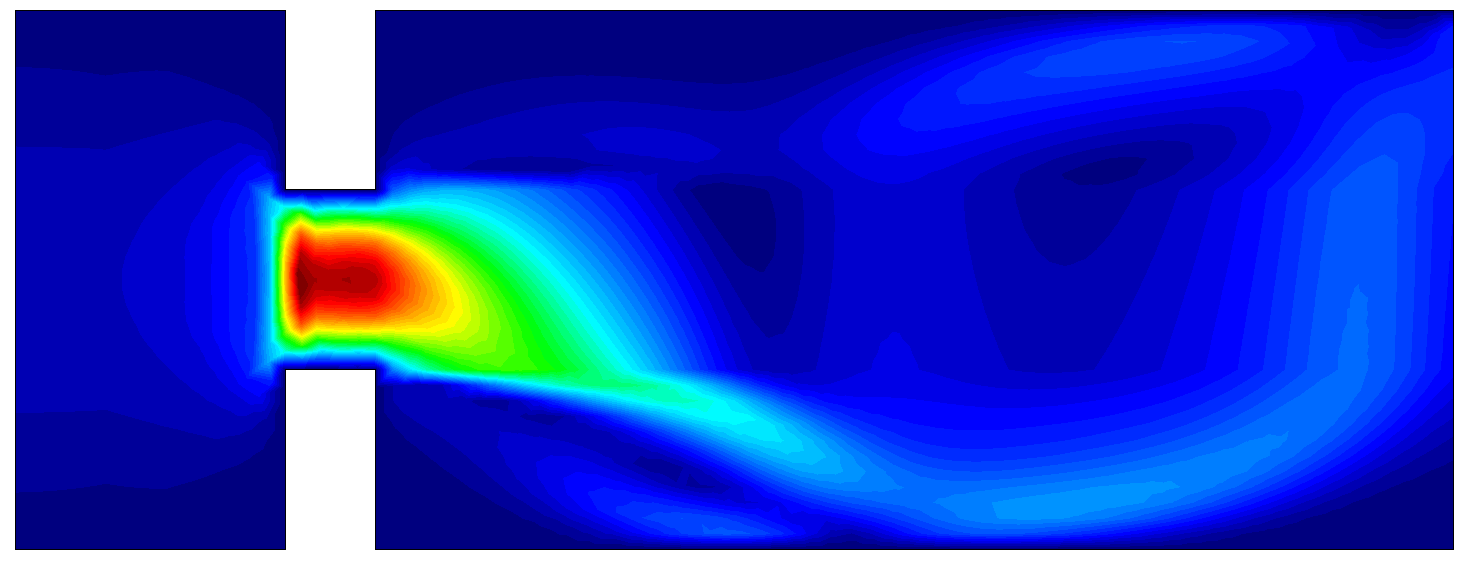}
    \caption{velocity mode 1}
  \end{subfigure}
  \hfill
  \begin{subfigure}[b]{0.245\textwidth}
    \includegraphics[width=\linewidth]{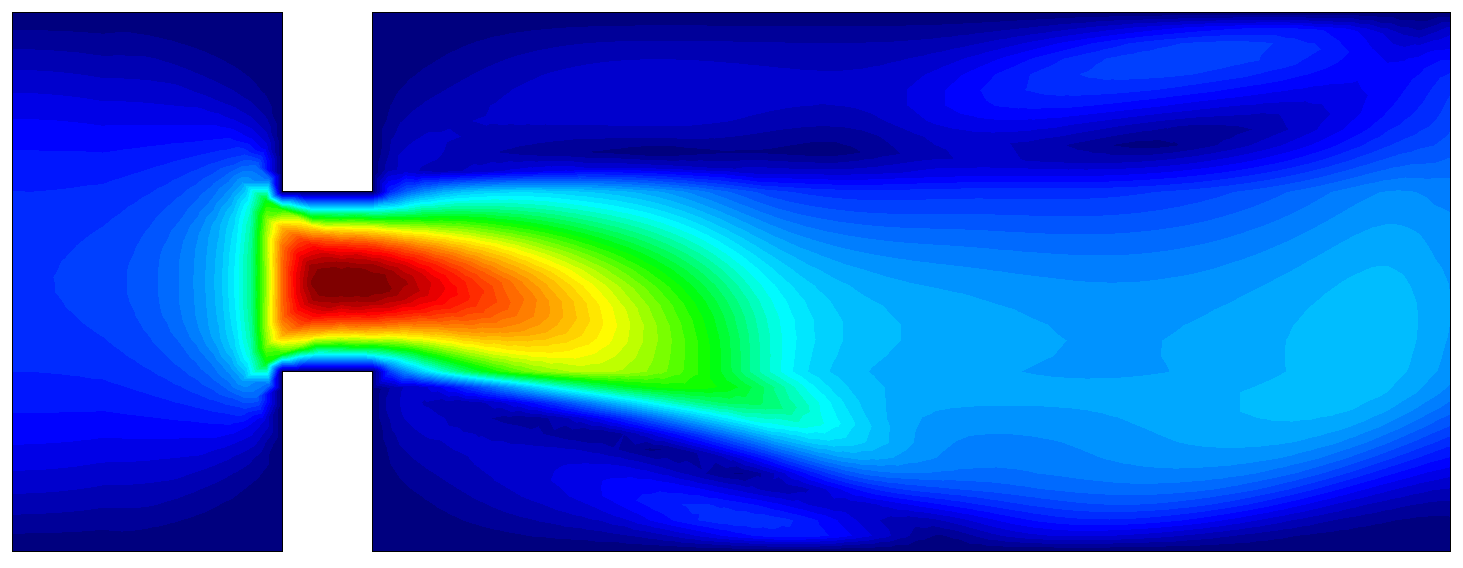}
    \caption{velocity mode 2}
  \end{subfigure}
  \hfill
  \begin{subfigure}[b]{0.245\textwidth}
    \includegraphics[width=\linewidth]{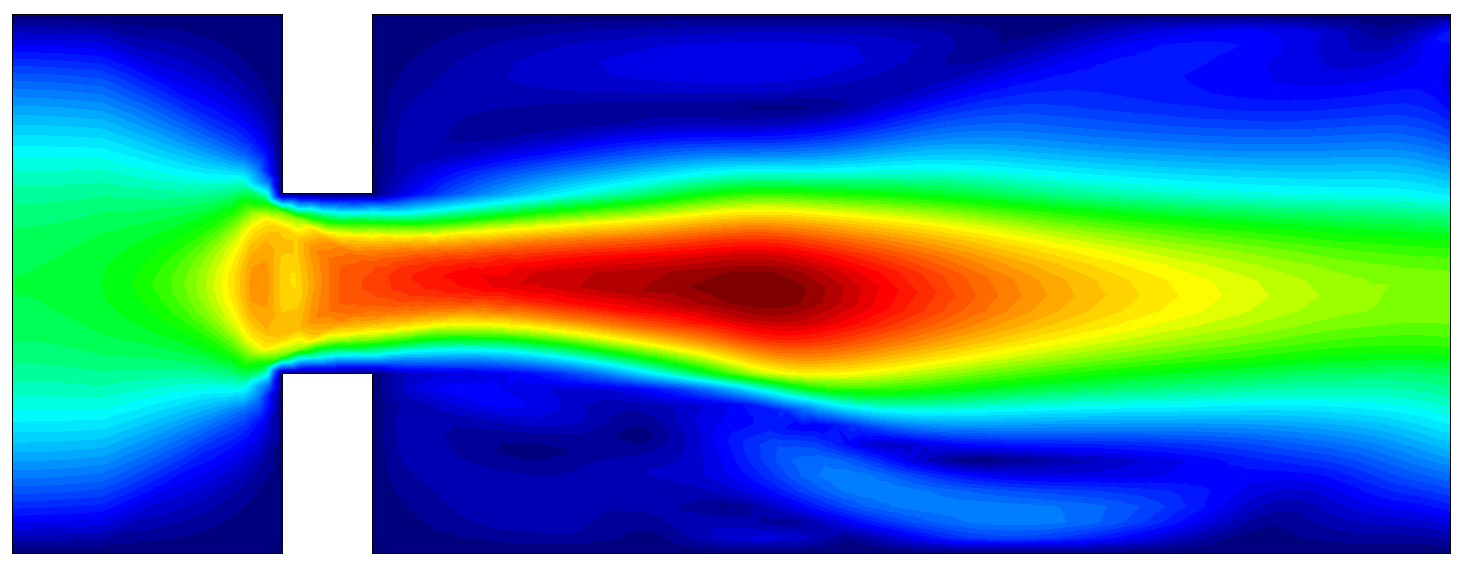}
    \caption{velocity mode 3}
  \end{subfigure}
  \hfill
  \begin{subfigure}[b]{0.245\textwidth}
    \includegraphics[width=\linewidth]{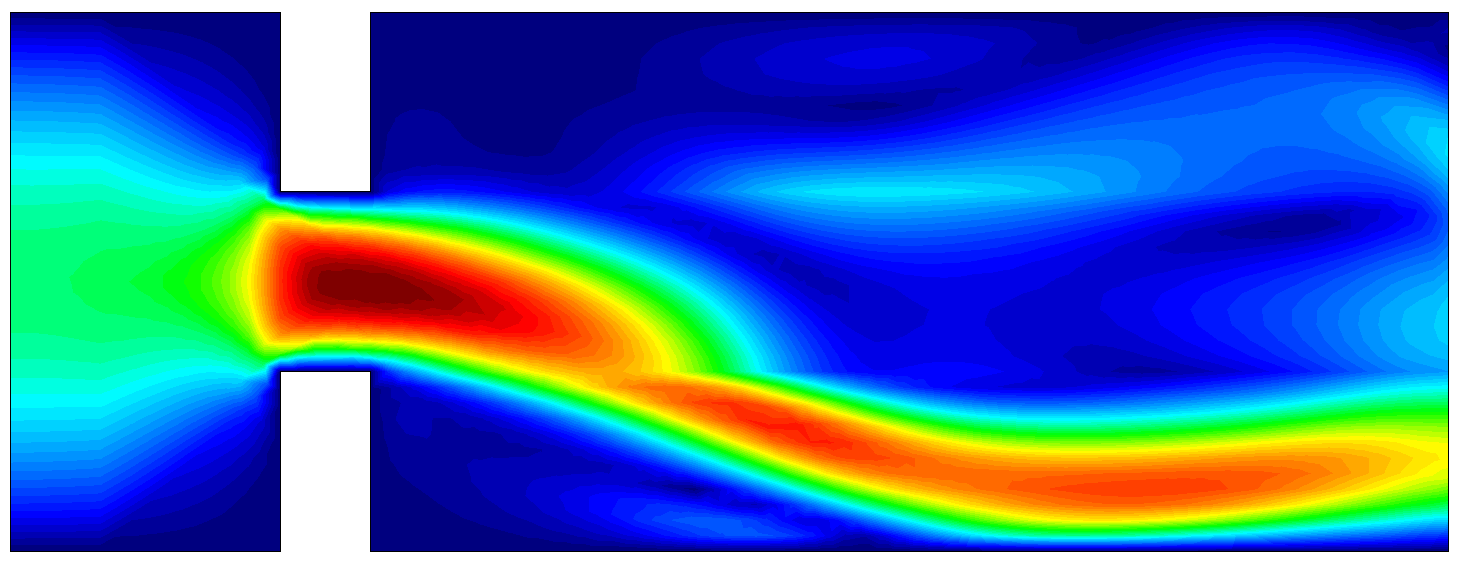}
    \caption{velocity mode 4}
  \end{subfigure}

  \begin{subfigure}[b]{0.245\textwidth}
    \includegraphics[width=\linewidth]{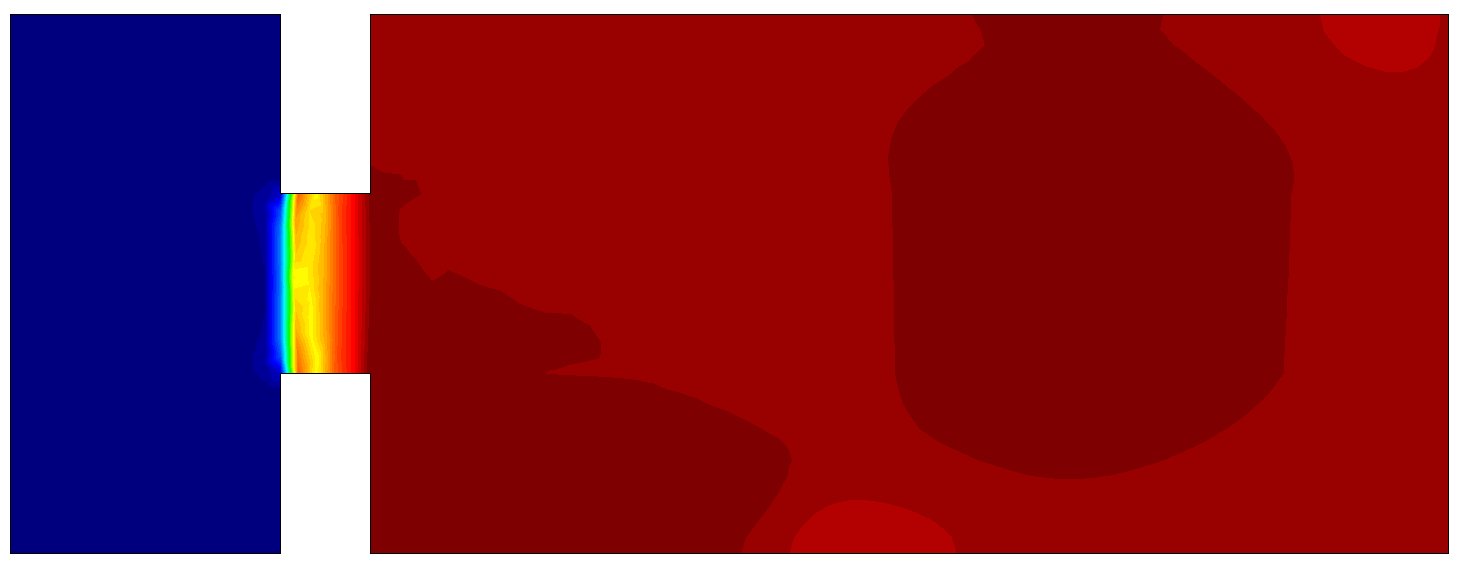}
    \caption{pressure mode 1}
  \end{subfigure}
  \hfill
  \begin{subfigure}[b]{0.245\textwidth}
    \includegraphics[width=\linewidth]{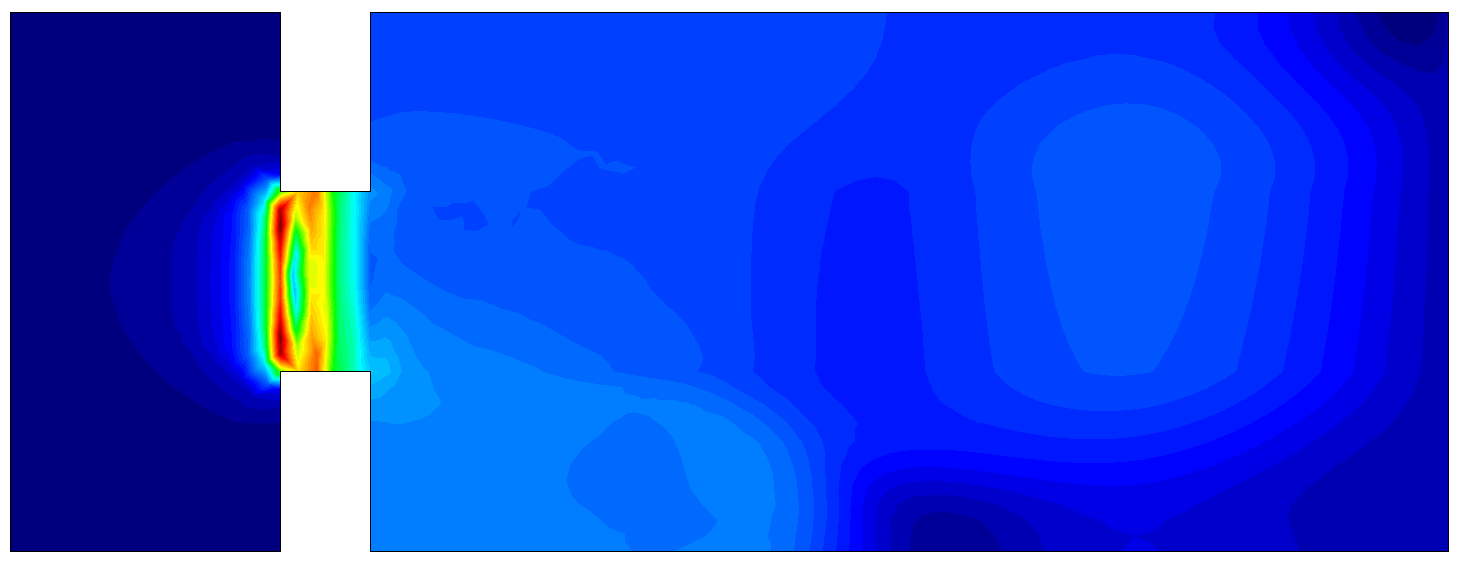}
    \caption{pressure mode 2}
  \end{subfigure}
  \hfill
  \begin{subfigure}[b]{0.245\textwidth}
    \includegraphics[width=\linewidth]{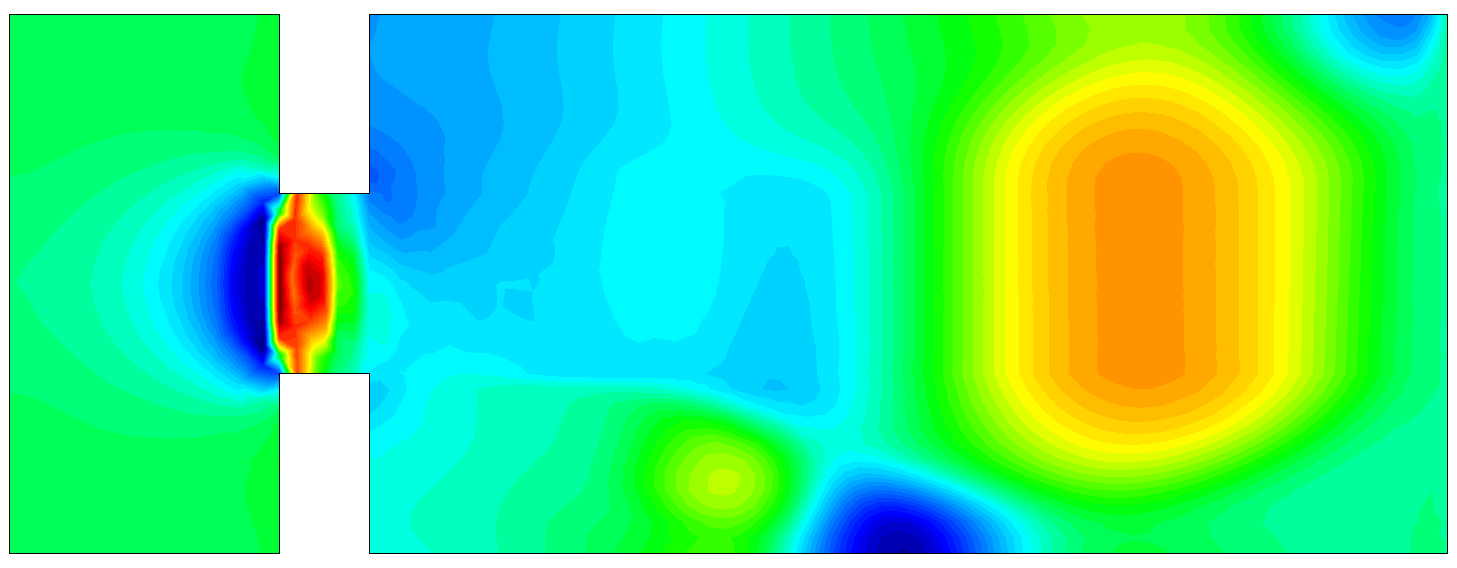}
    \caption{pressure mode 3}
  \end{subfigure}
  \hfill
  \begin{subfigure}[b]{0.245\textwidth}
    \includegraphics[width=\linewidth]{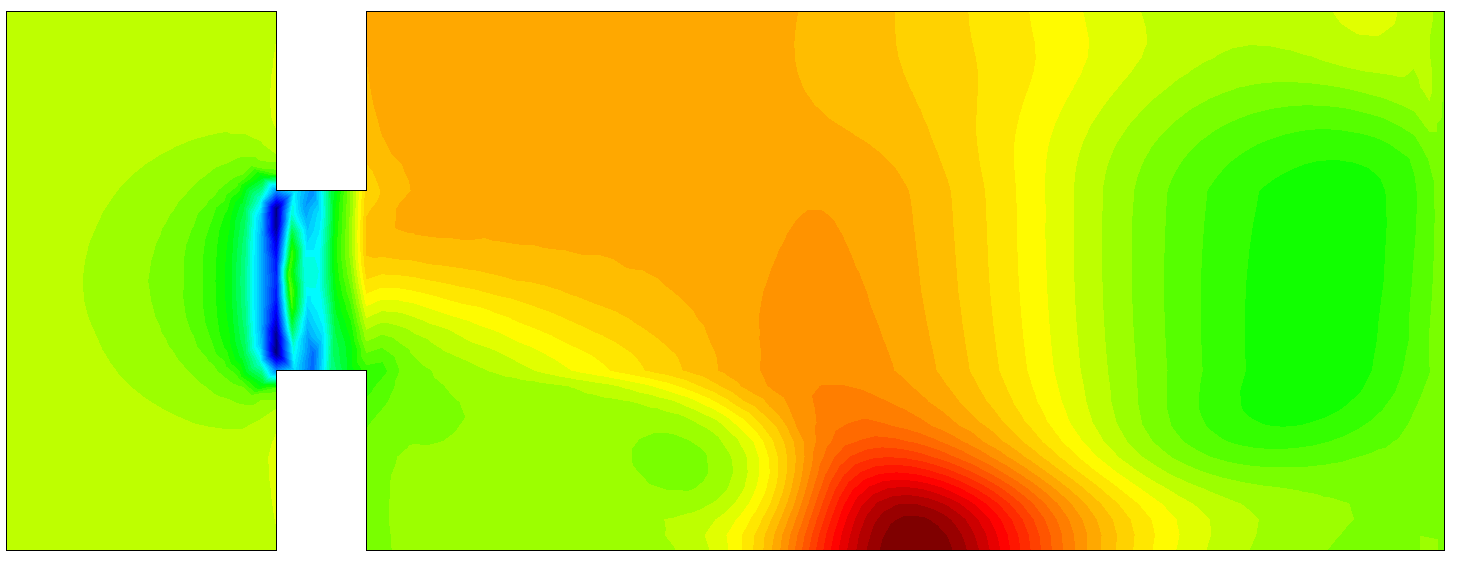}
    \caption{pressure mode 4}
  \end{subfigure}

  \caption{First four velocity and pressure modes for the $\boldsymbol{\varphi}_{\text{\tiny AFFINE}}$ mapping}
  \label{fig: Example 1 modes affine mapping}
\end{figure}

\begin{figure}[H]
  \centering
  \begin{subfigure}[b]{0.245\textwidth}
    \includegraphics[width=\linewidth]{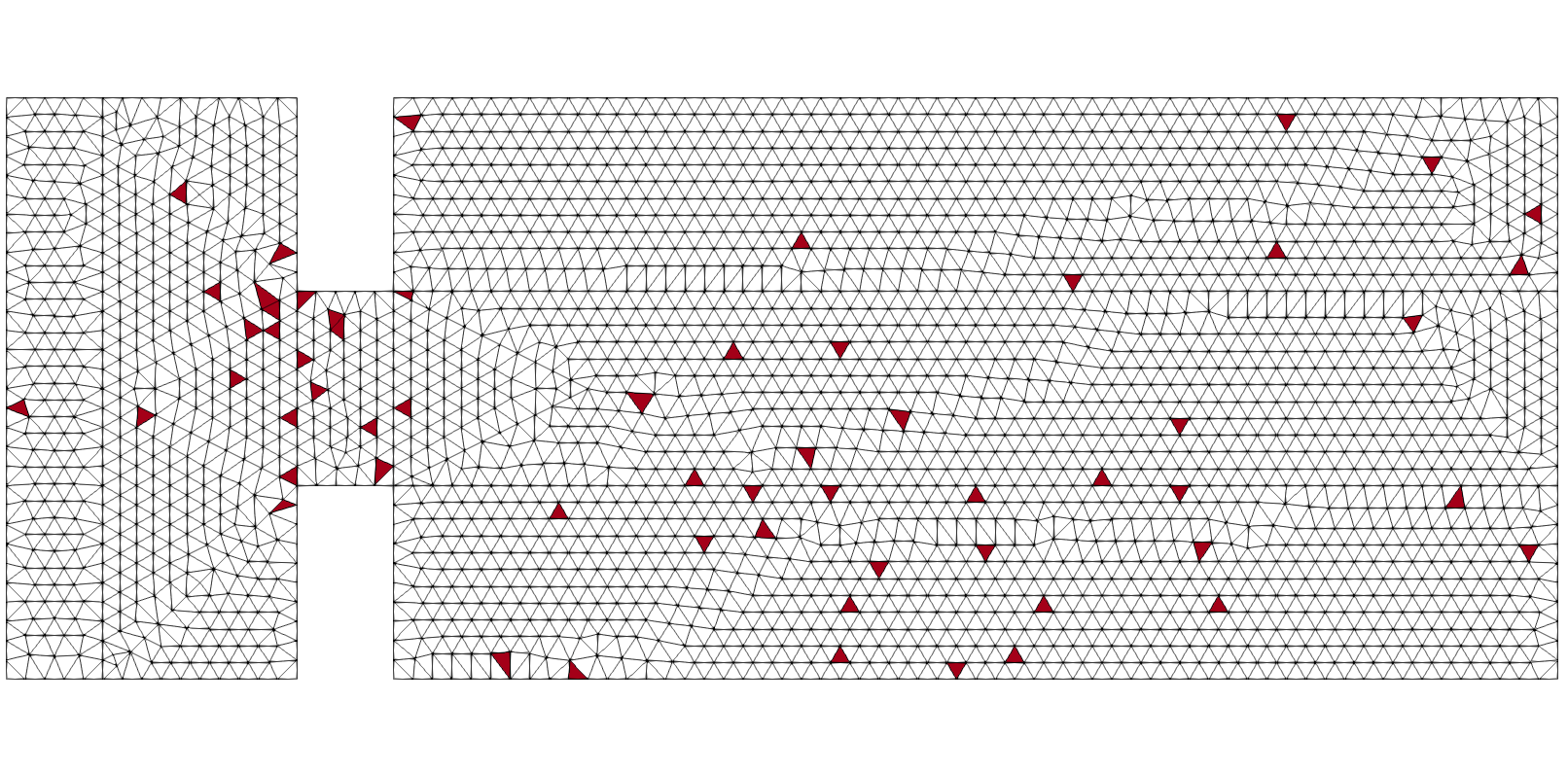}
    \caption{ $\epsilon_{\text{\tiny SOL}} = 1e-3, \epsilon_{\text{\tiny RES}} = 1e-3, $ }
    %\label{fig:HROM_elemns_a}
  \end{subfigure}
  \hfill
  \begin{subfigure}[b]{0.245\textwidth}
    \includegraphics[width=\linewidth]{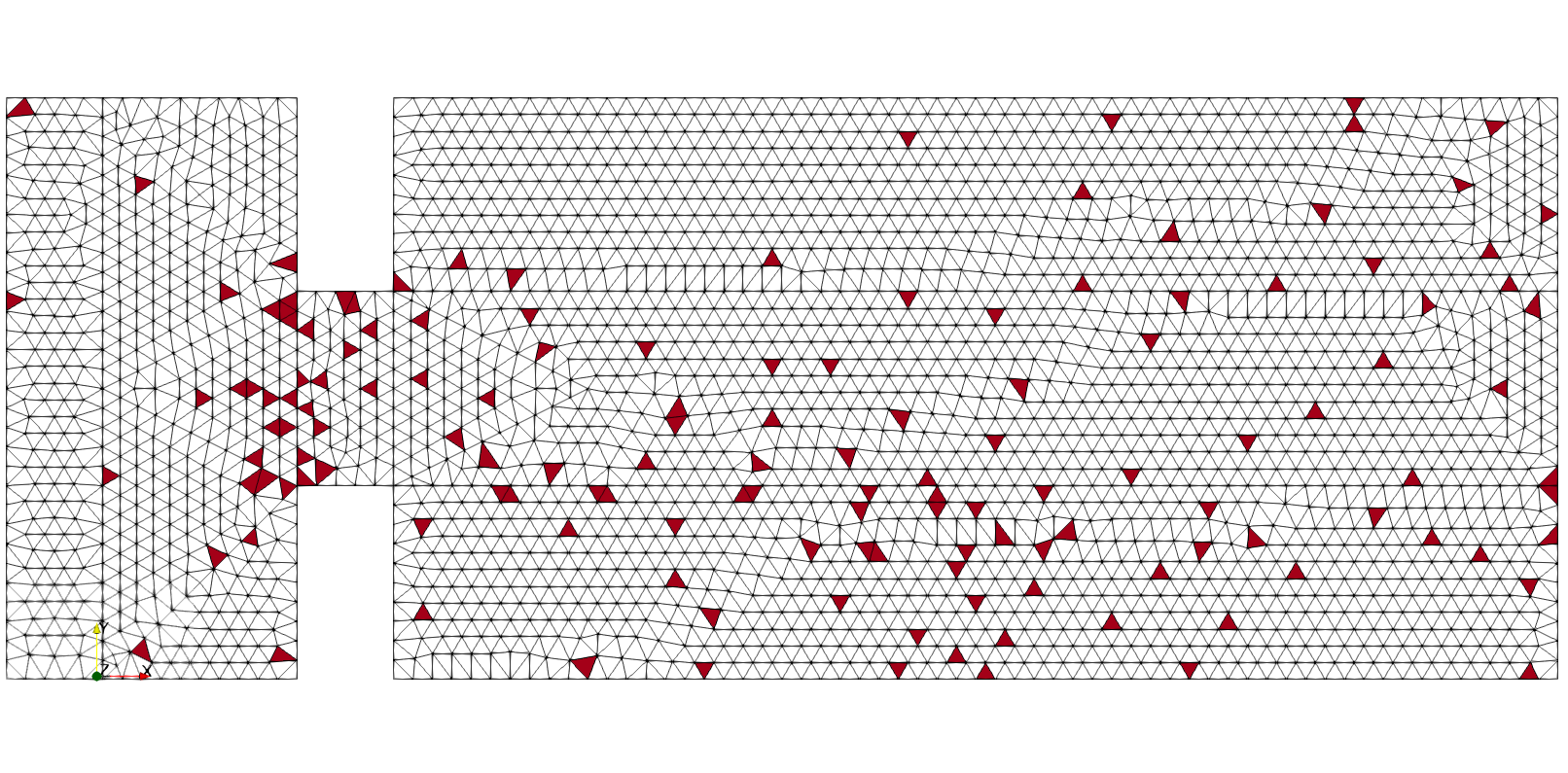}
    \caption{ $\epsilon_{\text{\tiny SOL}} = 1e-3, \epsilon_{\text{\tiny RES}} = 1e-4, $ }
    %\label{fig:HROM_elemns_b}
  \end{subfigure}
  \hfill
  \begin{subfigure}[b]{0.245\textwidth}
    \includegraphics[width=\linewidth]{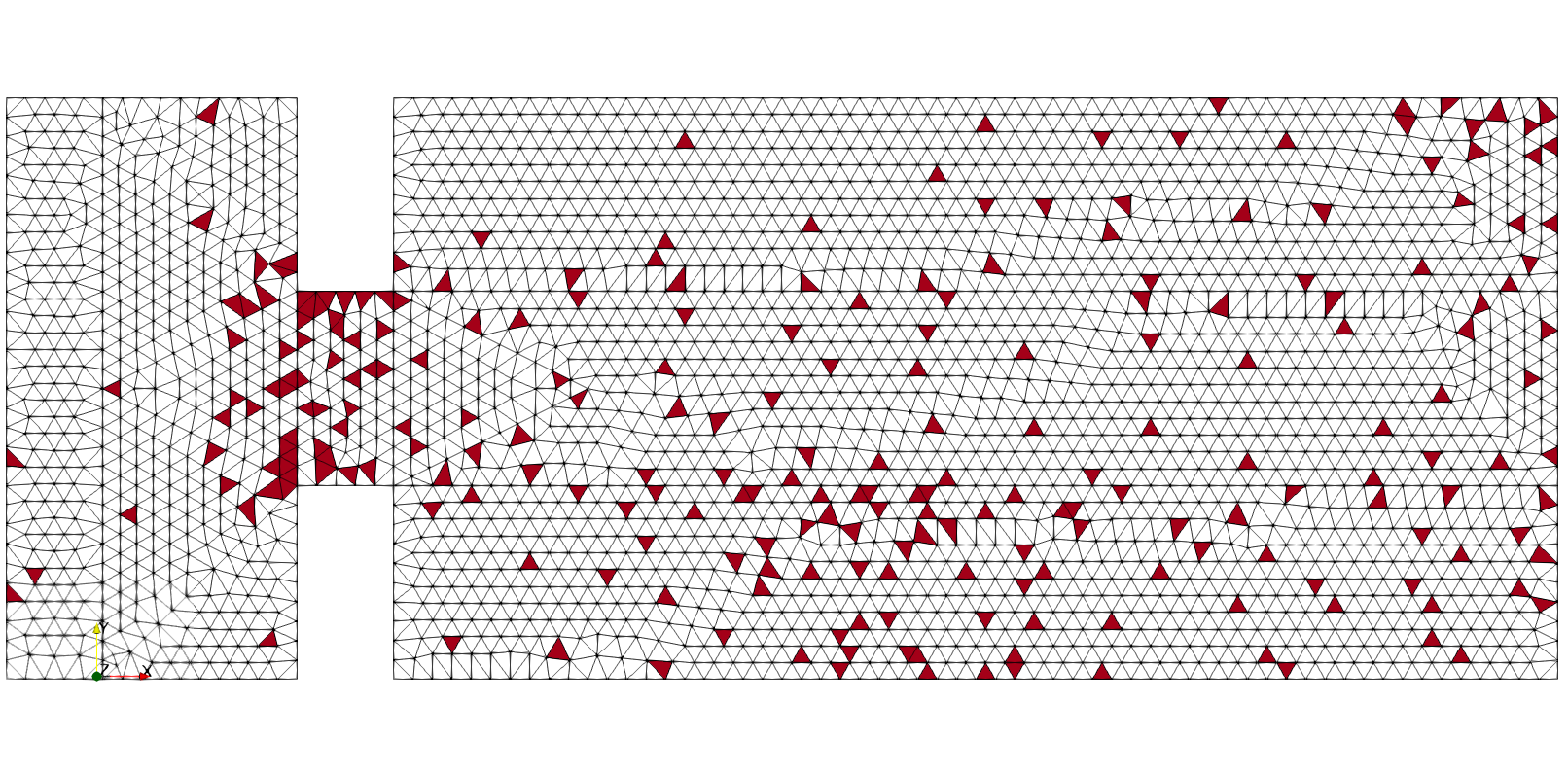}
    \caption{ $\epsilon_{\text{\tiny SOL}} = 1e-3, \epsilon_{\text{\tiny RES}} = 1e-5, $ }
    %\label{fig:HROM_elemns_c}
  \end{subfigure}
  \hfill
  \begin{subfigure}[b]{0.245\textwidth}
    \includegraphics[width=\linewidth]{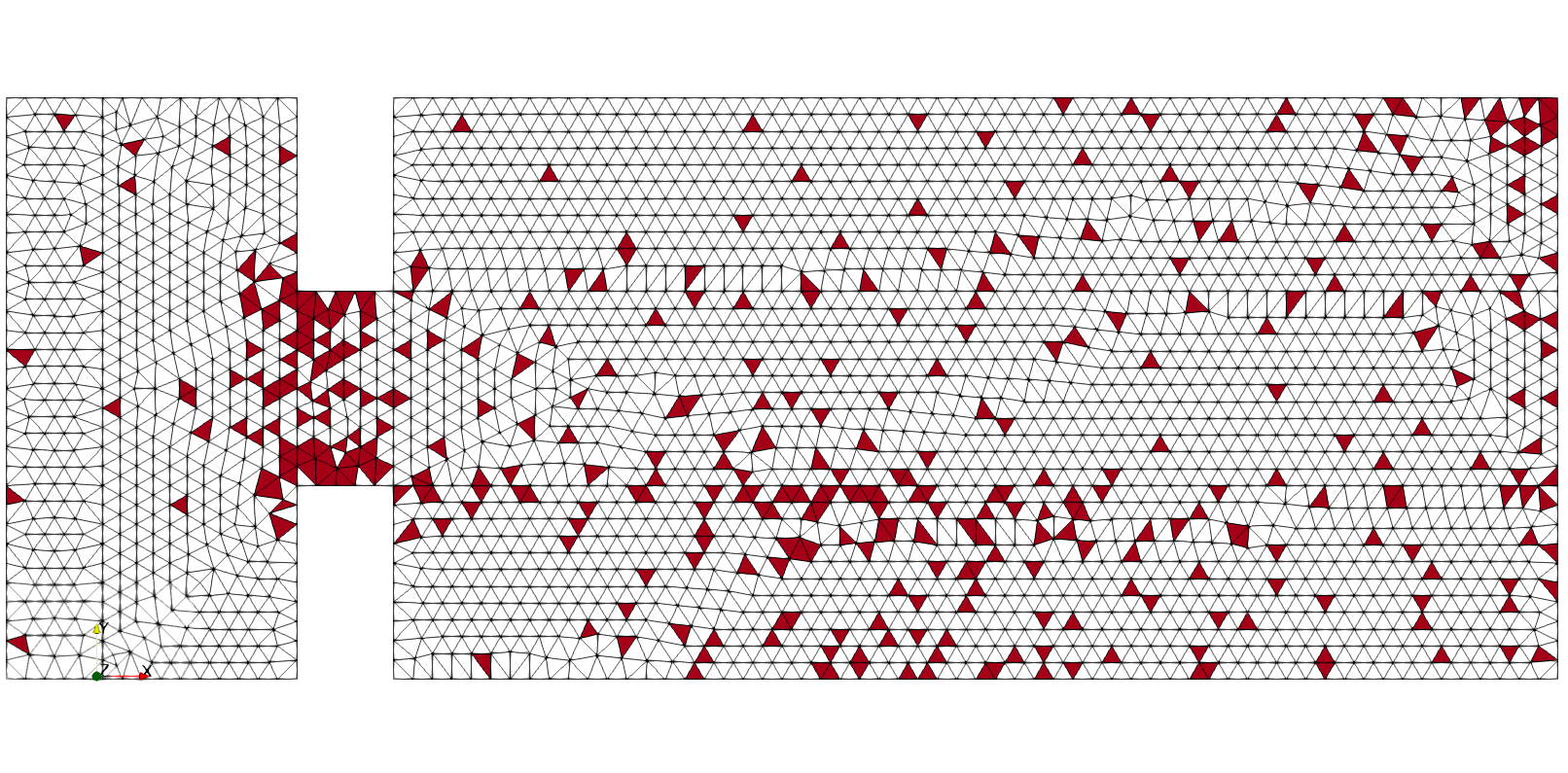}
    \caption{ $\epsilon_{\text{\tiny SOL}} = 1e-3, \epsilon_{\text{\tiny RES}} = 1e-6, $ }
    %\label{fig:HROM_elemns_d}
  \end{subfigure}

  \begin{subfigure}[b]{0.245\textwidth}
    \includegraphics[width=\linewidth]{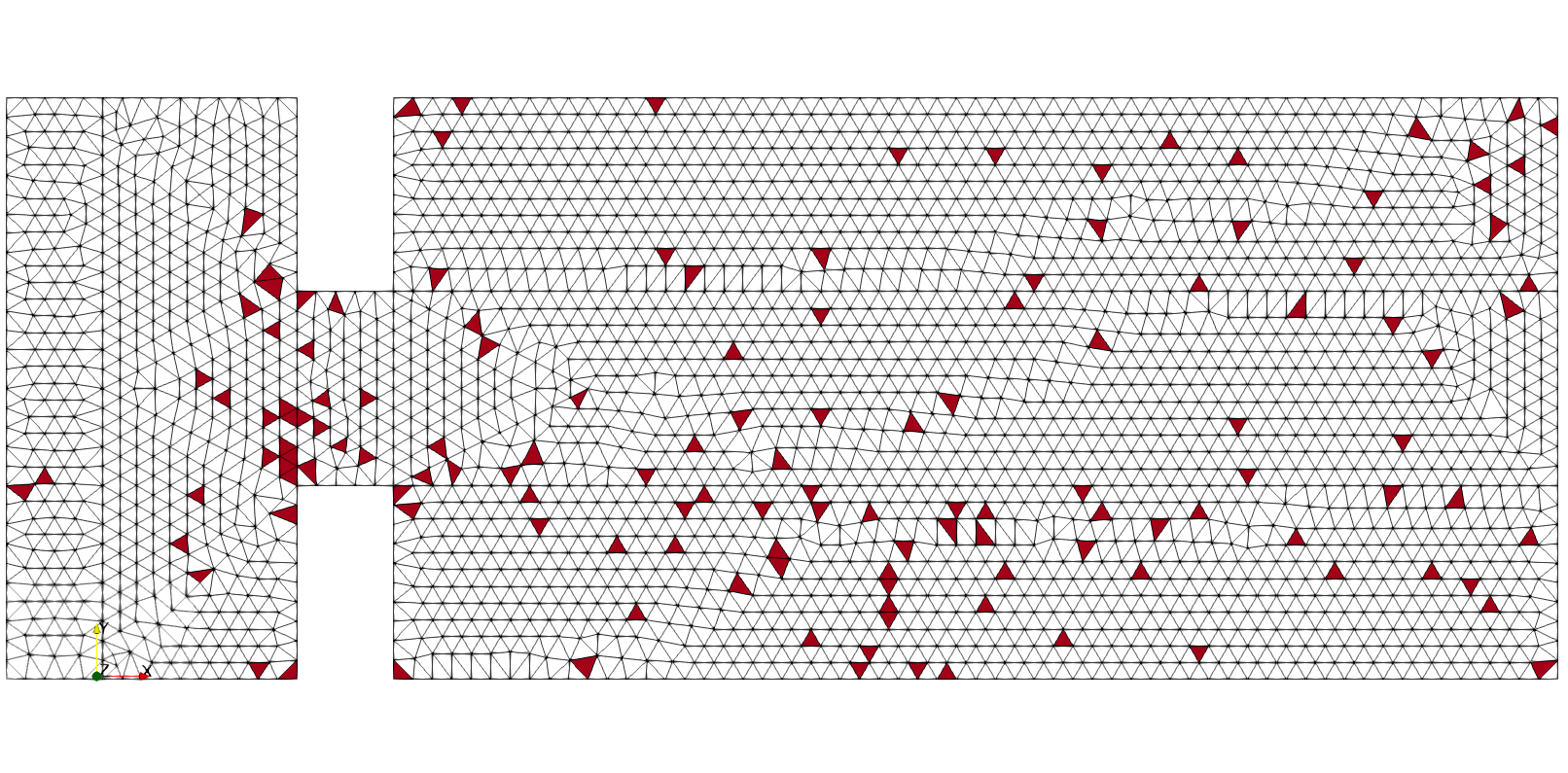}
    \caption{ $\epsilon_{\text{\tiny SOL}} = 1e-4, \epsilon_{\text{\tiny RES}} = 1e-3, $ }
    %\label{fig:HROM_elemns_e}
  \end{subfigure}
  \hfill
  \begin{subfigure}[b]{0.245\textwidth}
    \includegraphics[width=\linewidth]{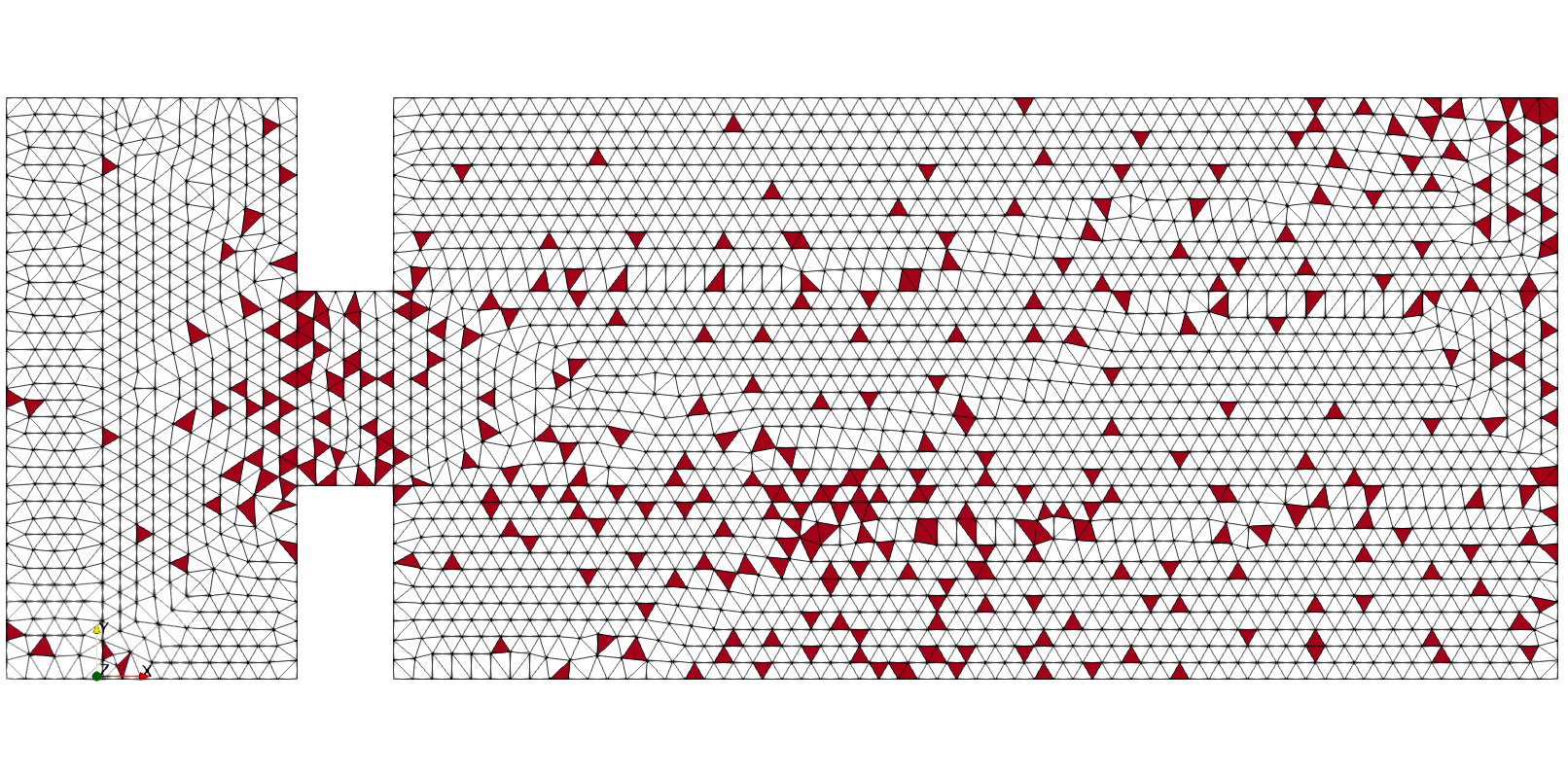}
    \caption{ $\epsilon_{\text{\tiny SOL}} = 1e-4, \epsilon_{\text{\tiny RES}} = 1e-4, $ }
    %\label{fig:HROM_elemns_f}
  \end{subfigure}
  \hfill
  \begin{subfigure}[b]{0.245\textwidth}
    \includegraphics[width=\linewidth]{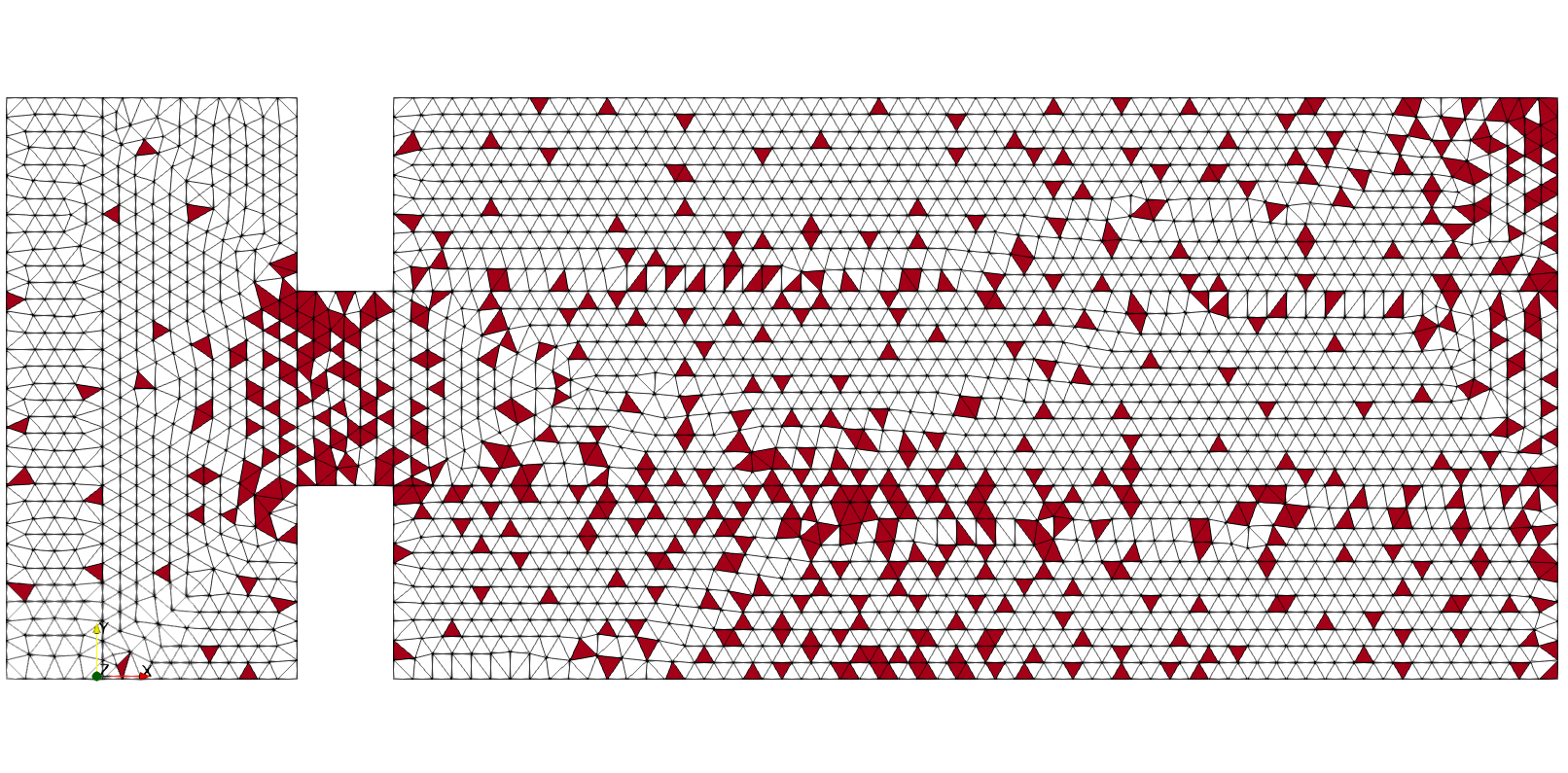}
    \caption{ $\epsilon_{\text{\tiny SOL}} = 1e-4, \epsilon_{\text{\tiny RES}} = 1e-5, $ }
    %\label{fig:HROM_elemns_g}
  \end{subfigure}
  \hfill
  \begin{subfigure}[b]{0.245\textwidth}
    \includegraphics[width=\linewidth]{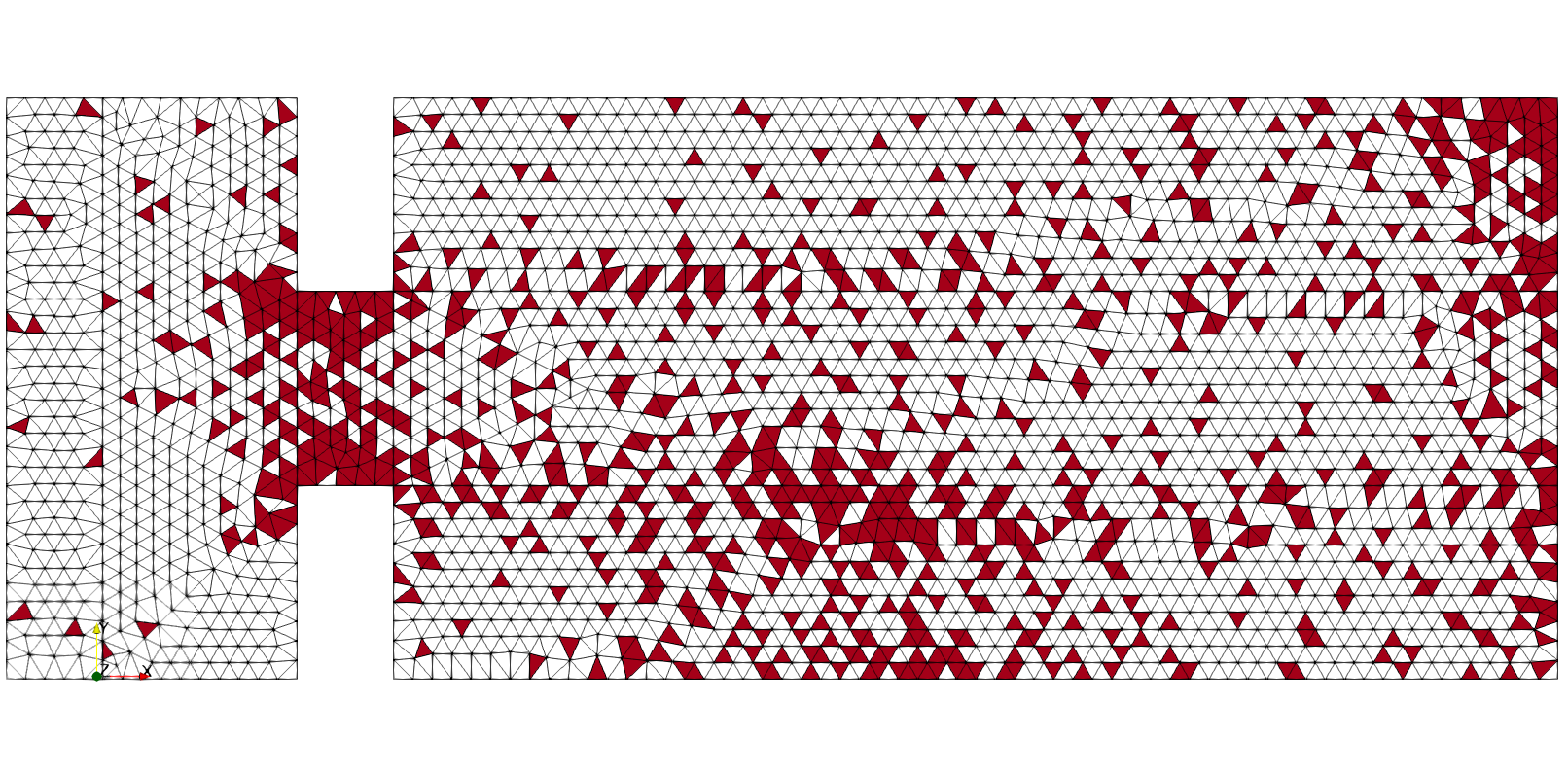}
    \caption{ $\epsilon_{\text{\tiny SOL}} = 1e-4, \epsilon_{\text{\tiny RES}} = 1e-6, $ }
    %\label{fig:HROM_elemns_h}
  \end{subfigure}

  \begin{subfigure}[b]{0.245\textwidth}
    \includegraphics[width=\linewidth]{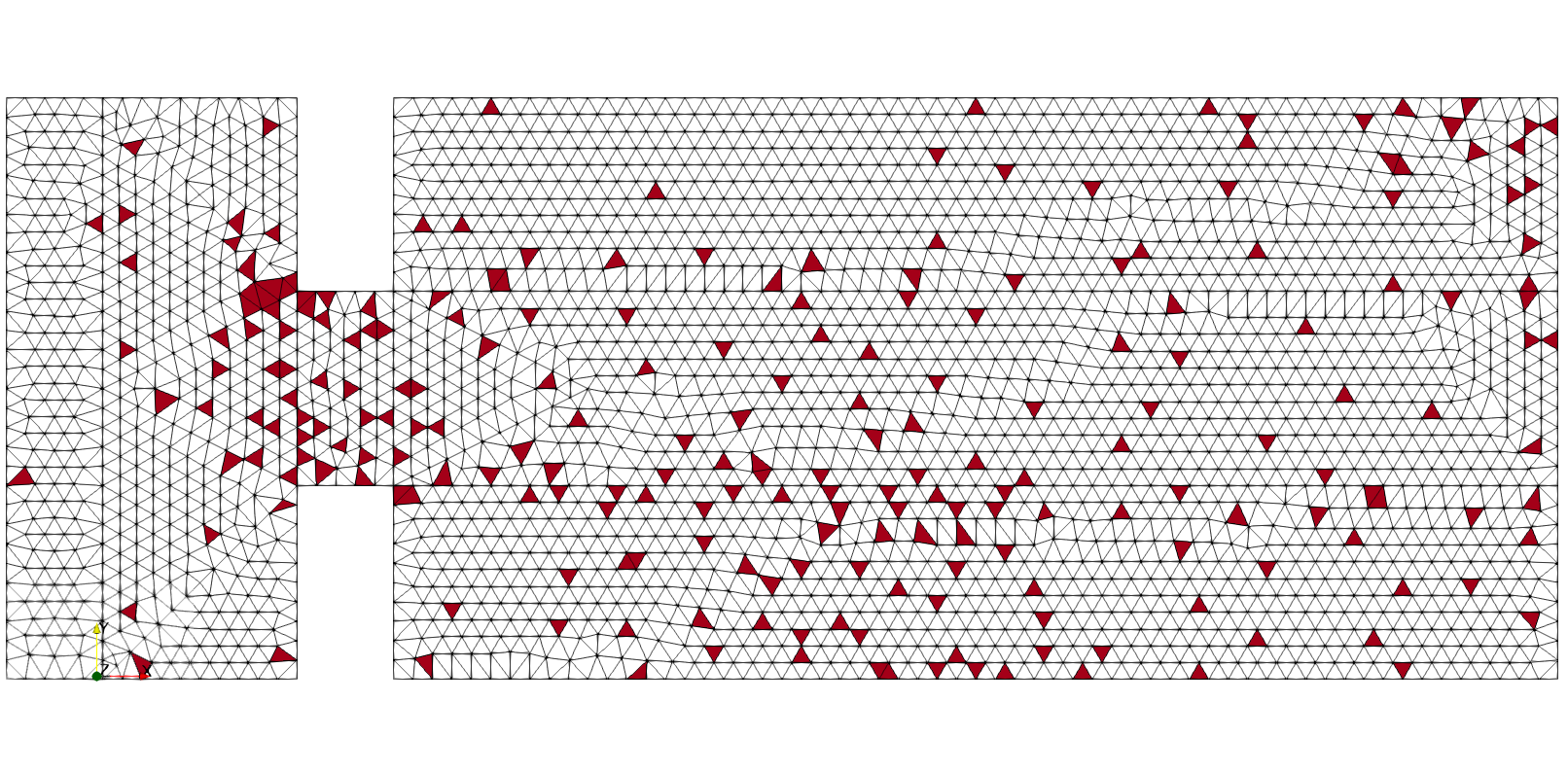}
    \caption{ $\epsilon_{\text{\tiny SOL}} = 1e-5, \epsilon_{\text{\tiny RES}} = 1e-3, $ }
    %\label{fig:HROM_elemns_i}
  \end{subfigure}
  \hfill
  \begin{subfigure}[b]{0.245\textwidth}
    \includegraphics[width=\linewidth]{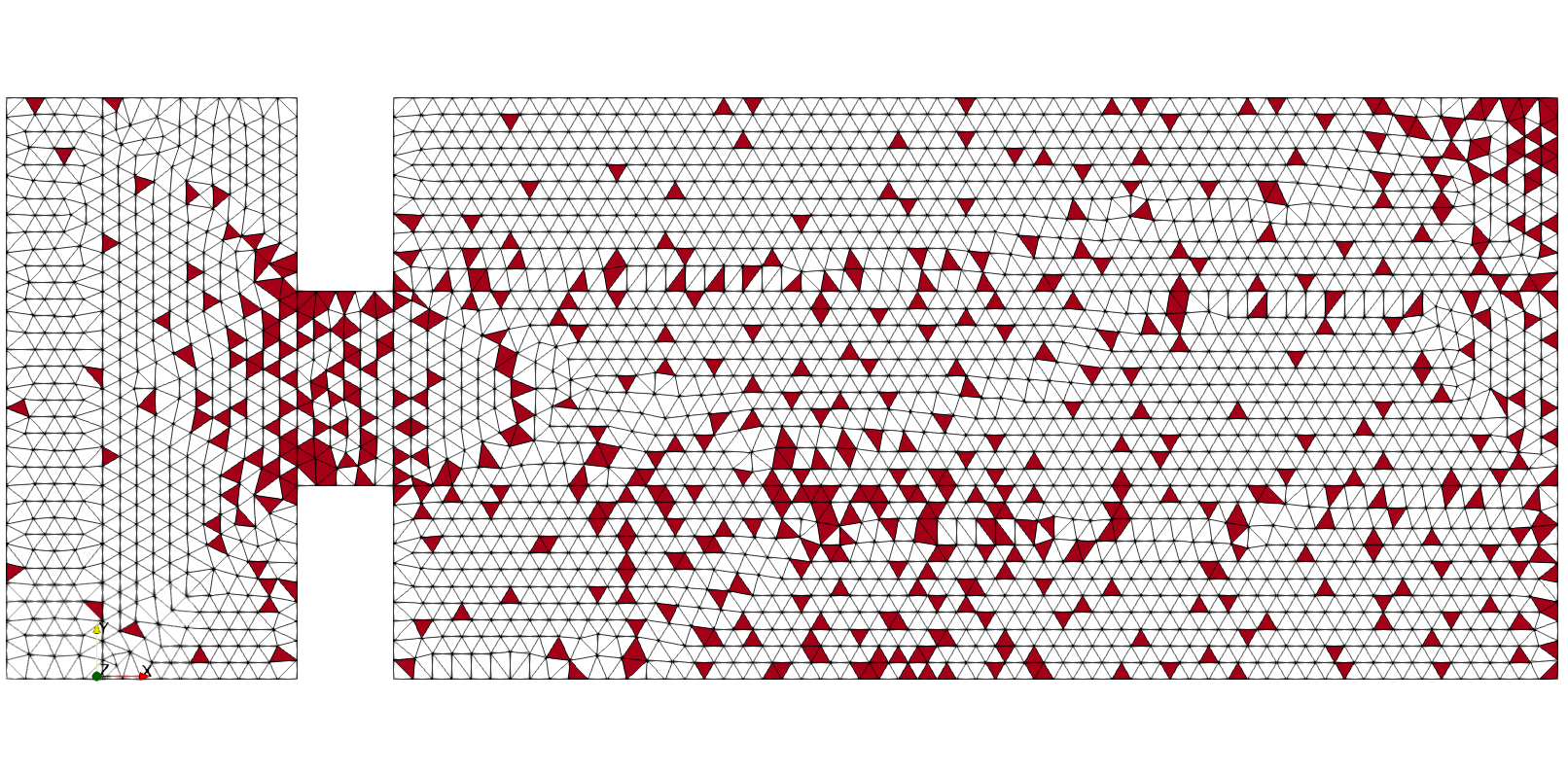}
    \caption{ $\epsilon_{\text{\tiny SOL}} = 1e-5, \epsilon_{\text{\tiny RES}} = 1e-4, $ }
    %\label{fig:HROM_elemns_j}
  \end{subfigure}
  \hfill
  \begin{subfigure}[b]{0.245\textwidth}
    \includegraphics[width=\linewidth]{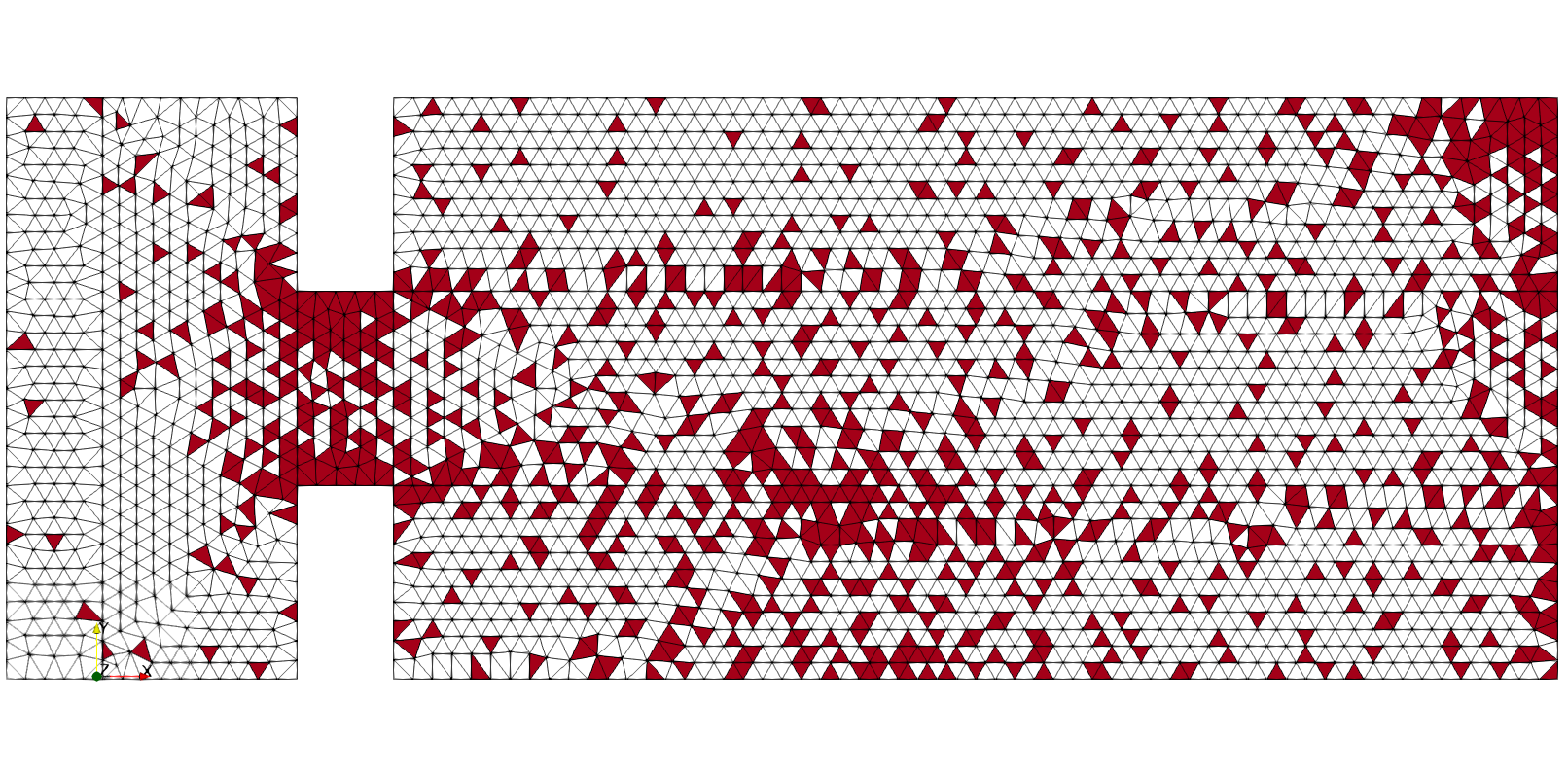}
    \caption{ $\epsilon_{\text{\tiny SOL}} = 1e-5, \epsilon_{\text{\tiny RES}} = 1e-5, $ }
    %\label{fig:HROM_elemns_k}
  \end{subfigure}
  \hfill
  \begin{subfigure}[b]{0.245\textwidth}
    \includegraphics[width=\linewidth]{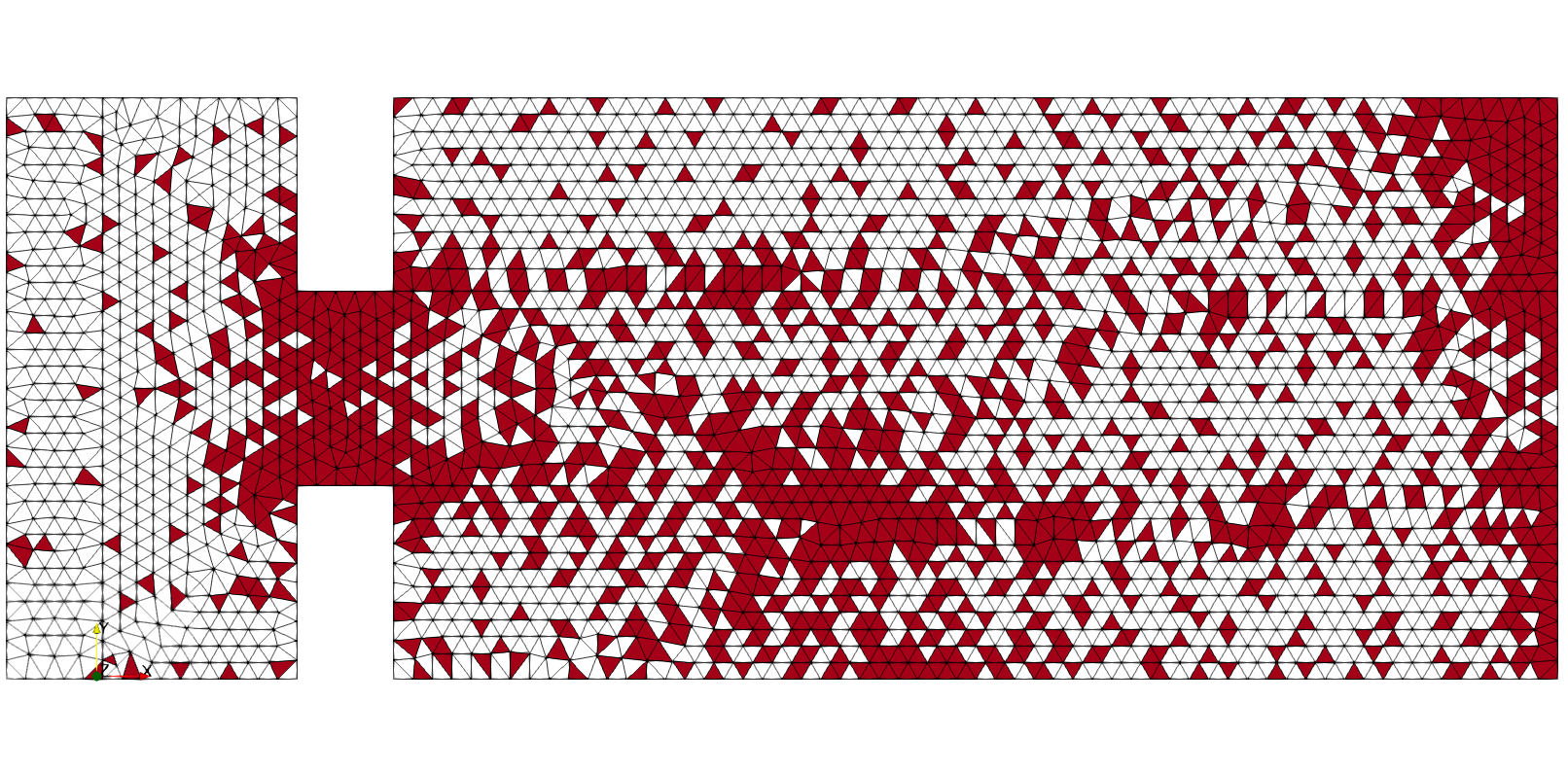}
    \caption{ $\epsilon_{\text{\tiny SOL}} = 1e-5, \epsilon_{\text{\tiny RES}} = 1e-6, $ }
    %\label{fig:HROM_elemns_l}
  \end{subfigure}

  \begin{subfigure}[b]{0.245\textwidth}
    \includegraphics[width=\linewidth]{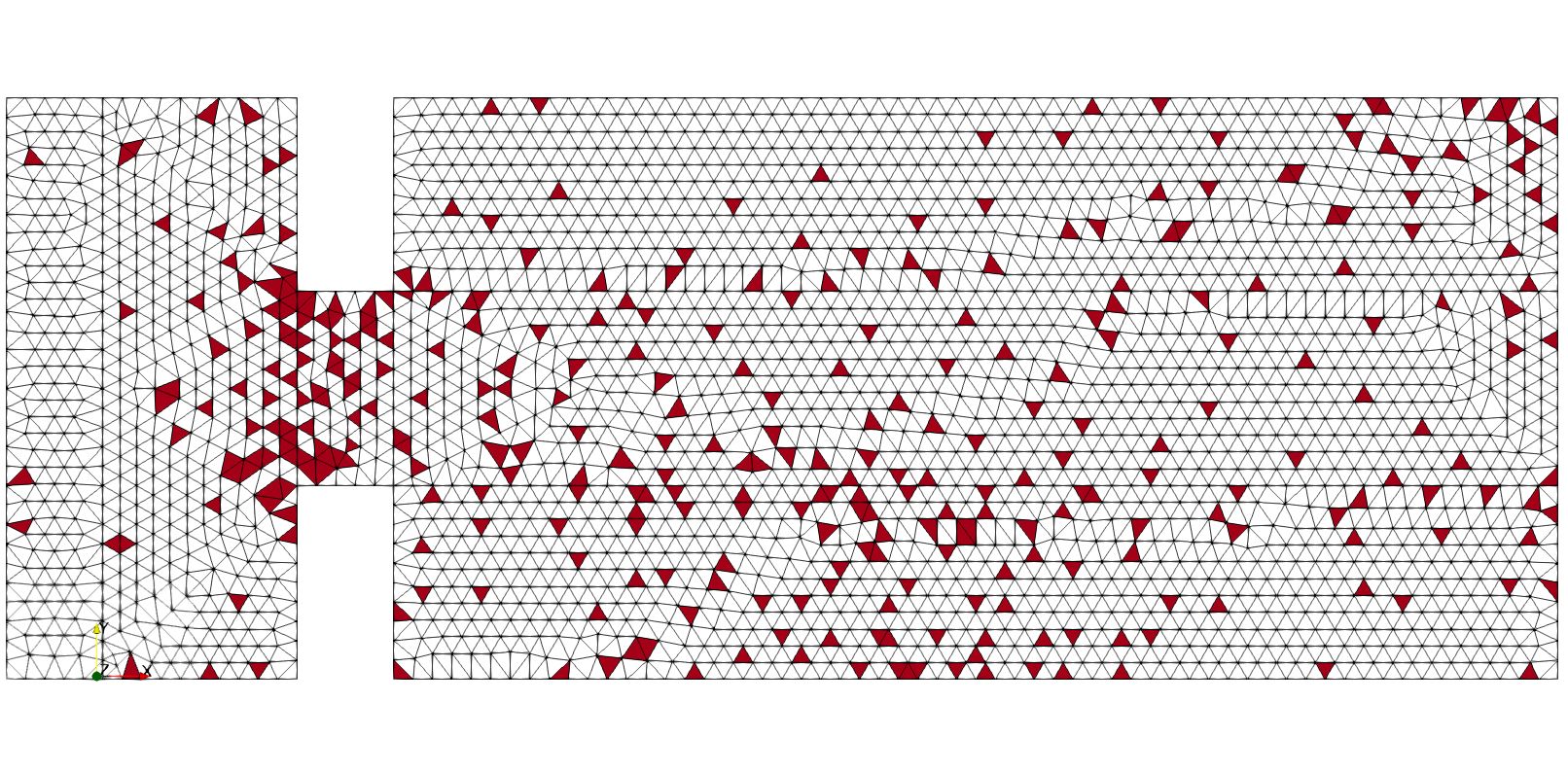}
    \caption{ $\epsilon_{\text{\tiny SOL}} = 1e-6, \epsilon_{\text{\tiny RES}} = 1e-3, $ }
    %\label{fig:HROM_elemns_m}
  \end{subfigure}
  \hfill
  \begin{subfigure}[b]{0.245\textwidth}
    \includegraphics[width=\linewidth]{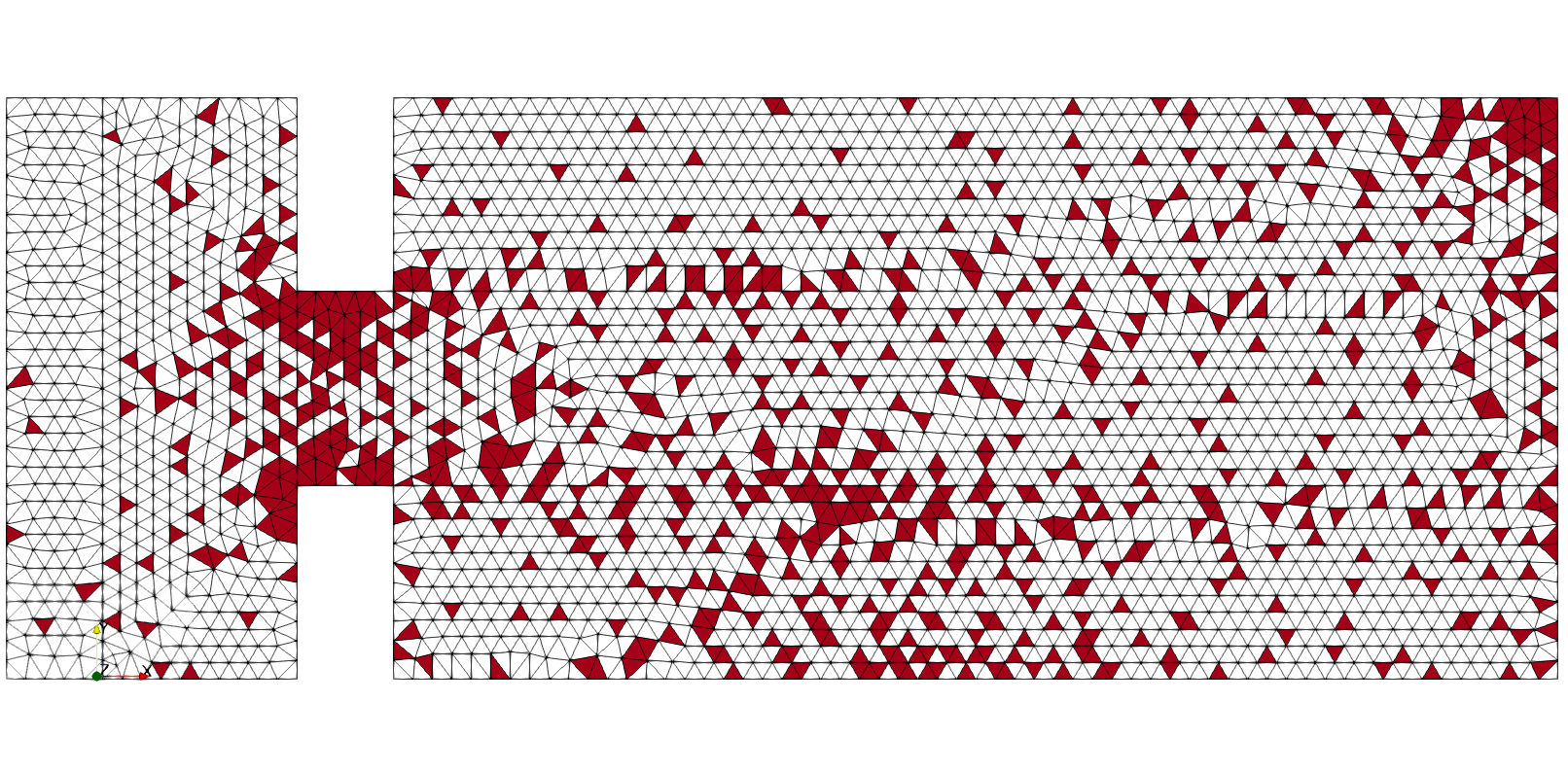}
    \caption{ $\epsilon_{\text{\tiny SOL}} = 1e-6, \epsilon_{\text{\tiny RES}} = 1e-4, $ }
    %\label{fig:HROM_elemns_n}
  \end{subfigure}
  \hfill
  \begin{subfigure}[b]{0.245\textwidth}
    \includegraphics[width=\linewidth]{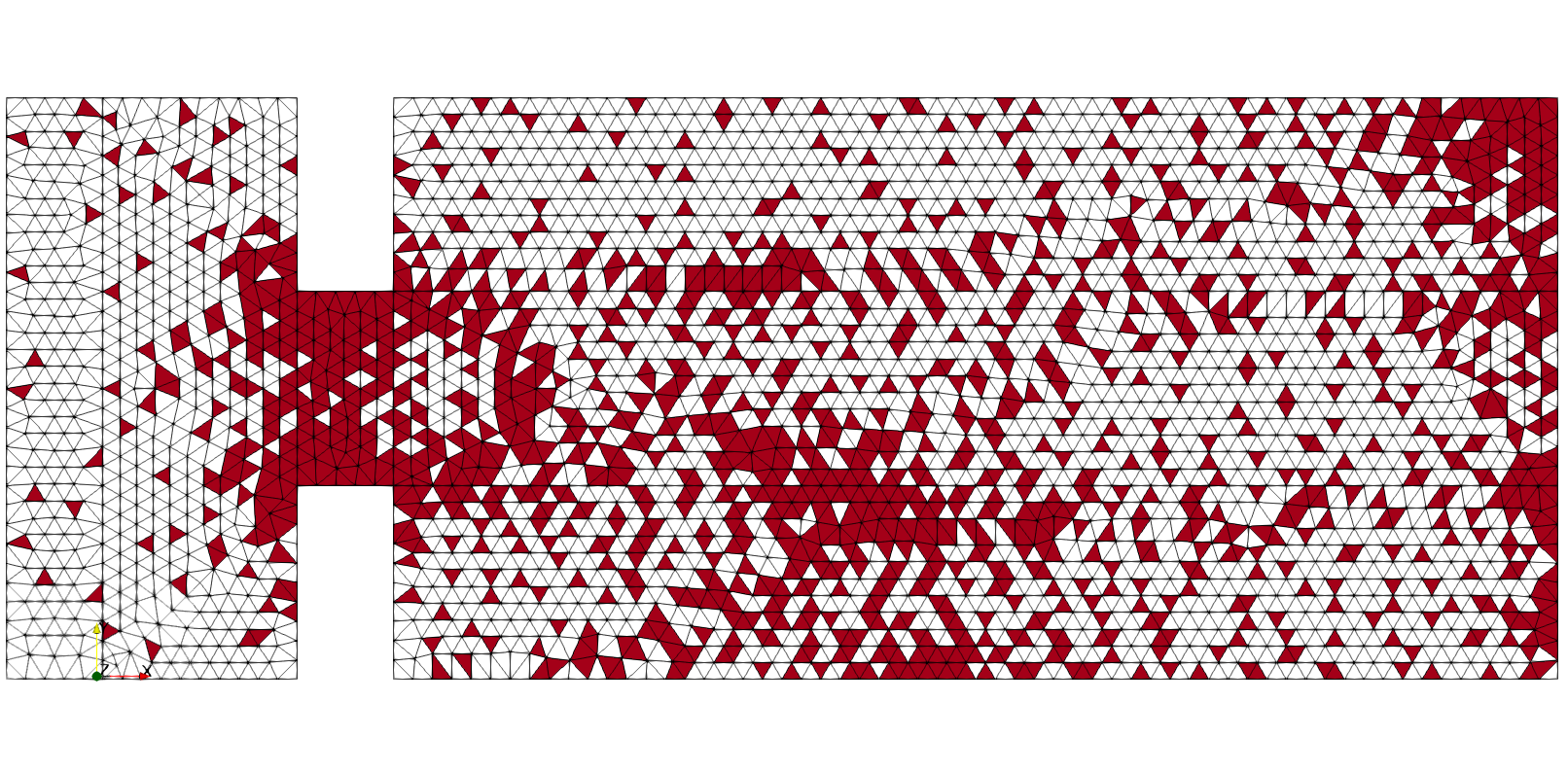}
    \caption{ $\epsilon_{\text{\tiny SOL}} = 1e-6, \epsilon_{\text{\tiny RES}} = 1e-5, $ }
    %\label{fig:HROM_elemns_o}
  \end{subfigure}
  \hfill
  \begin{subfigure}[b]{0.245\textwidth}
    \includegraphics[width=\linewidth]{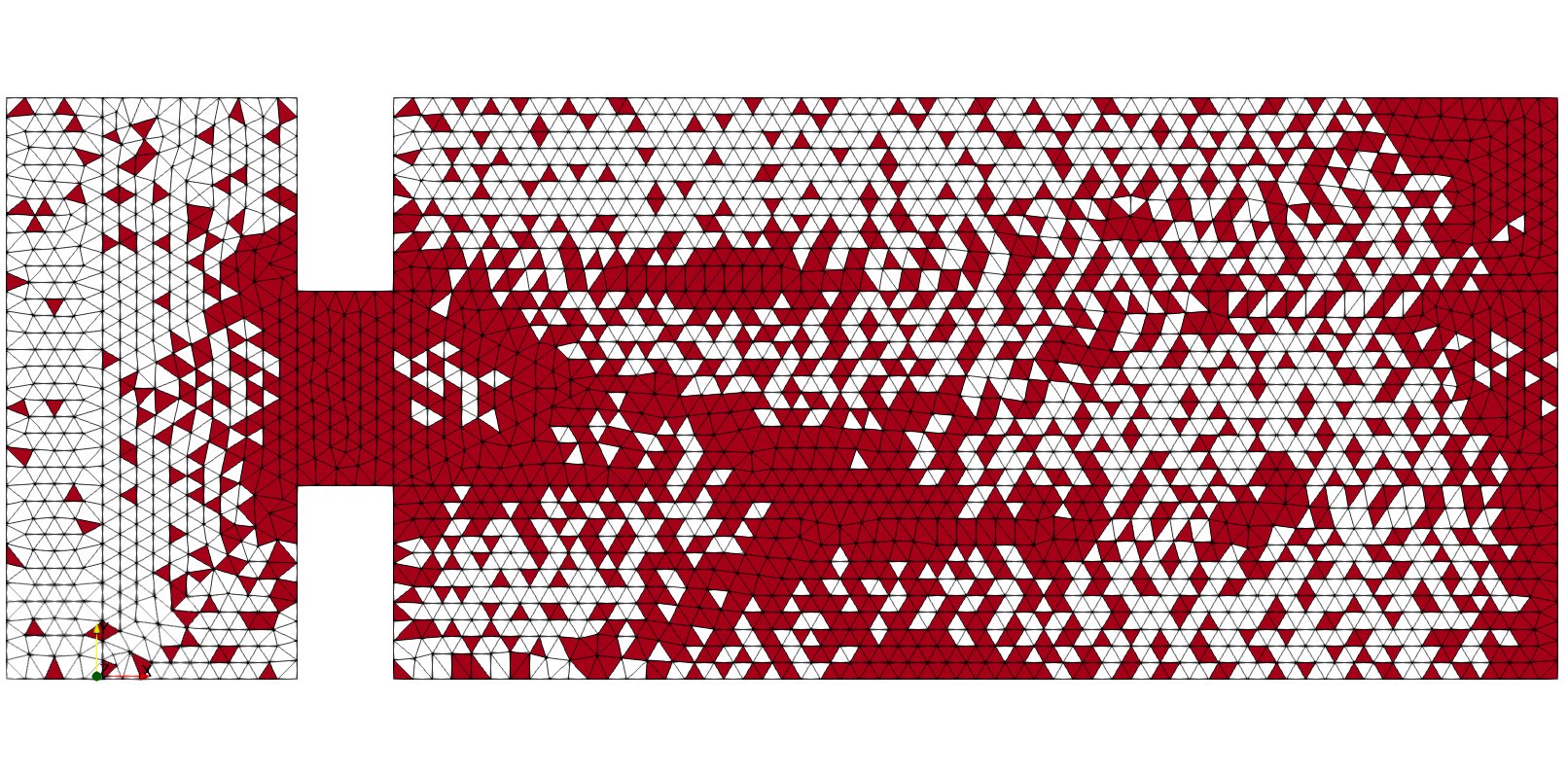}
    \caption{ $\epsilon_{\text{\tiny SOL}} = 1e-6, \epsilon_{\text{\tiny RES}} = 1e-6, $ }
    %\label{fig:HROM_elemns_p}
  \end{subfigure}

  \caption{Hyper-reduced elements selected for the $\boldsymbol{\varphi}_{\text{\tiny AFFINE}}$ mapping}
  \label{fig: Example 1 HROM elements affine}
\end{figure}

\append{The pressure and velocity modes for the $\varphi_{\text{\tiny AFFINE}}$ mapping are illustrated in Fig. \ref{fig: Example 1 modes affine mapping}. These modes capture the prevalence in Trajectory 1 of a solution with a jet attaching to the lower wall. Similarly, Fig. \ref{fig: Example 1 HROM elements affine} shows the spatial distribution of the elements selected by the ECM algorithm for the 16 HROMs considered in Example 1 for the affine mapping. Notably, the selected elements tend to accumulate in the lower part of the geometry, aligning with the observed patterns in the POD modes.}

\subsubsection{\append{Example 1. Nonaffine Mapping}}

\begin{figure}[H]
  \centering
  \begin{subfigure}[b]{0.245\textwidth}
    \includegraphics[width=\linewidth]{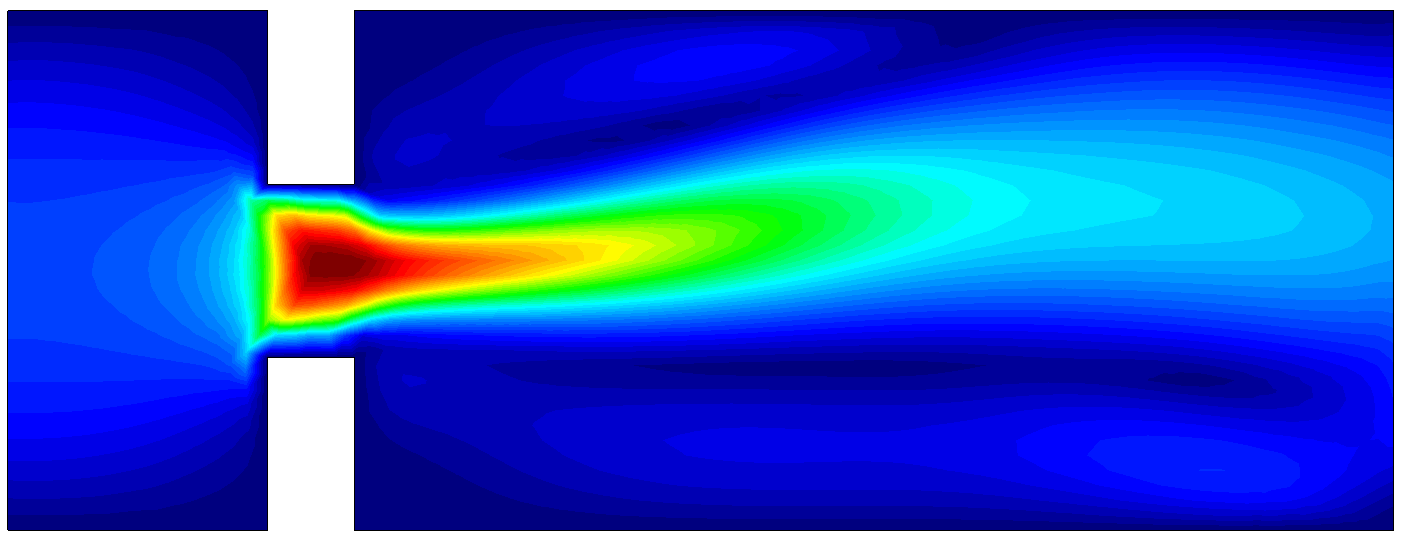}
    \caption{velocity mode 1}
  \end{subfigure}
  \hfill
  \begin{subfigure}[b]{0.245\textwidth}
    \includegraphics[width=\linewidth]{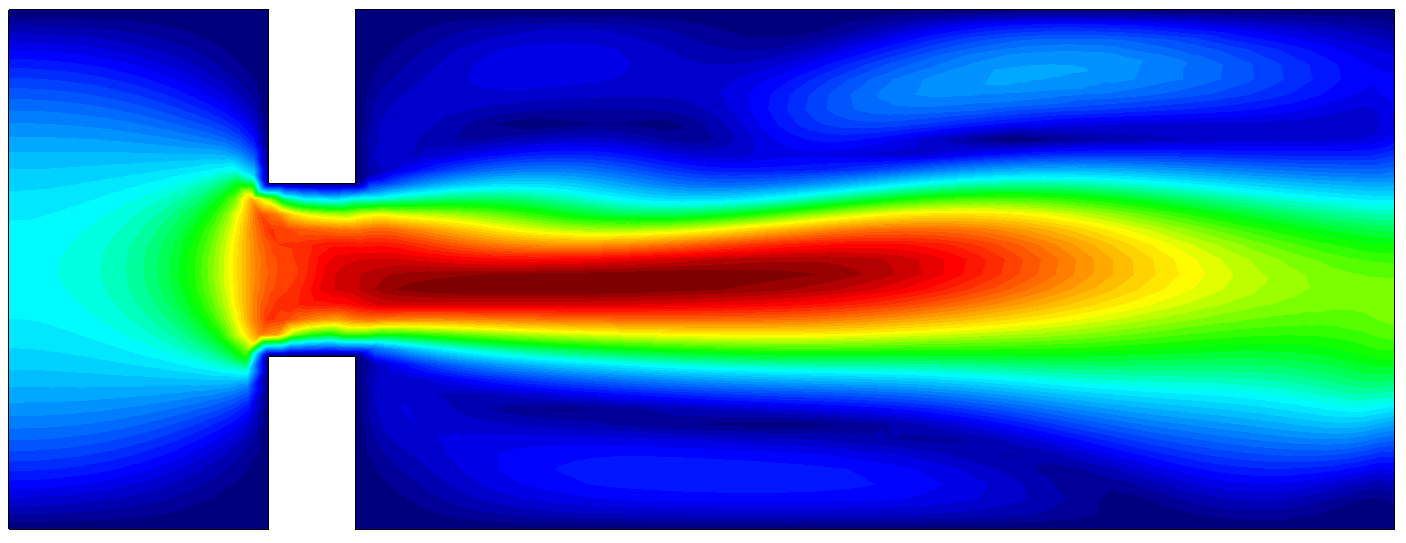}
    \caption{velocity mode 2}
  \end{subfigure}
  \hfill
  \begin{subfigure}[b]{0.245\textwidth}
    \includegraphics[width=\linewidth]{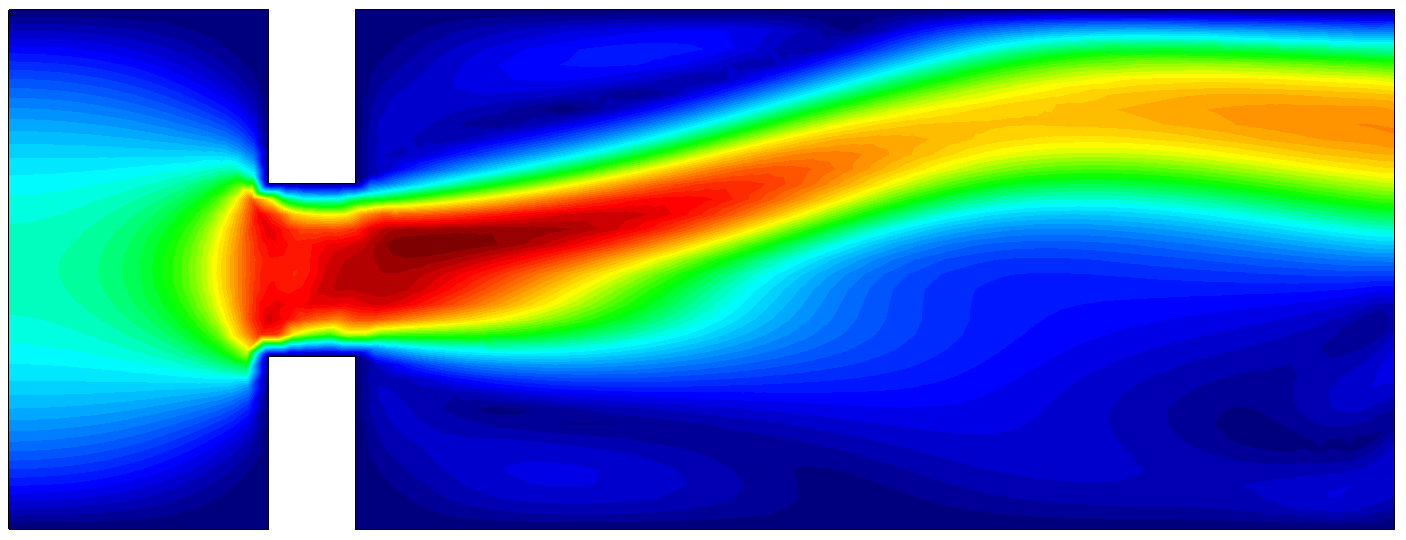}
    \caption{velocity mode 3}
  \end{subfigure}
  \hfill
  \begin{subfigure}[b]{0.245\textwidth}
    \includegraphics[width=\linewidth]{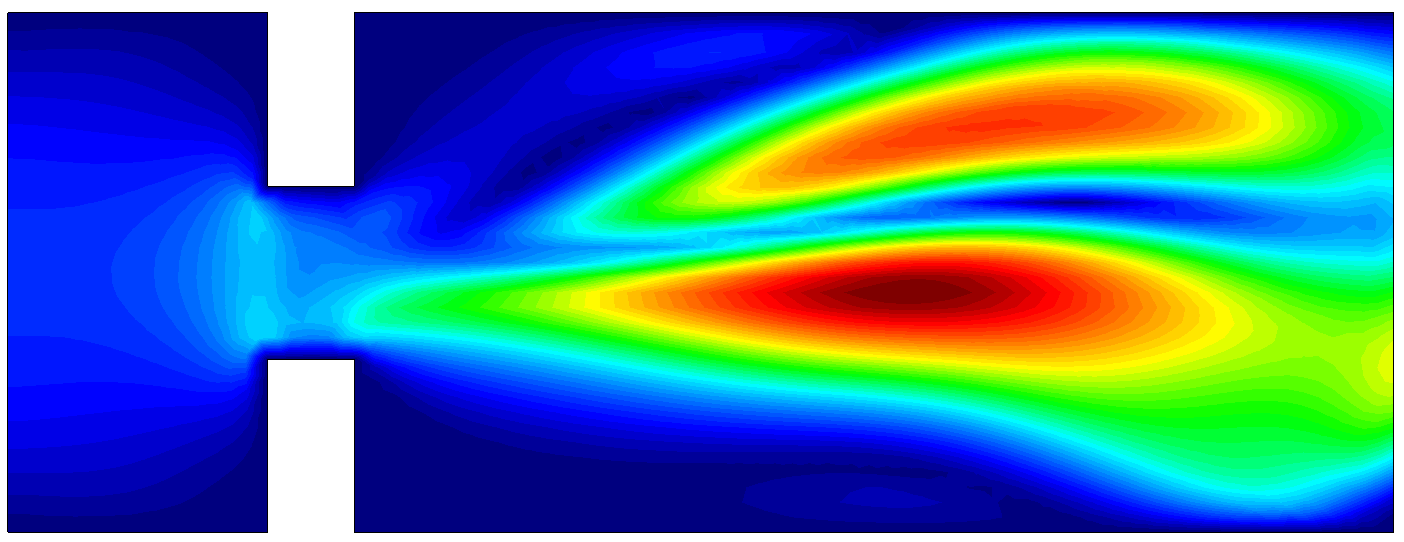}
    \caption{velocity mode 4}
  \end{subfigure}

  \begin{subfigure}[b]{0.245\textwidth}
    \includegraphics[width=\linewidth]{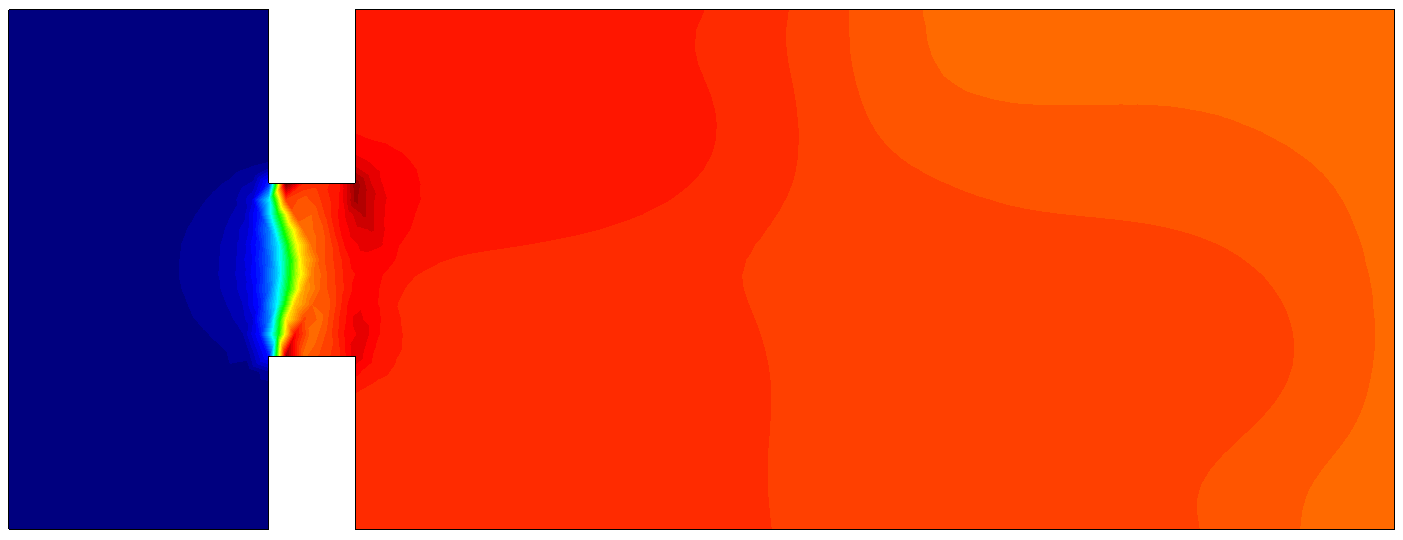}
    \caption{pressure mode 1}
  \end{subfigure}
  \hfill
  \begin{subfigure}[b]{0.245\textwidth}
    \includegraphics[width=\linewidth]{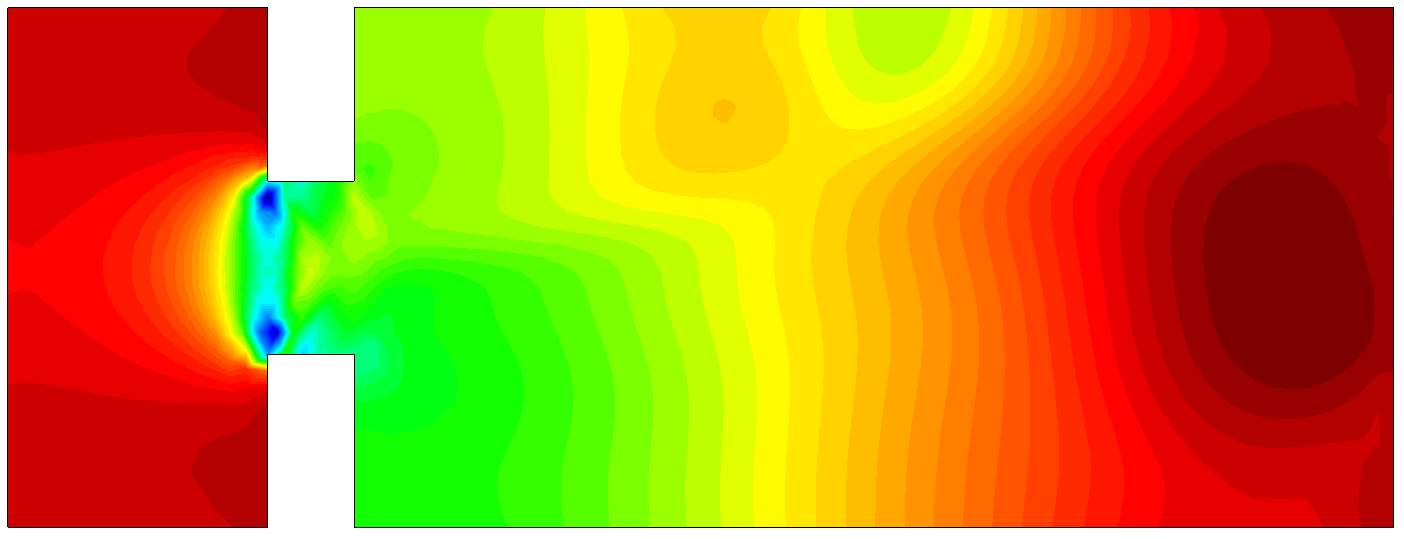}
    \caption{pressure mode 2}
  \end{subfigure}
  \hfill
  \begin{subfigure}[b]{0.245\textwidth}
    \includegraphics[width=\linewidth]{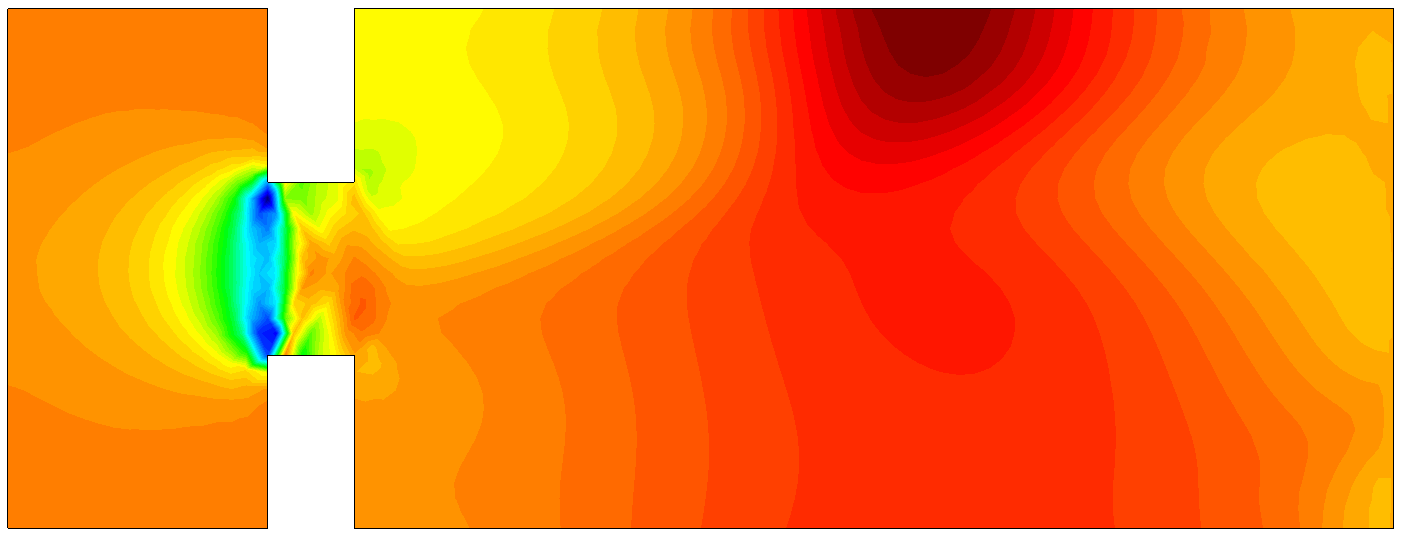}
    \caption{pressure mode 3}
  \end{subfigure}
  \hfill
  \begin{subfigure}[b]{0.245\textwidth}
    \includegraphics[width=\linewidth]{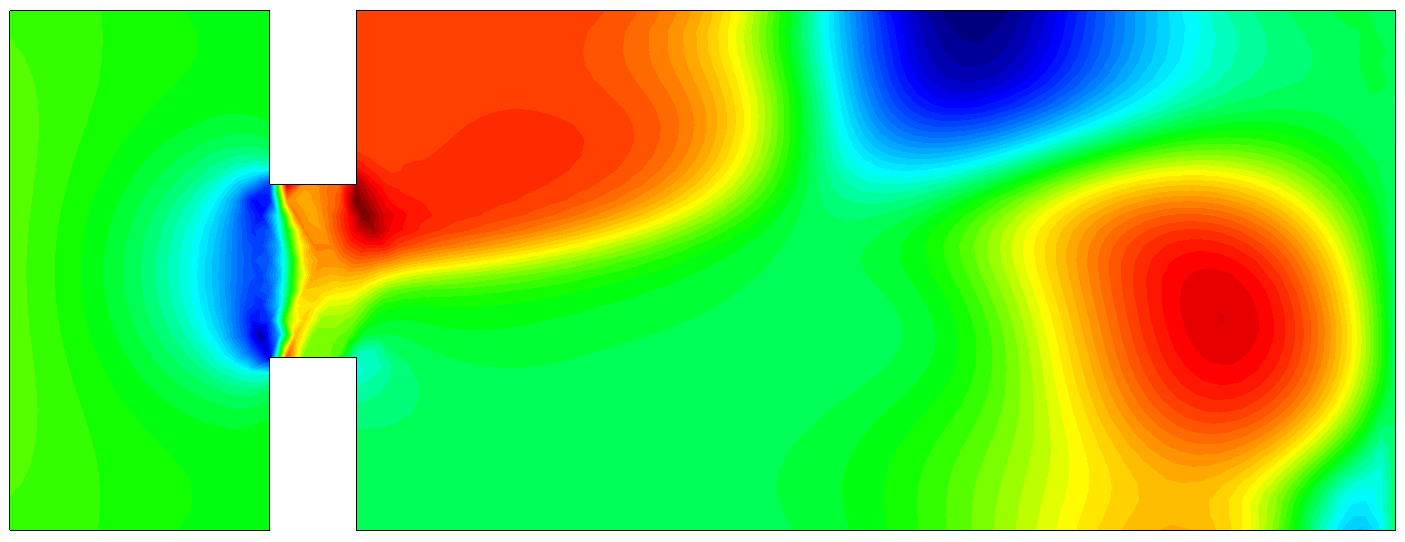}
    \caption{pressure mode 4}
  \end{subfigure}
  \caption{First four velocity and pressure modes for the $\boldsymbol{\varphi}_{\text{\tiny FFD+RBF}}$ mapping}
  \label{fig: Example 1 modes nonlinear mapping}
\end{figure}

\begin{figure}[H]
  \centering
  \begin{subfigure}[b]{0.245\textwidth}
    \includegraphics[width=\linewidth]{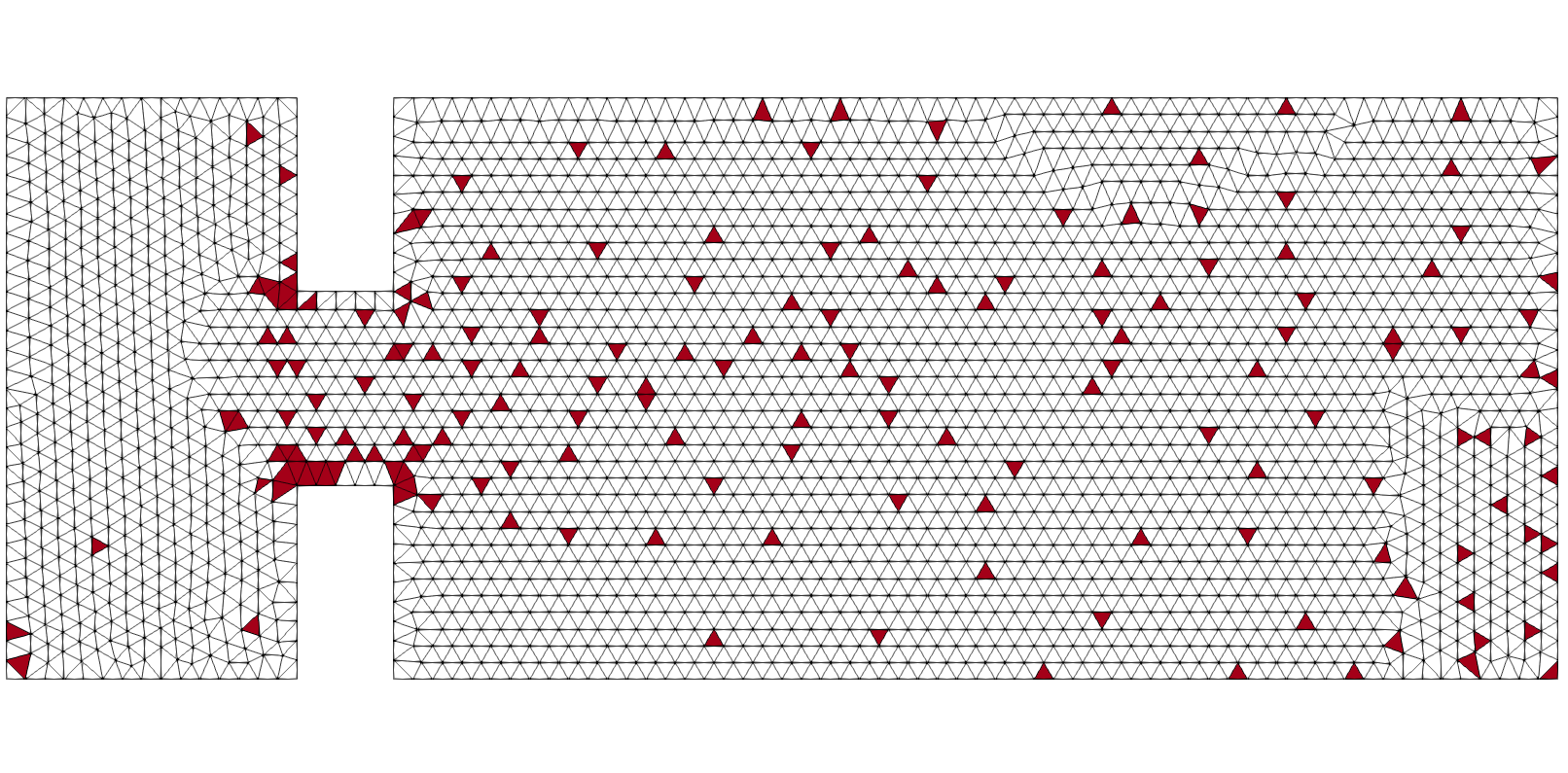}
    \caption{ $\epsilon_{\text{\tiny SOL}} = 1e-3, \epsilon_{\text{\tiny RES}} = 1e-3, $ }
    %\label{fig:HROM_elemns_ffd_rbf__a}
  \end{subfigure}
  \hfill
  \begin{subfigure}[b]{0.245\textwidth}
    \includegraphics[width=\linewidth]{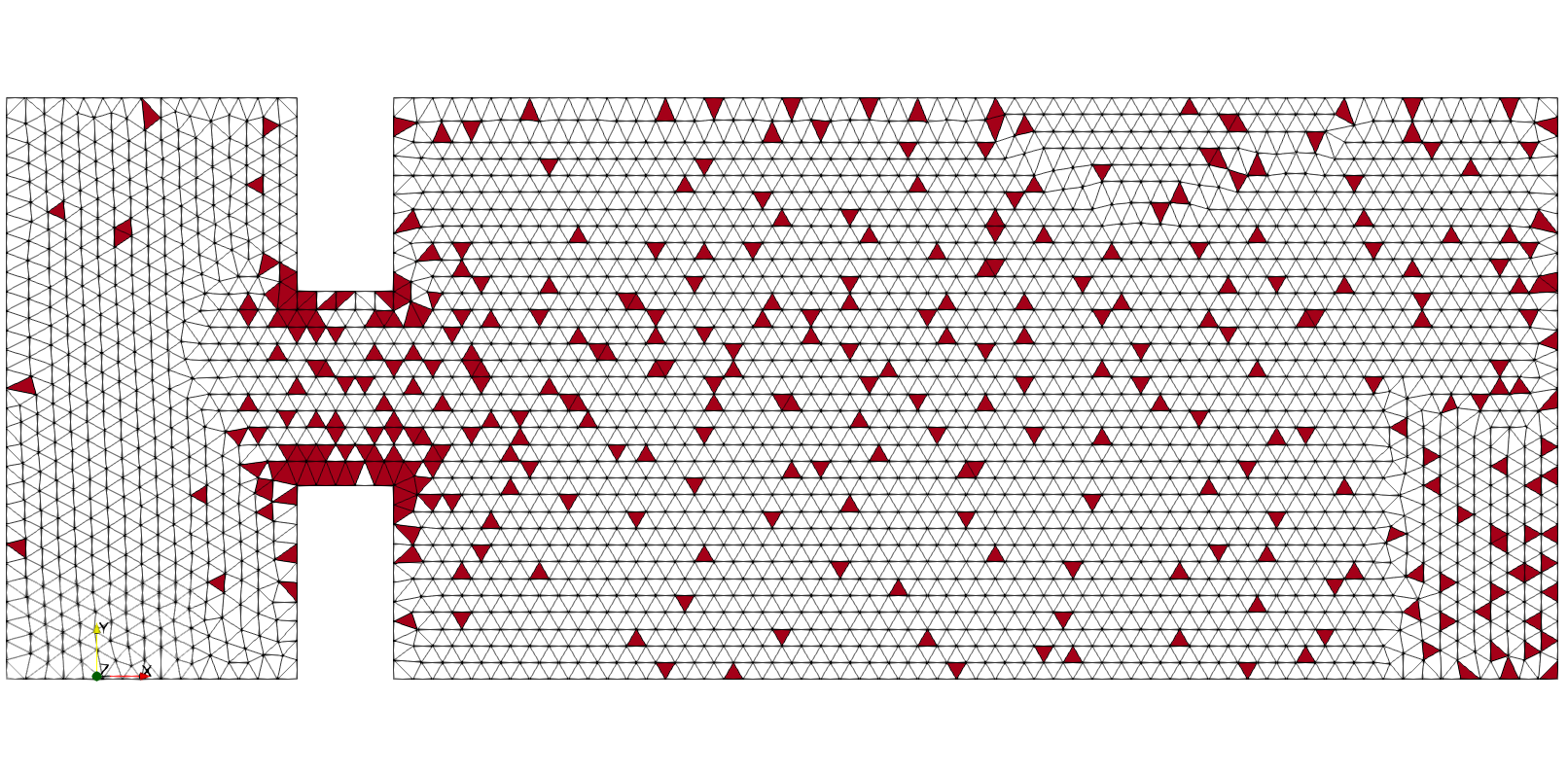}
    \caption{ $\epsilon_{\text{\tiny SOL}} = 1e-3, \epsilon_{\text{\tiny RES}} = 1e-4, $ }
    %\label{fig:HROM_elemns_ffd_rbf__b}
  \end{subfigure}
  \hfill
  \begin{subfigure}[b]{0.245\textwidth}
    \includegraphics[width=\linewidth]{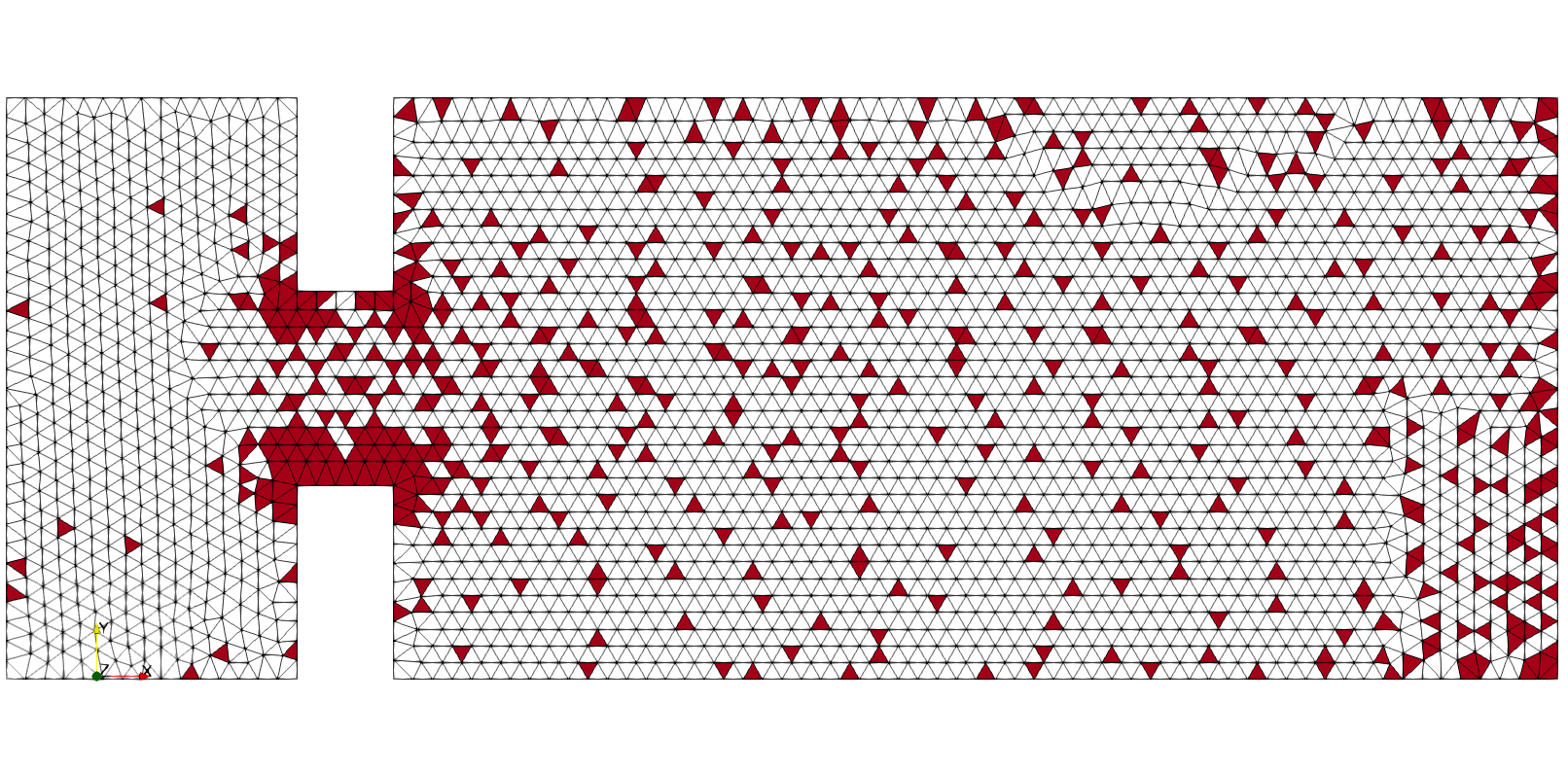}
    \caption{ $\epsilon_{\text{\tiny SOL}} = 1e-3, \epsilon_{\text{\tiny RES}} = 1e-5, $ }
    %\label{fig:HROM_elemns_ffd_rbf__c}
  \end{subfigure}
  \hfill
  \begin{subfigure}[b]{0.245\textwidth}
    \includegraphics[width=\linewidth]{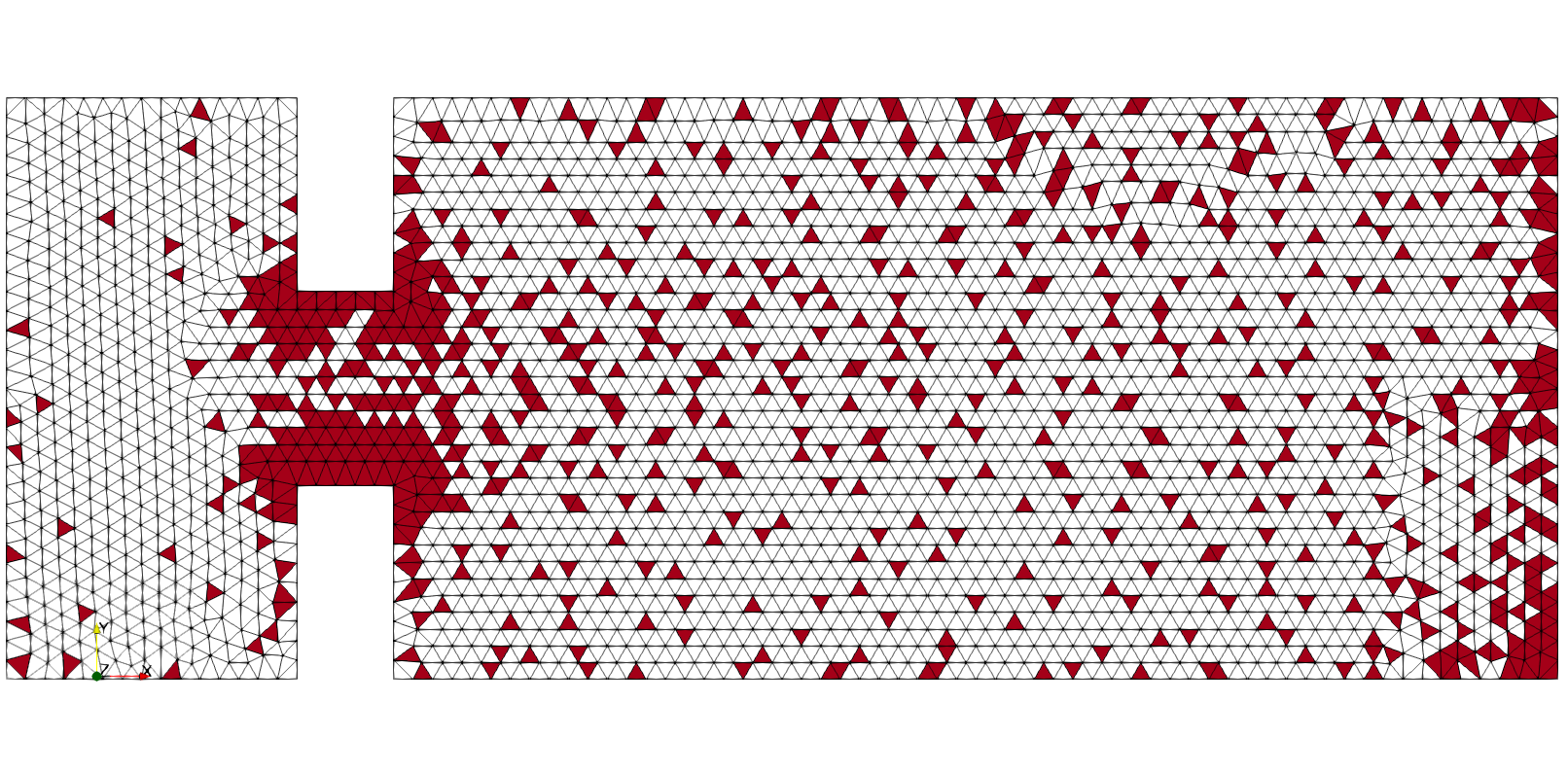}
    \caption{ $\epsilon_{\text{\tiny SOL}} = 1e-3, \epsilon_{\text{\tiny RES}} = 1e-6, $ }
    %\label{fig:HROM_elemns_ffd_rbf__d}
  \end{subfigure}

  \begin{subfigure}[b]{0.245\textwidth}
    \includegraphics[width=\linewidth]{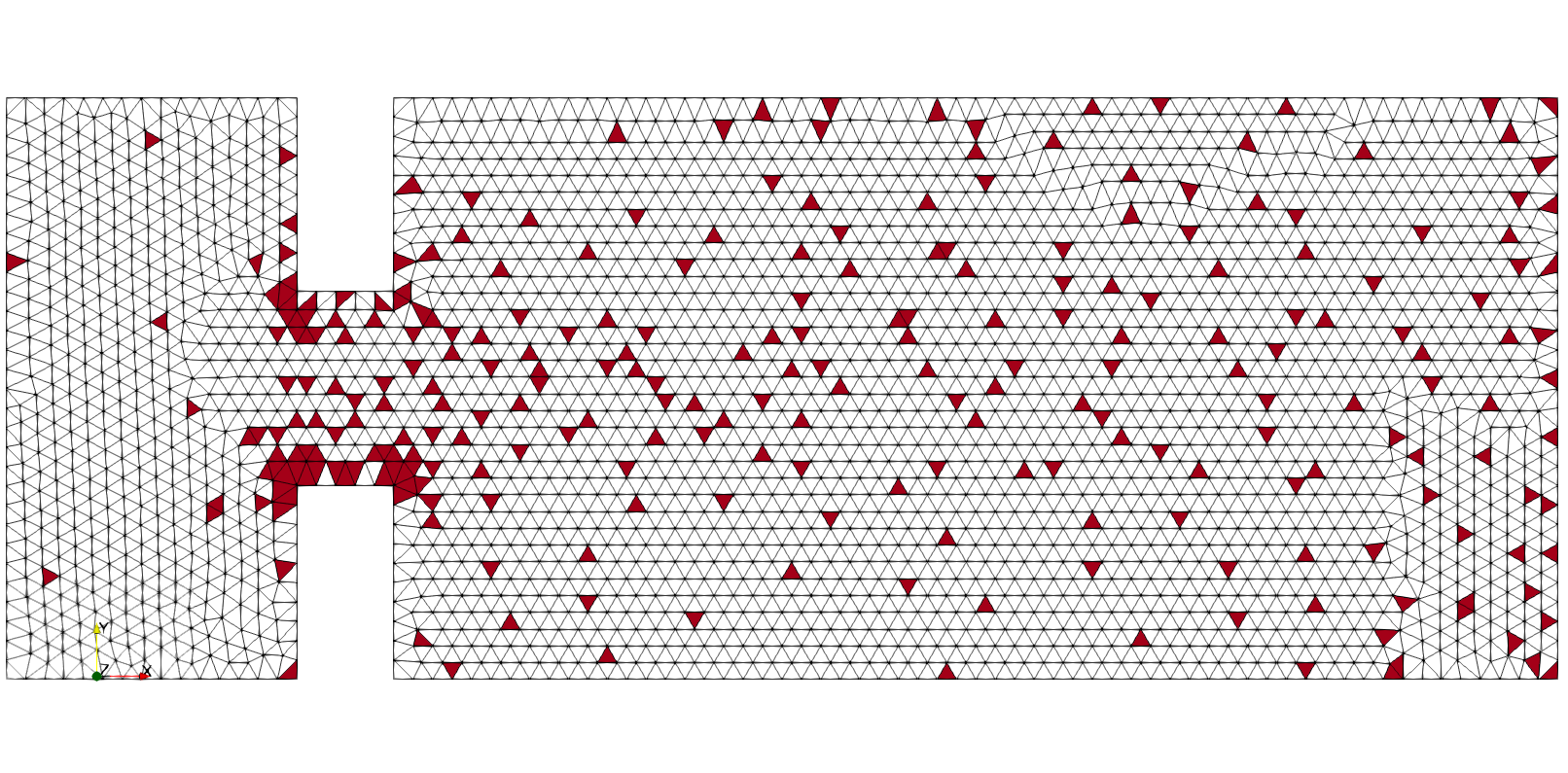}
    \caption{ $\epsilon_{\text{\tiny SOL}} = 1e-4, \epsilon_{\text{\tiny RES}} = 1e-3, $ }
    %\label{fig:HROM_elemns_ffd_rbf__e}
  \end{subfigure}
  \hfill
  \begin{subfigure}[b]{0.245\textwidth}
    \includegraphics[width=\linewidth]{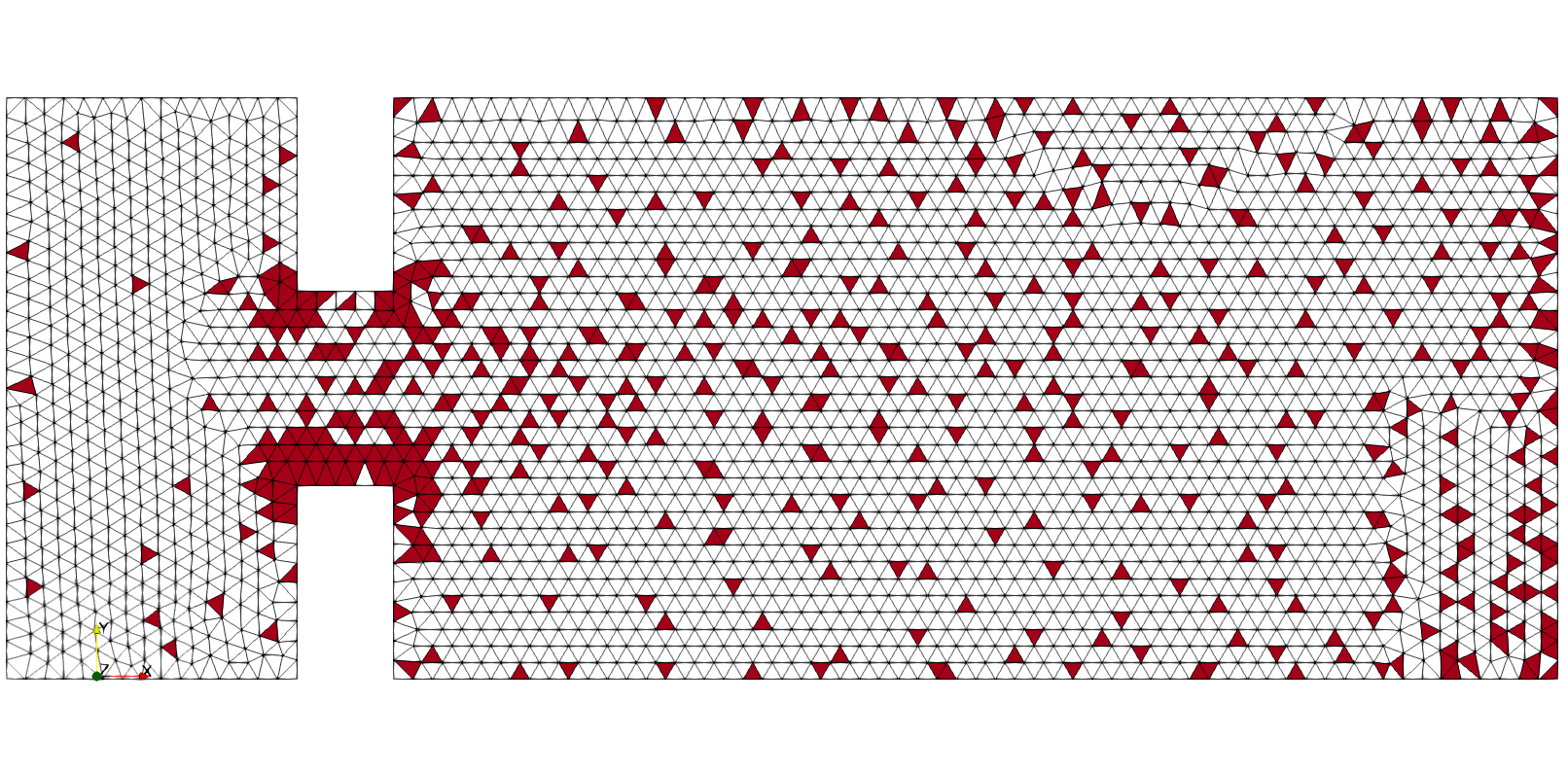}
    \caption{ $\epsilon_{\text{\tiny SOL}} = 1e-4, \epsilon_{\text{\tiny RES}} = 1e-4, $ }
    %\label{fig:HROM_elemns_ffd_rbf__f}
  \end{subfigure}
  \hfill
  \begin{subfigure}[b]{0.245\textwidth}
    \includegraphics[width=\linewidth]{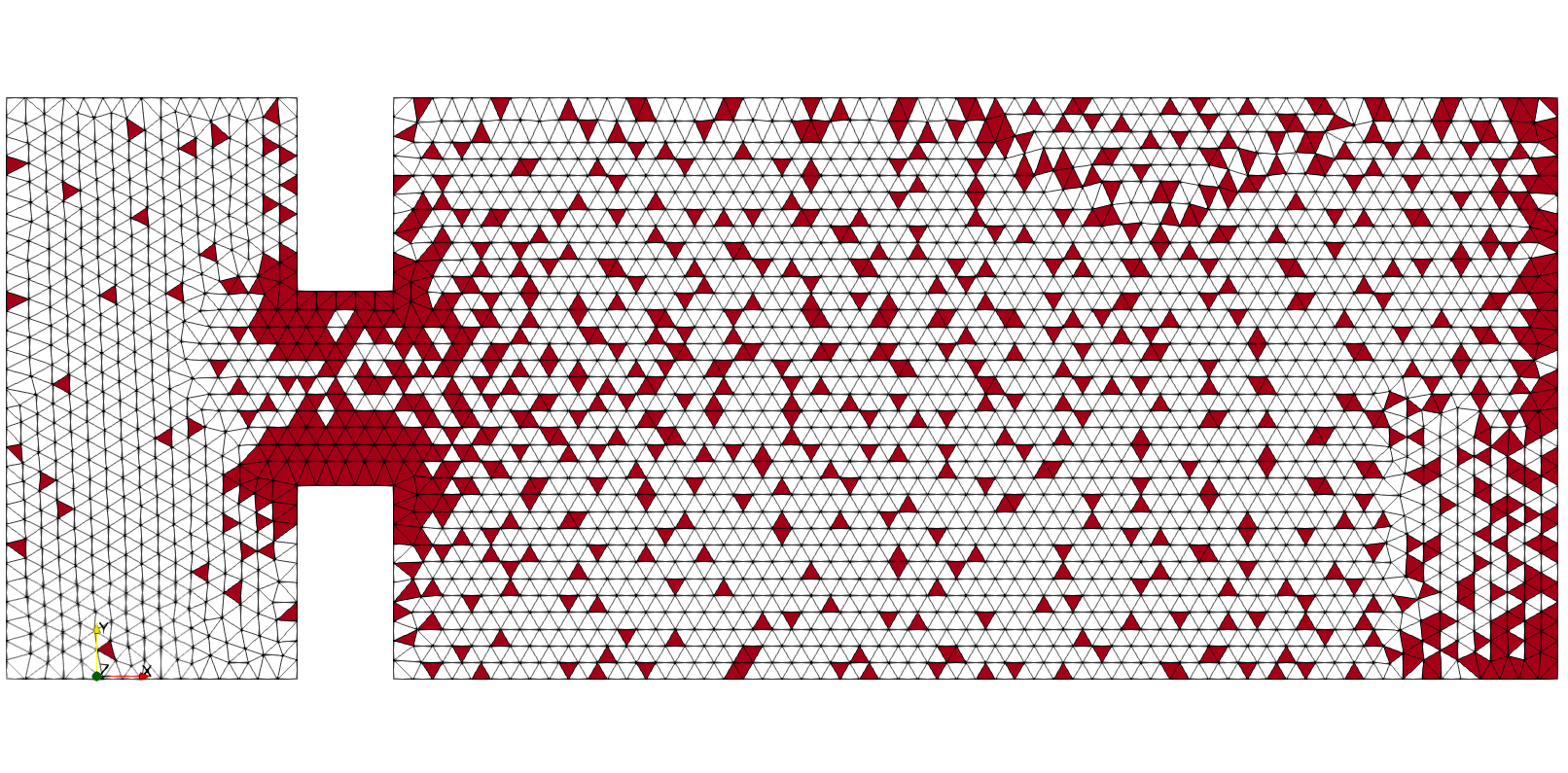}
    \caption{ $\epsilon_{\text{\tiny SOL}} = 1e-4, \epsilon_{\text{\tiny RES}} = 1e-5, $ }
    %\label{fig:HROM_elemns_ffd_rbf__g}
  \end{subfigure}
  \hfill
  \begin{subfigure}[b]{0.245\textwidth}
    \includegraphics[width=\linewidth]{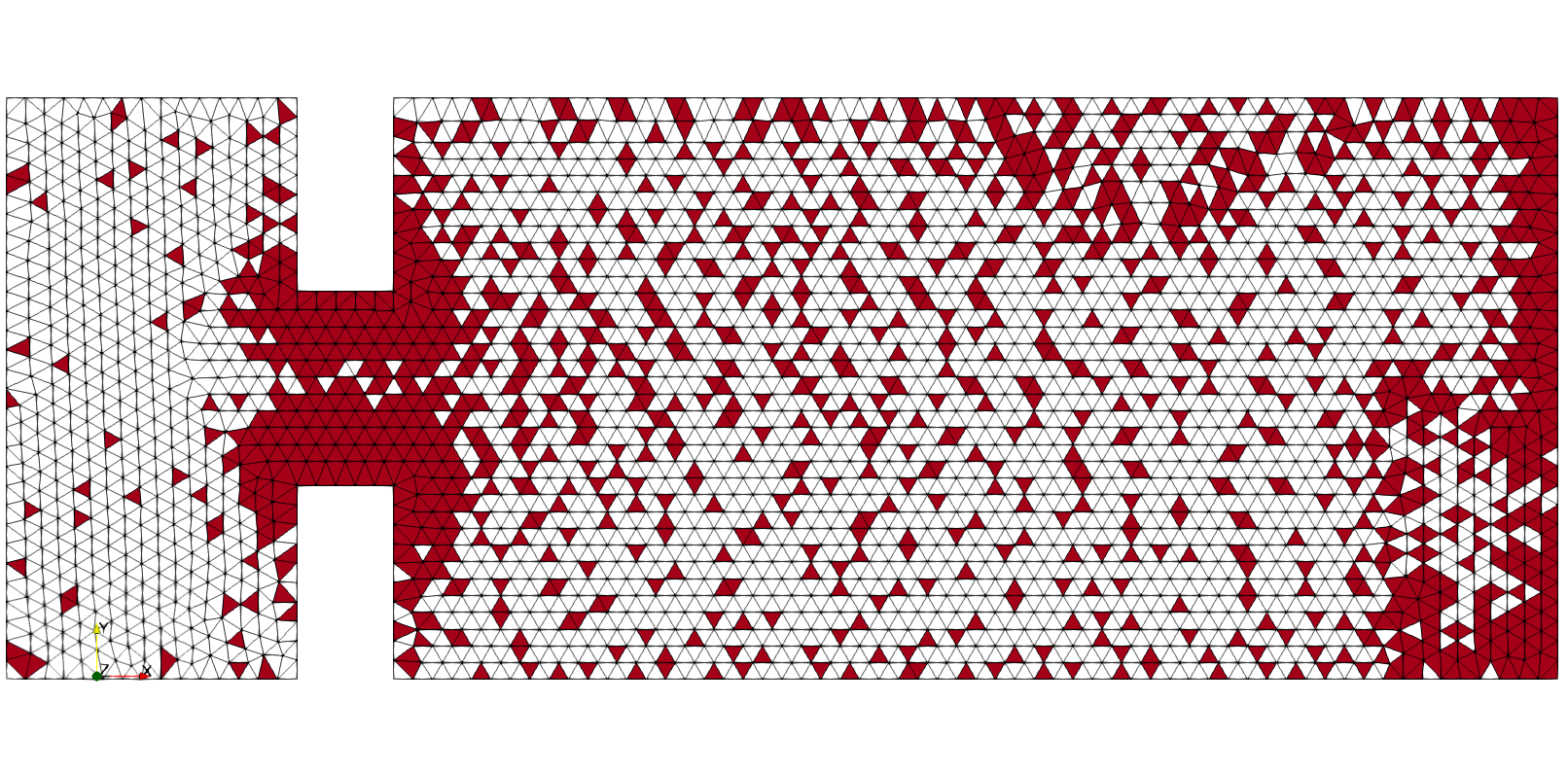}
    \caption{ $\epsilon_{\text{\tiny SOL}} = 1e-4, \epsilon_{\text{\tiny RES}} = 1e-6, $ }
    %\label{fig:HROM_elemns_ffd_rbf__h}
  \end{subfigure}

  \begin{subfigure}[b]{0.245\textwidth}
    \includegraphics[width=\linewidth]{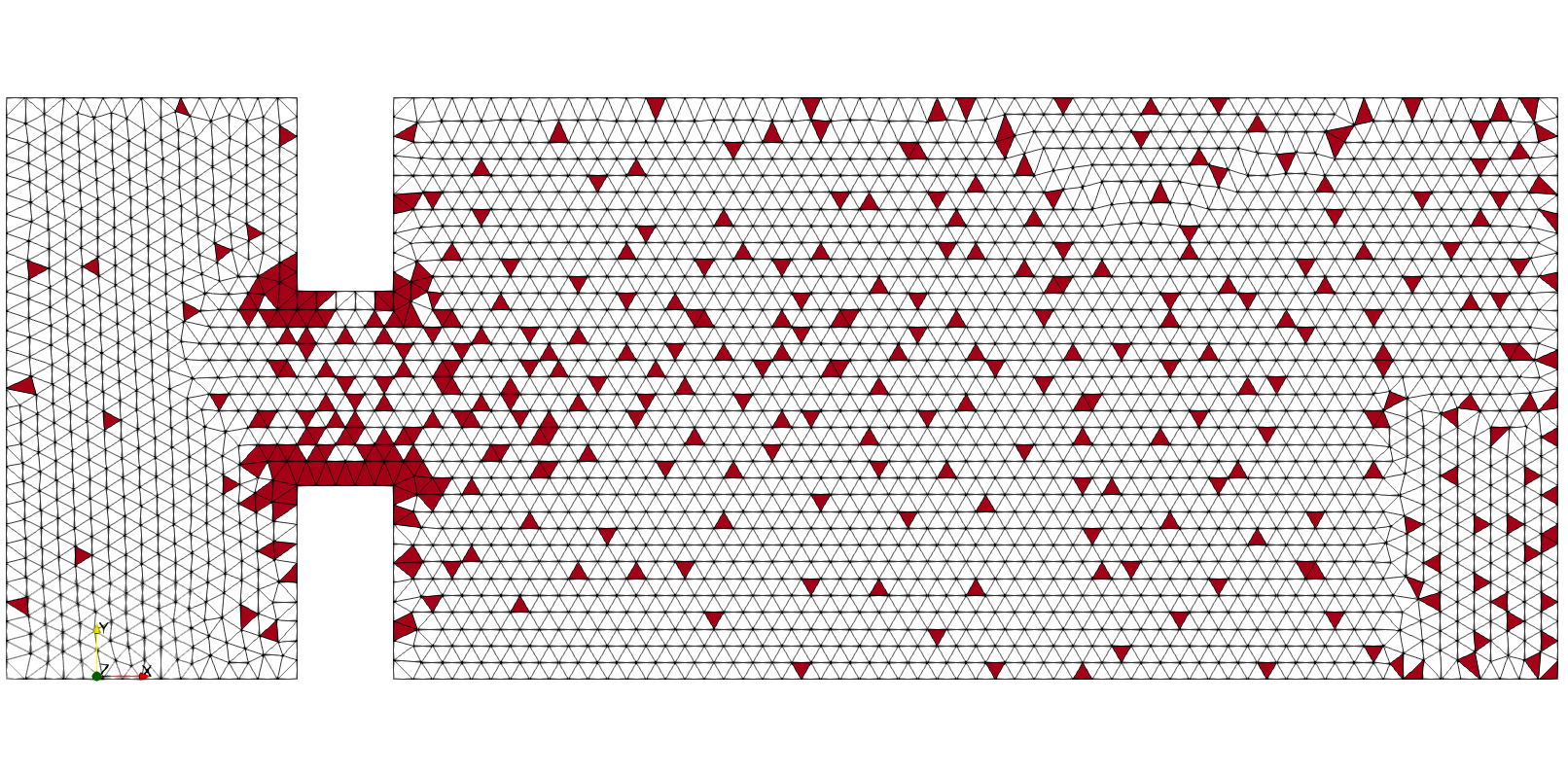}
    \caption{ $\epsilon_{\text{\tiny SOL}} = 1e-5, \epsilon_{\text{\tiny RES}} = 1e-3, $ }
    %\label{fig:HROM_elemns_ffd_rbf__i}
  \end{subfigure}
  \hfill
  \begin{subfigure}[b]{0.245\textwidth}
    \includegraphics[width=\linewidth]{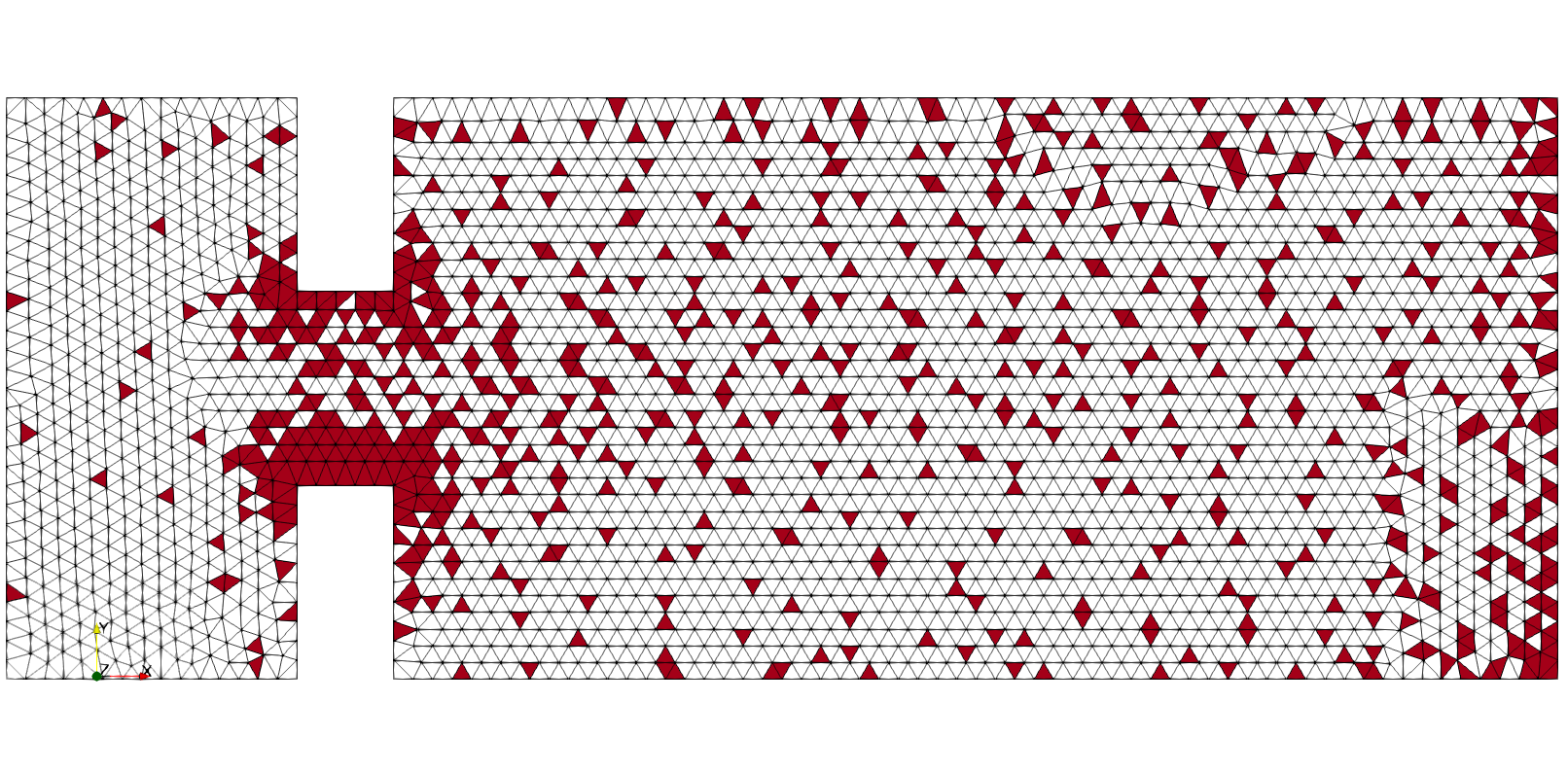}
    \caption{ $\epsilon_{\text{\tiny SOL}} = 1e-5, \epsilon_{\text{\tiny RES}} = 1e-4, $ }
    %\label{fig:HROM_elemns_ffd_rbf__j}
  \end{subfigure}
  \hfill
  \begin{subfigure}[b]{0.245\textwidth}
    \includegraphics[width=\linewidth]{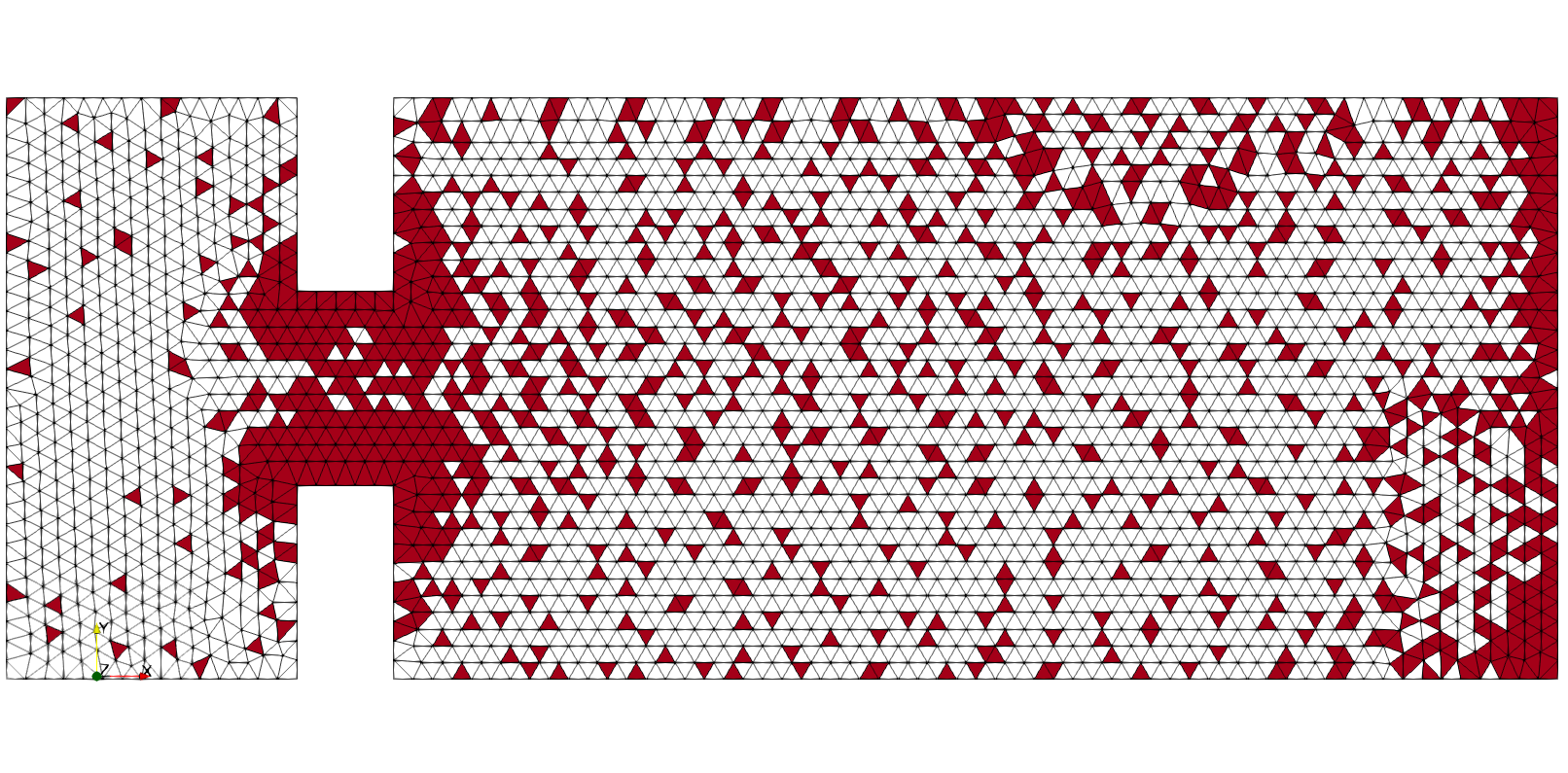}
    \caption{ $\epsilon_{\text{\tiny SOL}} = 1e-5, \epsilon_{\text{\tiny RES}} = 1e-5, $ }
    %\label{fig:HROM_elemns_ffd_rbf__k}
  \end{subfigure}
  \hfill
  \begin{subfigure}[b]{0.245\textwidth}
    \includegraphics[width=\linewidth]{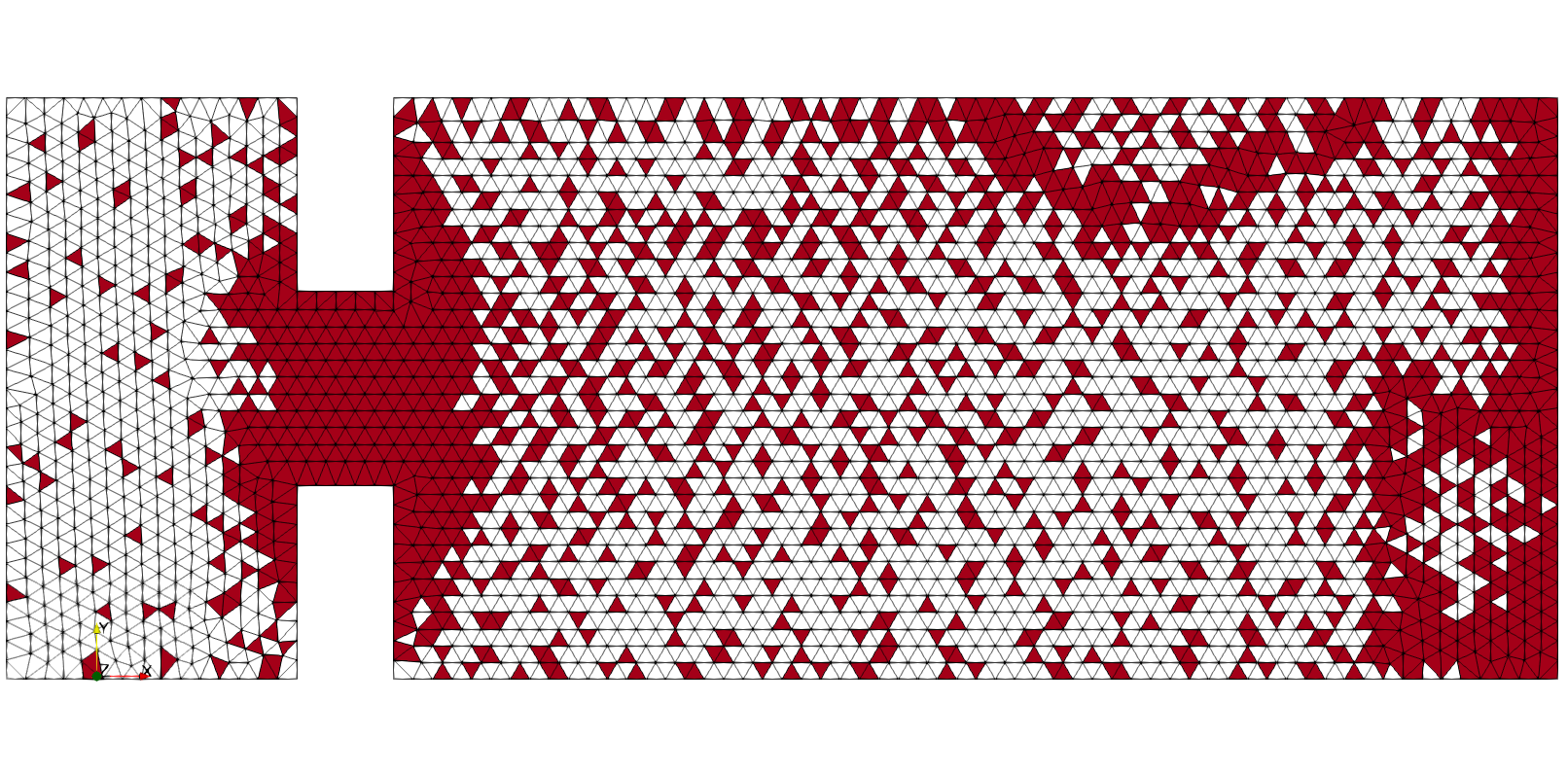}
    \caption{ $\epsilon_{\text{\tiny SOL}} = 1e-5, \epsilon_{\text{\tiny RES}} = 1e-6, $ }
    %\label{fig:HROM_elemns_ffd_rbf__l}
  \end{subfigure}

  \begin{subfigure}[b]{0.245\textwidth}
    \includegraphics[width=\linewidth]{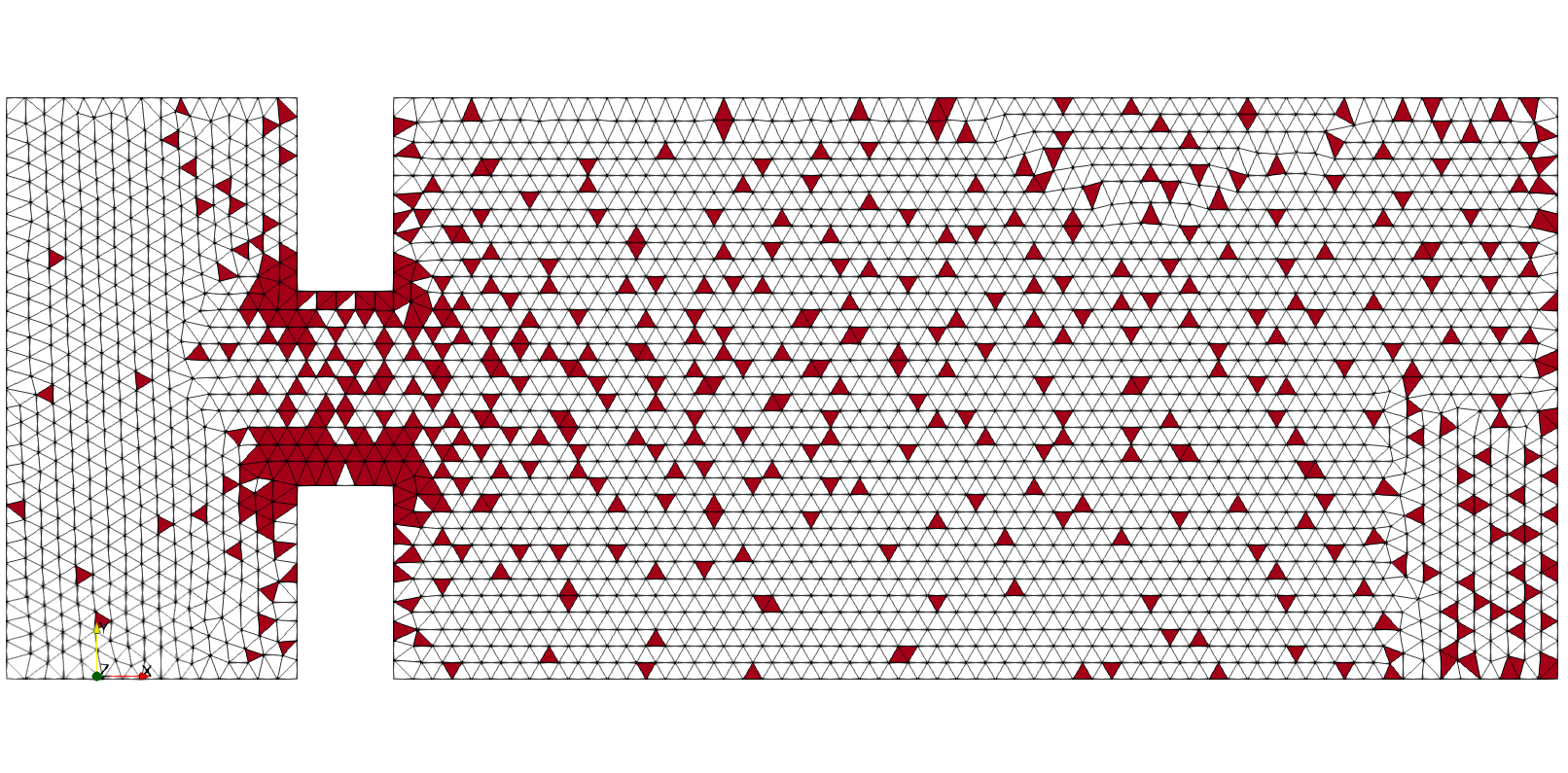}
    \caption{ $\epsilon_{\text{\tiny SOL}} = 1e-6, \epsilon_{\text{\tiny RES}} = 1e-3, $ }
    %\label{fig:HROM_elemns_ffd_rbf__m}
  \end{subfigure}
  \hfill
  \begin{subfigure}[b]{0.245\textwidth}
    \includegraphics[width=\linewidth]{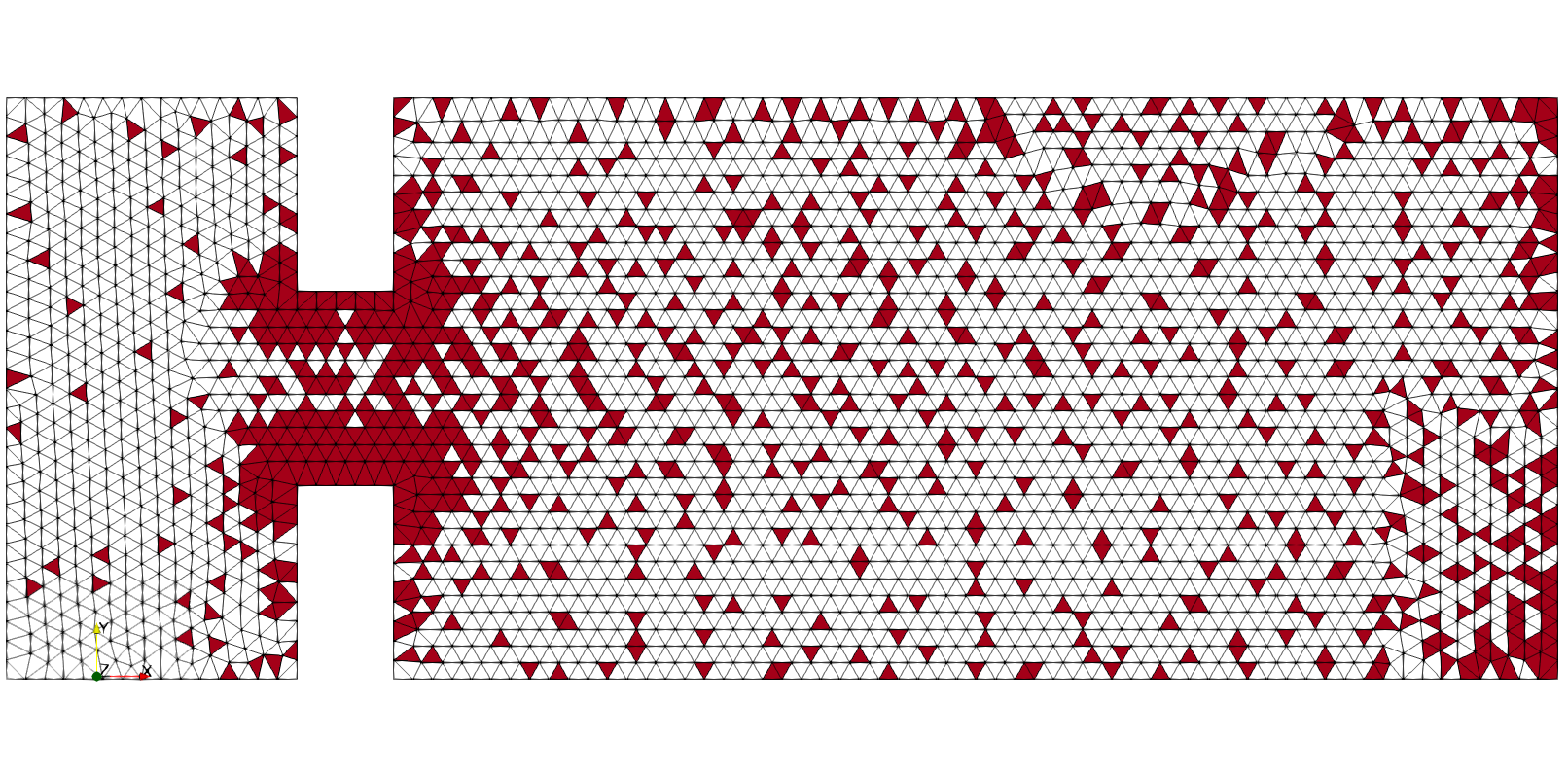}
    \caption{ $\epsilon_{\text{\tiny SOL}} = 1e-6, \epsilon_{\text{\tiny RES}} = 1e-4, $ }
    %\label{fig:HROM_elemns_ffd_rbf__n}
  \end{subfigure}
  \hfill
  \begin{subfigure}[b]{0.245\textwidth}
    \includegraphics[width=\linewidth]{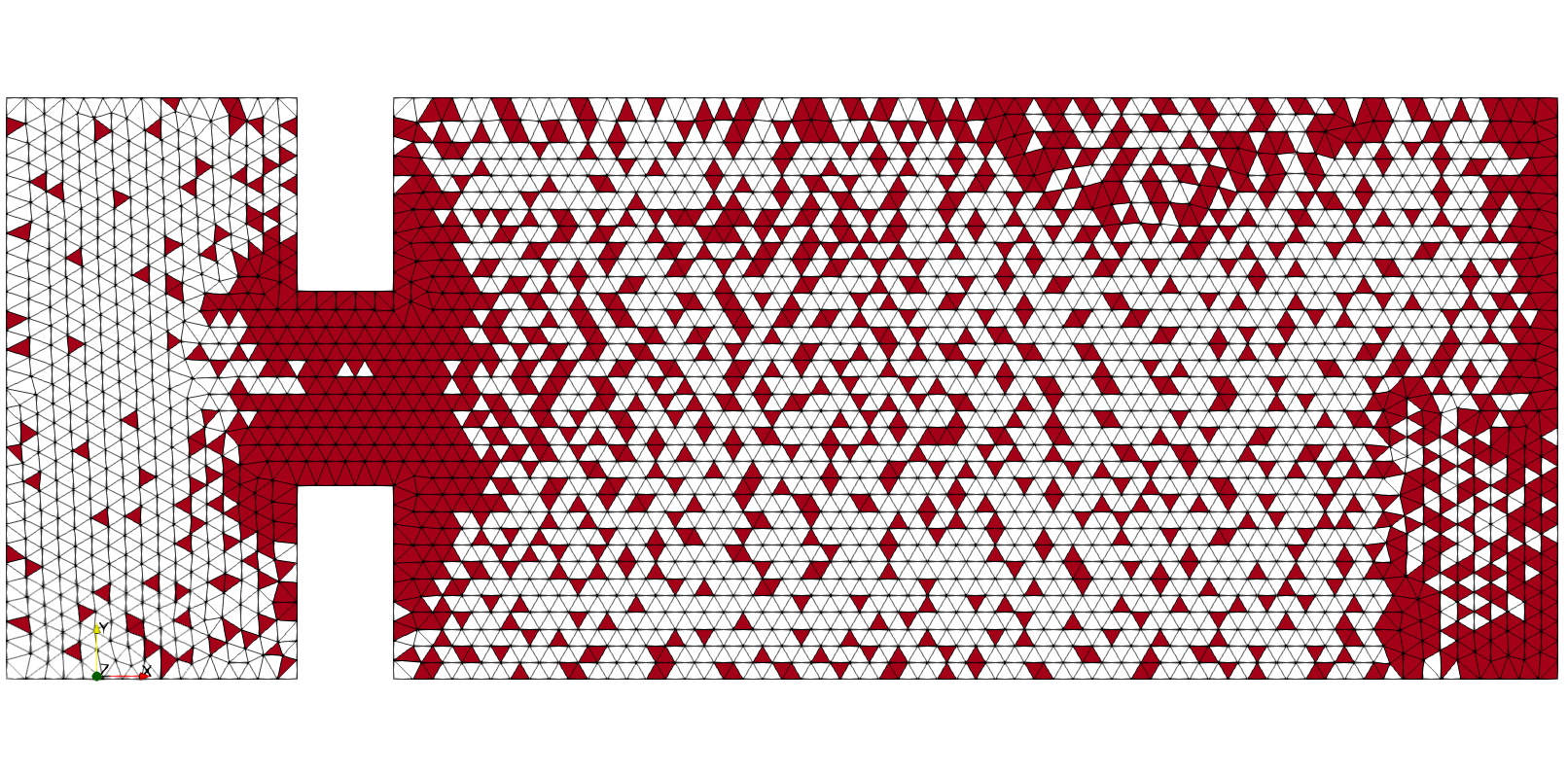}
    \caption{ $\epsilon_{\text{\tiny SOL}} = 1e-6, \epsilon_{\text{\tiny RES}} = 1e-5, $ }
    %\label{fig:HROM_elemns_ffd_rbf__o}
  \end{subfigure}
  \hfill
  \begin{subfigure}[b]{0.245\textwidth}
    \includegraphics[width=\linewidth]{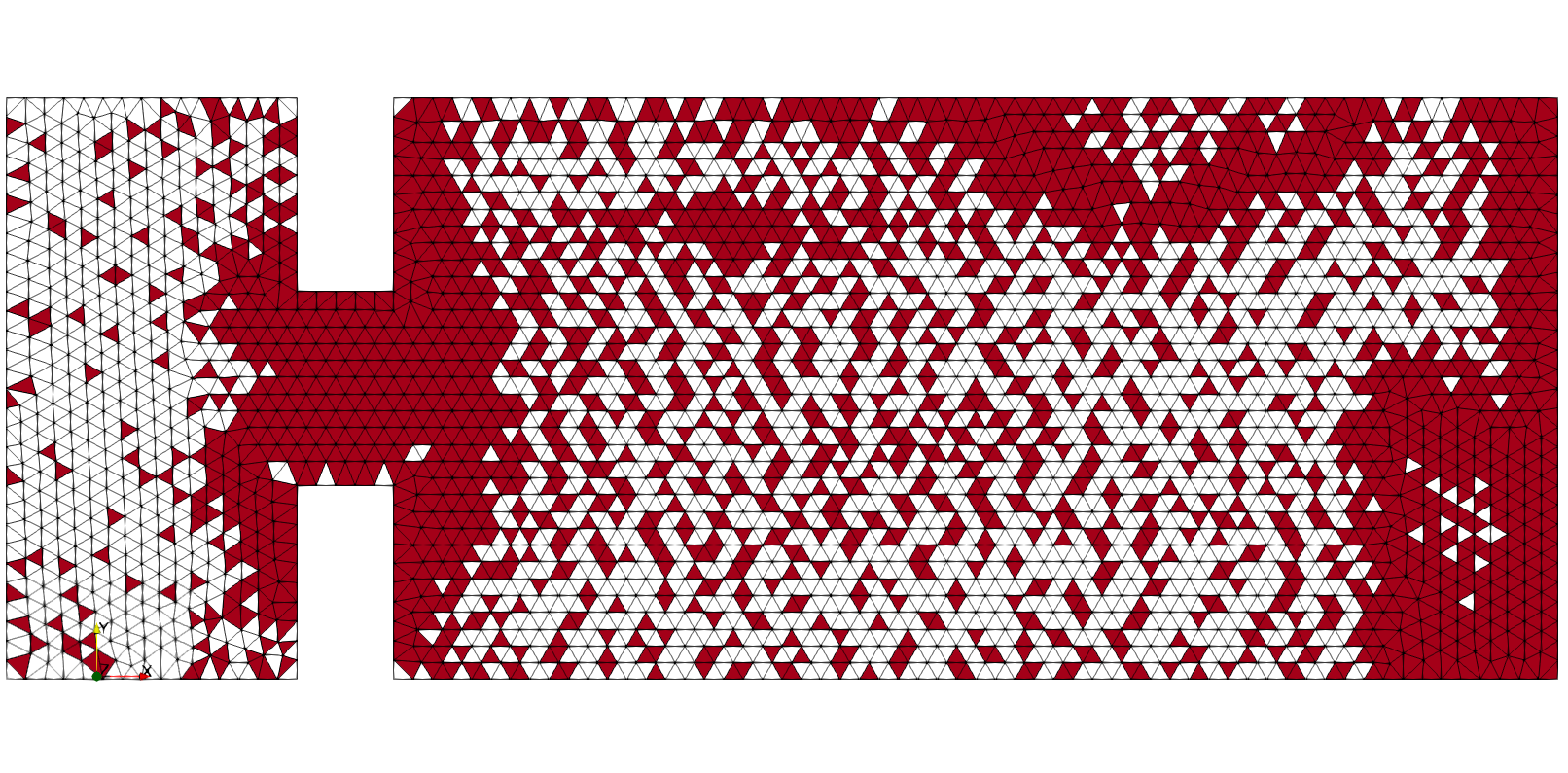}
    \caption{ $\epsilon_{\text{\tiny SOL}} = 1e-6, \epsilon_{\text{\tiny RES}} = 1e-6, $ }
    %\label{fig:HROM_elemns_ffd_rbf__p}
  \end{subfigure}
  \caption{Hyper-reduced elements selected for the $\boldsymbol{\varphi}_{\text{\tiny FFD+RBF}}$ mapping}
  \label{fig: Example 1 HROM elements nonlinear}
\end{figure}

\append{Fig. \ref{fig: Example 1 modes nonlinear mapping} show the pressure and velocity modes for the $\varphi_{\text{\tiny FFD+RBF}}$ mapping. In this scenario, the modes for the nonaffine mapping represent the prevalence in Trajectory 1 of a solution where the jet attaches to the upper wall. Similarly, Fig. \ref{fig: Example 1 HROM elements nonlinear} shows the spatial distribution of elements selected by the ECM algorithm for the 16 HROMs built for the nonaffine mapping. Once again, it is evident that the selected elements tend to accumulate in the upper part of the geometry, aligning with the observed patterns in the FOM solution for the training trajectory and POD modes.}

\subsubsection{\append{Example 2. Affine Mapping}}

\begin{figure}[H]
  \centering
  \begin{subfigure}[b]{0.245\textwidth}
    \includegraphics[width=\linewidth]{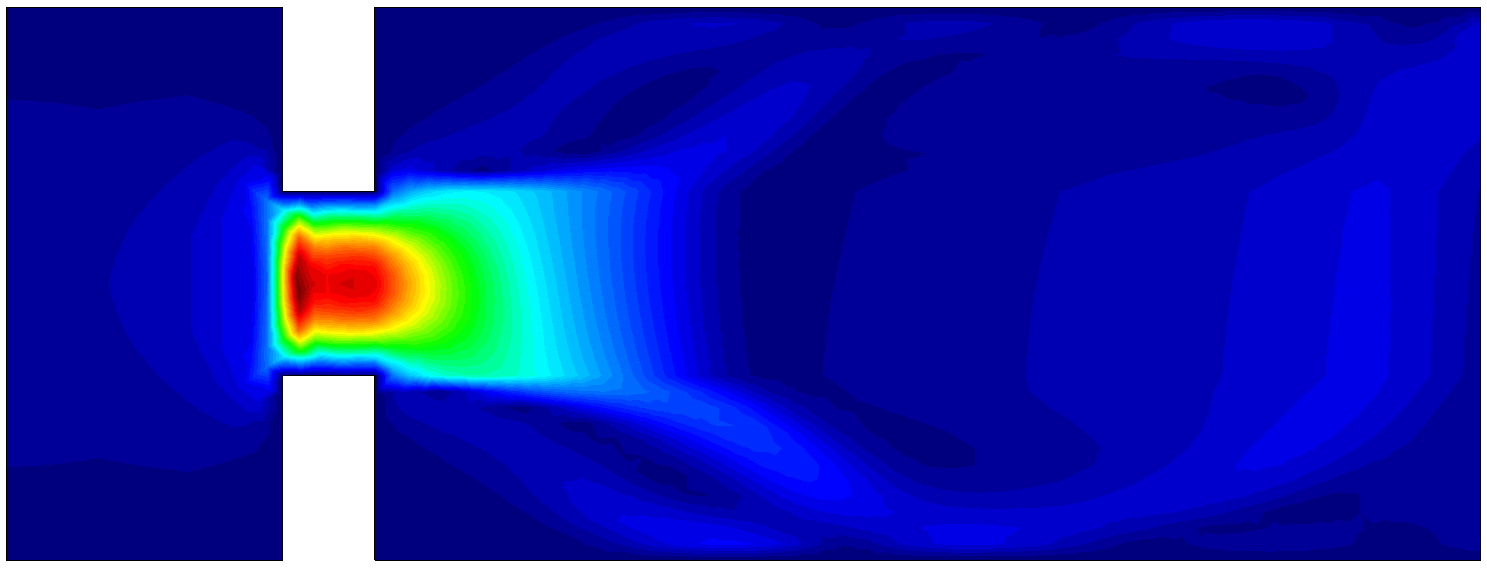}
    \caption{velocity mode 1}
  \end{subfigure}
  \hfill
  \begin{subfigure}[b]{0.245\textwidth}
    \includegraphics[width=\linewidth]{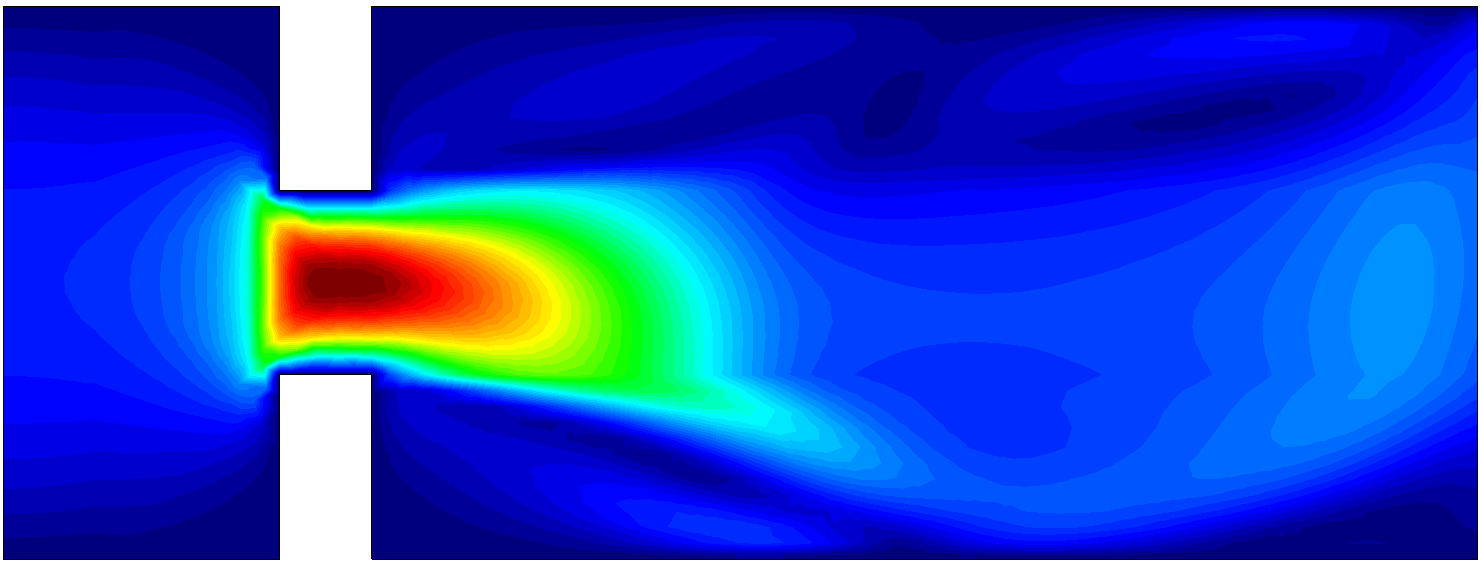}
    \caption{velocity mode 2}
  \end{subfigure}
  \hfill
  \begin{subfigure}[b]{0.245\textwidth}
    \includegraphics[width=\linewidth]{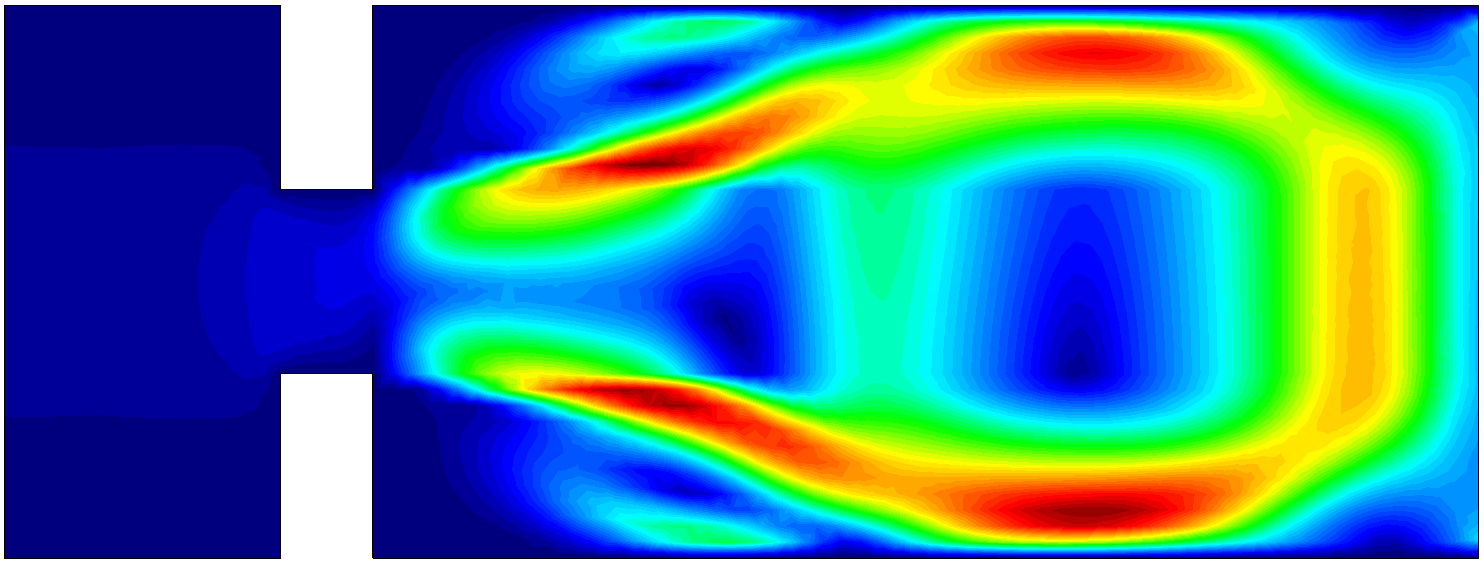}
    \caption{velocity mode 3}
  \end{subfigure}
  \hfill
  \begin{subfigure}[b]{0.245\textwidth}
    \includegraphics[width=\linewidth]{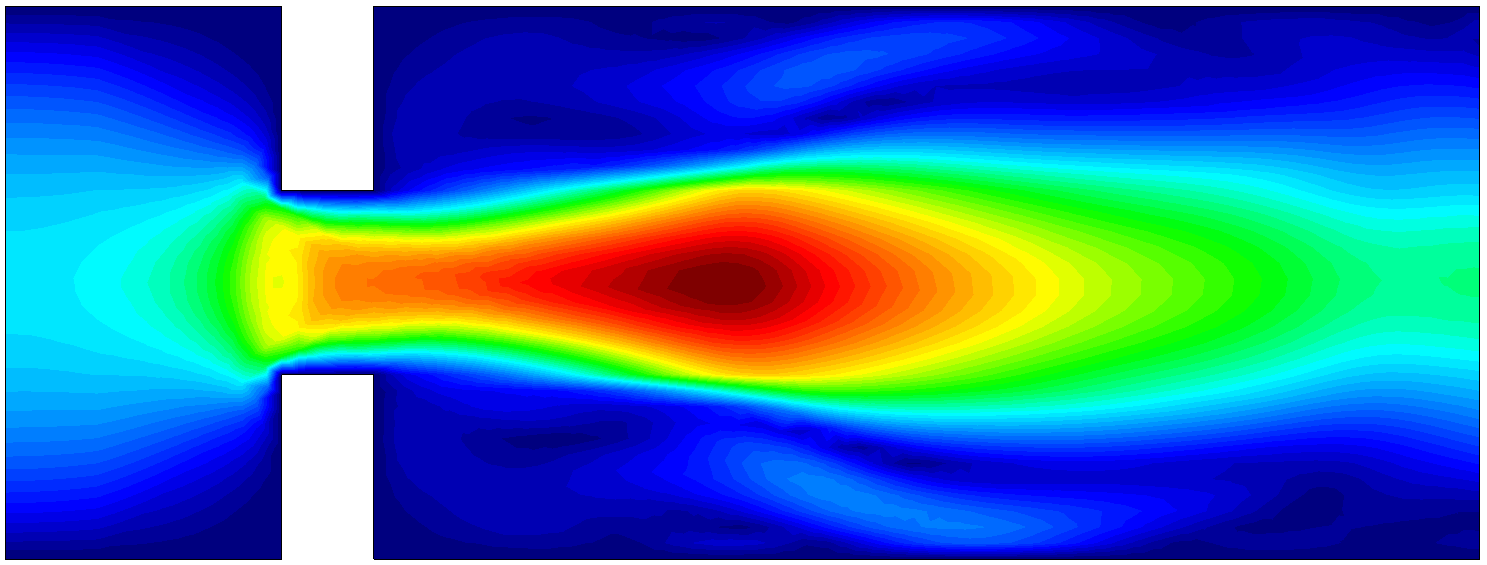}
    \caption{velocity mode 4}
  \end{subfigure}

  \begin{subfigure}[b]{0.245\textwidth}
    \includegraphics[width=\linewidth]{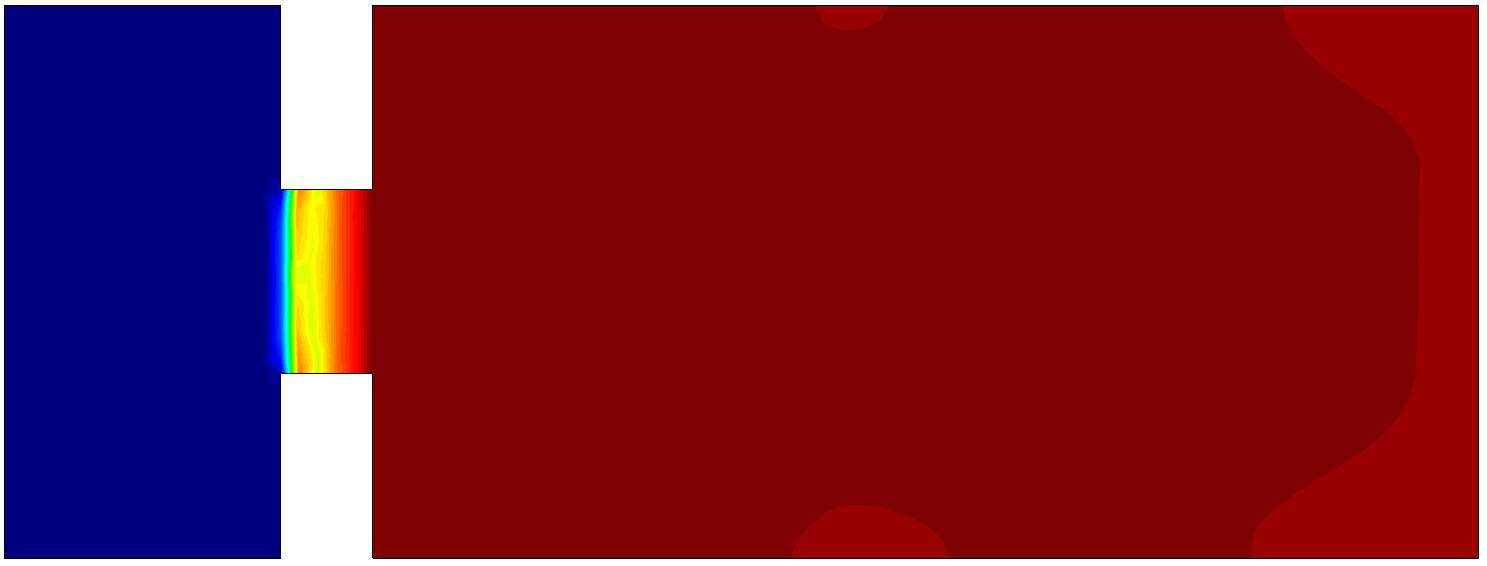}
    \caption{pressure mode 1}
  \end{subfigure}
  \hfill
  \begin{subfigure}[b]{0.245\textwidth}
    \includegraphics[width=\linewidth]{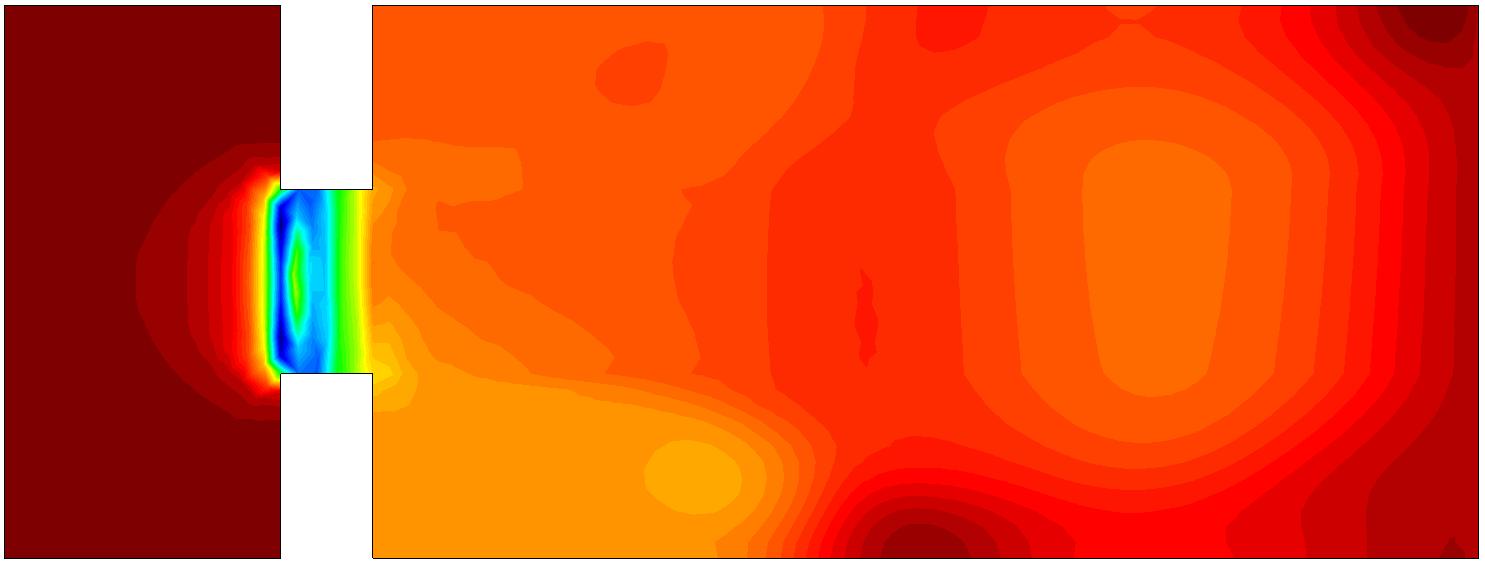}
    \caption{pressure mode 2}
  \end{subfigure}
  \hfill
  \begin{subfigure}[b]{0.245\textwidth}
    \includegraphics[width=\linewidth]{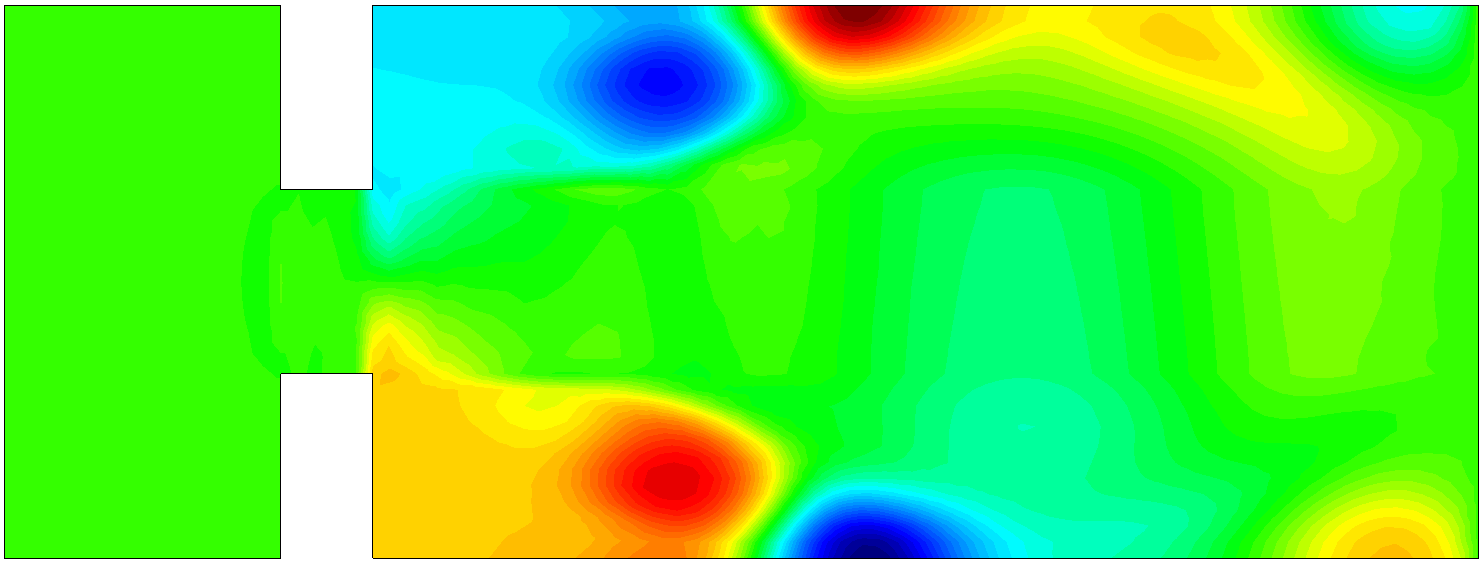}
    \caption{pressure mode 3}
  \end{subfigure}
  \hfill
  \begin{subfigure}[b]{0.245\textwidth}
    \includegraphics[width=\linewidth]{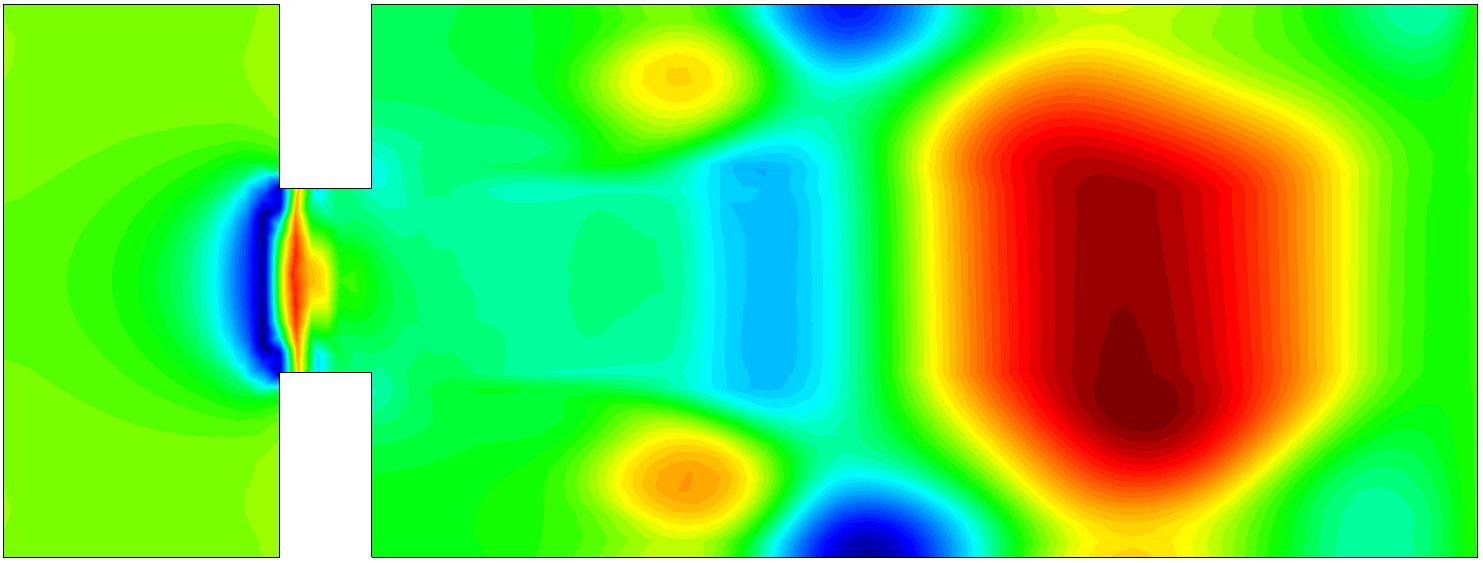}
    \caption{pressure mode 4}
  \end{subfigure}

  \caption{First four velocity and pressure modes for the $\boldsymbol{\varphi}_{\text{\tiny AFFINE}}$ mapping}
  \label{fig:modes affine mapping ex3}
\end{figure}

\begin{figure}[H]
  \centering
  \begin{subfigure}[b]{0.245\textwidth}
    \includegraphics[width=\linewidth]{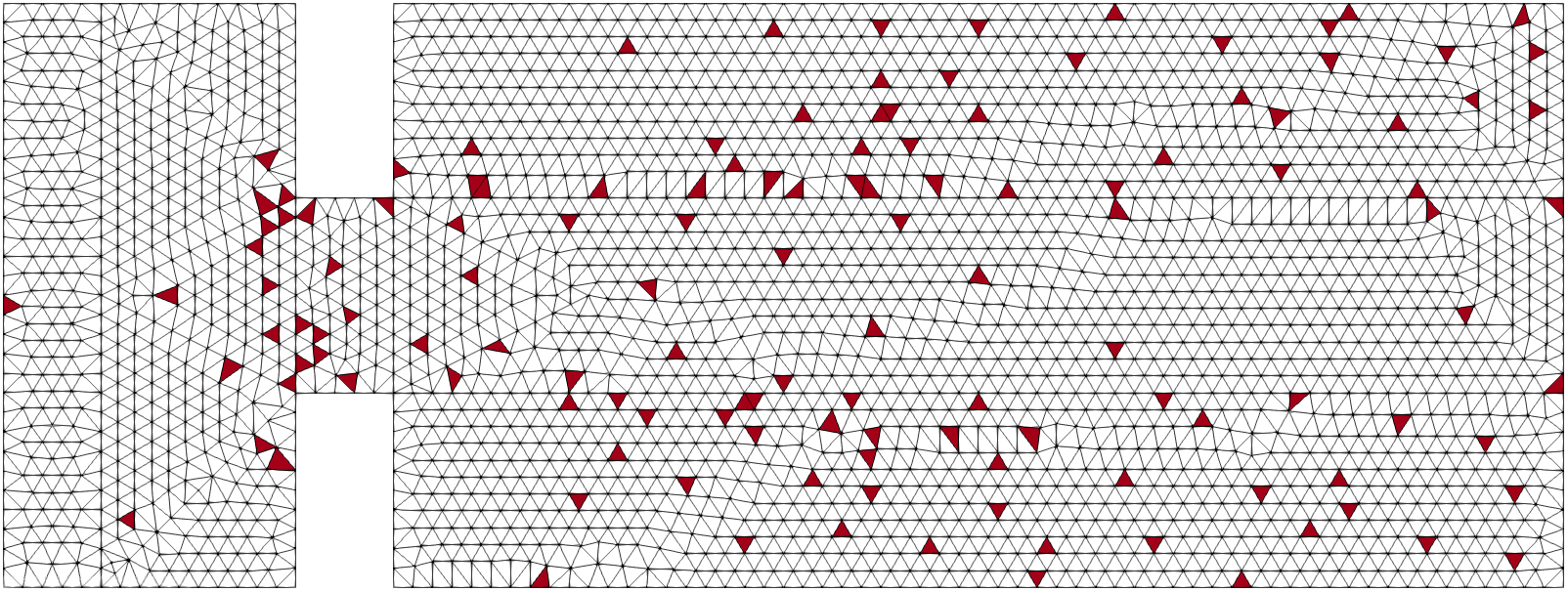}
    \caption{ $\epsilon_{\text{\tiny SOL}} = 1e-3, \epsilon_{\text{\tiny RES}} = 1e-3, $ }
    %\label{fig: ex3HROM_elemns_a}
  \end{subfigure}
  \hfill
  \begin{subfigure}[b]{0.245\textwidth}
    \includegraphics[width=\linewidth]{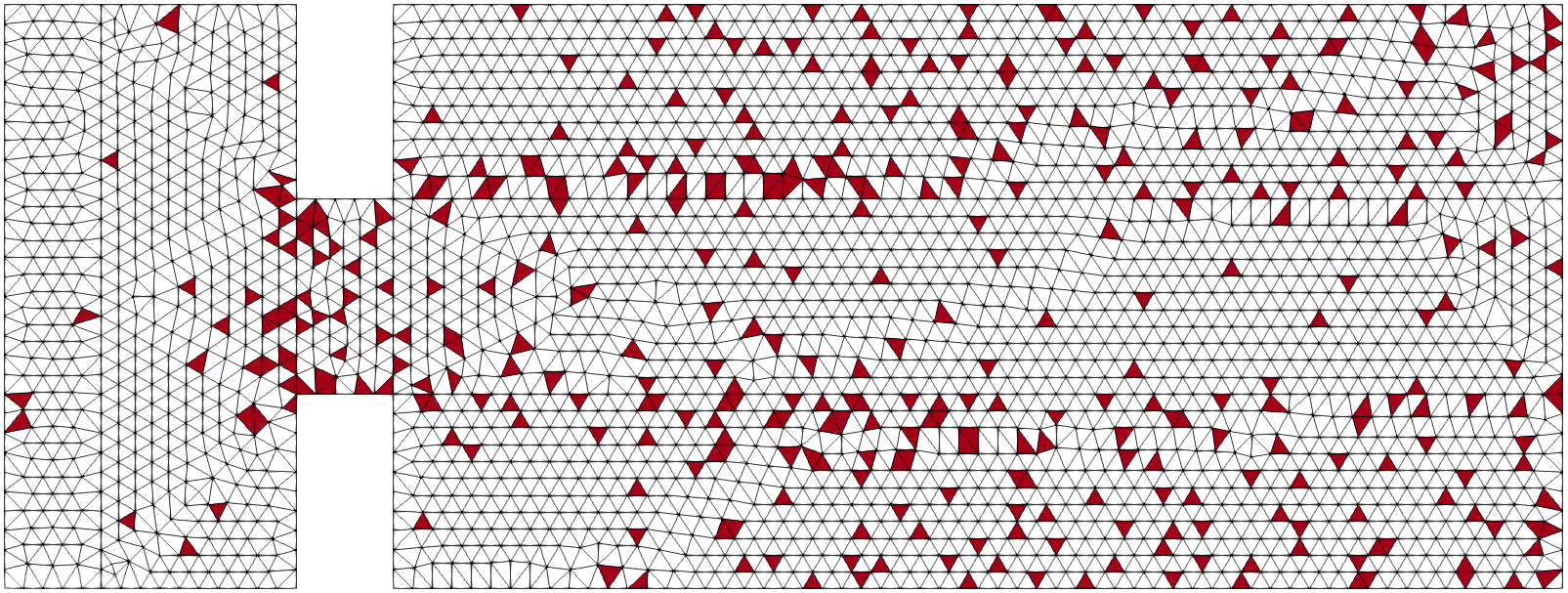}
    \caption{ $\epsilon_{\text{\tiny SOL}} = 1e-3, \epsilon_{\text{\tiny RES}} = 1e-4, $ }
    %\label{fig: ex3HROM_elemns_b}
  \end{subfigure}
  \hfill
  \begin{subfigure}[b]{0.245\textwidth}
    \includegraphics[width=\linewidth]{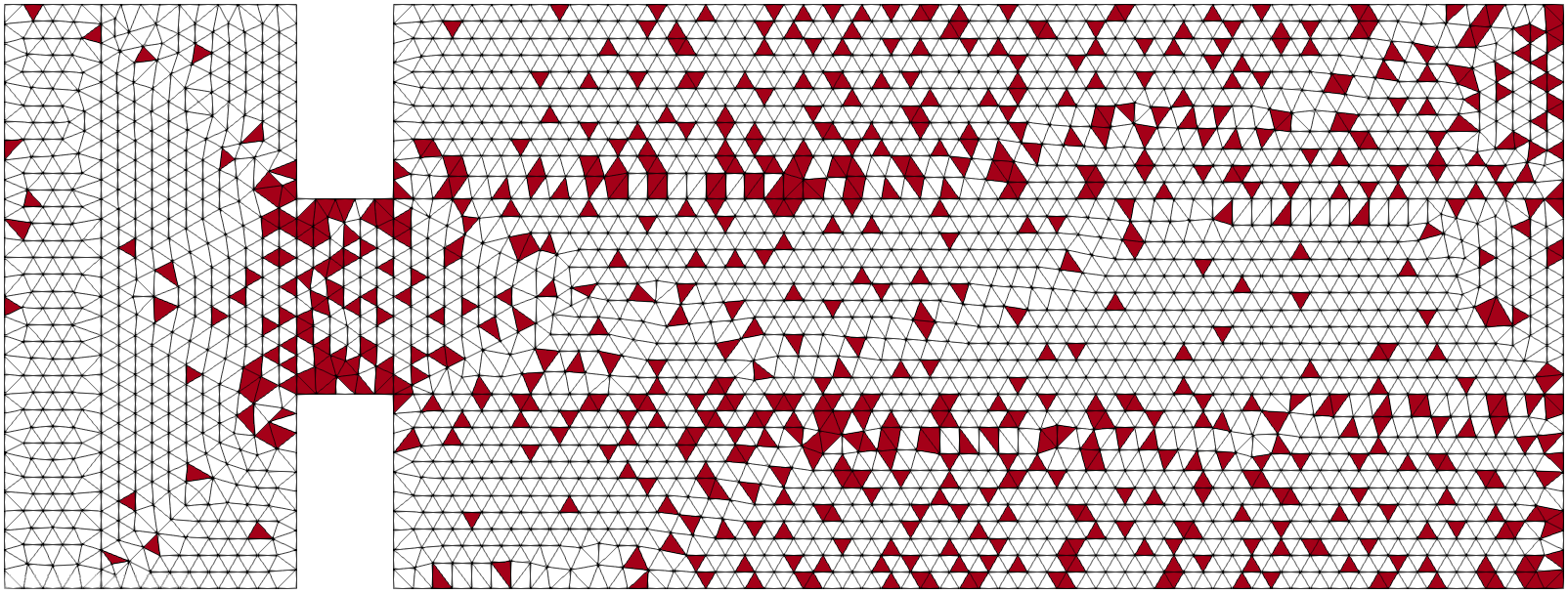}
    \caption{ $\epsilon_{\text{\tiny SOL}} = 1e-3, \epsilon_{\text{\tiny RES}} = 1e-5, $ }
    %\label{fig: ex3HROM_elemns_c}
  \end{subfigure}
  \hfill
  \begin{subfigure}[b]{0.245\textwidth}
    \includegraphics[width=\linewidth]{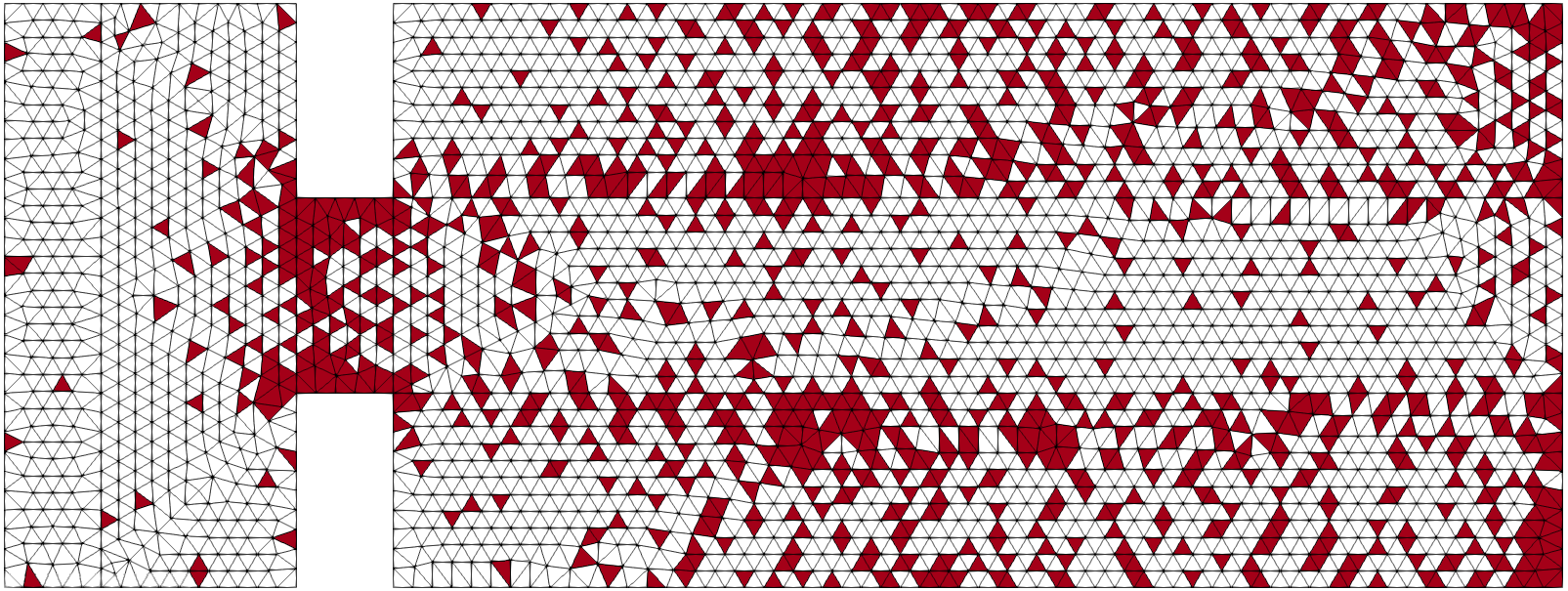}
    \caption{ $\epsilon_{\text{\tiny SOL}} = 1e-3, \epsilon_{\text{\tiny RES}} = 1e-6, $ }
    %\label{fig: ex3HROM_elemns_d}
  \end{subfigure}

  \begin{subfigure}[b]{0.245\textwidth}
    \includegraphics[width=\linewidth]{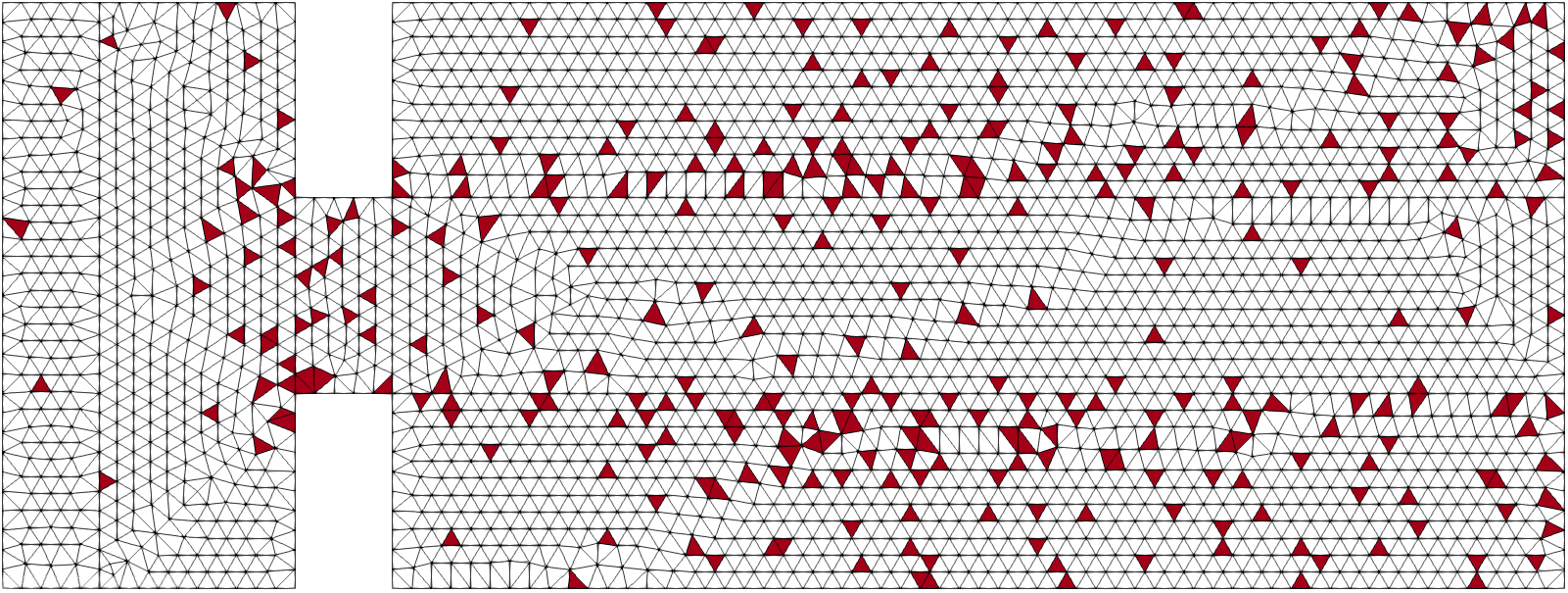}
    \caption{ $\epsilon_{\text{\tiny SOL}} = 1e-4, \epsilon_{\text{\tiny RES}} = 1e-3, $ }
    %\label{fig: ex3HROM_elemns_e}
  \end{subfigure}
  \hfill
  \begin{subfigure}[b]{0.245\textwidth}
    \includegraphics[width=\linewidth]{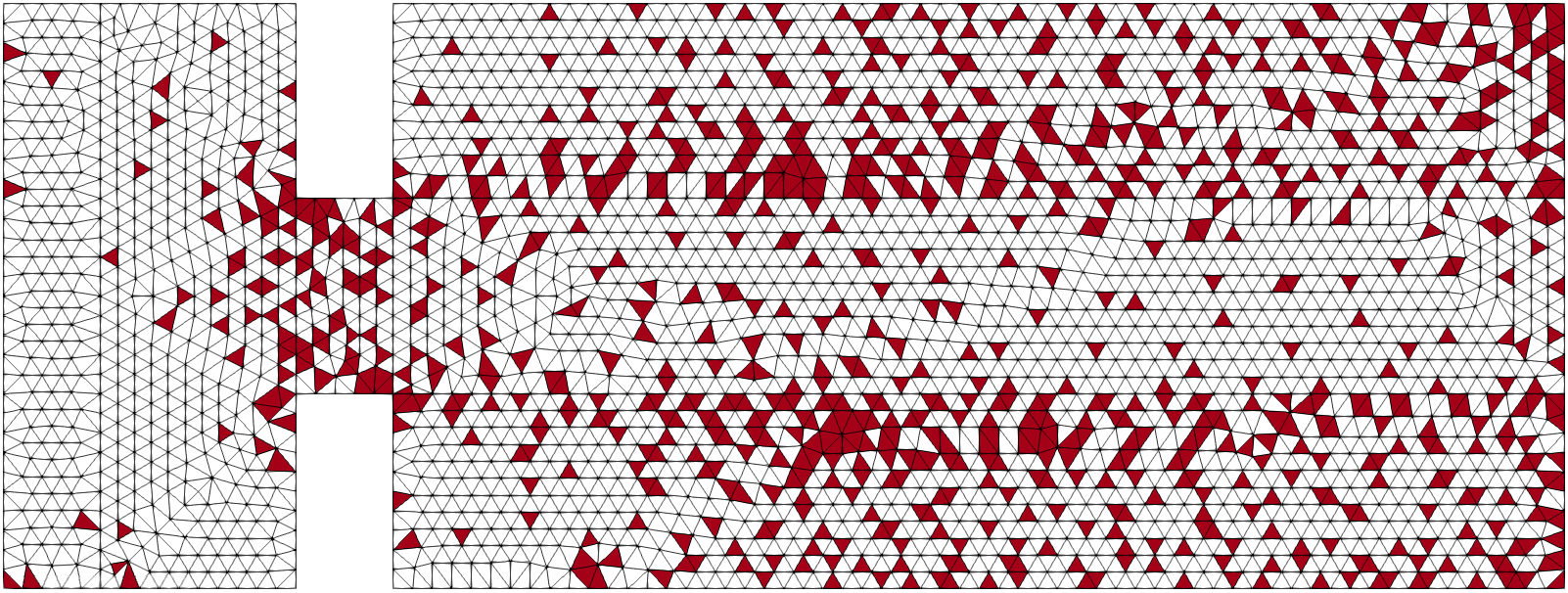}
    \caption{ $\epsilon_{\text{\tiny SOL}} = 1e-4, \epsilon_{\text{\tiny RES}} = 1e-4, $ }
    %\label{fig: ex3HROM_elemns_f}
  \end{subfigure}
  \hfill
  \begin{subfigure}[b]{0.245\textwidth}
    \includegraphics[width=\linewidth]{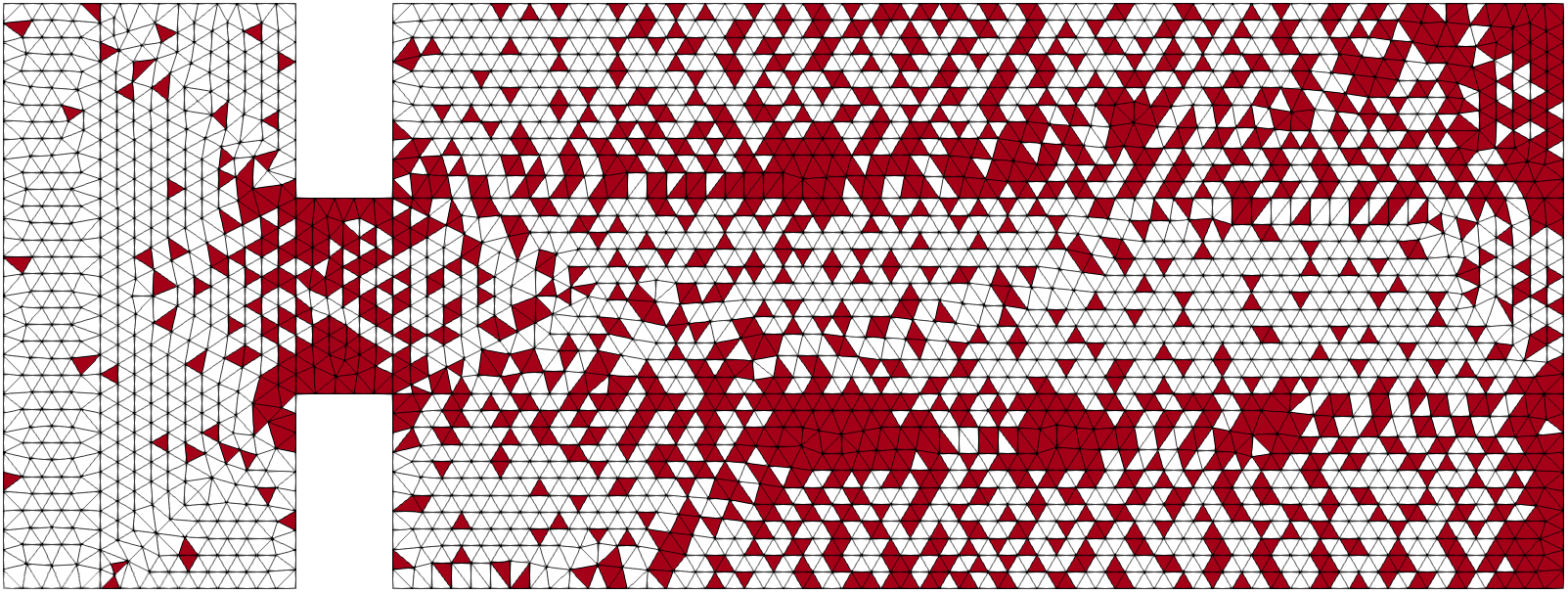}
    \caption{ $\epsilon_{\text{\tiny SOL}} = 1e-4, \epsilon_{\text{\tiny RES}} = 1e-5, $ }
    %\label{fig: ex3HROM_elemns_g}
  \end{subfigure}
  \hfill
  \begin{subfigure}[b]{0.245\textwidth}
    \includegraphics[width=\linewidth]{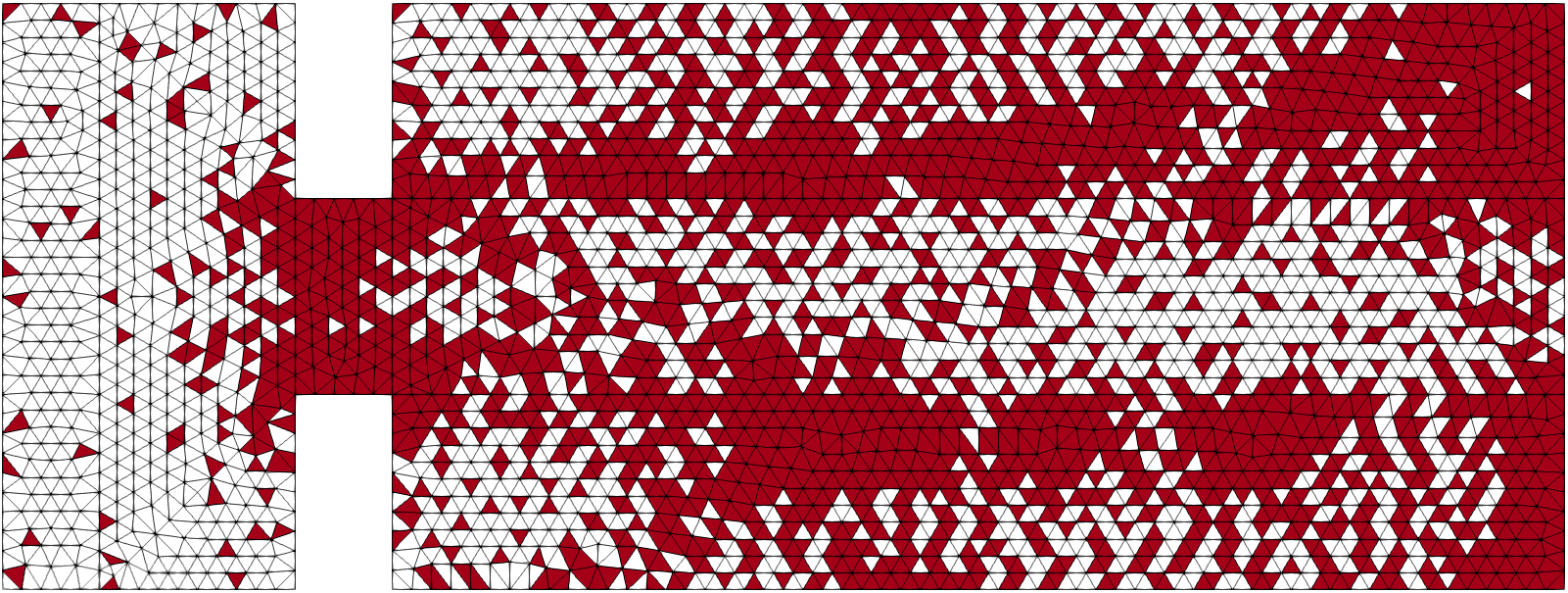}
    \caption{ $\epsilon_{\text{\tiny SOL}} = 1e-4, \epsilon_{\text{\tiny RES}} = 1e-6, $ }
    %\label{fig: ex3HROM_elemns_h}
  \end{subfigure}

  \begin{subfigure}[b]{0.245\textwidth}
    \includegraphics[width=\linewidth]{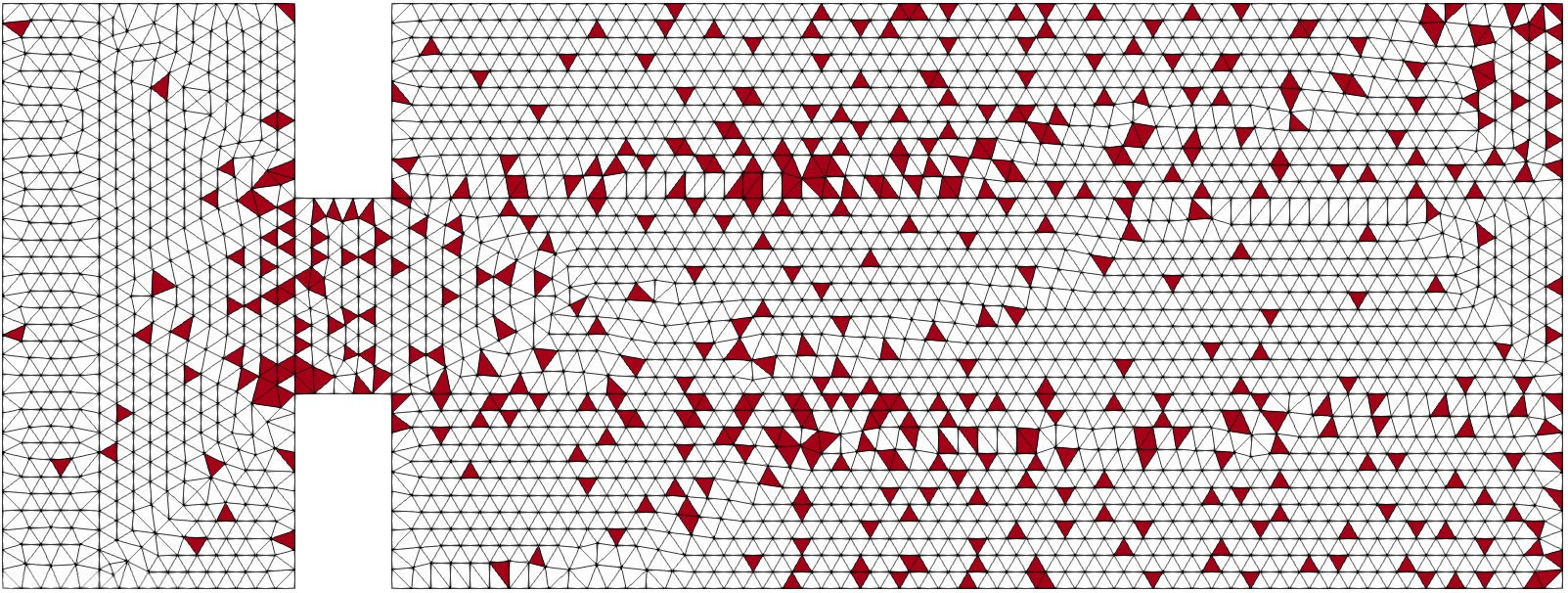}
    \caption{ $\epsilon_{\text{\tiny SOL}} = 1e-5, \epsilon_{\text{\tiny RES}} = 1e-3, $ }
    %\label{fig: ex3HROM_elemns_i}
  \end{subfigure}
  \hfill
  \begin{subfigure}[b]{0.245\textwidth}
    \includegraphics[width=\linewidth]{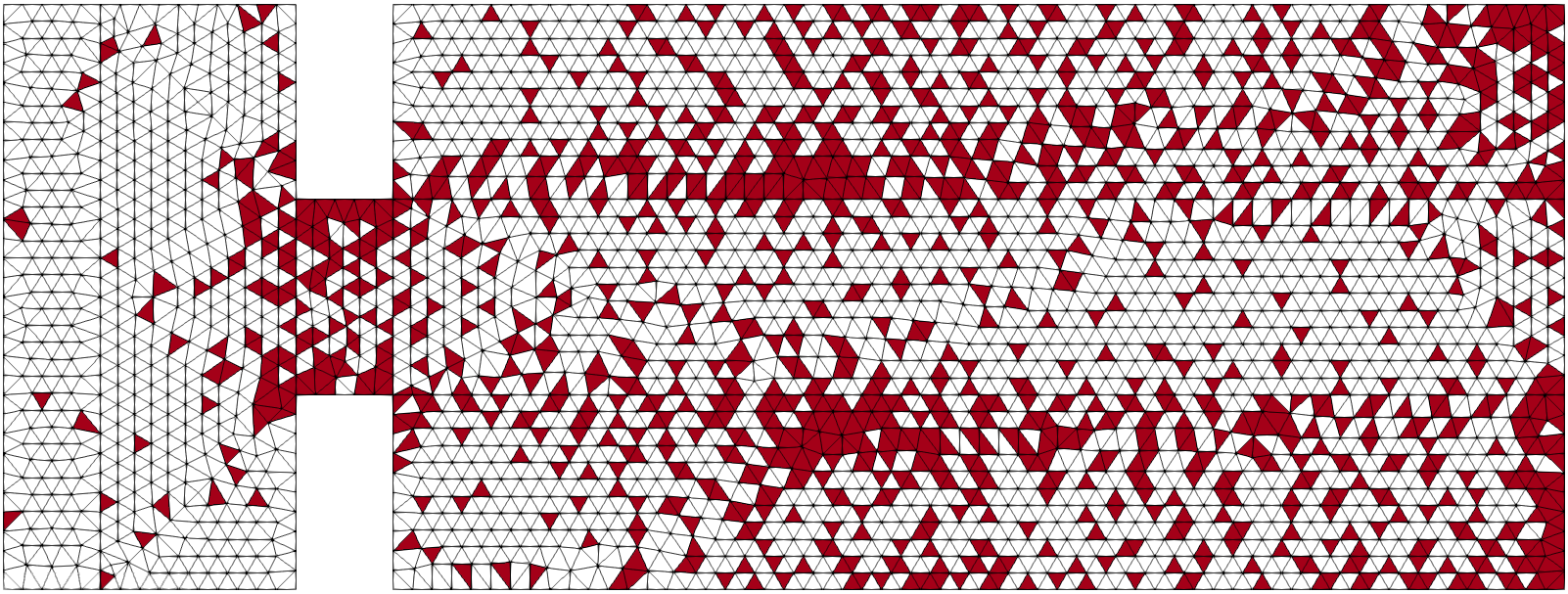}
    \caption{ $\epsilon_{\text{\tiny SOL}} = 1e-5, \epsilon_{\text{\tiny RES}} = 1e-4, $ }
    %\label{fig: ex3HROM_elemns_j}
  \end{subfigure}
  \hfill
  \begin{subfigure}[b]{0.245\textwidth}
    \includegraphics[width=\linewidth]{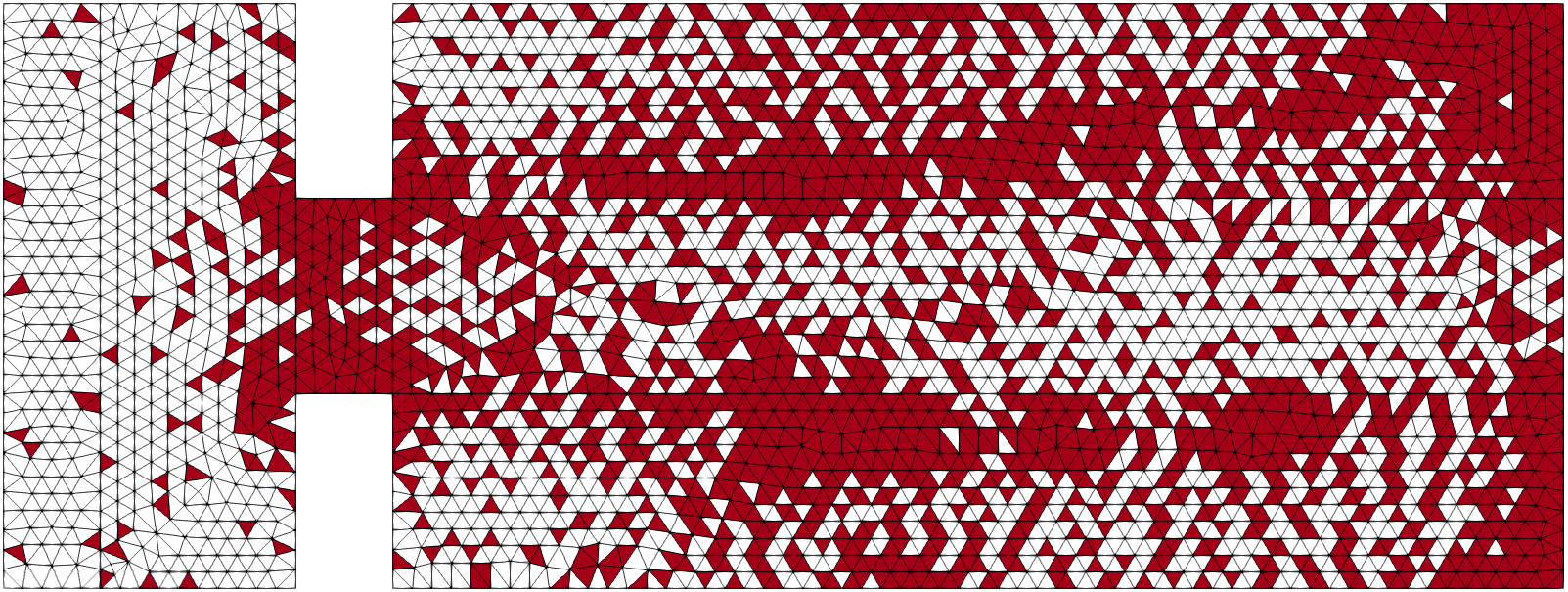}
    \caption{ $\epsilon_{\text{\tiny SOL}} = 1e-5, \epsilon_{\text{\tiny RES}} = 1e-5, $ }
    %\label{fig: ex3HROM_elemns_k}
  \end{subfigure}
  \hfill
  \begin{subfigure}[b]{0.245\textwidth}
    \includegraphics[width=\linewidth]{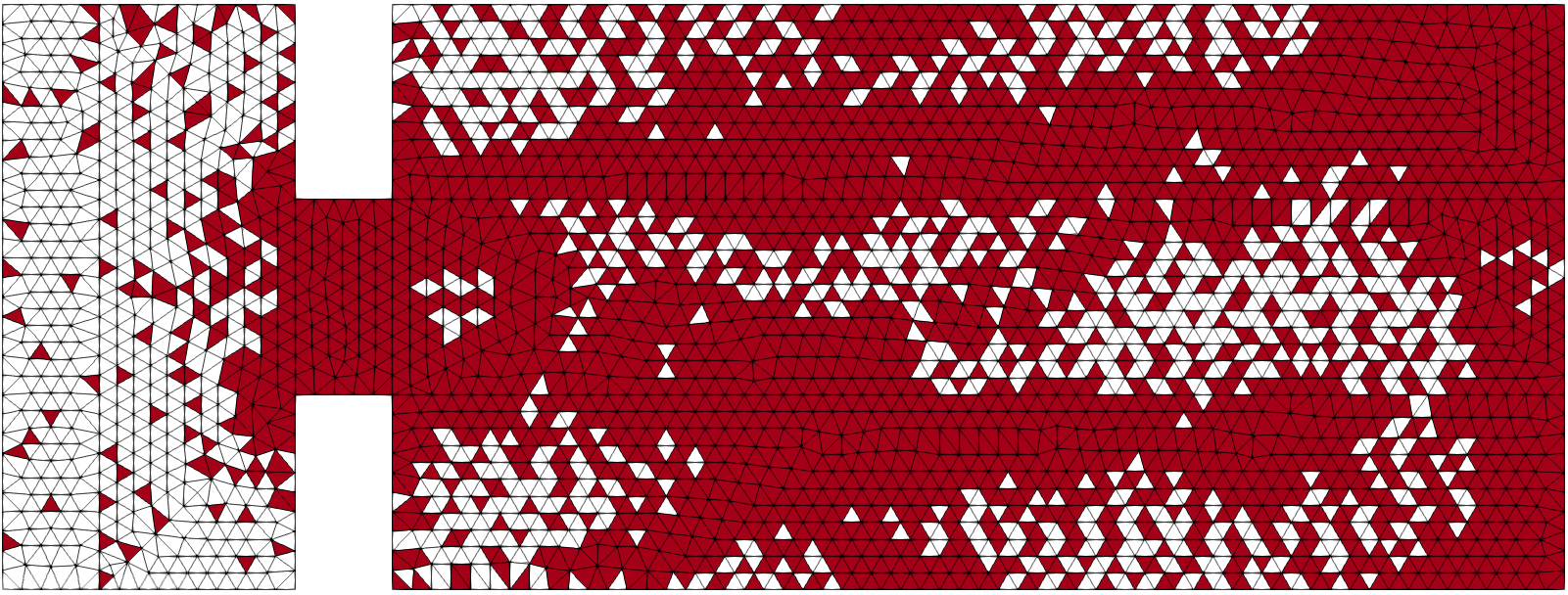}
    \caption{ $\epsilon_{\text{\tiny SOL}} = 1e-5, \epsilon_{\text{\tiny RES}} = 1e-6, $ }
    %\label{fig: ex3HROM_elemns_l}
  \end{subfigure}

  \begin{subfigure}[b]{0.245\textwidth}
    \includegraphics[width=\linewidth]{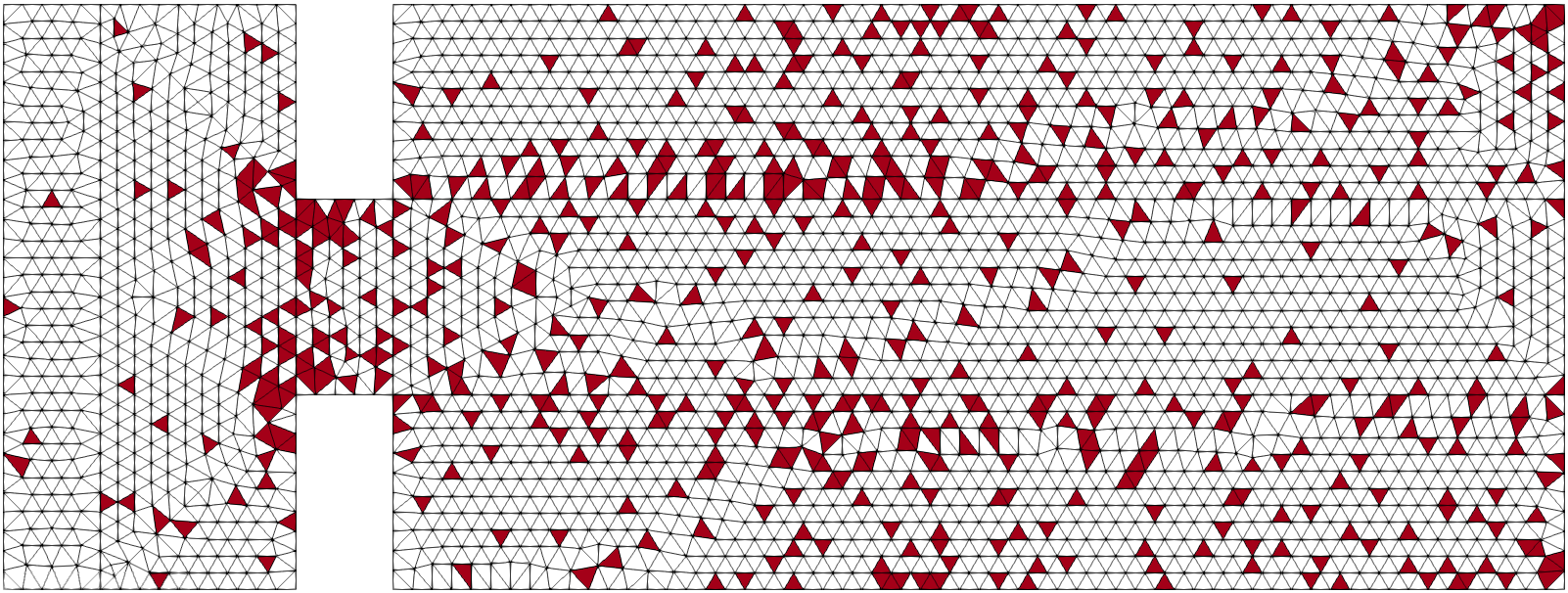}
    \caption{ $\epsilon_{\text{\tiny SOL}} = 1e-6, \epsilon_{\text{\tiny RES}} = 1e-3, $ }
    %\label{fig: ex3HROM_elemns_m}
  \end{subfigure}
  \hfill
  \begin{subfigure}[b]{0.245\textwidth}
    \includegraphics[width=\linewidth]{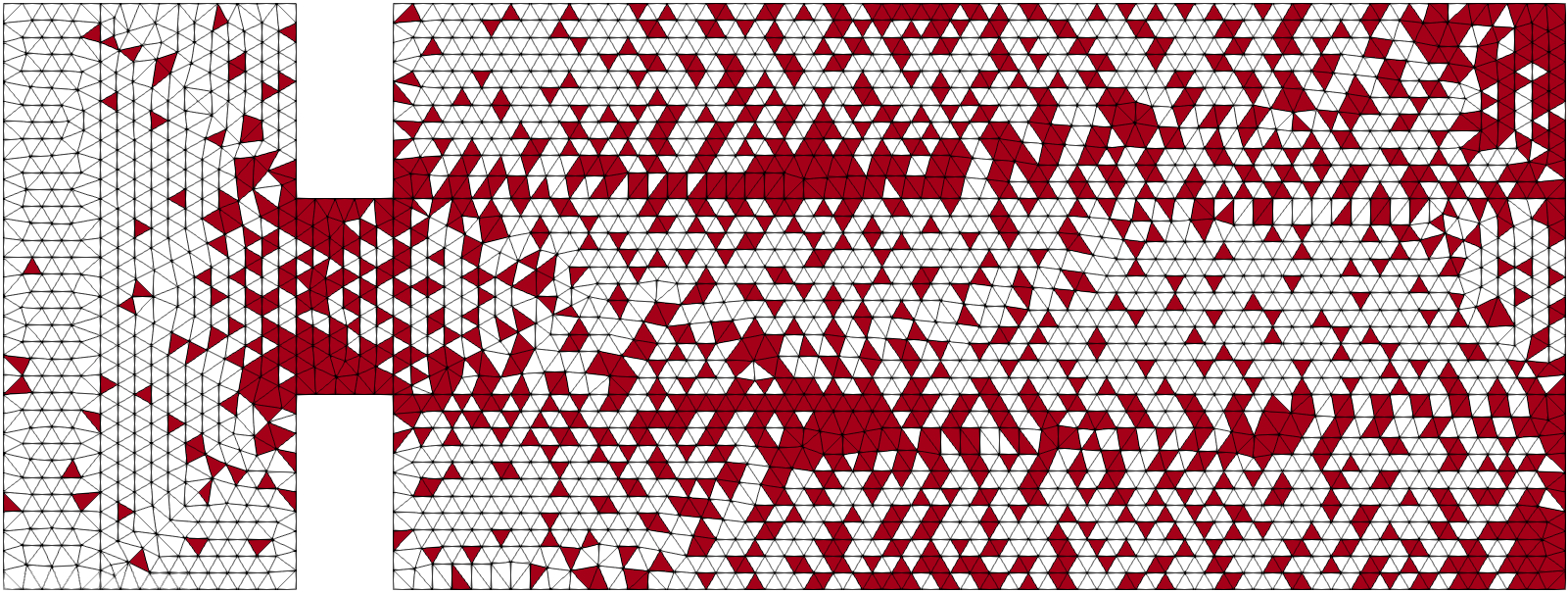}
    \caption{ $\epsilon_{\text{\tiny SOL}} = 1e-6, \epsilon_{\text{\tiny RES}} = 1e-4, $ }
    %\label{fig: ex3HROM_elemns_n}
  \end{subfigure}
  \hfill
  \begin{subfigure}[b]{0.245\textwidth}
    \includegraphics[width=\linewidth]{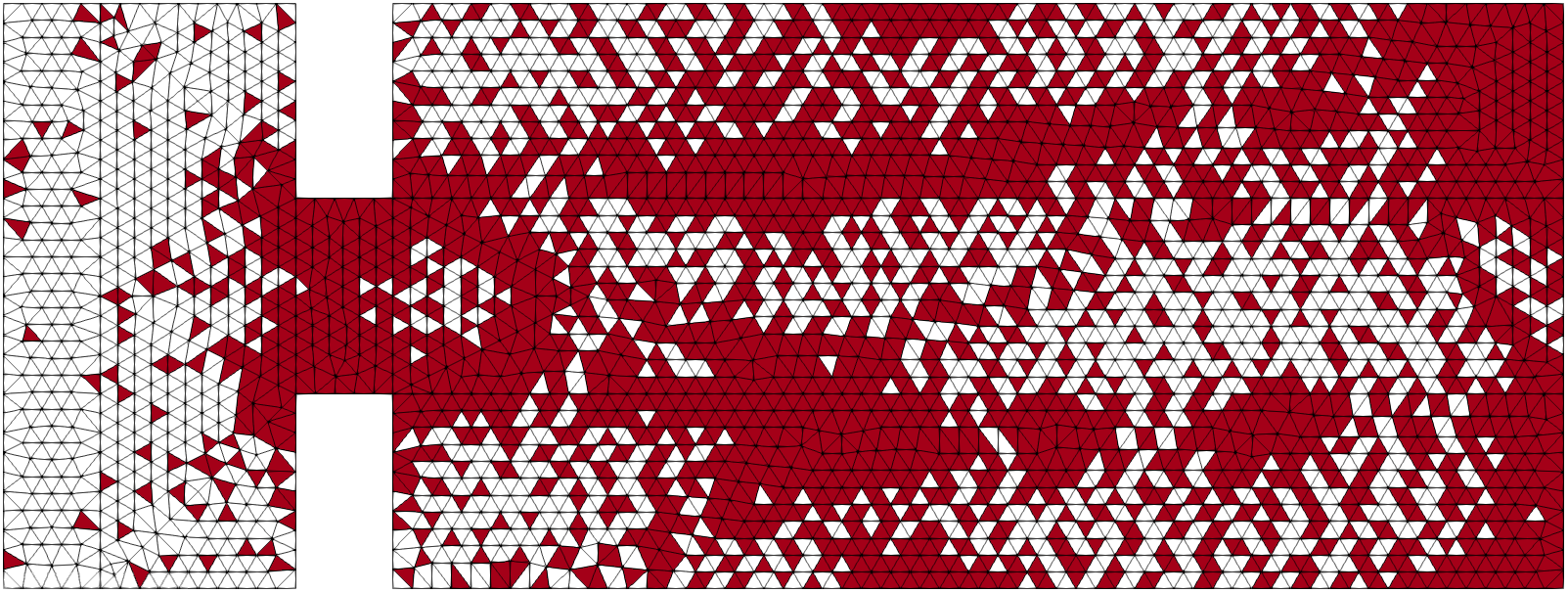}
    \caption{ $\epsilon_{\text{\tiny SOL}} = 1e-6, \epsilon_{\text{\tiny RES}} = 1e-5, $ }
    %\label{fig: ex3HROM_elemns_o}
  \end{subfigure}
  \hfill
  \begin{subfigure}[b]{0.245\textwidth}
    \includegraphics[width=\linewidth]{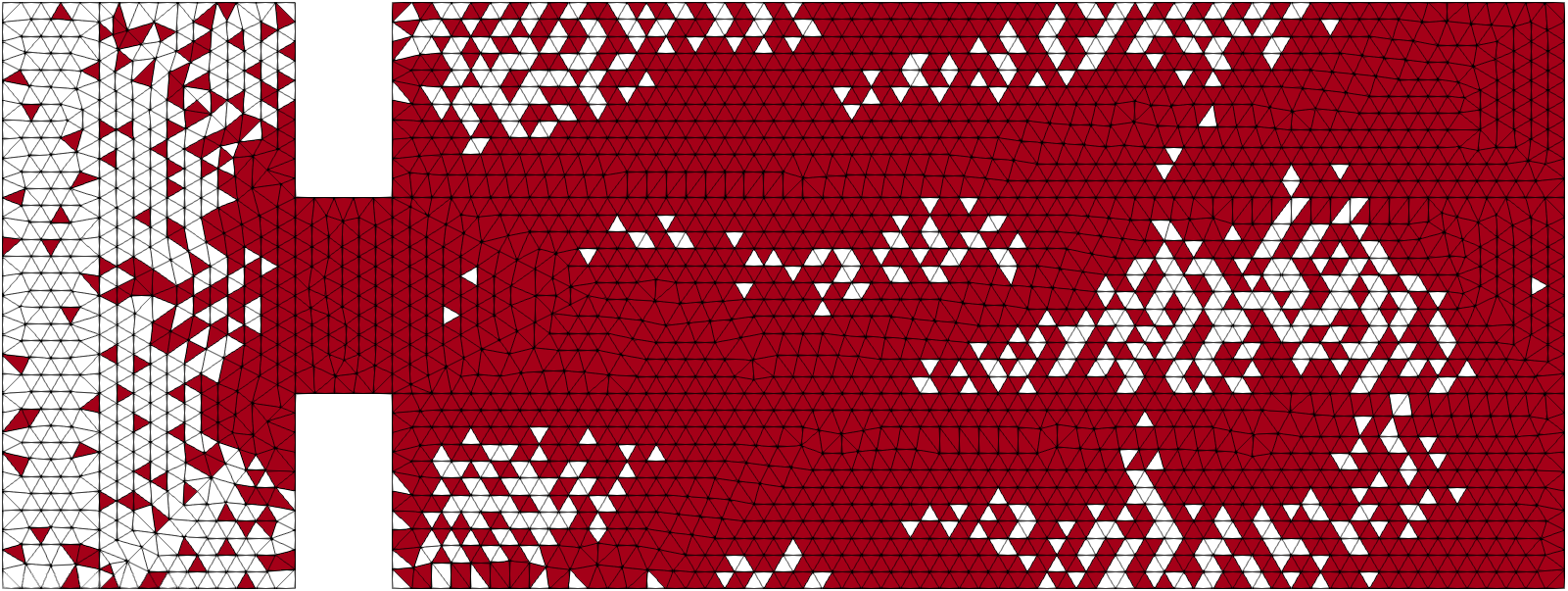}
    \caption{ $\epsilon_{\text{\tiny SOL}} = 1e-6, \epsilon_{\text{\tiny RES}} = 1e-6, $ }
    %\label{fig: ex3HROM_elemns_p}
  \end{subfigure}

  \caption{Hyper-reduced elements selected for the $\boldsymbol{\varphi}_{\text{\tiny AFFINE}}$ mapping}
  \label{fig: HROM elements affine ex3}
\end{figure}

\append{Example 2 considers Trajectory 1 and Trajectory 2 as training trajectories. The solutions of these trajectories involve jets attaching to both the upper and lower walls. Fig. \ref{fig:modes affine mapping ex3} shows the velocity and pressure POD modes for the affine mapping. These modes exhibit symmetric patterns that do not favor one wall over the other. Consistent with these observations, the elements selected by the ECM algorithm, as illustrated in Fig. \ref{fig: HROM elements affine ex3}, also demonstrate a more uniform distribution of selected elements across the domain.}

\subsubsection{\append{Example 2. Nonaffine Mapping}}

\begin{figure}[H]
  \centering
  \begin{subfigure}[b]{0.245\textwidth}
    \includegraphics[width=\linewidth]{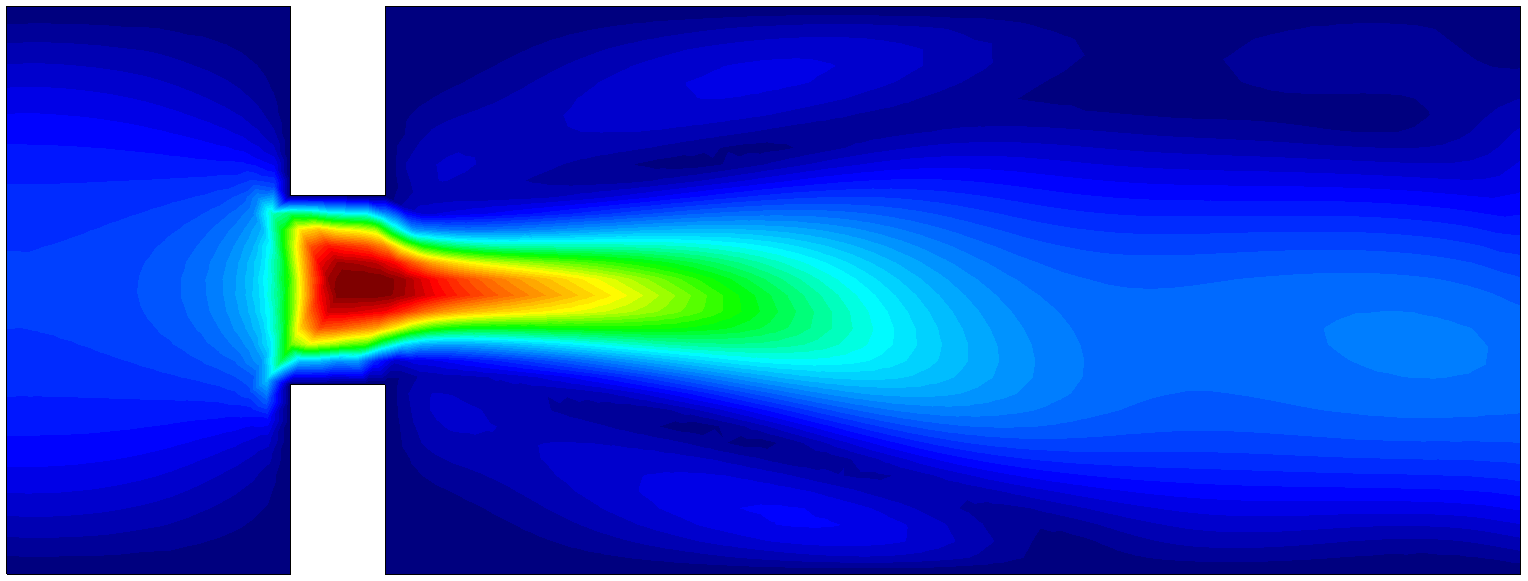}
    \caption{velocity mode 1}
  \end{subfigure}
  \hfill
  \begin{subfigure}[b]{0.245\textwidth}
    \includegraphics[width=\linewidth]{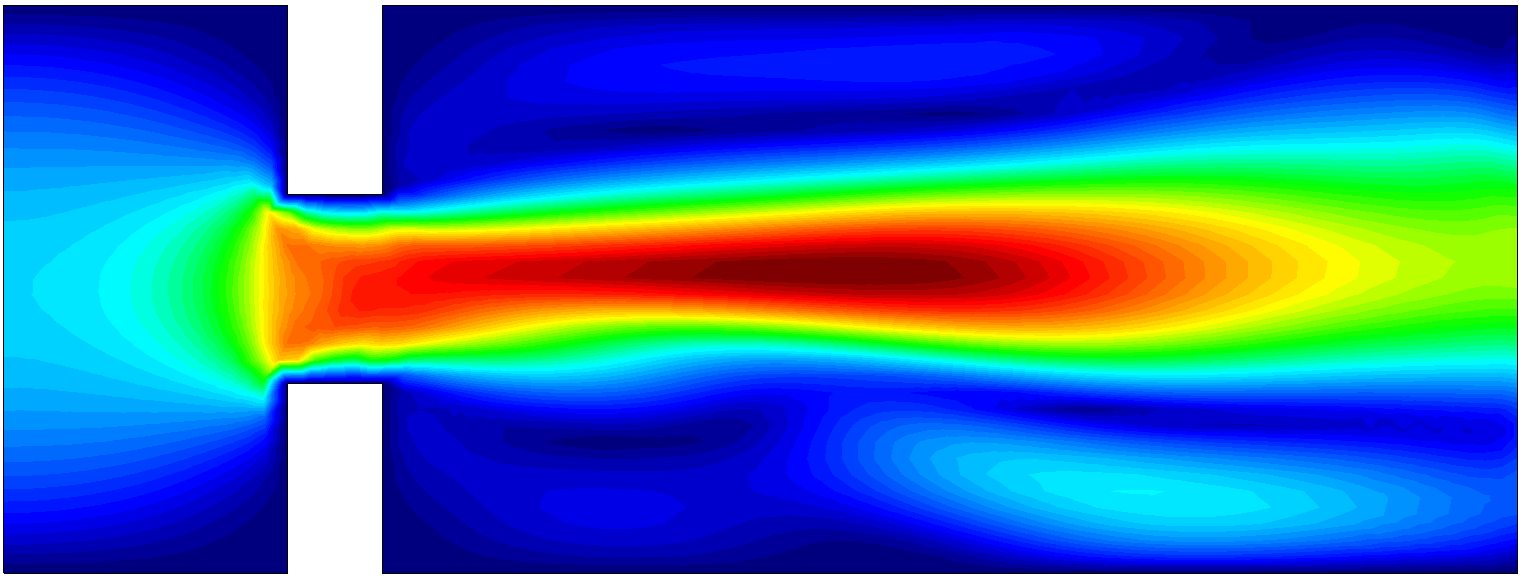}
    \caption{velocity mode 2}
  \end{subfigure}
  \hfill
  \begin{subfigure}[b]{0.245\textwidth}
    \includegraphics[width=\linewidth]{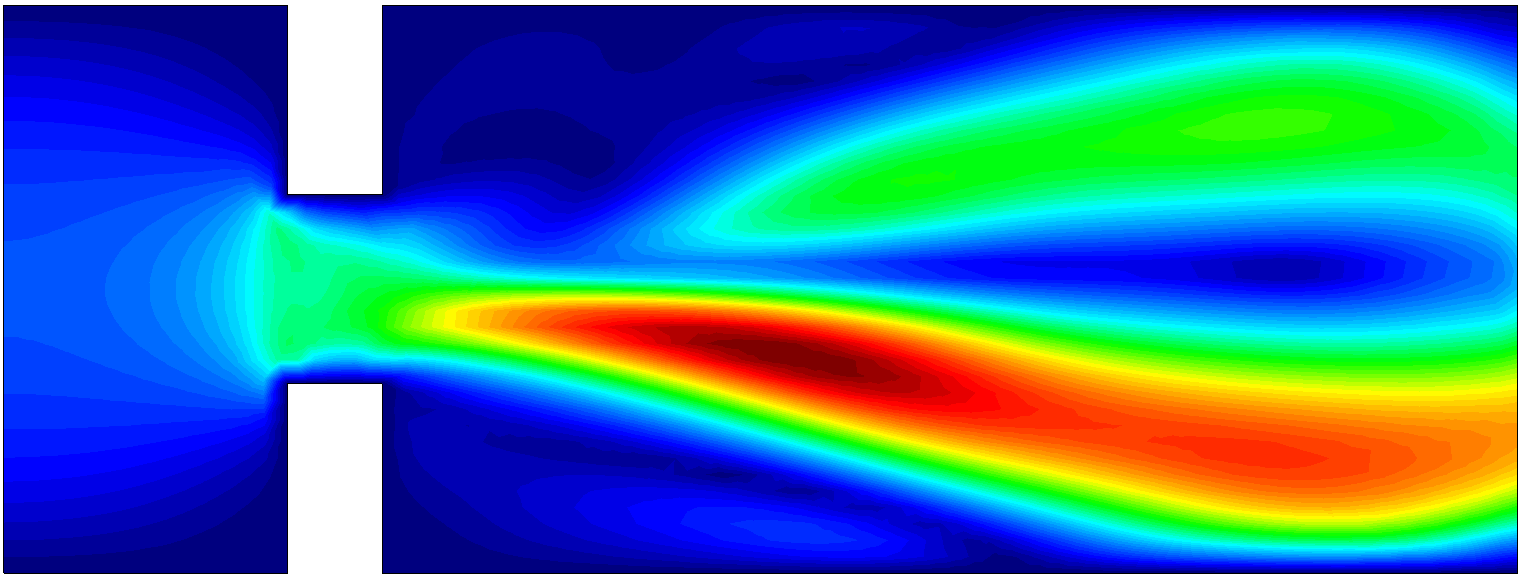}
    \caption{velocity mode 3}
  \end{subfigure}
  \hfill
  \begin{subfigure}[b]{0.245\textwidth}
    \includegraphics[width=\linewidth]{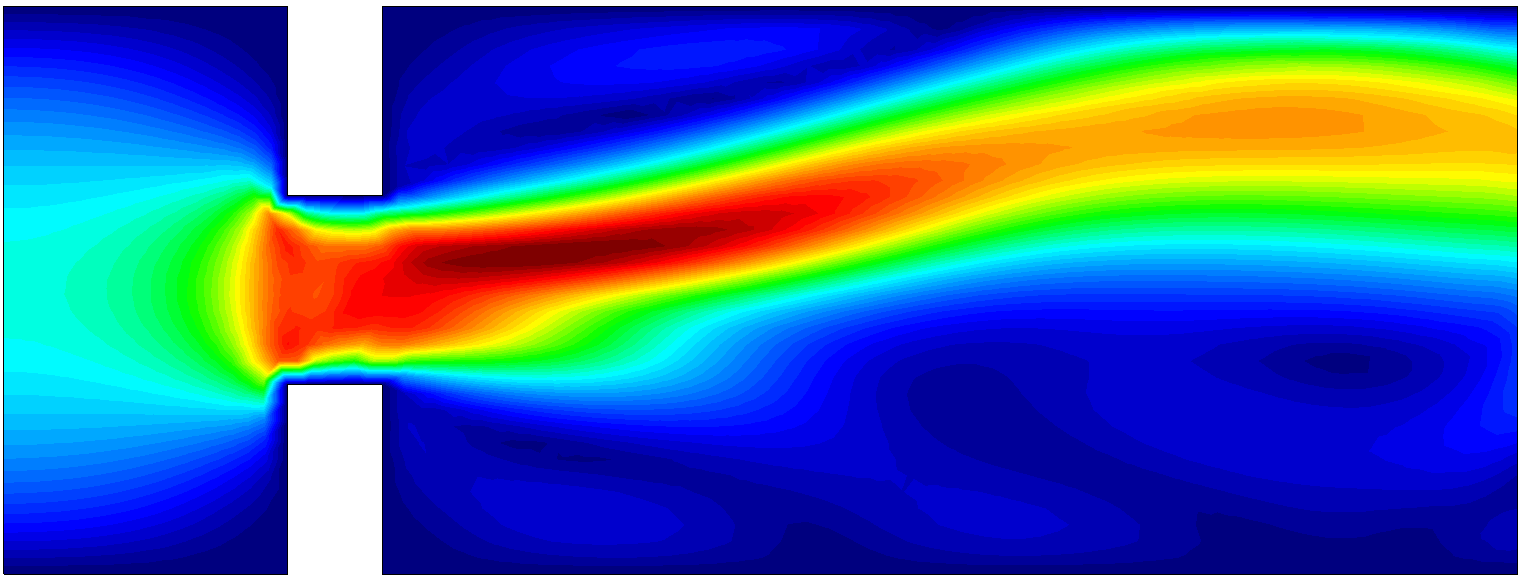}
    \caption{velocity mode 4}
  \end{subfigure}

  \begin{subfigure}[b]{0.245\textwidth}
    \includegraphics[width=\linewidth]{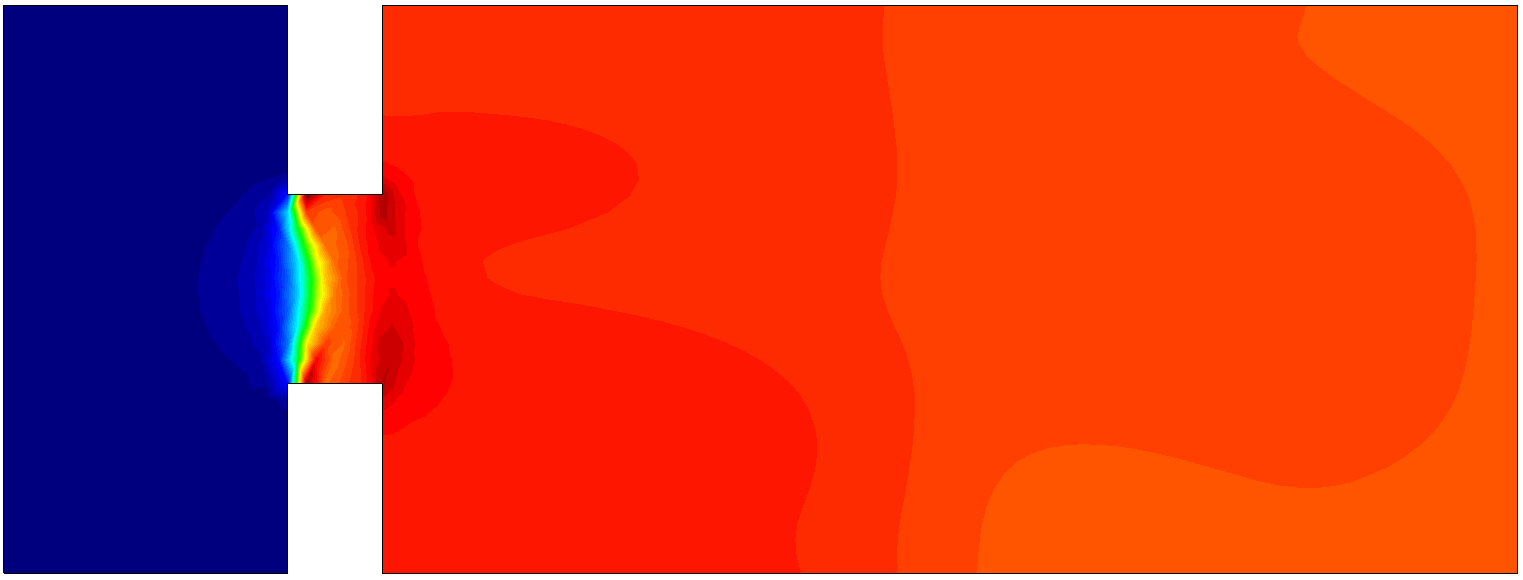}
    \caption{pressure mode 1}
  \end{subfigure}
  \hfill
  \begin{subfigure}[b]{0.245\textwidth}
    \includegraphics[width=\linewidth]{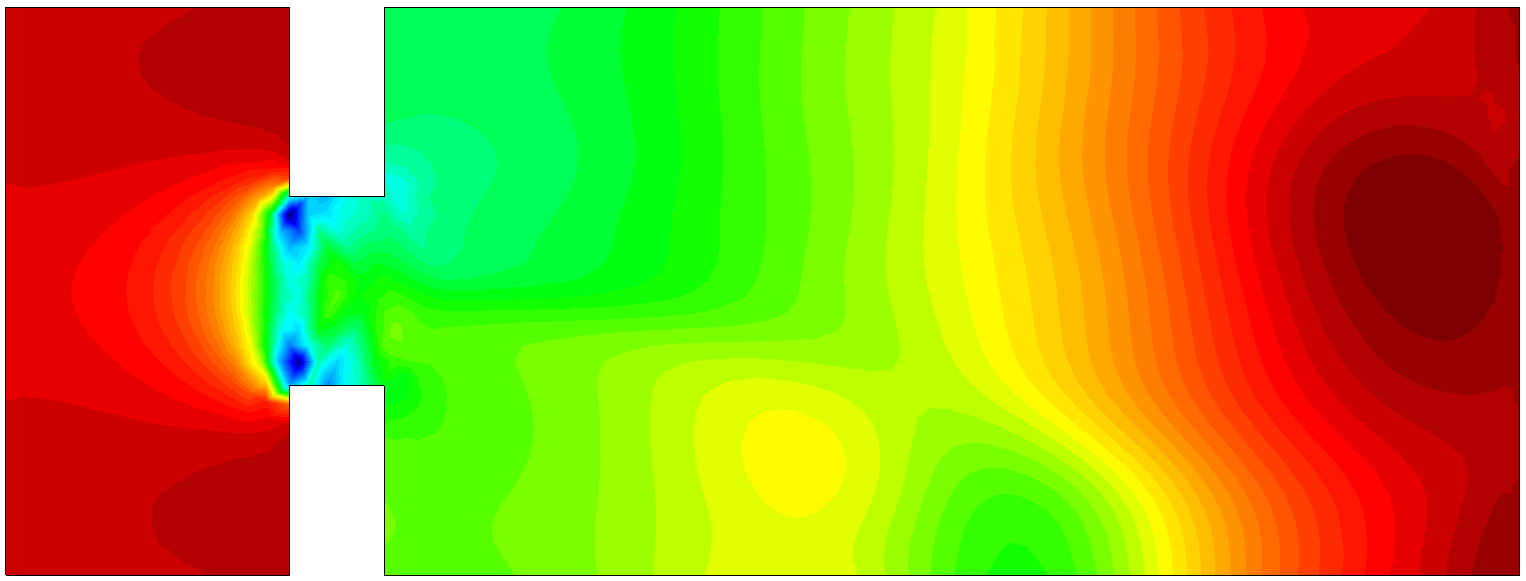}
    \caption{pressure mode 2}
  \end{subfigure}
  \hfill
  \begin{subfigure}[b]{0.245\textwidth}
    \includegraphics[width=\linewidth]{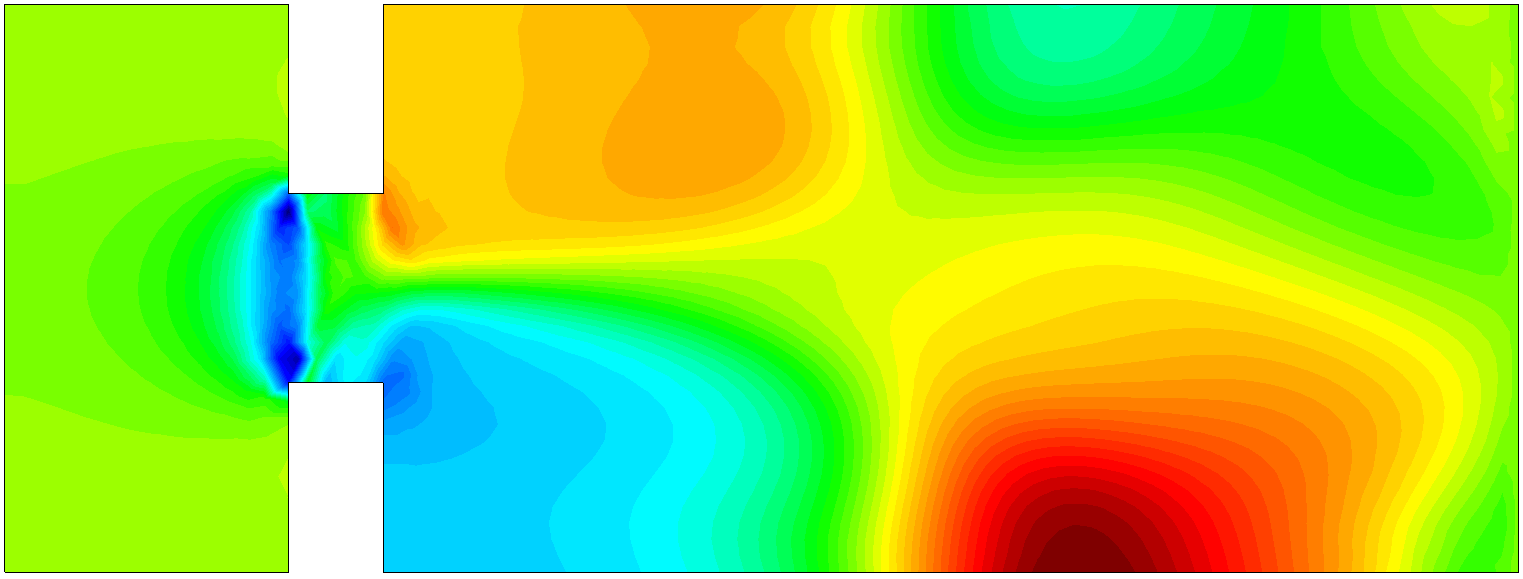}
    \caption{pressure mode 3}
  \end{subfigure}
  \hfill
  \begin{subfigure}[b]{0.245\textwidth}
    \includegraphics[width=\linewidth]{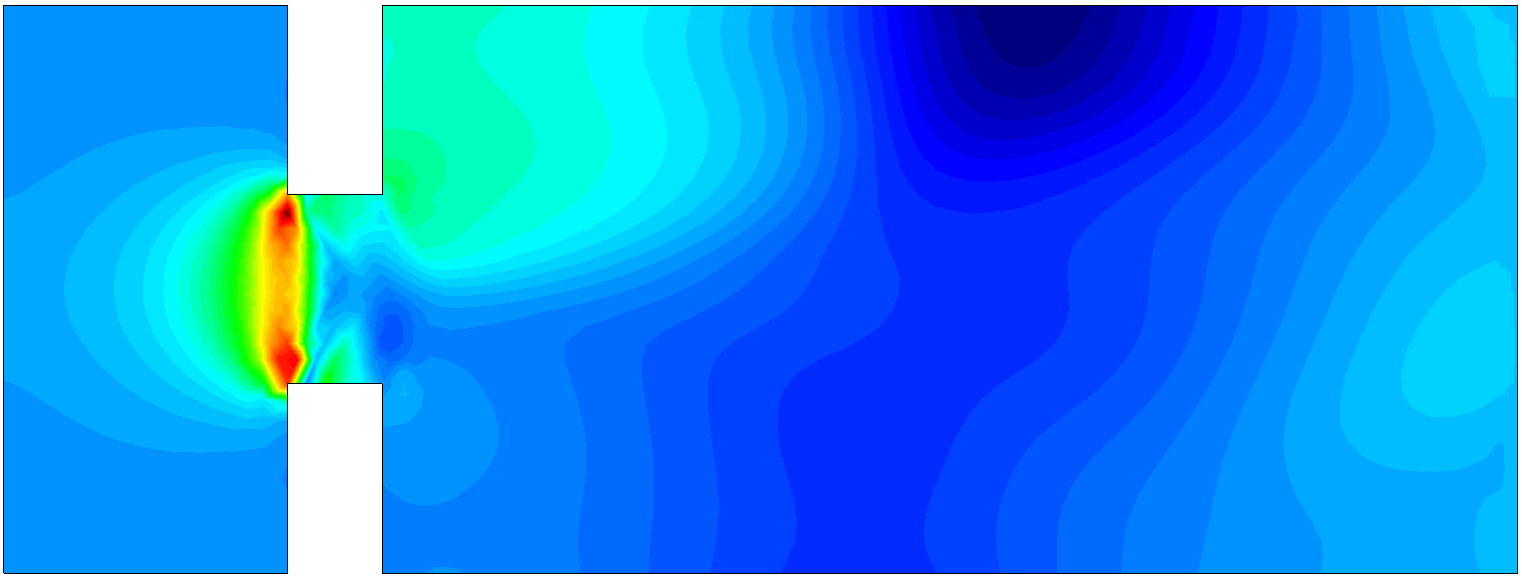}
    \caption{pressure mode 4}
  \end{subfigure}

  \caption{First four velocity and pressure modes for the $\boldsymbol{\varphi}_{\text{\tiny FFD+RBF}}$ mapping}
  \label{fig:modes ffd rbf mapping ex3}
\end{figure}

\begin{figure}[H]
  \centering
  \begin{subfigure}[b]{0.245\textwidth}
    \includegraphics[width=\linewidth]{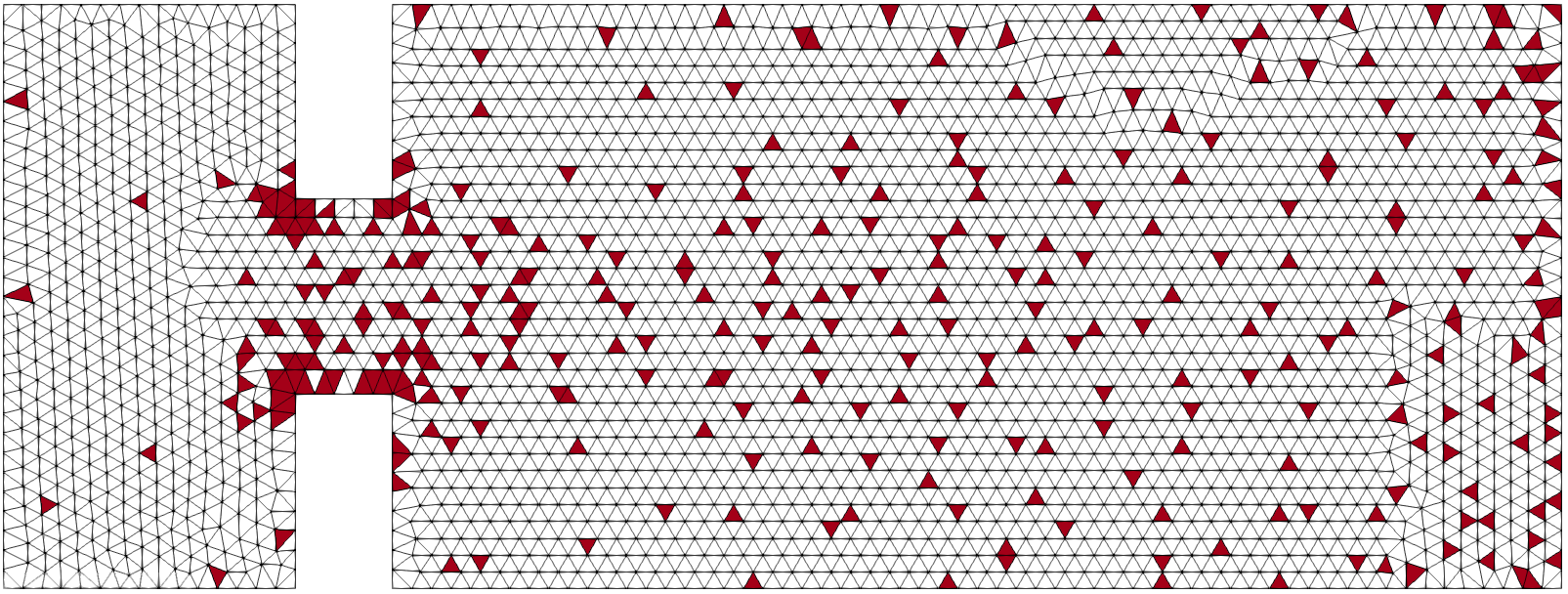}
    \caption{ $\epsilon_{\text{\tiny SOL}} = 1e-3, \epsilon_{\text{\tiny RES}} = 1e-3, $ }
    %\label{fig: example 3 HROM_elemns_ffd_rbf__a}
  \end{subfigure}
  \hfill
  \begin{subfigure}[b]{0.245\textwidth}
    \includegraphics[width=\linewidth]{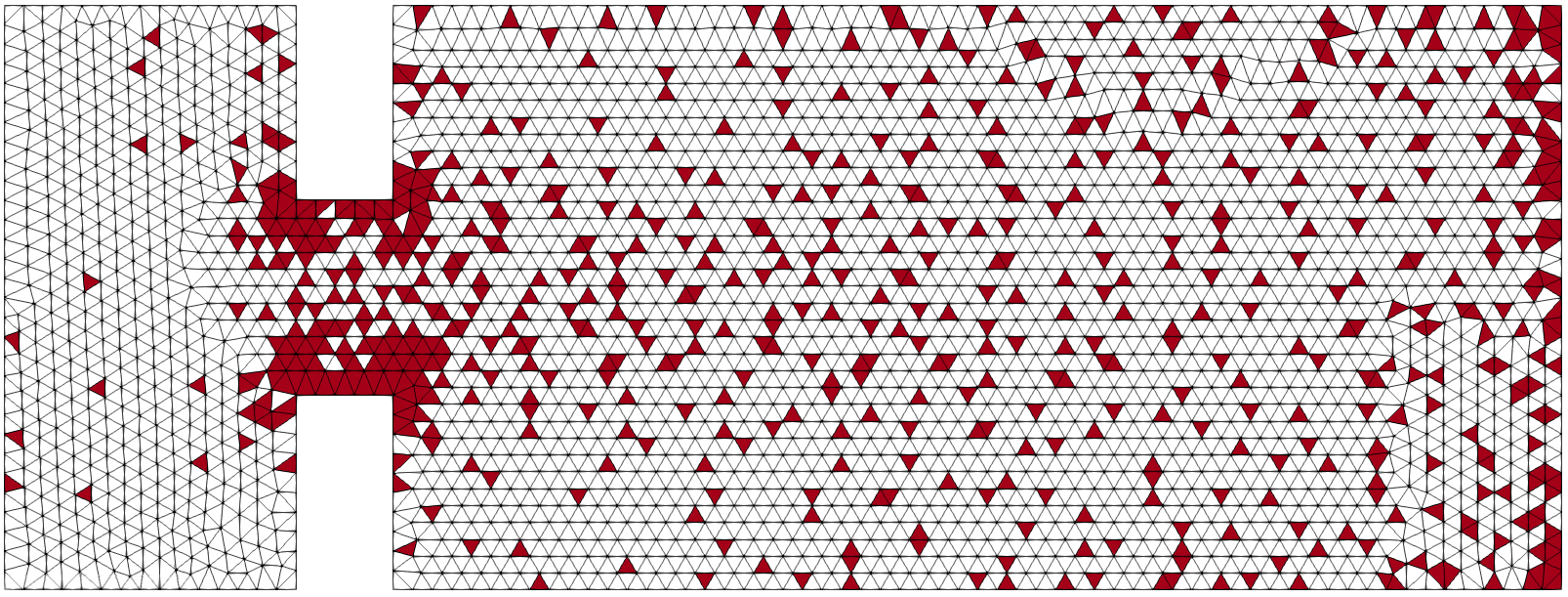}
    \caption{ $\epsilon_{\text{\tiny SOL}} = 1e-3, \epsilon_{\text{\tiny RES}} = 1e-4, $ }
    %\label{fig: example 3 HROM_elemns_ffd_rbf__b}
  \end{subfigure}
  \hfill
  \begin{subfigure}[b]{0.245\textwidth}
    \includegraphics[width=\linewidth]{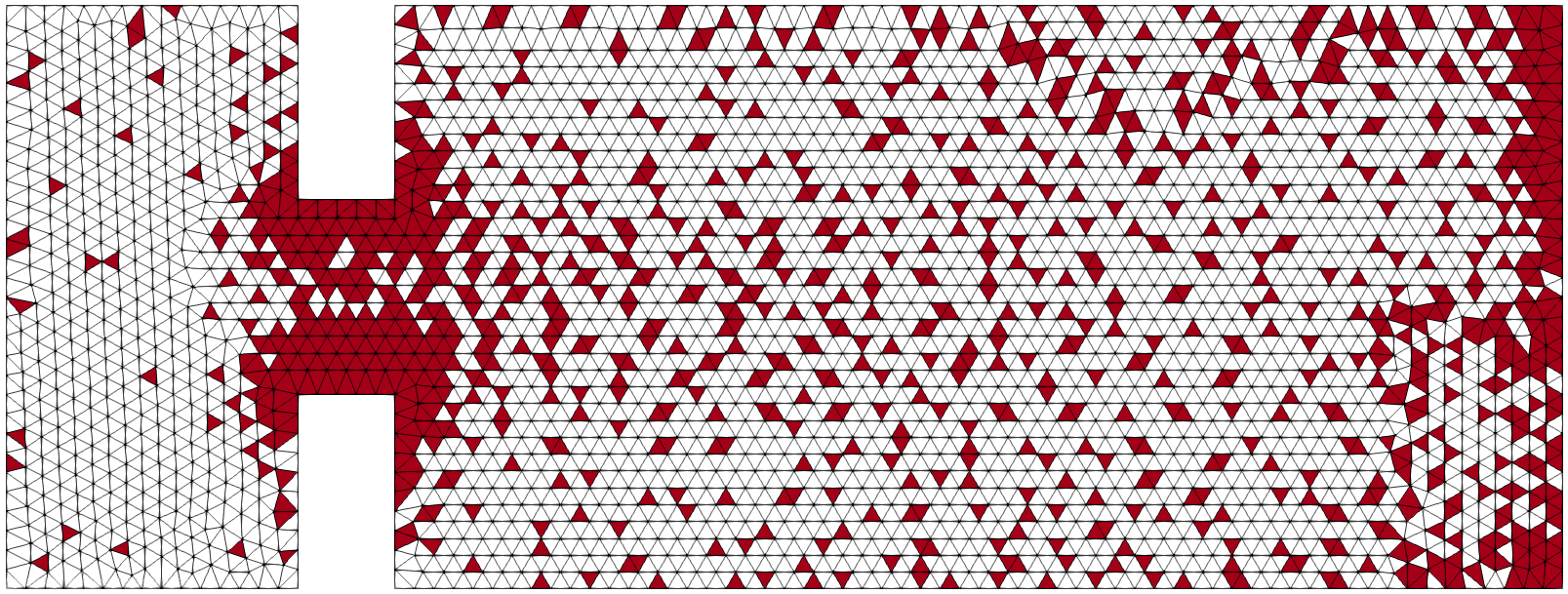}
    \caption{ $\epsilon_{\text{\tiny SOL}} = 1e-3, \epsilon_{\text{\tiny RES}} = 1e-5, $ }
    %\label{fig: example 3 HROM_elemns_ffd_rbf__c}
  \end{subfigure}
  \hfill
  \begin{subfigure}[b]{0.245\textwidth}
    \includegraphics[width=\linewidth]{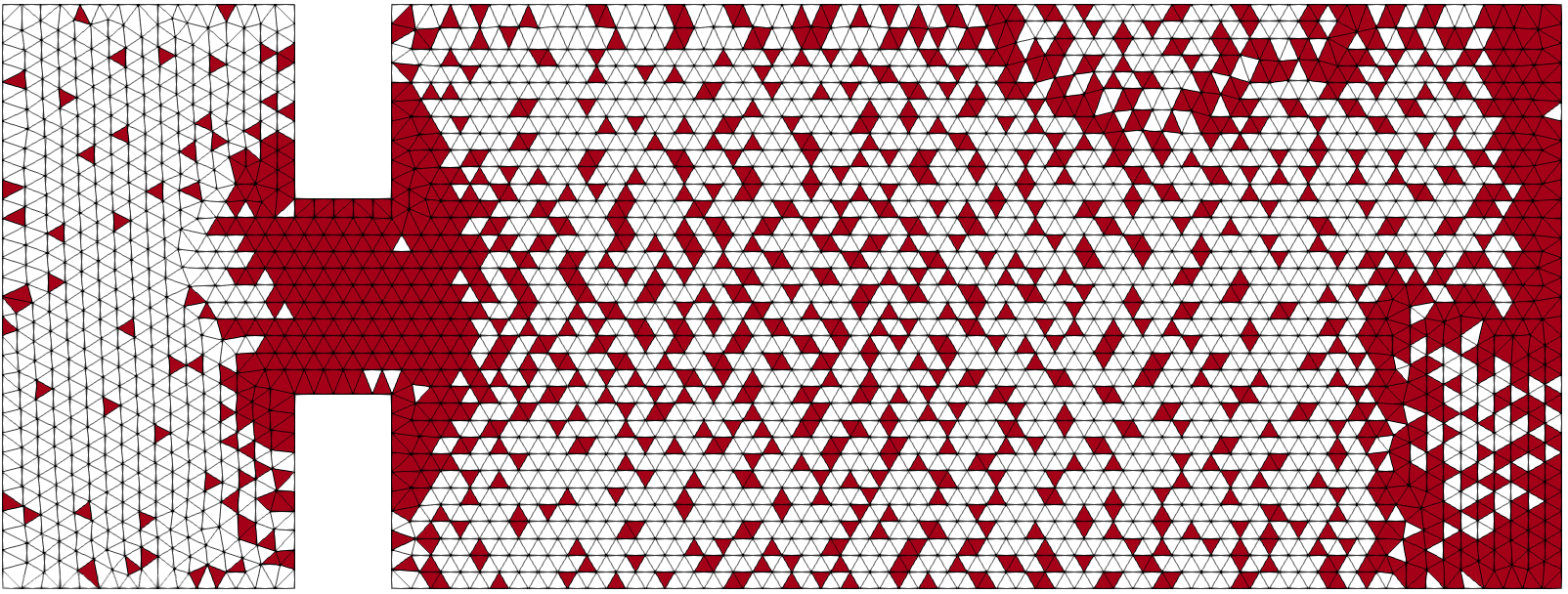}
    \caption{ $\epsilon_{\text{\tiny SOL}} = 1e-3, \epsilon_{\text{\tiny RES}} = 1e-6, $ }
    %\label{fig: example 3 HROM_elemns_ffd_rbf__d}
  \end{subfigure}

  \begin{subfigure}[b]{0.245\textwidth}
    \includegraphics[width=\linewidth]{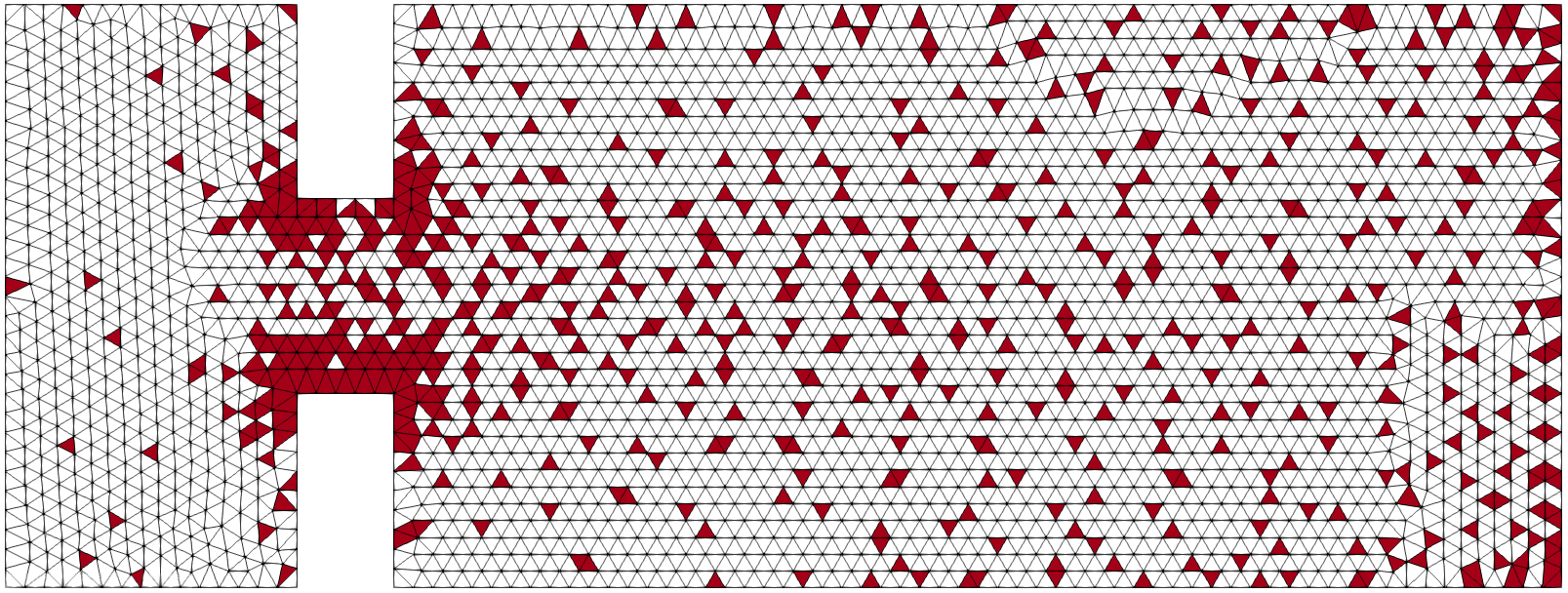}
    \caption{ $\epsilon_{\text{\tiny SOL}} = 1e-4, \epsilon_{\text{\tiny RES}} = 1e-3, $ }
    %\label{fig: example 3 HROM_elemns_ffd_rbf__e}
  \end{subfigure}
  \hfill
  \begin{subfigure}[b]{0.245\textwidth}
    \includegraphics[width=\linewidth]{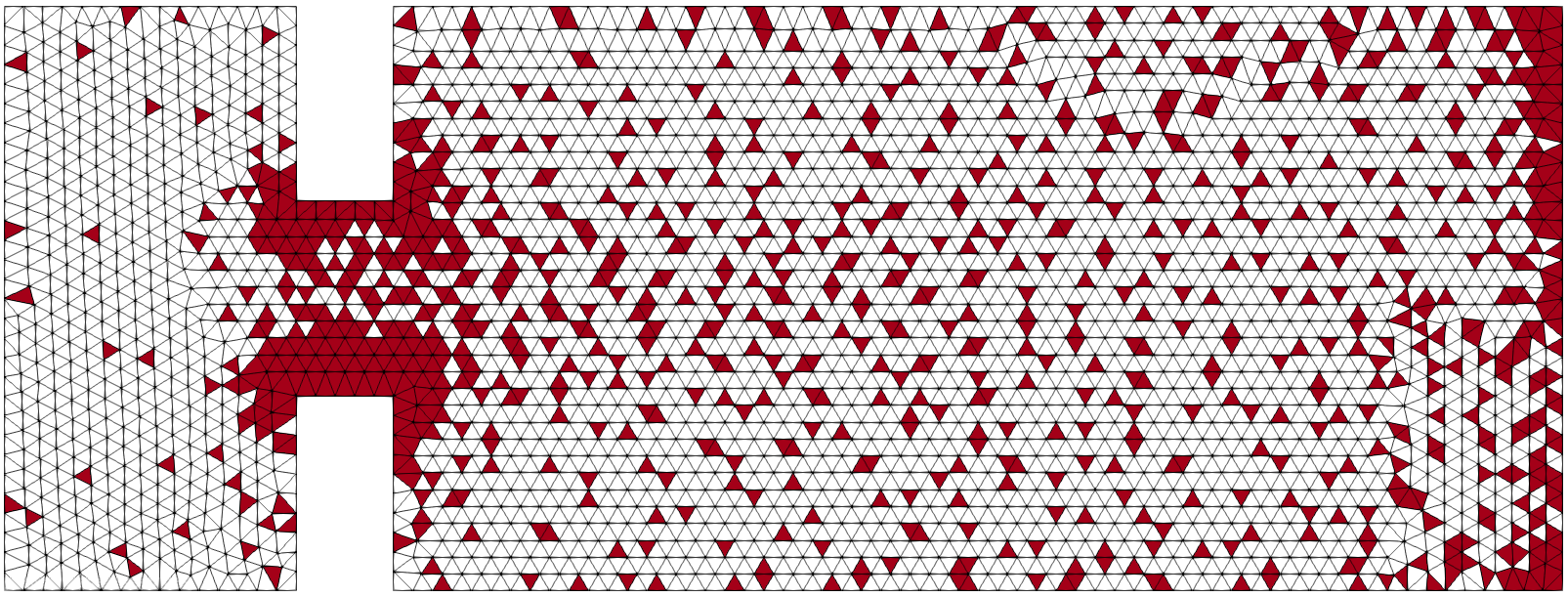}
    \caption{ $\epsilon_{\text{\tiny SOL}} = 1e-4, \epsilon_{\text{\tiny RES}} = 1e-4, $ }
    %\label{fig: example 3 HROM_elemns_ffd_rbf__f}
  \end{subfigure}
  \hfill
  \begin{subfigure}[b]{0.245\textwidth}
    \includegraphics[width=\linewidth]{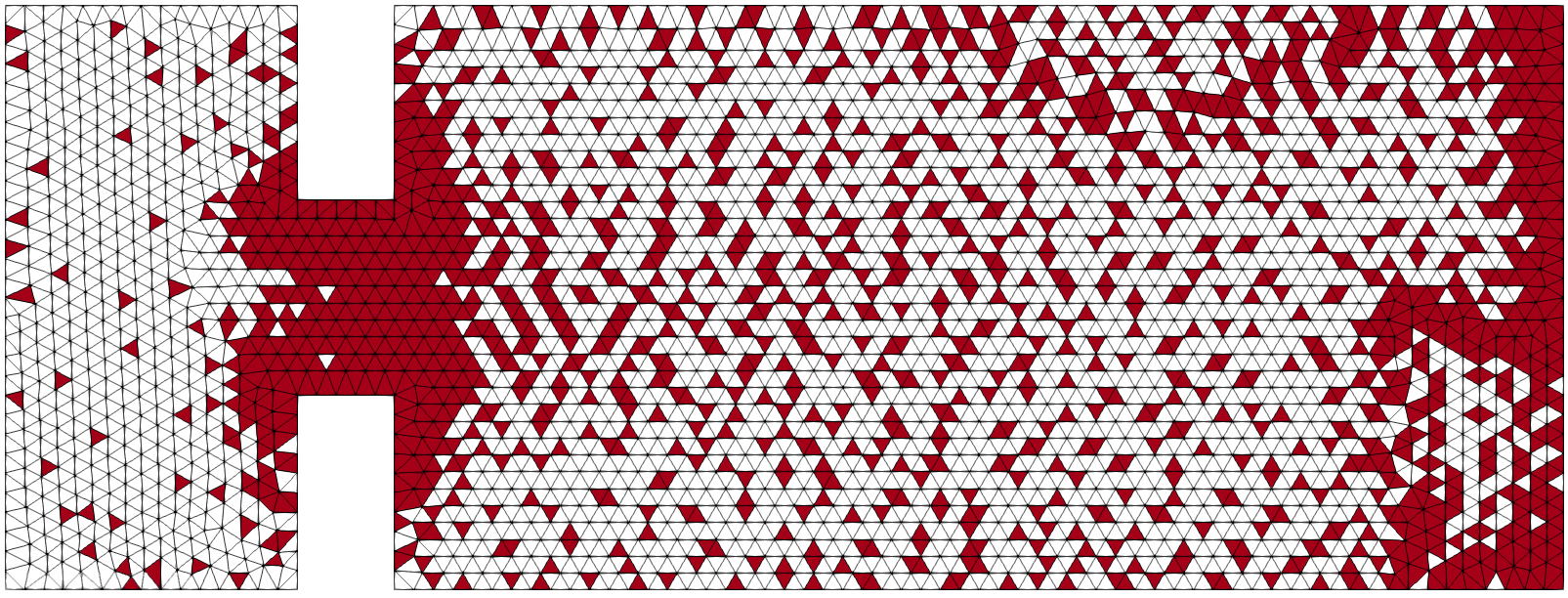}
    \caption{ $\epsilon_{\text{\tiny SOL}} = 1e-4, \epsilon_{\text{\tiny RES}} = 1e-5, $ }
    %\label{fig: example 3 HROM_elemns_ffd_rbf__g}
  \end{subfigure}
  \hfill
  \begin{subfigure}[b]{0.245\textwidth}
    \includegraphics[width=\linewidth]{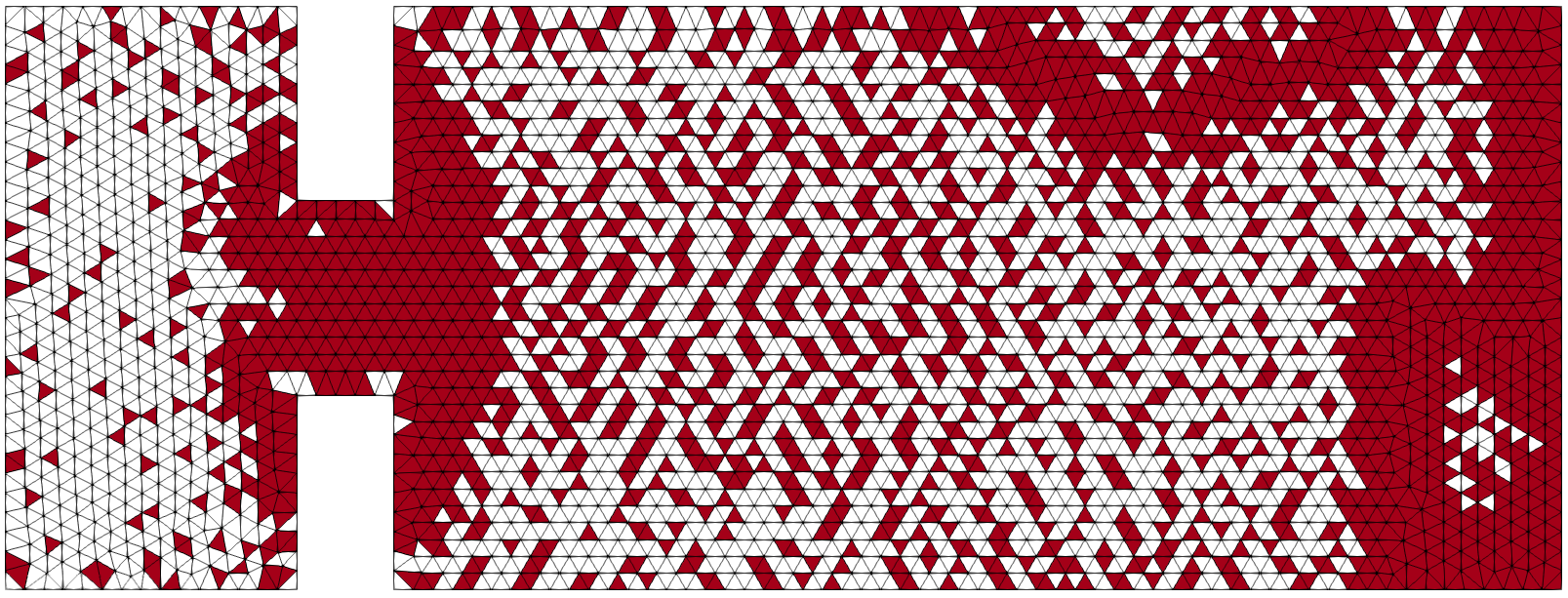}
    \caption{ $\epsilon_{\text{\tiny SOL}} = 1e-4, \epsilon_{\text{\tiny RES}} = 1e-6, $ }
    %\label{fig: example 3 HROM_elemns_ffd_rbf__h}
  \end{subfigure}

  \begin{subfigure}[b]{0.245\textwidth}
    \includegraphics[width=\linewidth]{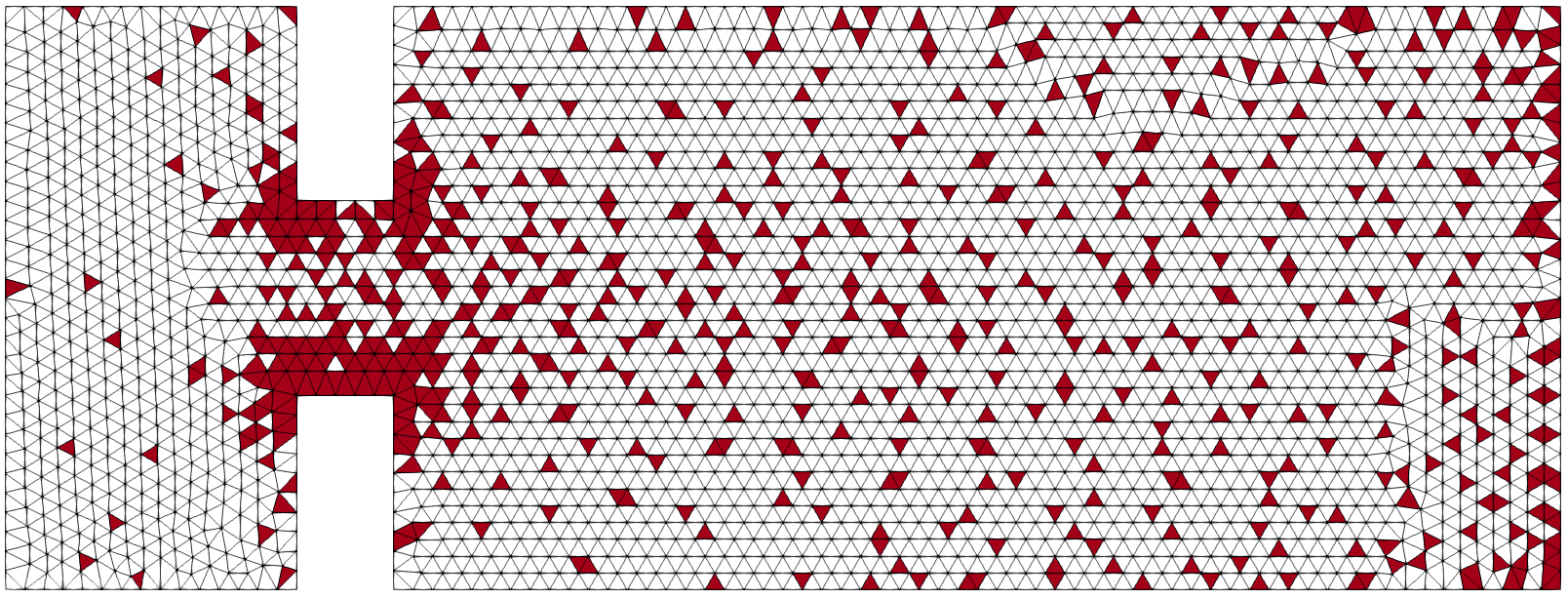}
    \caption{ $\epsilon_{\text{\tiny SOL}} = 1e-5, \epsilon_{\text{\tiny RES}} = 1e-3, $ }
    %\label{fig: example 3 HROM_elemns_ffd_rbf__i}
  \end{subfigure}
  \hfill
  \begin{subfigure}[b]{0.245\textwidth}
    \includegraphics[width=\linewidth]{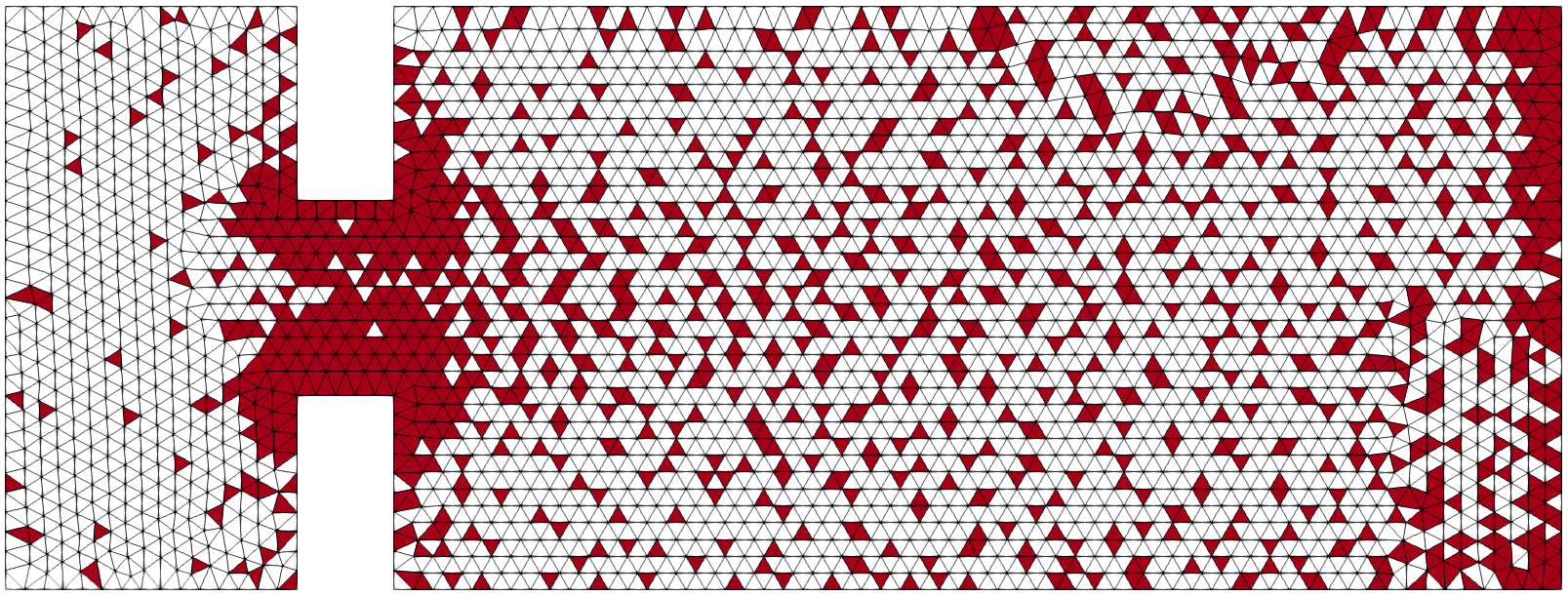}
    \caption{ $\epsilon_{\text{\tiny SOL}} = 1e-5, \epsilon_{\text{\tiny RES}} = 1e-4, $ }
    %\label{fig: example 3 HROM_elemns_ffd_rbf__j}
  \end{subfigure}
  \hfill
  \begin{subfigure}[b]{0.245\textwidth}
    \includegraphics[width=\linewidth]{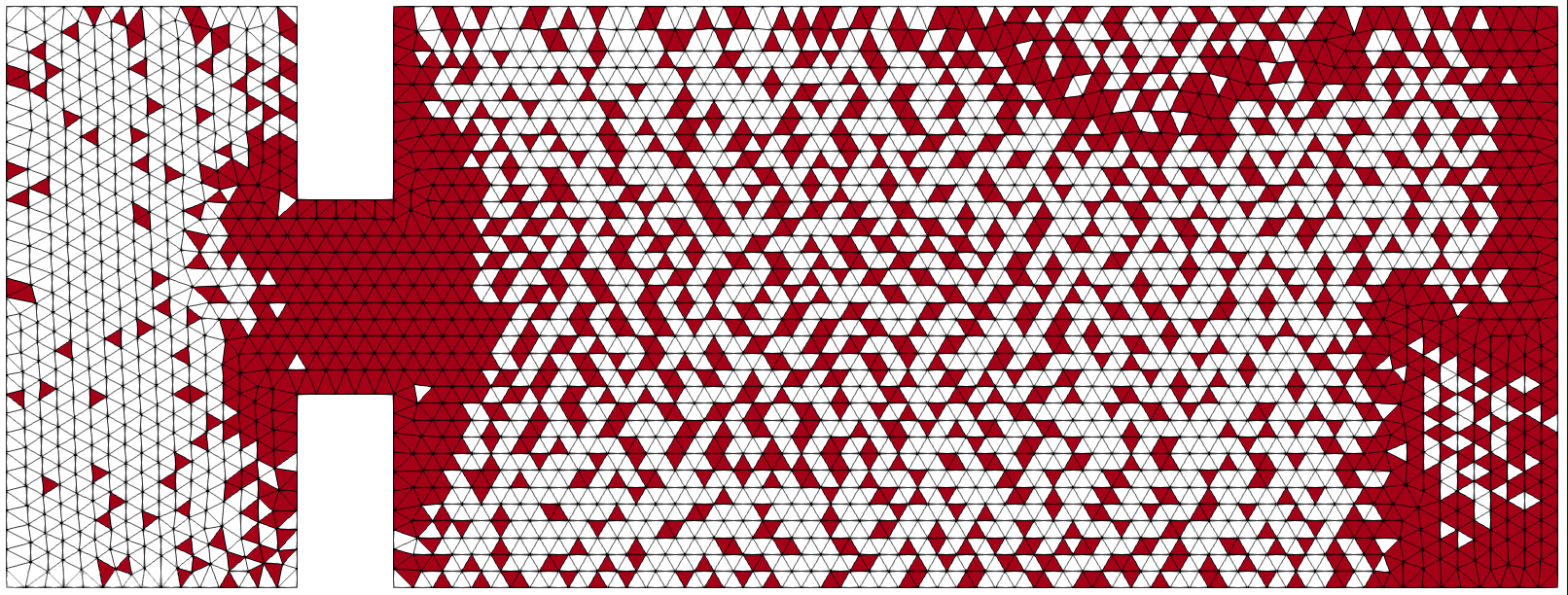}
    \caption{ $\epsilon_{\text{\tiny SOL}} = 1e-5, \epsilon_{\text{\tiny RES}} = 1e-5, $ }
    %\label{fig: example 3 HROM_elemns_ffd_rbf__k}
  \end{subfigure}
  \hfill
  \begin{subfigure}[b]{0.245\textwidth}
    \includegraphics[width=\linewidth]{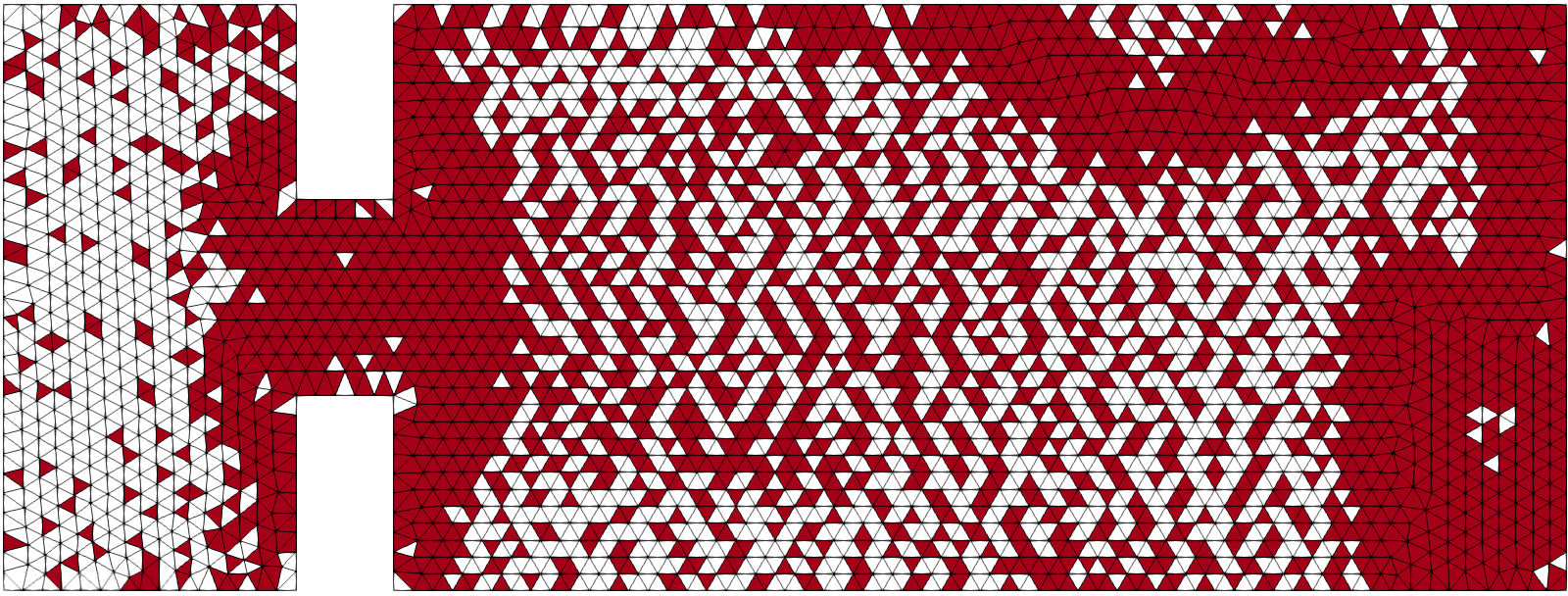}
    \caption{ $\epsilon_{\text{\tiny SOL}} = 1e-5, \epsilon_{\text{\tiny RES}} = 1e-6, $ }
    %\label{fig: example 3 HROM_elemns_ffd_rbf__l}
  \end{subfigure}

  \begin{subfigure}[b]{0.245\textwidth}
    \includegraphics[width=\linewidth]{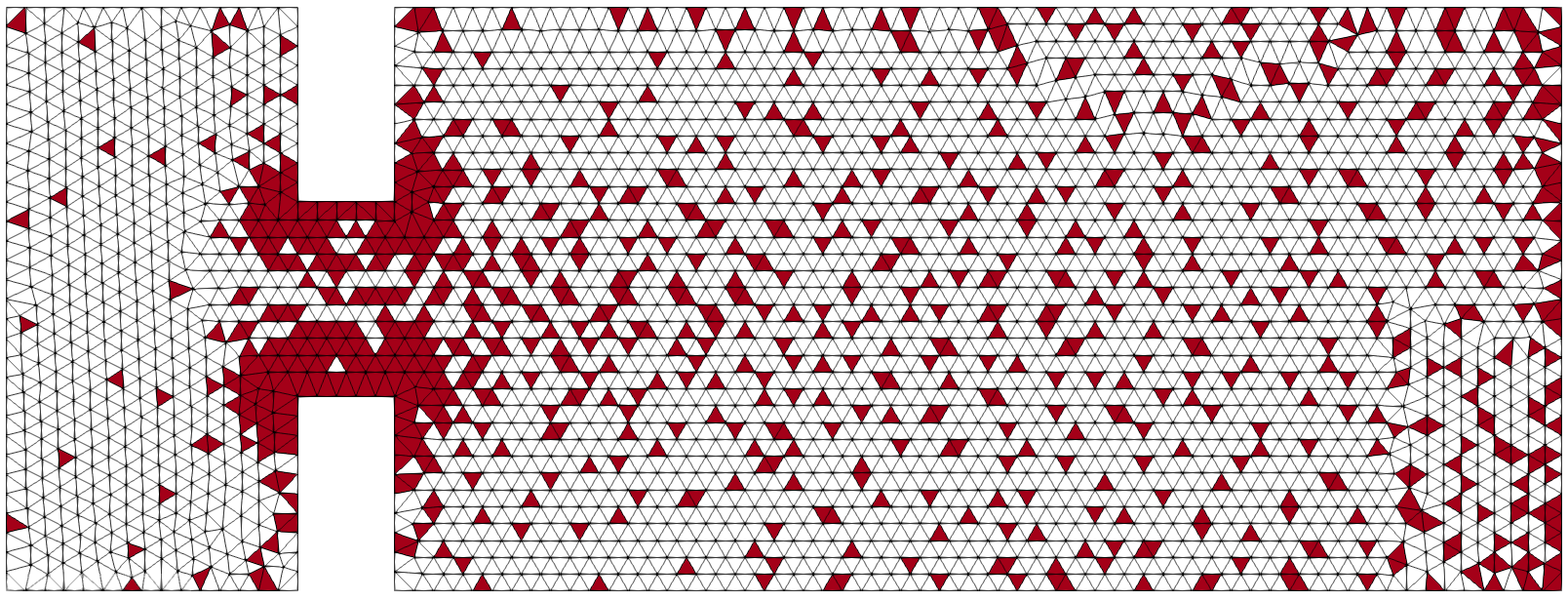}
    \caption{ $\epsilon_{\text{\tiny SOL}} = 1e-6, \epsilon_{\text{\tiny RES}} = 1e-3, $ }
    %\label{fig: example 3 HROM_elemns_ffd_rbf__m}
  \end{subfigure}
  \hfill
  \begin{subfigure}[b]{0.245\textwidth}
    \includegraphics[width=\linewidth]{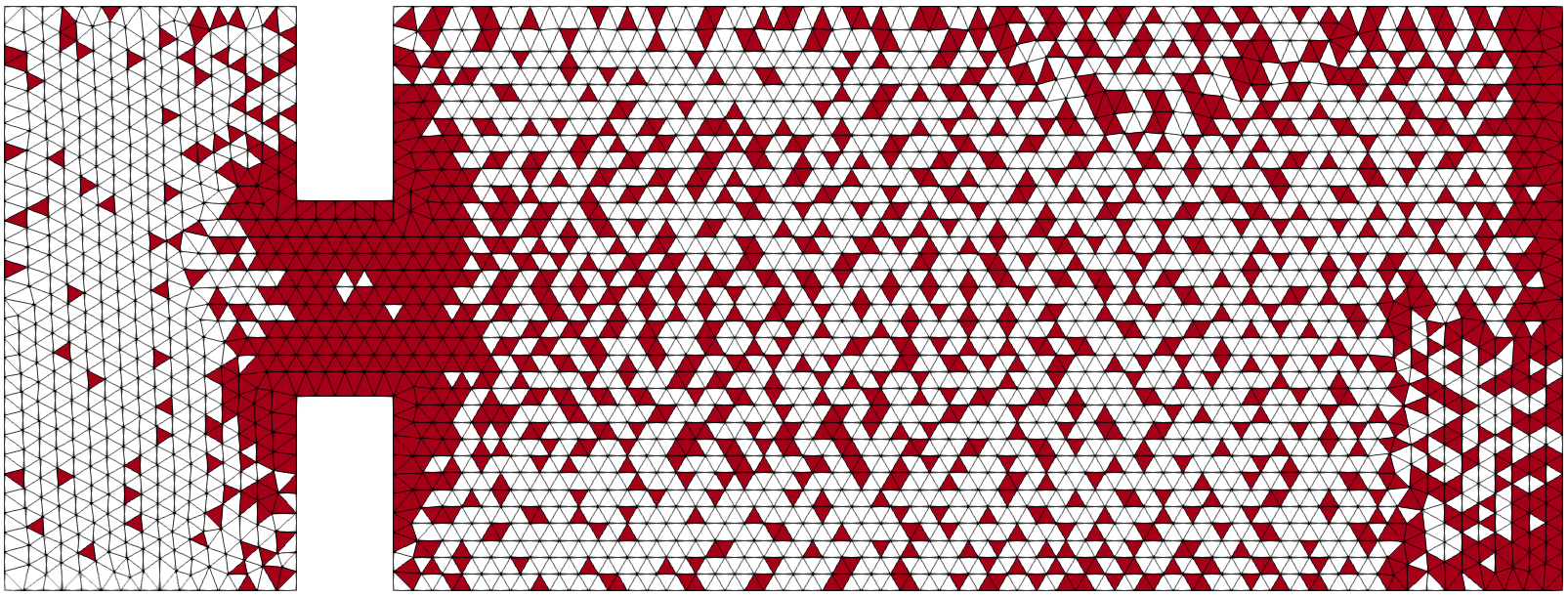}
    \caption{ $\epsilon_{\text{\tiny SOL}} = 1e-6, \epsilon_{\text{\tiny RES}} = 1e-4, $ }
    %\label{fig: example 3 HROM_elemns_ffd_rbf__n}
  \end{subfigure}
  \hfill
  \begin{subfigure}[b]{0.245\textwidth}
    \includegraphics[width=\linewidth]{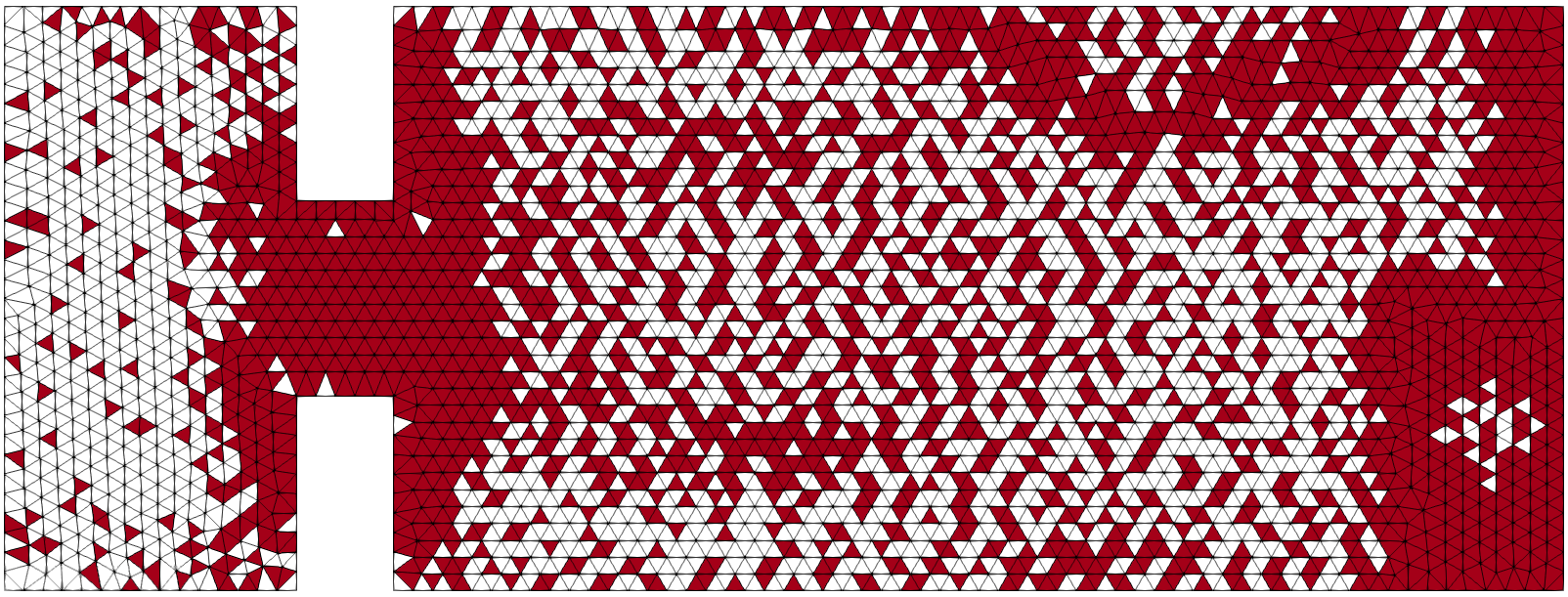}
    \caption{ $\epsilon_{\text{\tiny SOL}} = 1e-6, \epsilon_{\text{\tiny RES}} = 1e-5, $ }
    %\label{fig: example 3 HROM_elemns_ffd_rbf__o}
  \end{subfigure}
  \hfill
  \begin{subfigure}[b]{0.245\textwidth}
    \includegraphics[width=\linewidth]{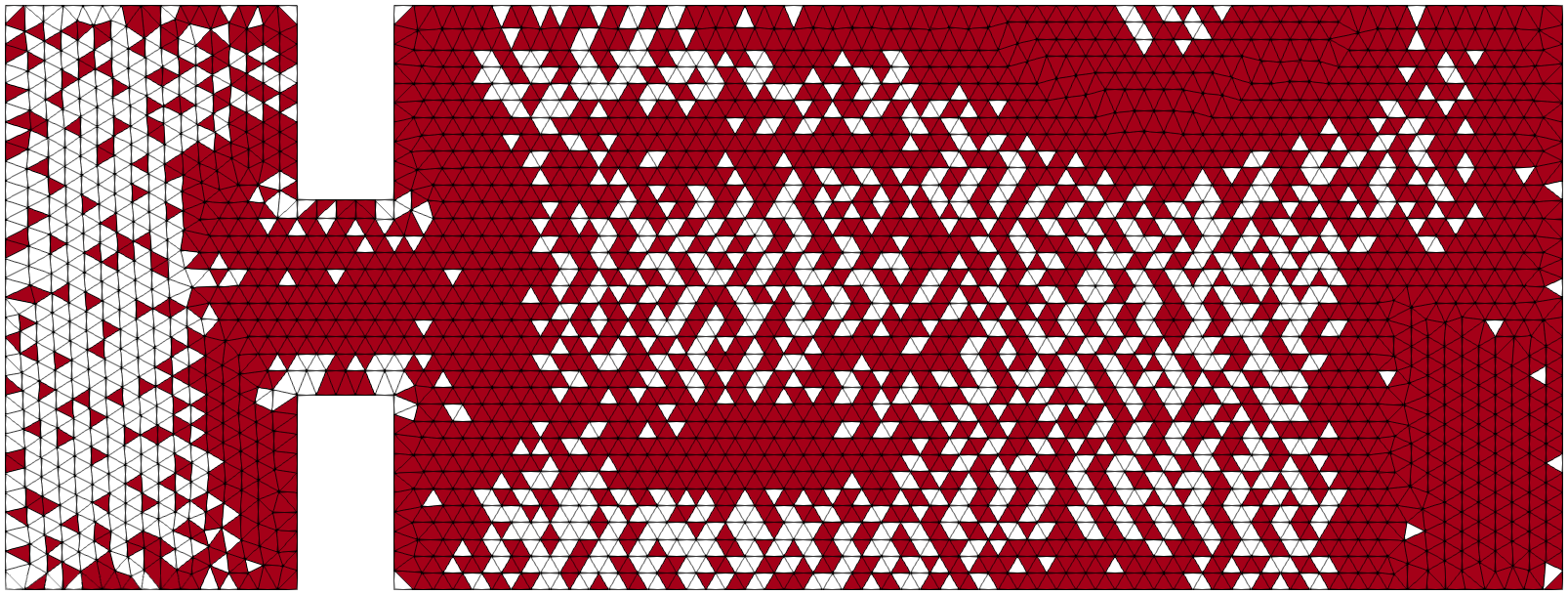}
    \caption{ $\epsilon_{\text{\tiny SOL}} = 1e-6, \epsilon_{\text{\tiny RES}} = 1e-6, $ }
    %\label{fig: example 3 HROM_elemns_ffd_rbf__p}
  \end{subfigure}

  \caption{Hyper-reduced elements selected for the $\boldsymbol{\varphi}_{\text{\tiny FFD+RBF}}$ mapping}
  \label{fig: example 3  HROM elements ffd rbf}
\end{figure}

\append{Similarly, for the nonaffine mapping, Figs. \ref{fig:modes ffd rbf mapping ex3} and \ref{fig: example 3  HROM elements ffd rbf} demonstrate a more equilibrated pattern, which, however, is not as symmetric as in the case of the affine mapping. The pattern of selected elements' distribution is still much more uniform than for Example 1, showing some slightly asymmetric patterns on the upper and lower walls.}

\appendix
\end{document}